\documentclass[preprint]{elsarticle}
\journal{Physics Reports}

\usepackage{color}
\usepackage[utf8]{inputenc}
\usepackage{multirow}

\usepackage{hyperref}
\definecolor{nicered}{rgb}{0.7,0.1,0.1}
\definecolor{nicegreen}{rgb}{0.1,0.5,0.1}
\hypersetup{colorlinks,citecolor= nicegreen,linkcolor= nicered}

\usepackage{braket}
\usepackage{slashed}
\usepackage{multirow}
\usepackage{array}
\newcolumntype{P}[1]{>{\centering\arraybackslash}p{#1}}

\newcommand{\E}[1]{\times 10^{#1}}
\newcommand{\e}[1]{\,\mathrm{#1}}
\newcommand{\mc}[1]{\mathcal{#1}}
\newcommand{\mrm}[1]{\mathrm{#1}}
\newcommand{\re}[0]{\mrm{Re}}
\newcommand{\im}[0]{\mrm{Im}}
\newcommand{\mlq}[0]{m_\mrm{LQ}}
\newcommand{\qq}{(q^2)}

\newcommand{\aem}[0]{\alpha_\mrm{em}}

\usepackage[T1]{fontenc}
\usepackage{amsmath} 
\usepackage{amssymb}

\begin{document}

\title{Physics of leptoquarks in precision experiments and at particle colliders}

\author[ST]{I.~Dor\v sner\corref{cor1}} 
\ead{dorsner@fesb.hr}

\author[LJ_1,LJ_2]{S.~Fajfer} 
\ead{svjetlana.fajfer@ijs.si} 

\author[CH_1,SJ]{A.~Greljo} 
\ead{admir@physik.uzh.ch} 

\author[LJ_1,LJ_2]{J.F.~Kamenik} 
\ead{jernej.kamenik@ijs.si} 

\author[LJ_1,LJ_2]{N.~Ko\v snik} 
\ead{nejc.kosnik@ijs.si}

\cortext[cor1]{Corresponding author}
\address[ST]{University of Split, Faculty of Electrical Engineering, Mechanical Engineering and Naval Architecture in Split (FESB), Ru\dj era Bo\v skovi\' ca 32, 21000 Split, Croatia}
\address[CH_1]{Physik-Institut, Universit\"at Z\"urich, CH-8057 Z\"urich, Switzerland}
\address[SJ]{Faculty of Science, University of Sarajevo, Zmaja od Bosne 33-35, \\ 71000 Sarajevo, Bosnia and Herzegovina}
\address[LJ_1]{Department of Physics, University of Ljubljana, Jadranska 19, 1000 Ljubljana, Slovenia}
\address[LJ_2]{Jo\v zef Stefan Institute, Jamova 39, P.\ O.\ Box 3000, 1001
  Ljubljana, Slovenia}

\begin{abstract}
We present a comprehensive review of physics effects generated by  leptoquarks (LQs), i.e., hypothetical particles that can turn quarks into leptons and vice versa, of either scalar or vector nature. These considerations include discussion of possible completions of the Standard Model that contain LQ fields. The main focus of the review is on those LQ scenarios that are not problematic with regard to proton stability. We accordingly concentrate on the phenomenology of light leptoquarks that is relevant for precision experiments and particle colliders. Important constraints on LQ interactions with matter are derived from precision low-energy observables such as electric dipole moments, ($g-2$) of charged leptons, atomic parity violation, neutral meson mixing, Kaon, $B$, and $D$ meson decays, etc. We provide a general analysis of indirect constraints on the strength of LQ interactions with the quarks and leptons to make statements that are as model independent as possible. We address complementary constraints that originate from electroweak precision measurements, top, and Higgs physics. The Higgs physics analysis we present covers not only the most recent but also expected results from the Large Hadron Collider (LHC). We finally discuss direct LQ searches. Current experimental situation is summarized and self-consistency of assumptions that go into existing accelerator-based searches is discussed. A progress in making next-to-leading order predictions for both pair and single LQ productions at colliders is also outlined.
\end{abstract}
\maketitle

\tableofcontents

\section{Introduction}
\label{sec:INTRODUCTION}
Leptoquarks (LQs), also referred to as lepto-quarks, are hypothetical particles that emerge in a particularly natural fashion in extensions of the Standard Model (SM) that unify matter~\cite{Pati:1973uk}. They have one special property that sets them apart from all other elementary particles. Namely, LQs can turn quarks into leptons and vice versa~\cite{Pati:1974yy}. They thus represent a unique source of new physics that can be and has been extensively tested. LQ discovery would be a tentative signal of matter unification. This would nicely dovetail with the observed unification of weak and electromagnetic interactions.

Even though the notion of LQs has been on the market for more than forty years there is only a handful of reviews that cover the diverse aspects of LQ phenomenology~\cite{Davidson:1993qk,Hewett:1997ce,Nath:2006ut}. Admittedly there are several comprehensive studies of specific facets of LQ properties~\cite{Shanker:1981mj,Shanker:1982nd,Buchmuller:1986iq,Buchmuller:1986zs,Hewett:1987yg,Leurer:1993em,Leurer:1993qx} that reflect the richness of the subject. However, there is not a single contemporary review that covers the physics of light LQs. In our view, this is the region of parameter space where most of the recent experimental and theoretical progress has been made. It is thus our opinion that the time is right to remedy this situation. We accordingly provide a comprehensive yet succinct review of past and recent advances in this field of research.

Let us for the benefit of the reader stress what one should not expect to find in this review. We do not intend to provide comprehensive table(s) of quantitative constraints on the strengths of either single or particular combinations of multiple couplings of leptoquarks to the SM matter as a function of the LQ mass. That sort of analysis is very model dependent and tends to be outdated with even moderate improvements of experimental constraints. We instead provide tools that enable one to derive relevant constraints within a well-defined model with relative ease. As for the direct searches for LQs, we provide the latest lower limits on the LQ masses together with the assumptions that were used to extract them.

The outline of the review is as follows. We list possible scalar and vector LQ states and present the associated nomenclature that one can find in the literature in this section. We also address the issue of baryon ($B$) and lepton ($L$) number violation with the aim to single out those LQ scenarios that are (not) problematic with regard to matter stability. This is followed by a discussion in Section~\ref{sec:MODELS} of possible completions of the SM that contain fields of an LQ nature. From that point on we concentrate on the physics of light LQs. 

A large number of important constraints on LQ interactions with matter are extracted from precision low-energy observables in Section~\ref{sec:FLAVOR}. There are, moreover, many processes where the physics of LQs might provide observable correlated effects. Recent notable examples include: (i) $B \rightarrow K^* \,\mu^+\, \mu^-$ decays, where the LHCb Collaboration reported deviations in their measurements of angular observables from the corresponding SM theoretical estimates; (ii) lepton flavor universality ratios of $B \rightarrow K \, \mu^+\, \mu^-$ and $B \rightarrow K\, e^+\, e^-$ decays, the most precise LHCb measurements which currently also deviate from precise SM predictions; (iii) lepton flavor universality ratios of semitauonic $b$ to $c$ decays, as reported by Belle, Babar and LHCb exhibit persisting deviations from SM expectations. In a number of theoretical explanations TeV scale LQs were considered as likely candidates to account for these anomalies. We address these issues within the context of a general analysis of indirect constraints on LQ Yukawa interactions trying to make statements that are as model independent as possible.

We proceed by addressing complementary constraints that originate from top quark phenomenology, electroweak precision measurements and Higgs physics in Section~\ref{sec:EWK&HIGGS}. The latter is done in view of not only the most recent results but also those expected from the LHC.

Finally, we discuss direct searches for LQs at colliders, in Section~\ref{sec:COLLIDERS}. There is a relatively large number of new constraints that has been extracted from the $pp$ collisions at the LHC. However, the collider results have not originated only from the LHC but also from PETRA, TRISTAN, LEP, Tevatron, and HERA. We summarize past and current experimental situation and discuss the self-consistency of assumptions that go into existing accelerator-based searches. For example, one of them is a common {\it ansatz} (with a few notable exceptions) that LQs decay into a particular generation of quarks and leptons without any flavor mixings. We also present, in Section~\ref{sec:COLLIDERS}, the leading order and the progress in making next-to-leading order predictions for both pair and single LQ production. 

 
Closing remarks are put forth in Section~\ref{sec:CONCLUSIONS}.


\subsection{Nomenclature}
\label{subsec:NOMENCLATURE}
LQs are fields that simultaneously couple to a quark and a lepton. In local quantum field theories they can be of either scalar (spin-zero) or vector (spin-one) nature. Since the number of representations of quark and lepton fields in the SM is finite and small it is easy to classify all possible LQ states. The SM fermion multiplets are $L_L^{i} \equiv ({\bf 1},{\bf 2},-1/2)^{i} =(\nu_L^i \qquad e_L^i)^T$, $e_R^{i} \equiv ({\bf 1},{\bf 1},-1)^{i}$, $Q_L^{i} \equiv ({\bf 3},{\bf 2},1/6)^{i} =(u_L^i \qquad  d_L^i)^T$, $u_R^{i} \equiv ({\bf 3},{\bf 1},2/3)^{i}$, and $d_R^{i} \equiv ({\bf 3},{\bf 1},-1/3)^{i}$, where the numbers within brackets represent the SM gauge group $SU(3) \times SU(2) \times U(1)$ transformation properties. For example, a state denoted as $({\bf 3},{\bf 2},1/6)$ transforms as triplet (doublet) of $SU(3)$ ($SU(2)$) with the $U(1)$ hyper-charge of $1/6$. Superscript $i(=1,2,3)$ is a flavor index and subscripts $L$ and $R$ denote left- and right-chiral fermion fields, respectively. Superscript $T$ will always stand for transposition. It is in the $SU(2)$ group space of the SM in this particular instance. We take quarks (leptons) to have a baryon (lepton) number $B=1/3$ ($L=1$) in accordance with the usual convention. This allows us to have a well-defined fermion number $F=3B+L$ for all leptoquarks.

The hyper-charge normalization is defined through $\hat{Q}=I^3+Y$, where $\hat{Q}$ is the electric charge operator that yields eigenvalues we denote with $Q$ in units of absolute value of the electron charge, $I^3$ stands for the diagonal generator of $SU(2)$, and $Y$ represents $U(1)$ hyper-charge operator. The electric charge of $d_R \equiv (\mathbf{3},\mathbf{1},-1/3)$ is, for example, $Q=0+(-1/3)=-1/3$, where $d_R$ is a right-chiral down-type quark. 

At least two neutrinos are conclusively massive. However, their Dirac and/or Majorana nature is not yet experimentally ascertained. One might accordingly add to the SM fermion content one or more electrically neutral fields that could take on the role of right-chiral neutrinos. We denote these hypothetical fermions with $\nu_R (\equiv ({\bf 1},{\bf 1},0))$. If these states are added one could have more LQ states than there would be in the SM model with purely left-chiral neutrinos. We include this possibility to ensure the generality of our considerations.

We adopt a notation that was introduced in Ref.~\cite{Buchmuller:1986zs} to denote individual leptoquark multiplets under the SM gauge group and provide a list of all possible LQs in Table~\ref{tab:LQs}. There are, all in all, six scalar ($S_3$, $R_2$, $\tilde{R}_2$, $\tilde{S}_1$, $S_1$, and $\bar{S}_1$) and six vector ($U_3$, $V_2$, $\tilde{V}_2$, $\tilde{U}_1$, $U_1$, and $\bar{U}_1$) LQ multiplets if one uses transformations under the SM gauge group as the classification criterion. In the first column of Table~\ref{tab:LQs} we explicitly specify the SM transformation properties that can be easily understood on purely group theoretical grounds as follows. 
\begin{table}[tbp]
\centering
\begin{tabular}{|c|cccc|}
\hline
$(SU(3),SU(2),U(1))$ &  Spin & Symbol & Type & $F$ \\
\hline 
$(\overline{\mathbf{3}},\mathbf{3},1/3)$ & 0 & $S_3$ & $LL$\,$(S^L_1)$ & $-2$ \\
$(\mathbf{3},\mathbf{2},7/6)$ & 0 & $R_2$ & $RL$\,$(S^L_{1/2})$, $LR$\,$(S^R_{1/2})$ & $0$ \\
$(\mathbf{3},\mathbf{2},1/6)$ & 0 & $\tilde{R}_2$ & $RL$\,$(\tilde{S}^L_{1/2})$, $\overline{LR}$\,$(\tilde{S}^{\overline{L}}_{1/2})$ & $0$ \\
$(\overline{\mathbf{3}},\mathbf{1},4/3)$ & 0 & $\tilde{S}_1$ & $RR$\,$(\tilde{S}^R_{0})$ & $-2$ \\
$(\overline{\mathbf{3}},\mathbf{1},1/3)$ & 0 & $S_1$ & $LL$\,$(S^L_0)$, $RR$\,$(S^R_0)$, $\overline{RR}$\,$(S^{\overline{R}}_0)$ & $-2$ \\
$(\overline{\mathbf{3}},\mathbf{1},-2/3)$ & 0 & $\bar{S}_1$ & $\overline{RR}$\,$(\bar{S}^{\overline{R}}_0)$ & $-2$ \\
\hline
\hline
$(\mathbf{3},\mathbf{3},2/3)$ & 1 & $U_3$ & $LL$\,$(V^L_1)$ & $0$ \\
$(\overline{\mathbf{3}},\mathbf{2},5/6)$ & 1 & $V_2$ & $RL$\,$(V^L_{1/2})$, $LR$\,$(V^R_{1/2})$ & $-2$ \\
$(\overline{\mathbf{3}},\mathbf{2},-1/6)$ & 1 & $\tilde{V}_2$ & $RL$\,$(\tilde{V}^L_{1/2})$, $\overline{LR}$\,$(\tilde{V}^{\overline{R}}_{1/2})$ & $-2$ \\
$(\mathbf{3},\mathbf{1},5/3)$ & 1 & $\tilde{U}_1$ & $RR$\,$(\tilde{V}^R_0)$ & $0$ \\
$(\mathbf{3},\mathbf{1},2/3)$ & 1 & $U_1$ & $LL$\,$(V^L_0)$, $RR$\,$(V^R_0)$, $\overline{RR}$\,$(V^{\overline{R}}_0)$ & $0$ \\
$(\mathbf{3},\mathbf{1},-1/3)$ & 1 & $\bar{U}_1$ & $\overline{RR}$\,$(\bar{V}^{\overline{R}}_0)$ & $0$ \\
\hline 
\end{tabular}
\caption{\label{tab:LQs} List of scalar and vector LQs. See text for details.}
\end{table} 

All quarks (leptons) are triplets (singlets) with regard to $SU(3)$. This means that all LQs should transform as 3-dimensional representations of $SU(3)$ in order for quark--lepton--LQ contraction(s) to be invariant under the $SU(3)$ group transformations. In group theoretical terms one uses a singlet ($\mathbf{1}$) of $SU(3)$ in decomposition of the direct product $\mathbf{3} \otimes \mathbf{1} \otimes \overline{\mathbf{3}} \equiv \mathbf{8} \oplus \mathbf{1}$ that corresponds to quark--lepton--LQ contraction(s). The dimensionality of $SU(3)$ representation is thus the same for all LQs. They are all triplets of $SU(3)$. One consequence of this property is that some LQs can couple to a quark-quark pair while not a single LQ can couple to a lepton--lepton pair.

The LQ dimensionality assignment under $SU(2)$ is less trivial if compared with the $SU(3)$ case since both quarks and leptons are either singlets ($\mathbf{1}$) or doublets ($\mathbf{2}$) of $SU(2)$. Quark--lepton contractions can then be of triplet, doublet, and singlet nature in $SU(2)$ space since $\mathbf{2} \otimes \mathbf{2} \equiv \mathbf{3} \oplus \mathbf{1}$, $\mathbf{2} \otimes \mathbf{1} \equiv \mathbf{2}$, and $\mathbf{1} \otimes \mathbf{1} \equiv \mathbf{1}$. These representations need to be contracted with LQs that have the same $SU(2)$ dimensionality in order to create gauge invariant terms. LQs can thus be triplets, doublets, and singlets of $SU(2)$. The dimensionality under $SU(2)$ can accordingly be used to distinguish between different LQs. This corresponds to a subscript of the LQ symbol in the third column of Table~\ref{tab:LQs}. If LQ multiplets have the same dimensionality under $SU(2)$ but differ in hyper-charge $U(1)$ one needs to introduce additional markings --- a tilde or a bar --- above the LQ symbol. For example, there are three scalar multiplets that are singlets under $SU(2)$. We accordingly denote them as $S_1$, $\tilde{S}_1$, and $\bar{S}_1$. This particular notation was introduced in Ref.~\cite{Buchmuller:1986zs} for all LQs except for $\bar{S}_1$ and $\bar{U}_1$ states. The only leptons that $\bar{S}_1$ and $\bar{U}_1$ can couple to are right-chiral neutrinos that are not part of the SM field content. This is the reason why $\bar{S}_1$ and $\bar{U}_1$ were omitted from the analysis presented in Ref.~\cite{Buchmuller:1986zs}. We note in passing that $S_3$ and $U_3$ were  initially denoted as $\vec{S}_3$ and $\vec{U}_3$, respectively~\cite{Buchmuller:1986zs}. 

In more recent studies a symbol assigned to a given scalar (vector) LQ is simply replaced with $S_j$ ($V_j$)~\cite{Djouadi:1989md}, where $j$ is the highest eigenvalue of $I^3$ generator in a given $SU(2)$ representation. The following correspondence prescribes transition from one notation to the other: $S_3 \leftrightarrow S_1$, $R_2 \leftrightarrow S_{1/2}$, $\tilde{R}_2 \leftrightarrow \tilde{S}_{1/2}$, $\tilde{S}_1 \leftrightarrow \tilde{S}_{0}$, $S_1 \leftrightarrow S_{0}$, $\bar{S}_1 \leftrightarrow \bar{S}_{0}$, $U_3 \leftrightarrow V_1$, $V_2 \leftrightarrow V_{1/2}$, $\tilde{V}_2 \leftrightarrow \tilde{V}_{1/2}$, $\tilde{U}_1 \leftrightarrow \tilde{V}_{0}$, $U_1 \leftrightarrow U_{0}$, and $\bar{U}_1 \leftrightarrow \bar{V}_{0}$. Some authors denote $S_j$ and $V_j$ with $S_j^\dagger$ and $V_j^\dagger$, respectively. See, for example, section on LQs in Ref.~\cite{Agashe:2014kda}.

The LQ hyper-charge is easily deduced from group theory since it is an additive quantity. For example, let us determine the hyper-charge of a scalar LQ that transforms as a triplet of both $SU(3)$ and $SU(2)$. This state can only be contracted with quark and lepton doublets in a gauge invariant way. The hyper-charge of the quark--lepton contraction of doublet representations $L_L$ and $Q_L$ is $(-1/2)+1/6=-1/3$. This then implies that an LQ that contracts with $L_L$ and $Q_L$ in a gauge invariant manner has hyper-charge $1/3$, i.e., $({\bf 1},{\bf 1},0) \in ({\bf 1},{\bf 2},-1/2) \otimes ({\bf 3},{\bf 2},1/6) \otimes (\overline{\mathbf{3}},\mathbf{3},1/3)$. 

There are six scalar and six vector LQ multiplets under the SM gauge group. These comprise ten different LQ states each if we disregard transformations under $SU(3)$. There are, however, only four possible (electric) charge assignments LQs can have since they couple the up-type quarks ($Q=\pm 2/3$) or the down-type quarks ($Q=\pm 1/3$) to the charged leptons ($Q=\pm 1$) or the neutral leptons ($Q=0$). The possible LQ charges are thus $Q=\pm 5/3$, $Q=\pm 4/3$, $Q=\pm 2/3$, and $Q=\pm 1/3$. The upshot of this is that some fields from different LQ multiplets can, in principle, couple to the same quark--lepton combination. These would very often provide indistinguishable (final state) signatures at modern colliders. The couplings of the leptoquark multiplets to the SM fermions are such that the $|F| = 2$ ($F = 0$) leptoquarks decay to quark--lepton (antiquark--lepton) pairs. The LQ fermion number $F=3B+L$ is shown in the fifth column of Table~\ref{tab:LQs}. 

We put forward one mnemonic that concerns the LQ multiplet assignment. Note that $S_3=(\overline{\mathbf{3}},\mathbf{3},1/3)$ ($U_3=(\overline{\mathbf{3}},\mathbf{3},2/3)$) comprises three states with electric charges $Q=4/3$, $Q=1/3$, and $Q=-2/3$ ($Q=5/3$, $Q=2/3$, and $Q=-1/3$). These, conveniently enough, are exactly the electric charges and hyper-charges of three scalar (vector) LQs that are singlets of $SU(2)$, i.e., $\tilde{S}_1$, $S_1$, and $\bar{S}_1$ ($\tilde{U}_1$, $U_1$, and $\bar{U}_1$). Since the $SU(3)$ charge assignment is the same for all these LQs this fact allows for an easy reconstruction of gauge transformation properties of three scalar (vector) singlets using the quantum numbers of $S_3$ ($U_3$).

LQs are often referred to as being of the first, second or third generation type. The idea behind this nomenclature is to name the LQ according to the generation of the charged lepton and the quark it couples to. This terminology can be ambiguous. For example, an LQ that couples to an electron and a down quark certainly qualifies to be of the first generation type. But the same would be true of an LQ that couples to an electron and an up quark. We will use this nomenclature with care from time to time. 

Let us now specify all possible renormalizable interactions of scalar  and vector LQs with fermions. These interactions, excluding the ones related to the right-chiral neutrinos, have been specified in Ref.~\cite{Nieves:1981tv}. The right-chiral neutrino interactions, for the case of scalar LQs, have been discussed in Ref.~\cite{Davies:1990sc}. LQs (presumably) couple to gauge fields and the Higgs scalar of the SM. We relegate the list of all leptoquark--Higgs couplings to Appendix~\ref{sec:APPENDIX_A}. For a full list of the LQ couplings to photon, $Z$ boson, $W$ bosons, and gluon fields see Appendix~\ref{sec:APPENDIX_C}. For more details on the LQ couplings to gauge fields we refer the reader to Refs.~\cite{Blumlein:1992ej,Blumlein:1996qp}. If more than one type of LQ multiplet is added to the SM the gauge symmetry would allow interaction between these multiplets and the SM Higgs scalar. The invariant LQ couplings with the Higgs scalar that correspond to such an arrangement can be found in Ref.~\cite{Hirsch:1996qy}.

We will generically use $y$ and $z$ ($x$ and $w$) to denote coupling matrices of scalar (vector) leptoquarks with the quark--lepton and quark--quark pairs, respectively. The dimension of the $SU(2)$ representation of the leptoquark multiplet will correspond to a subscript of these matrices. We will also introduce a superscript to denote chiralities of the fermions the LQ couples to. Our convention will be such that the first index in the superscripts of $y$ and $x$ is associated with the quark chirality. Note that for scalar (vector) LQs the coupling matrices to quark-quark pairs will always be of $x^{LL}$ or $x^{RR}$ ($w^{LR}$ or $w^{RL}$) type. We provide for convenience of the reader a summary of all leptoquark--quark--lepton couplings in Appendix~\ref{sec:APPENDIX_B} for scalar and vector leptoquarks in Eqs.~\eqref{eq:main_scalars} and~\eqref{eq:main_vectors}, respectively. 


\subsection{$S_3= (\overline{\mathbf{3}},\mathbf{3},1/3)$}

There are two operators that describe how scalar $S_3$ interacts with matter. They are 
\begin{align}
\label{eq:main_S_3}
\mathcal{L} \supset &+ y^{LL}_{3\,ij}\bar{Q}_{L}^{C\,i,a} \epsilon^{ab} (\tau^k S^k_{3})^{bc} L_{L}^{j,c}+z^{LL}_{3\,ij}\bar{Q}_{L}^{C\,i,a} \epsilon^{ab} ((\tau^k S^k_{3})^\dagger)^{bc} Q_{L}^{j,c}+\textrm{h.c.},
\end{align}
where $\tau^k$, $k=1,2,3$, are Pauli matrices, $i,j=1,2,3$ ($a,b=1,2$) are flavor ($SU(2)$) indices, $ \epsilon^{ab}=(i \tau^2)^{ab}$, and $S^k_{3}$ are components of $S_{3}$ in $SU(2)$ space. We always assume summation over repeated indices. Contraction $(\tau^k S^k_{3})$ can be written as $(\vec{\tau} \cdot \vec{S}_{3})$. This is the reason why $S_3$ was initially denoted as $\vec{S}_{3}$ in Ref.~\cite{Buchmuller:1986zs}. It can be treated as an (iso)vector of weak isospin. We opt not to display $SU(3)$ indices. Superscript $C$ stands for charge conjugation operation. Our notation is such that for the fermion field $\psi$ we write $\psi_{L,R}=P_{L,R}\psi$, $\overline{\psi}=\psi^\dagger \gamma^0$, and $\psi^C=C\overline{\psi}^T$, where $P_{L,R}=(1 \mp \gamma^5)/2$ and $C=i\gamma^2 \gamma^0$. We introduce the gamma matrices $\gamma^0$, $\gamma^1$, $\gamma^2$, and $\gamma^3$, where $\gamma^5=i\gamma^0 \gamma^1 \gamma^2 \gamma^3$. $y^{LL}_{3\,ij}$ are elements of an arbitrary complex $3 \times 3$ Yukawa coupling matrix. The $z^{LL}_3$ matrix, on the other hand, is antisymmetric in flavor space, i.e., $z^{LL}_{3\,ij}=-z^{LL}_{3\,ji}$~\cite{Davies:1990sc}. Matrices $y$ and $z$ describe the strength of LQ interaction with the quark--lepton and quark--quark pairs, respectively. 

It is customary to refer to the quark--lepton--LQ operator via the chirality of both fermions. If that is adopted one would refer to the first operator in Eq.~\eqref{eq:main_S_3} as an $LL$ operator, where convention dictates that the chirality of the quark precedes that of the lepton. We specify the LQ chirality properties in the fourth column of Table~\ref{tab:LQs}. One can also read off the operator chirality from the superscripts of $y$ and $x$ matrices for scalar and vector LQs, respectively. In fact, the chirality properties of the LQ couplings to leptons of the SM have also been used to classify LQs. The notation in question stipulates~\cite{Schrempp:1991vp} that all scalar (vector) LQs are denoted with $S^C_j$ ($V^C_j$), where $C$ denotes the chirality of the lepton and $j$ represents the highest eigenvalue of $I^3$ generator of $SU(2)$ of a given LQ multiplet. If LQs have the same chirality and same $j$ but differ in hyper-charge $U(1)$ one introduces an additional tilde or a bar above the symbol. We present this alternative notation in the fourth column of Table~\ref{tab:LQs} next to the corresponding chirality of the LQ couplings. Note that within the SM framework this notation corresponds to fourteen different possibilities for LQs to couple to the quark--lepton pair. Seven of these are due to the scalar LQ couplings. There are four more possibilities if one includes right-chiral neutrinos that are equally shared between scalars and vectors. If the LQs couple exclusively to either left- or right-chiral quarks they are referred to as left- and right-type LQs~\cite{Hirsch:1996qy}, respectively. $S_3$ clearly has only left-type LQ couplings to quarks.

The second operator in Eq.~\eqref{eq:main_S_3} corresponds to a ``diquark'' coupling of $S_3$. If an LQ does not possess ``diquark'' couplings due to the SM quantum number assignment we refer to it as a genuine LQ~\cite{Dorsner:2009cu}. Matrices associated with the ``diquark'' couplings are $z$ and $w$ for scalar and vector LQs, respectively. In the absence of the ``diquark'' couplings one could also assign $B$ and $L$ numbers to a given LQ~\cite{Nieves:1981tv}. Interactions of genuine LQs could thus automatically conserve both $B$ and $L$ at renormalizable level. One should, however, check that the assignment in question is not in conflict with all possible renormalizable interactions between a given LQ and the SM fields~\cite{Arnold:2012sd}. We postpone the discussion on baryon and lepton numbers of LQs to the end of this section. 

To proceed we define $S^{4/3}_3=(S^1_3-i S^2_3)/\sqrt{2}$, $S^{-2/3}_3=(S^1_3+i S^2_3)/\sqrt{2}$, and $S^{1/3}_3=S^3_3$, where the rational superscripts denote electric charges of $S_3$ components. This allows us to write
\begin{align}
\nonumber
\mathcal{L} \supset &-(y^{LL}_{3}U)_{ij}\bar{d}_{L}^{C\,i} S^{1/3}_{3} \nu_{L}^{j}-\sqrt{2} y^{LL}_{3\,ij}\bar{d}_{L}^{C\,i} S^{4/3}_{3} e_{L}^{j}+\\
\nonumber
&+\sqrt{2} (V^Ty^{LL}_{3}U)_{ij}\bar{u}_{L}^{C\,i} S^{-2/3}_{3} \nu_{L}^{j}-(V^Ty^{LL}_{3})_{ij}\bar{u}_{L}^{C\,i} S^{1/3}_{3} e_{L}^{j}-\\
\nonumber
&-(z^{LL}_{3}V^\dagger)_{ij}\bar{d}_{L}^{C\,i} S^{1/3\,*}_{3} u_{L}^{j}-\sqrt{2}z^{LL}_{3\,ij}\bar{d}_{L}^{C\,i} S^{-2/3\,*}_{3} d_{L}^{j}+\\
\label{eq:main_S_3_a}
&+\sqrt{2}(V^T z^{LL}_3 V^\dagger)_{ij}\bar{u}_{L}^{C\,i} S^{4/3\,*}_{3} u_{L}^{j}-(V^Tz^{LL}_3)_{ij}\bar{u}_{L}^{C\,i} S^{1/3\,*}_{3} d_{L}^{j}+\textrm{h.c.},
\end{align}
where $U$ represents a Pontecorvo--Maki--Nakagawa--Sakata (PMNS) unitary mixing matrix and $V$ is a Cabibbo--Kobayashi--Maskawa (CKM) mixing matrix. Note that $S_3$ has purely $LL$ LQ couplings. 

The specific form of the flavor structures is derived when starting from a flavor basis which coincides with the mass-ordered mass eigenstate basis of the down-type quarks and charged leptons.
All fields in Eq.~\eqref{eq:main_S_3_a} are however specified in the mass eigenstate basis. The transition from the chosen flavor to the mass eigenstate basis is accomplished through the following transformations: $u_{L}^{j} \rightarrow (V^\dagger)_{jk} u_{L}^{k}$, $d_{L}^{j} \rightarrow d_{L}^{j}$, $e_{L}^{j} \rightarrow e_{L}^{j}$, and $\nu_{L}^{j} \rightarrow (U)_{jk} \nu_{L}^{k}$. We implicitly assume that the  {\it individual} rotations in the left-chiral sector are not physical. This might not be the case in all circumstances. 


\subsection{$R_2=(\mathbf{3},\mathbf{2},7/6)$}
\label{R_2}
Yukawa couplings of $R_2$ to the SM fermions are
\begin{align}
\label{eq:main_R_2}
\mathcal{L} \supset &-y^{RL}_{2\,ij}\bar{u}_{R}^{i} R_{2}^{a}\epsilon^{ab}L_{L}^{j,b}+y^{LR}_{2\,ij}\bar{e}_{R}^{i} R_{2}^{a\,*}Q_{L}^{j,a} +\textrm{h.c.}.
\end{align}
$y^{RL}_{2}$ and $y^{LR}_{2}$ in Eq.~\eqref{eq:main_R_2} are arbitrary complex $3 \times 3$ Yukawa matrices. In the mass eigenstate basis we have 
\begin{align}
\nonumber
\mathcal{L} \supset  &  - y^{RL}_{2\,ij} \bar{u}^i_R e^j_L R_2^{5/3}+  (y^{RL}_2
  U)_{ij} \bar{u}^i_R 
  \nu^j_L R_2^{2/3}+\\
 \label{eq:main_R_2_a}
 &+(y^{LR}_2 V^\dagger)_{ij} \bar{e}^i_R 
  u^j_L R_2^{5/3\,*} +y^{LR}_{2\,ij} \bar{e}^i_R d^j_L R_2^{2/3\,*} + \textrm{h.c.},
\end{align}
where the LQ superscript denotes electric charge of $R_2$ components. There are two independent Yukawa couplings for both $R_2^{5/3}$ and $R_2^{2/3}$. One is of $RL$ and the other of $LR$ type. If LQ couples to both left- and right-chiral quarks it is referred to as a non-chiral LQ. $R_2$ is thus a non-chiral LQ of the genuine kind.

We have already outlined the ambiguity attached to the first, second, and third generation LQ nomenclature. There is yet another aspect to this issue. Namely, LQ multiplets that transform non-trivially under $SU(2)$ can have one charged component to be of the first, second or third generation type but that would not be the case with the other charged component(s) of the same multiplet. This can be nicely illustrated with this particular LQ. For example, $R_2^{2/3}$ could be referred to as first generation LQ if $y^{LR}_{2\,ij}=y \delta_{1i}\delta_{1j}$ and $y^{RL}_{2\,ij}=0$, $i,j=1,2,3$, where $\delta_{ij}$ is the Kronecker delta. However, $R_2^{5/3}$ would then be an LQ that simultaneously couples an electron to all up-type quarks. We implicitly assume that all unitary rotations in the flavor space of the right-chiral sector and the {\it individual} rotations of the left-chiral sector are not experimentally accessible.

\subsection{$\tilde{R}_2=(\mathbf{3},\mathbf{2},1/6)$}
\label{R_2_tilda}
There are two renormalizable terms that describe interactions of $\tilde{R}_2$ with matter. These are
\begin{align}
\label{eq:main_t_R_2}
\mathcal{L} \supset &-\tilde{y}^{RL}_{2\,ij}\bar{d}_{R}^{i}\tilde{R}_{2}^{a}\epsilon^{ab}L_{L}^{j,b}+\tilde{y}^{\overline{LR}}_{2\,ij}\bar{Q}_{L}^{i,a} \tilde{R}_{2}^{a}\nu_{R}^{j}+\textrm{h.c.}.
\end{align}
We assume that there are three right-chiral neutrinos $\nu_R^{j}$, $j=1,2,3$. $\tilde{y}^{RL}_{2\,ij}$ and $\tilde{y}^{\overline{LR}}_{2\,ij}$ are then elements of arbitrary complex $3 \times 3$ Yukawa coupling matrices. Since the presence of right-chiral neutrinos is questionable we bar all symbols associated with them including the superscript of the associated Yukawa coupling matrix. For example, the second term in Eq.~\eqref{eq:main_t_R_2} is of $\overline{LR}$ type. This can be read off from Table~\ref{tab:LQs} allowing for an easy removal of the relevant coupling(s) if necessary. The list of scalar LQs that also includes those that couple to (three) $\nu_R$ states with the associated list of fermion bilinears has been provided in Ref.~\cite{Davies:1990sc}. 

Note that $\tilde{R}_{2}$ actually comprises two LQs. One has $Q=2/3$ and the other has $Q=-1/3$. The $SU(2)$ contraction yields
\begin{align}
\nonumber
\mathcal{L} \supset &-\tilde{y}^{RL}_{2\,ij}\bar{d}_{R}^{i}e_{L}^{j}\tilde{R}_{2}^{2/3}+(\tilde{y}^{RL}_2 U)_{ij}\bar{d}_{R}^{i}\nu_{L}^{j}\tilde{R}_{2}^{-1/3}+\\
\label{eq:main_t_R_2_a}
&+(V \tilde{y}^{\overline{LR}}_2)_{ij} \bar{u}_{L}^{i} \nu_{R}^{j} \tilde{R}_{2}^{2/3}+\tilde{y}^{\overline{LR}}_{2\,ij}\bar{d}_{L}^{i}\nu_{R}^{j}\tilde{R}_{2}^{-1/3}+\textrm{h.c.},
\end{align}
where the LQ superscript denotes electric charge of a given $SU(2)$ doublet component of $\tilde{R}_{2}$. The important point to notice is that the Yukawa couplings of $\tilde{R}_{2}^{2/3}$ and $\tilde{R}_{2}^{-1/3}$ that are associated with $\tilde{y}^{RL}_{2}$ ($\tilde{y}^{\overline{LR}}_{2}$) are related through PMNS (CKM) parameters. These are well-measured physical parameters. (For the latest experimental measurements of the mixing parameters see, for example, Ref.~\cite{Agashe:2014kda}.) $\tilde{R}_{2}$ is clearly a genuine LQ with $F=0$. In fact, all genuine LQs have $F=0$.


\subsection{$\tilde{S}_1=(\overline{\mathbf{3}},\mathbf{1},4/3)$}

The couplings of $\tilde{S}_1$ to matter read
\begin{align}
\label{eq:main_t_S_1}
\mathcal{L} \supset &+ \tilde{y}^{RR}_{1\,ij}\bar{d}_{R}^{C\,i} \tilde{S}_{1} e_{R}^{j}+\tilde{z}^{RR}_{1\,ij}\bar{u}_{R}^{C\,i} \tilde{S}^*_{1} u_{R}^{j}+\textrm{h.c.}.
\end{align}
This LQ has purely $RR$ couplings. One can observe that $\tilde{z}^{RR}_{1}$ is an antisymmetric matrix in any basis.


\subsection{$S_1=(\overline{\mathbf{3}},\mathbf{1},1/3)$}

There are five possible operators that describe how $S_1$ interacts with fermions. These are
\begin{align}
\nonumber
\mathcal{L} \supset &+y^{LL}_{1\,ij}\bar{Q}_{L}^{C\,i,a} S_{1} \epsilon^{ab}L_{L}^{j,b}+y^{RR}_{1\,ij}\bar{u}_{R}^{C\,i} S_{1} e_{R}^{j}+y^{\overline{RR}}_{1\,ij}\bar{d}_{R}^{C\,i} S_{1} \nu_{R}^{j}+\\
\label{eq:main_S_1}
&+z^{LL}_{1\,ij}\bar{Q}_{L}^{C\,i,a} S^*_{1} \epsilon^{ab}Q_{L}^{j,b}+z^{RR}_{1\,ij}\bar{u}_{R}^{C\,i} S^*_{1} d_{R}^{j}+\textrm{h.c.},
\end{align}
where $z^{LL}_{1}$ is a symmetric matrix in flavor space~\cite{Davies:1990sc}, i.e., $z^{LL}_{1\,ij}=z^{LL}_{1\,ji}$, whereas all other matrices are {\it a priori} completely arbitrary. 

The operators of Eq.~\eqref{eq:main_S_1}, after the contraction in the $SU(2)$ space, yield the following terms:
\begin{align}
\nonumber
\mathcal{L} \supset &-(y^{LL}_1 U)_{ij} \bar{d}_{L}^{C\,i} S_{1} \nu_{L}^{j}+(V^T y^{LL}_1)_{ij}\bar{u}_{L}^{C\,i} S_{1} e_{L}^{j}+y^{RR}_{1\,ij}\bar{u}_{R}^{C\,i} S_{1} e_{R}^{j}+\\
\nonumber
&+y^{\overline{RR}}_{1\,ij}\bar{d}_{R}^{C\,i} S_{1} \nu_{R}^{j}
+(V^T z^{LL}_1)_{ij}\bar{u}_{L}^{C\,i} S^*_{1} d_{L}^{j}-(z^{LL}_1 V^\dagger)_{ij}\bar{d}_{L}^{C\,i} S^*_{1} u_{L}^{j}+\\
\label{eq:main_S_1_a}
&+z^{RR}_{1\,ij}\bar{u}_{R}^{C\,i} S^*_{1} d_{R}^{j}+\textrm{h.c.}.
\end{align}
This LQ allows for $LL$, $RR$, and $\overline{RR}$ operators. $S_1$ is often referred to simply as a color triplet behind the ``infamous'' doublet--triplet splitting problem that is frequently discussed in the field of grand unified model building.


\subsection{$\bar{S}_1=(\overline{\mathbf{3}},\mathbf{1},-2/3)$}

$\bar{S}_1$ couples to one or more states that transform as $\nu_{R}$. Accordingly its symbol is barred. We assume, for convenience of notation, that there are three right-chiral neutrinos in what follows. The relevant couplings are
\begin{align}
\label{eq:main_b_S_1}
\mathcal{L} \supset &+\bar{y}^{\overline{RR}}_{1\,ij}\bar{u}_{R}^{C\,i} \bar{S}_1 \nu_{R}^{j}+\bar{z}^{RR}_{1\,ij}\bar{d}_{R}^{C\,i} \bar{S}_1^* d_{R}^{j}+\textrm{h.c.}.
\end{align}

This LQ has purely $\overline{RR}$ couplings. Since it does not couple to charged leptons it can never be referred to as first, second or third generation LQ. Note that $\bar{z}^{RR}_{1}$ is an antisymmetric matrix in any (flavor) basis.

 
\subsection{$U_3=(\mathbf{3},\mathbf{3},2/3)$}

$U_3$ is a genuine vector LQ with $LL$ type of couplings. The relevant operator reads
\begin{align}
\label{eq:main_U_3}
\mathcal{L} \supset &+ x^{LL}_{3\,ij}\bar{Q}_{L}^{i,a} \gamma^\mu (\tau^k U^k_{3,\mu})^{ab}  L_{L}^{j,b} +\textrm{h.c.},
\end{align}
where $U^k_{3}$ are components of $U_{3}$ in $SU(2)$ space and $\mu(=0,1,2,3)$ is an index in the Lorentz space. We define $U^{5/3}_3=(U^1_3-i U^2_3)/\sqrt{2}$, $U^{-1/3}_3=(U^1_3+i U^2_3)/\sqrt{2}$, and $U^{2/3}_3=U^3_3$, where the rational superscripts denote electric charges of $U_3$ components. This allows us to write
\begin{align}
\nonumber
\mathcal{L} \supset &-x^{LL}_{3\,ij}\bar{d}_{L}^{i} \gamma^\mu U^{2/3}_{\mu}  e_{L}^{j}+(V x^{LL}_3 U)_{ij}\bar{u}_{L}^{i} \gamma^\mu U^{2/3}_{\mu}  \nu_{L}^{j}+\\
\label{eq:main_U_3_a}
&+\sqrt{2}(x^{LL}_3 U)_{ij}\bar{d}_{L}^{i} \gamma^\mu U^{-1/3}_{\mu}  \nu_{L}^{j}+\sqrt{2}(V x^{LL}_3)_{ij}\bar{u}_{L}^{i} \gamma^\mu U^{5/3}_{\mu}  e_{L}^{j}+\textrm{h.c.}.
\end{align}

Vector LQs are usually associated with carriers of new interactions. This implies that $x^{LL}_{3\,ij}$ in Eq.~\eqref{eq:main_U_3} can be identified with gauge couplings that measure the strength of relevant interactions. These are flavor blind. If that is assumed one could accordingly write $x^{LL}_{3\,ij}=x \delta_{ij}$, where $x$ would represents the gauge coupling in question. We again remind the reader that we assume that the right-chiral sector rotations are not physical. The same assumption applies to {\it individual} rotations in the left-chiral sector. If that is not the case one needs to insert unitary transformations of all chiral states where appropriate. 


\subsection{$V_2=(\overline{\mathbf{3}},\mathbf{2},5/6)$}

$V_2$ is vector LQ of non-chiral nature. It has the following couplings with the SM fermions 
\begin{align}
\nonumber
\mathcal{L} \supset &+x^{RL}_{2\,ij}\bar{d}_{R}^{C\,i} \gamma^\mu V^{a}_{2,\mu} \epsilon^{ab}L_{L}^{j,b} + x^{LR}_{2\,ij}\bar{Q}_{L}^{C\,i,a} \gamma^\mu \epsilon^{ab} V^b_{2,\mu} e_{R}^{j}+\\
\label{eq:main_V_2}
&+ w^{LR}_{2\,ij}\bar{Q}_{L}^{C\,i,a} \gamma^\mu V^{a\,*}_{2,\mu} u_{R}^{j}+\textrm{h.c.}.
\end{align}
These couplings, in the mass eigenstate basis, read 
\begin{align}
\nonumber
\mathcal{L} \supset&-(x^{RL}_2 U)_{ij}\bar{d}_{R}^{C\,i} \gamma^\mu V^{1/3}_{2,\mu} \nu_{L}^{j}+x^{RL}_{2\,ij}\bar{d}_{R}^{C\,i} \gamma^\mu V^{4/3}_{2,\mu} e_{L}^{j}+\\
\nonumber
&+ (V^T x^{LR}_{2})_{ij}\bar{u}_{L}^{C\,i} \gamma^\mu  V^{1/3}_{2,\mu} e_{R}^{j}-x^{LR}_{2\,ij}\bar{d}_{L}^{C\,i} \gamma^\mu  V^{4/3}_{2,\mu} e_{R}^{j}+\\
\label{eq:main_V_2_a}
&+ (V^T w^{LR}_2)_{ij}\bar{u}_{L}^{C\,i} \gamma^\mu  V^{4/3\,*}_{2,\mu} u_{R}^{j}+w^{LR}_{2\,ij}\bar{d}_{L}^{C\,i} \gamma^\mu  V^{1/3\,*}_{2,\mu} u_{R}^{j}+\textrm{h.c.},
\end{align}
where $V^{4/3}_{2,\mu}$ and $V^{1/3}_{2,\mu}$ are two components of $V_2$. These are often referred to as $X$ and $Y$ gauge bosons, respectively.

\subsection{$\tilde{V}_2=(\overline{\mathbf{3}},\mathbf{2},-1/6)$}

$\tilde{V}_2$ is vector LQ with $RL$ and $\overline{LR}$ couplings:
\begin{align}
\nonumber
\mathcal{L} \supset&+\tilde{x}^{RL}_{2\,ij}\bar{u}_{R}^{C\,i} \gamma^\mu \tilde{V}^{b}_{2,\mu} \epsilon^{ab}L_{L}^{j,a} +\tilde{x}^{\overline{LR}}_{2\,ij}\bar{Q}_{L}^{C\,i,a} \gamma^\mu \epsilon^{ab} \tilde{V}^{b}_{2,\mu}  \nu_{R}^{j}+\\
\label{eq:main_t_V_2}
&+\tilde{w}^{RL}_{2\,ij}\bar{d}_{R}^{C\,i} \gamma^\mu \tilde{V}^{a\,*}_{2,\mu} Q_{L}^{j,a} +\textrm{h.c.}.
\end{align}
After the $SU(2)$ decomposition, i.e., $\tilde{V}_2 \rightarrow (\tilde{V}^{1/3},\,\tilde{V}^{-2/3})$, we get 
\begin{align}
\nonumber
\mathcal{L} \supset&-\tilde{x}^{RL}_{2\,ij}\bar{u}_{R}^{C\,i} \gamma^\mu \tilde{V}^{1/3}_{2,\mu} e_{L}^{j} +(\tilde{x}^{RL}_2 U)_{ij}\bar{u}_{R}^{C\,i} \gamma^\mu \tilde{V}^{-2/3}_{2,\mu} \nu_{L}^{j}+\\
\nonumber
&+(V^T \tilde{x}^{\overline{LR}}_2)_{ij}\bar{u}_{L}^{C\,i} \gamma^\mu \tilde{V}^{-2/3}_{2,\mu} \nu_{R}^{j} -\tilde{x}^{\overline{LR}}_{2\,ij}\bar{d}_{L}^{C\,i} \gamma^\mu \tilde{V}^{1/3}_{2,\mu} \nu_{R}^{j}+\\
\label{eq:main_t_V_2_a}
&+\tilde{w}^{RL}_{2\,ij}\bar{d}_{R}^{C\,i} \gamma^\mu \tilde{V}^{-2/3\,*}_{2,\mu} d_{L}^{j} +(\tilde{w}^{RL}_2 V^\dagger)_{ij}\bar{d}_{R}^{C\,i} \gamma^\mu \tilde{V}^{1/3\,*}_{2,\mu} u_{L}^{j}+\textrm{h.c.}.
\end{align}
It is very common in the grand unified theory model building community to refer to $\tilde{V}^{1/3}_{2,\mu}$ and $\tilde{V}^{-2/3}_{2,\mu}$ as $X'$ and $Y'$ gauge bosons, respectively.


\subsection{$\tilde{U}_1=(\mathbf{3},\mathbf{1},5/3)$}

$\tilde{U}_1$ is a genuine vector LQ with $RR$ couplings. The couplings are 
\begin{align}
\label{eq:main_t_U_1}
\mathcal{L} \supset&+\tilde{x}^{RR}_{1\,ij}\bar{u}_{R}^{i} \gamma^\mu \tilde{U}_{1,\mu} e_{R}^{j} +\textrm{h.c.}.
\end{align}


\subsection{$U_1=(\mathbf{3},\mathbf{1},2/3)$}

This is a genuine non-chiral vector LQ that has the following couplings:
\begin{align}
\label{eq:main_U_1}
\mathcal{L} \supset&+x^{LL}_{1\,ij}\bar{Q}_{L}^{i,a} \gamma^\mu U_{1,\mu} L_{L}^{j,a} + x^{RR}_{1\,ij}\bar{d}_{R}^{i} \gamma^\mu U_{1,\mu} e_{R}^{j}+ x^{\overline{RR}}_{1\,ij}\bar{u}_{R}^{i} \gamma^\mu U_{1,\mu} \nu_{R}^{j}+\textrm{h.c.}.
\end{align}
In the mass eigenstate basis one obtains
\begin{align}
\nonumber
\mathcal{L} \supset&+(V x^{LL}_1 U)_{ij}\bar{u}_{L}^{i} \gamma^\mu U_{1,\mu} \nu_{L}^{j} +x^{LL}_{1\,ij}\bar{d}_{L}^{i} \gamma^\mu U_{1,\mu} e_{L}^{j}+ \\
\label{eq:main_U_1_a}
&+x^{RR}_{1\,ij}\bar{d}_{R}^{i} \gamma^\mu U_{1,\mu} e_{R}^{j}+ x^{\overline{RR}}_{1\,ij}\bar{u}_{R}^{i} \gamma^\mu U_{1,\mu} \nu_{R}^{j}+\textrm{h.c.}.
\end{align}

$U_1$ is part of the gauge sector of the Pati--Salam model~\cite{Pati:1974yy}. It is thus the very first LQ that was studied in the literature. 


\subsection{$\bar{U}_1=(\mathbf{3},\mathbf{1},-1/3)$}

$\bar{U}_1$ couples exclusively to the right-chiral neutrinos. It has only the $\overline{RR}$ type of couplings. These read 
\begin{align}
\label{eq:main_b_U_1}
\mathcal{L} \supset&+x^{\overline{RR}}_{1\,ij}\bar{d}_{R}^{i} \gamma^\mu \bar{U}_{1,\mu} \nu_{R}^{j} +\textrm{h.c.}.
\end{align}


\subsection{$B$ and $L$ considerations}

This is a review on LQs in precision experiments and at particle colliders. We will thus not delve too deep into the issues of $B$ and $L$ conservation unless absolutely necessary. The idea instead is to succinctly present main reasons why the observed level of matter stability puts stringent constraints on the way leptoquarks couple to the SM fermions. For exhaustive reviews on proton decay see, for example, Refs.~\cite{Langacker:1980js,Nath:2006ut}.

LQs have a well-defined fermion number $F=3B+L$ that is shown in the fifth column of Table~\ref{tab:LQs}. Again, the couplings of the leptoquark multiplets to the SM fermions are such that the $|F| = 2$ ($F = 0$) leptoquarks decay to quark--lepton (antiquark--lepton) pairs. In the absence of the ``diquark'' couplings one could, in principle, also assign $B$ and $L$ numbers to a given LQ in a self-consistent manner. This would allow one to concentrate on the low-energy phenomenology of these states without the pressing need to address the issue of matter stability. Note, however, that the LQ in question could couple to other fields in the SM in both the gauge and Lorentz invariant way that would violate the self-consistency of the $B$ and $L$ assignments. For example, even though $\tilde{R}_2$ does not possess ``diquark'' couplings it can couple to the Higgs field $H$ of the SM via the renormalizable operator $H\tilde{R}_2 \tilde{R}_2 \tilde{R}_2$. This would lead to the $p \rightarrow \pi^+ \,\pi^+ \,e^- \,\nu \,\nu$ process~\cite{Arnold:2012sd}. We will, for the time being, assume that this particular operator can be either neglected or altogether forbidden by (some) symmetry.

There are two scalar LQs --- $R_2$ and $\tilde{R}_2$ --- that do not possess ``diquark'' couplings. The same statement is valid for vector LQs $U_3$, $\tilde{U}_1$, $U_1$, and $\bar{U}_1$. These six genuine LQ multiplets can all be assigned a baryon number $B=1/3$ and lepton number $L=-1$ if one considers only couplings to fermions. One can easily single them out in Table~\ref{tab:LQs} as they all have $F=3B+L=0$. LQs with $F=-2$, on the other hand, can mediate proton decay if one does not forbid the ``diquark'' coupling(s) to make them purely LQs~\cite{Barr:1989fi}. Strictly speaking, these statements hold true within a framework comprising the SM and a single LQ state at renormalizable level. The situation might be much more involved in any other framework.

LQs that mediate proton decay and bound neutron decay at tree-level need to be very heavy and/or some of their couplings to matter need to be extremely small. Either way, they are irrelevant for low-energy phenomenology. The lower bounds on the masses of the $|F|=2$ vector LQs due to proton decay lifetime data are particularly severe. These states have been extensively investigated in the context of grand unified theories, where, again, $V^{4/3}_{2,\mu}$, $V^{1/3}_{2,\mu}$,  $\tilde{V}^{1/3}_{2,\mu}$, and $\tilde{V}^{-2/3}_{2,\mu}$ are referred to as $X$, $Y$, $X'$, and $Y'$ gauge bosons, respectively. 

\begin{figure}[tbp]
\centering
\begin{tabular}{cc}
\includegraphics[scale=0.75]{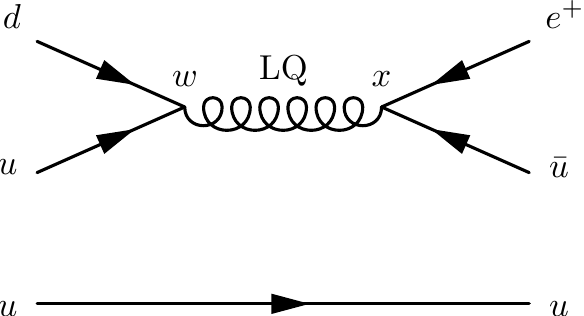}  & \includegraphics[scale=0.75]{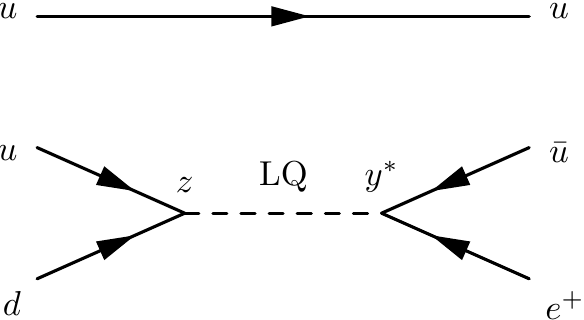}
\end{tabular}
\caption{\label{fig:PROTON_V_S} Schematic representation of tree-level proton decay $p \rightarrow \pi^0 \,e^+$ due to vector and scalar LQ mediation in the left and right panels, respectively. $x$ and $w$ ($y$ and $z$) represent relevant gauge (Yukawa) coupling strengths of vector (scalar) leptoquarks to the quark--quark and quark--lepton pairs, respectively.}
\end{figure}

Vector LQs couple to matter via gauge couplings. These couplings can be related to the measured gauge couplings and are thus well-known in most of the available models. The only thing that is {\it a priori} not determined are rotations in the left- and right-chiral sectors but these need to be unitary. (See, for example, Eqs.~\eqref{eq:main_V_2_a} and \eqref{eq:main_t_V_2_a}.) The lower bound on the LQ mass due to the matter stability constraints can be easily estimated on general grounds. Let us denote the mass of proton decay mediating vector LQ or any LQ for that matter to be $m_{\mathrm{LQ}}$. The relevant schematic diagram for two-body proton decay $p \rightarrow \pi^0\, e^+$ mediated by, for example, either $V^{1/3}_{2}$ or $\tilde{V}^{1/3}_{2}$ is given in the left panel of Fig.~\ref{fig:PROTON_V_S}, where we take $x$ and $w$ to represent generic values of gauge coupling strengths of vector leptoquarks to the quark--quark and quark--lepton pairs, respectively. We also promote all unitary transformations of fermion fields into identity matrices to make the most conservative estimate possible. 
The amplitude (decay width) associated with this process is proportional to $(x w)/m^2_{\mathrm{LQ}}$ ($(x w)^2/m^4_{\mathrm{LQ}}$). To get the dimensions right, one needs to introduce the only other relevant scale in the problem, i.e., the proton mass $m_p$. We thus have the following decay width ($\Gamma$) estimate for the proton due to the exchange of vector LQs
\begin{equation}
\Gamma \sim \frac{(x w)^2}{m^4_{\mathrm{LQ}}}m_p^5.
\end{equation}
If we take $x=w=1/2$, $m_p \approx 1$\,GeV and $\tau^{\mathrm{exp.}}_p >1.0\times10^{34}$\,years~\cite{Miura:2010zz} we get that $m_{\mathrm{LQ}} \gtrsim 10^{16}$\,GeV. One can certainly refine this constrain within a well-defined model but the scale we present is rather accurate. 

To establish a lower bound on the mass of the $|F|=2$ scalar LQs that mediates proton decay we proceed in the same manner. The relevant schematic diagram for two-body proton decay $p \rightarrow \pi^0 \,e^+$ mediated by, for example, scalar LQ $S_1$ is given in the right panel of Fig.~\ref{fig:PROTON_V_S}, where we take $y$ and $z$ to represent generic values of relevant Yukawa couplings of the leptoquark with the quark--quark and quark--lepton pairs, respectively. This time the decay width reads
\begin{equation}
\Gamma \sim \frac{|y|^2 |z|^2}{m^4_{\mathrm{LQ}}}m_p^5.
\end{equation}
If we now associate $|y|$ and $|z|$ with Yukawa coupling of the up quark ($|y|=|z| \approx 10^{-5}$) we get $m_{\mathrm{LQ}} \gtrsim 2 \times 10^{11}$\,GeV. Note that the limits on a scalar LQ mass are approximately a factor of $\sqrt{|y z|/|x w|} \approx 10^{-5}$ weaker with respect to the limits on the mass of vector LQs.

There are ways to avoid and/or lower these stringent bounds even within a framework that only allows for three generations of the SM fermions. One can, for example, use unitary transformations to partially suppress~\cite{Nandi:1982ew} and/or completely rotate away proton decay due to the vector LQ mediation for some~\cite{Dorsner:2004xa} or all available decay channels~\cite{Dorsner:2004jj} at the leading order. Proton decay through scalar LQ exchange involves dependence on Yukawa couplings. This makes it even more model dependent when compared with the vector LQ case. It can consequently be completely or partially suppressed with a suitable choice of the LQ couplings to matter~\cite{Nath:2006ut,Dorsner:2012nq}. Also, some scalar LQs generate suppressed contributions to proton decay due to intrinsic particularities of their flavor structure. This, for example, is the case with $\tilde{S}_1$ that has antisymmetric couplings to the up-type quark bilinears~\cite{Dong:2011rh,Dorsner:2012nq,Arnold:2013cva}. There are also numerous ways to forbid ``diquark'' coupling(s) altogether.

\begin{figure}[tbp]
\centering
\includegraphics[scale=0.9]{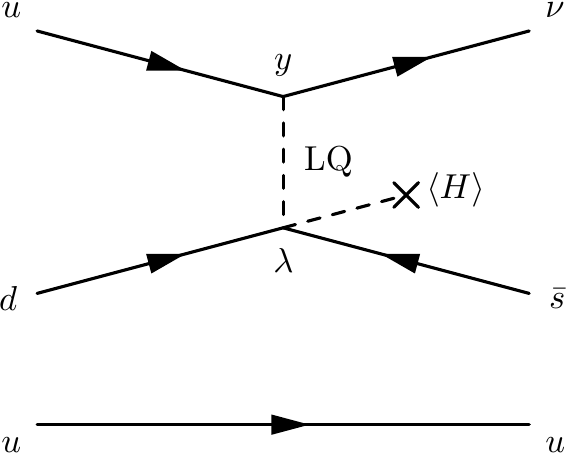}
\caption{\label{fig:PROTON_5D} Schematic representation of tree-level proton decay $p \rightarrow K^+ \,\nu$ due to LQ $R^{2/3}_2$ mediation. Coupling $\lambda$ represents dimensionless coefficient of dimension five operator (suppressed with appropriate scale), $y$ represents relevant Yukawa coupling(s), and $\langle H \rangle$ stands for the vacuum expectation value of the SM Higgs field.}
\end{figure}

It has been recently argued that even the SM extension with a single scalar LQ multiplet $R_2$ or $\tilde{R}_2$ could yield an unacceptably high level of baryon number violation if one allows operators of dimension five that are suppressed by the Planck scale and that feature couplings of order one~\cite{Arnold:2013cva}. For example, $R_2$ can couple to two down-type quarks and the Higgs field ($H$) via the dimension five operator to generate proton decay. See, for example, Fig.~\ref{fig:PROTON_5D} for a representative diagram. It has also been shown that this class of operators can be forbidden in a systematic way~\cite{Arnold:2013cva}, if needed.


\section{Leptoquark models}
\label{sec:MODELS}

LQs are fields that are a natural outcome of unification of quarks and leptons~\cite{Pati:1973uk} as initially proposed in the Pati--Salam model~\cite{Pati:1974yy}. It was clear from the very onset that they would turn quarks into leptons and accordingly generate qualitatively new physical effects. This notion has been further supported with a successful unification of quarks and leptons into simple groups of $SU(5)$~\cite{Georgi:1974sy} and $SO(10)$~\cite{Georgi:1974my,Fritzsch:1974nn}, where the exchange of vector LQs was shown to lead to yet to be observed physical processes of proton and bound neutron decay. 

Unification of leptons and quarks into the same representation requires unification of scalar LQ(s) with the SM Higgs boson if one resorts to simple groups. These scalar LQs very often tend to mediate proton decay. This is in fact the source of the so-called doublet--triplet splitting problem we refer to when we discuss Yukawa couplings of scalar LQ $S_1$ in Section~\ref{sec:INTRODUCTION}. Namely, one must address the fact that the SM Higgs boson, a doublet of $SU(2)$, has to be light whereas an associated LQ, a triplet of $SU(3)$, needs to be very heavy to satisfy experimental constraints on the proton lifetime. The mass splitting of this magnitude is highly unnatural especially since both the doublet and the triplet reside in the same representation. This proved to be very challenging and has led to an alternative way to unify fermions into a flipped $SU(5)$ framework~\cite{Barr:1981qv,DeRujula:1980qc,Derendinger:1983aj,Antoniadis:1987dx} that is based on an $SU(5) \times U(1)$ embedding of the SM fields. 

We will demonstrate with explicit models that the potential lightness of scalar LQs in the context of grand unified theories, if LQs are phenomenologically allowed to reside at low scale, is usually directly connected to viable gauge coupling unification. Naturally, it is necessary and sometimes difficult in view of preceding comments to insure that the light LQ scenario is compatible with proton (and bound neutron) stability.  

Introduction of supersymmetry (SUSY) has opened a door to a rich LQ phenomenology. A closer look at the SM fermion transformation properties shows that scalar partners of all quarks carry correct quantum numbers to couple quarks to leptons the way LQs do. The actual realization of these scenarios within SUSY framework, though, requires explicit R-parity violation to facilitate necessary couplings~\cite{Hall:1983id,Dawson:1985vr,Giudice:1997wb,Csaki:2011ge}. (It is possible to combine the idea of unification with R-parity violation. For the proposals on how to embed R-parity violating interactions in the grand unified theory frameworks see, for example, Refs.~\cite{Giudice:1997wb,Bajc:2015zja}.) This is an extremely reach subject that partially overlaps with the scope of present review. We do not, however, intend to cover physics of R-parity violation and instead point the reader to extensive reviews and studies on the subject \cite{Allanach:1999ic,Chemtob:2004xr,Barbier:2004ez,Dreiner:2006gu,FileviezPerez:2010ek,Csaki:2011ge,Dreiner:2012mx,Arvanitaki:2013yja}. Supersymmetric partners of leptoquarks are sometimes referred to as leptoquarkinos. The relevant phenomenology of the latter has been discussed in Ref.~\cite{Hewett:1988xc}.

LQs also appear within technicolor scenarios~\cite{Dimopoulos:1979es,Dimopoulos:1979sp,Farhi:1980xs,Georgi:1981xw} and composite models~\cite{Abbott:1981re,Schrempp:1984nj,Wudka:1985ef}, where one expects them to reside at a sufficiently low energy scale to be directly accessible with modern colliders. For the early studies of the possible signatures of technipions --- the particles of leptoquark nature --- within a technicolor scenario at hadron and/or $e^+ e^-$ colliders see Refs.~\cite{Eichten:1986eq,Lane:1991qh}. An example of how scalar LQs can arise as pseudo-Nambu Goldstone bosons of a spontaneously broken, approximate symmetry, within the Pati--Salam embedding can be found in Ref.~\cite{Gripaios:2009dq}. Phenomenology of the third generation LQs in models where electroweak symmetry breaking is driven by strong dynamics, and strategies on how to look for these states at the LHC have been discussed in Ref.~\cite{Gripaios:2010hv}. 

Phenomenology of light LQs has also been discussed within superstring theory that is compactified on a Calabi--Yau manifold~\cite{Angelopoulos:1986uq}. For the discussion of how the leptoquarks emerge from the superstring derived models see Ref.~\cite{Elwood:1997ds}. 

LQs have furthermore been proposed to transmit spontaneous charge and parity (CP) breaking to quarks and leptons of the SM in models that aim to solve the strong CP problem~\cite{Hall:1986fi}.

Finally, there are scenarios that are extensions of the SM that simply add one or more LQs and possibly some other fields to the experimentally observed ones in order to address certain phenomenological issue(s) such as the origin of neutrino mass, to name one example. We will discuss this type of scenarios towards the end of this section.


\subsection{Unification scenarios}

One can say with certainty that any unification leads to appearance of LQs. The real question from the theoretical point of view is whether these LQs can be or are predicted to be light enough to be relevant for the low-energy phenomenology. The possibility to embed light LQs into SUSY and/or non-SUSY grand unified theory scenarios has been discussed, for example, in Refs.~\cite{Senjanovic:1982ex,Buchmuller:1986iq,Frampton:1989fu,Frampton:1990hz,Frampton:1991ay,Rizzo:1991tc,Murayama:1991ah,Hewett:1997ba,Gershtein:1999gp,Dorsner:2005fq}. Our intention is to discuss models of gauge unification that predict light LQs. We will accordingly start with the LQs in $SU(5)$ gauge group since the language of $SU(5)$ is often used to simplify discussions of more complicated frameworks. Before we proceed we list in Table~\ref{tab:LQs_GUT} the most commonly used representations in $SU(5)$ that also accommodate LQs. The LQ embedding in the $SU(5)$ representations has already been presented and discussed, for example, in Ref.~\cite{Hewett:1997ce}. In what follows we denote representations through their dimensionality using conventions of Ref.~\cite{Slansky:1981yr}.
\begin{table}[tbp]
\centering
\begin{tabular}{|c|c|}
\hline
LEPTOQUARK &  $SU(5)$ \\
\hline 
$S_3 \equiv (\overline{\mathbf{3}},\mathbf{3},1/3) $ & $ \overline{\mathbf{45}},\overline{\mathbf{70}}$\\
$R_2 \equiv (\mathbf{3},\mathbf{2},7/6) $ & $ \overline{\mathbf{45}},\overline{\mathbf{50}}$\\
$\tilde{R}_2 \equiv (\mathbf{3},\mathbf{2},1/6) $ & $ \mathbf{10},\mathbf{15},\mathbf{40}$\\
$\tilde{S}_1 \equiv (\overline{\mathbf{3}},\mathbf{1},4/3) $ & $ \mathbf{45}$\\
$S_1 \equiv (\overline{\mathbf{3}},\mathbf{1},1/3)$ & $\overline{\mathbf{5}}, \overline{\mathbf{45}}, \overline{\mathbf{50}}, \overline{\mathbf{70}}$\\
$\bar{S}_1 \equiv (\overline{\mathbf{3}},\mathbf{1},-2/3) $ & $ \mathbf{10}, \mathbf{40}$\\
\hline
\hline
$U_3 \equiv (\mathbf{3},\mathbf{3},2/3) $ & $ \overline{\mathbf{35}}, \overline{\mathbf{40}}$\\
$V_2 \equiv (\overline{\mathbf{3}},\mathbf{2},5/6) $ & $ \mathbf{24}, \mathbf{75}$\\
$\tilde{V}_2 \equiv (\overline{\mathbf{3}},\mathbf{2},-1/6) $ & $ \overline{\mathbf{10}}, \overline{\mathbf{40}}$\\
$\tilde{U}_1 \equiv (\mathbf{3},\mathbf{1},5/3) $ & $ \mathbf{75}$\\
$U_1 \equiv (\mathbf{3},\mathbf{1},2/3) $ & $ \overline{\mathbf{10}}, \overline{\mathbf{40}}$\\
$\bar{U}_1 \equiv (\mathbf{3},\mathbf{1},-1/3) $ & $ \mathbf{5}, \mathbf{45}, \mathbf{50}, \mathbf{70}$\\
\hline 
\end{tabular}
\caption{\label{tab:LQs_GUT} List of $SU(5)$ representations (right column) that accommodate LQs (left column).}
\end{table} 

The SM fermions comprise three ten-dimensional representations ($\mathbf{10}_i$) and three five-dimensional representations ($\overline{\mathbf{5}}_j$), where $i,j(=1,2,3)$ are family indices.  
Namely, $\mathbf{10}_i \equiv (\mathbf{1},\mathbf{1},1)_i \oplus(\overline{\mathbf{3}},\mathbf{1},-2/3)_i
\oplus(\mathbf{3},\mathbf{2},1/6)_i=(e^C_i,u^C_i,Q_i)$ and
$\overline{\mathbf{5}}_j \equiv (\mathbf{1},\mathbf{2},-1/2)_j\oplus
(\overline{\mathbf{3}},\mathbf{1},1/3)_j=(L_j,d^C_j)$~\cite{Georgi:1974sy}. Note that all of the SM fermions in a given $SU(5)$ multiplet are left-chiral. (This allows us to drop the chirality subscript that we use throughout Section~\ref{sec:INTRODUCTION}.) All potentially relevant contractions of these representations in the $SU(5)$ group space read $\mathbf{10} \otimes \mathbf{10} =
\overline{\mathbf{5}} \oplus \overline{\mathbf{45}} \oplus \overline{\mathbf{50}}$, $\mathbf{10}
\otimes \overline{\mathbf{5}} = \mathbf{5}\oplus\mathbf{45}$, and $\overline{\mathbf{5}}
\otimes \overline{\mathbf{5}} = \overline{\mathbf{10}} \oplus \overline{\mathbf{15}}$. It is now easy to single out those scalar representations that can directly couple to two fermion fields at the renormalizable level. These are the $5$-, $10$-, $15$-, $45$-, and $50$-dimensional representations. On the other hand, the only scalar representations that contain fields capable of getting vacuum expectation value (VEV) without breaking electric charge and color are $5$-, $15$-, $24$-, and $45$-dimensional representations. Even though $24$-dimensional representation does not contain scalar LQs we note that vector LQs in $SU(5)$ reside in an adjoint representation. This simply means that the only vector LQ that always appears in any $SU(5)$ model is $V_2$. (Again, $V^{4/3}_{2}$ and $V^{1/3}_{2}$ --- two LQs comprising the SM multiplet $V_2$ --- are referred to as $X$ and $Y$ gauge bosons, respectively. See Eq.~\eqref{eq:main_V_2} for relevant couplings of these fields with matter.) Since at least one scalar representation with LQs needs to be present in the concrete model to generate electroweak symmetry breaking and fermion masses we can state with certainty that there is always at least one vector and one scalar LQ in any scenario that is based on the $SU(5)$ gauge group. 

If the $SU(5)$ scenario is realized with the minimal scalar content there would be two LQs in the theory and both would need to be superheavy. For example, the most stringent experimental limit on the mass of $V_2$ in the Georgi--Glashow model~\cite{Georgi:1974sy} originates from two-body proton decay $p \rightarrow \pi^0 e^+$. The flavor dependent part of the proton decay width in this channel in the minimal $SU(5)$ model is equal to $[ 1+ (1+|V_{11}|^2)^2]$, where $V_{11}$ is the $11$ element of the CKM mixing matrix. (In flipped $SU(5)$ the flavor dependent part of $p \rightarrow \pi^0 e^+$ process mediated by $\tilde{V}_2$ gauge boson is $|V_{11}|^4$~\cite{Barr:1981qv}.) In a more realistic setting with regard to fermion masses this term can be completely rotated away. However, one cannot simultaneously suppress all two-body proton decay channels in $SU(5)$~\cite{Nandi:1982ew}. (One can actually suppress all relevant channels in flipped $SU(5)$~\cite{Dorsner:2004jj}.) In particular, phenomenologically viable suppression in $SU(5)$ cannot drive flavor dependent term for proton decaying into an $s$-meson and a muon below a value of $2 |V_{13}|^2$~\cite{Dorsner:2004xa}, where $|V_{13}| \sim 0.004$ is the absolute value of the $13$ element of the CKM mixing matrix. Even with this level of suppression one would get a lower bound on the mass of $V_2$ to be of the order of $10^{13}$\,GeV. The point is that $V_2$ within the $SU(5)$ framework is irrelevant for the low-energy phenomenology except for proton decay. 

The lower bound on the scalar LQ mass can also be derived. If one takes the Georgi--Glashow model~\cite{Georgi:1974sy} but allows for realistic fermion masses through higher-dimensional operators it can be shown~\cite{Dorsner:2012uz} that the mass limit is generated by the experimental bound on the partial lifetime for $p \rightarrow K^+ \overline{\nu}$ channel. It depends only on one {\it a priori} unknown phase $\phi$ via a term proportional to $(m_u^2+m_d^2+2 m_u m_d \cos \phi)^{-1/4}$, where $m_u$ and $m_d$ are the up quark and the down quark masses, respectively. This result implies that the mass of the scalar LQ $S_1$ needs to be above $2.4 \times 10^{11}$\,GeV for the $SU(5)$ model to be viable with certainty. Here we present the limit using $\tau_{p \rightarrow K^+ \overline{\nu}}>5.9 \times 10^{33}$\,years~\cite{Abe:2014mwa}. (In the minimal flipped $SU(5)$ scenario the corresponding limit on the scalar LQ $S_1$ mass from $p \rightarrow K^+ \overline{\nu}$ channel is $10^{12}$\,GeV~\cite{Dorsner:2012nq}.) The main message of this discussion is that presence of light LQs in any scenario that is based on the $SU(5)$ gauge group requires additional scalar representations beside those of the minimal model. For the first study of the light scalar LQ phenomenology within the $SU(5)$ framework see Ref.~\cite{Buchmuller:1986iq}.

\begin{figure}[tbp]
\centering
\includegraphics[scale=1.2]{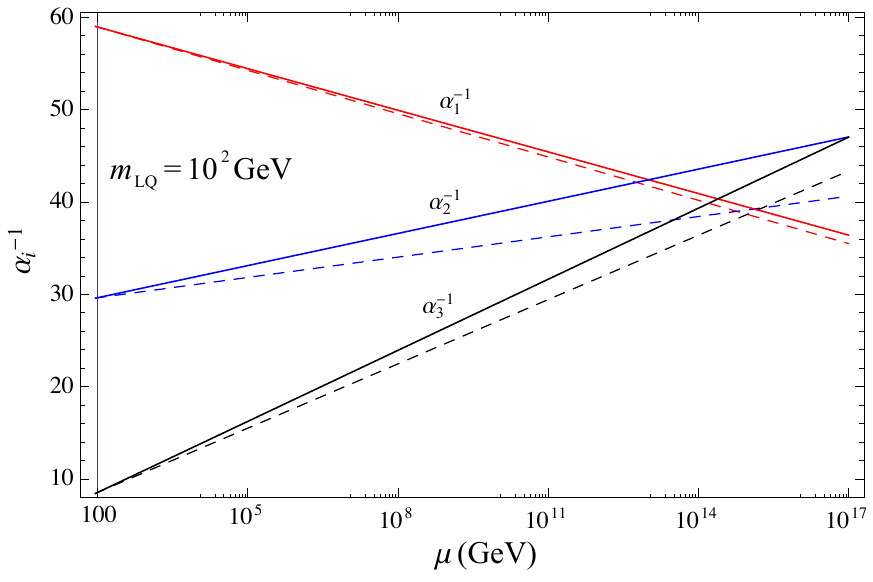}
\caption{\label{fig:MY} Running of gauge couplings with the SM particle content (solid lines) and with additional fields (dashed lines) comprising one scalar transforming as $(\mathbf{1},\mathbf{2},1/2)$ and two scalars transforming as $(\mathbf{3},\mathbf{2},1/6)$. Vertical line denotes the mass scale of additional fields.}
\end{figure}

Let us finally give an example of how (very) light scalar LQs might appear in $SU(5)$. We describe a model presented in Ref.~\cite{Murayama:1991ah}. The authors start with the Georgi--Glashow model and add to it one $5$-dimensional and two $10$-dimensional scalar representations. The Georgi--Glashow (GG) model~\cite{Georgi:1974sy} itself consists of matter fields --- $\mathbf{10}_i$ and $\overline{\mathbf{5}}_i$ ($i=1,2,3$) --- and scalars comprising one $24$- and one $5$-dimensional representation to start with. All vector bosons, on the other hand, reside in one $24$-dimensional representation. The VEV of the scalar adjoint breaks $SU(5)$ down to $SU(3) \times SU(2) \times U(1)$ and the fundamental scalar representation is there to provide electroweak symmetry breaking, i.e., $SU(3) \times SU(2) \times U(1) \rightarrow SU(3) \times U(1)_\mathrm{em}$. The authors of Ref.~\cite{Murayama:1991ah} proceed by assuming the validity of the ``desert hypothesis'' which stipulates that the fields are either superheavy or very light. The only light fields beside the SM fermions and gauge bosons are, by assumption, two multiplets in $5$-dimensional representation and two multiplets in $10$-dimensional scalar representation that transform as $(\mathbf{1},\mathbf{2},1/2)$ and $(\mathbf{3},\mathbf{2},1/6)$, respectively. This particular field content and these assumptions lead to unification of gauge couplings if the additional scalar fields reside at the electroweak scale. We show in Fig.~\ref{fig:MY} running of gauge couplings $\alpha_i=g_i^2/(4 \pi)$, $i=1,2,3$, within the SM framework (solid lines) with one Higgs scalar and confront it with the unification scenario when additional scalar fields are taken to be at the common scale of $10^2$\,GeV (dashed lines). $g_3$, $g_2$, and $g_1$ are couplings associated with $SU(3)$, $SU(2)$, and $U(1)$ gauge groups, respectively.
To produce Fig.~\ref{fig:MY} we use $\alpha_3 = 0.1176$, $\aem^{-1} = 127.906$, and $\sin^2 \theta_W = 0.23122$~\cite{Amsler:2008zzb}, where these (central) values are given at $m_Z$ in the $\overline{MS}$ scheme.

Admittedly, the model of Ref.~\cite{Murayama:1991ah} is not (any more) phenomenologically viable for multiple reasons. We, nevertheless, opt to present this scenario since it provides a kind of connection between the unification scale and the scalar LQ mass scale that persists in other models that predict light LQs. Scalar LQs simply need to be light if couplings are to unify. 

\begin{figure}[tbp]
\centering
\includegraphics[scale=0.8]{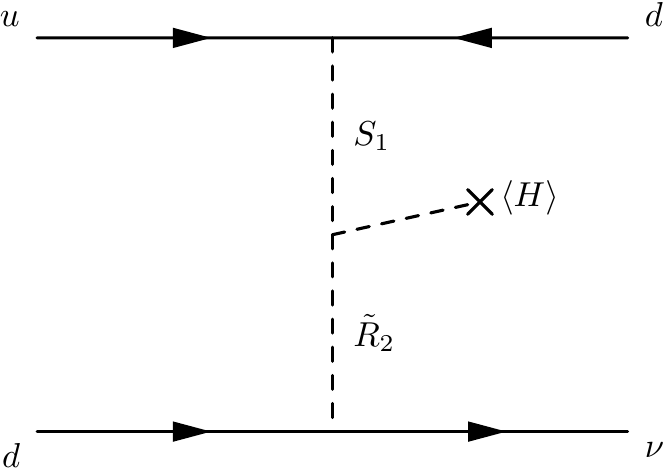}
\caption{\label{fig:p_decay_SU_5} Contribution to proton decay due to mixing between two color triplets and the Higgs scalar.}
\end{figure}

The model of Ref.~\cite{Murayama:1991ah} can be furthermore used to demonstrate how difficult it is to insure matter stability. Namely, there are multiple operators that should be suppressed if one is to prevent rapid proton decay. For example, the contraction $\epsilon_{\alpha \beta \gamma \delta \eta} \mathbf{10}^{\alpha \beta} \mathbf{10}^{\gamma \delta} \mathbf{10}^{\eta \nu} \overline{\mathbf{5}}_{\nu}$ of four scalar representations yields an operator that involves three LQs ($(\mathbf{3},\mathbf{2},1/6)=\tilde{R}_2$) and one Higgs field ($(\mathbf{1},\mathbf{2},1/2)=H$) that could eventually lead to $p \rightarrow \pi^+ \,\pi^+ \,e^- \,\nu \,\nu$ process~\cite{Arnold:2012sd}. (See discussion in Section~\ref{sec:INTRODUCTION} for details.) Here, $\epsilon_{\alpha \beta \gamma \delta \eta}$ is Levi-Civita tensor and $\alpha, \beta, \gamma, \delta, \eta, \nu(=1,\ldots,5)$ are $SU(5)$ indices. Moreover, if one allows for contraction $\overline{\mathbf{5}}_{\alpha} \mathbf{10}^{\alpha \beta} \overline{\mathbf{5}}_{\beta}$ of scalar representations $\mathbf{5}(=(H,S_1))$ with two $10$-dimensional scalar representations that differs from zero when there is more than one $\overline{\mathbf{5}}$ one can get contribution to proton decay that is shown in Fig.~\ref{fig:p_decay_SU_5}~\cite{Mohapatra:1986uf}.
It might be possible to forbid all problematic operators using additional symmetries that are external to the underlying gauge group. 

Note that additional representations in the previous example are introduced for the sole purpose of improving unification. There are, however, scenarios that have additional representations beside those in the GG model that do not only improve some of its shortcomings but also predict existence of light LQs in a self-consistent manner along the way. For example, the model presented in Ref.~\cite{Dorsner:2005fq} extends the GG model with only one $15$-dimensional scalar representation. The VEV of electrically neutral field in $\mathbf{15} (\equiv (\mathbf{1},\mathbf{3},1) \oplus(\mathbf{6},\mathbf{1},-2/3)
\oplus(\mathbf{3},\mathbf{2},1/6))$ generates experimentally observed neutrino masses through the type II see-saw mechanism~\cite{Lazarides:1980nt,Mohapatra:1980yp} while all the multiplets in $\mathbf{15}$ help establish gauge coupling unification. The operator responsible for neutrino masses is $y_{ij} \overline{\mathbf{5}}_{i\,\alpha} \mathbf{15}^{\alpha \beta} \overline{\mathbf{5}}_{j\,\beta}$, where $y$ is symmetric matrix in flavor space. One of the fields in $15$-dimensional scalar representation that transforms as $(\mathbf{3},\mathbf{2},1/6) \equiv \tilde{R}_2$ turns out to be rather light in order for unification to take place at sufficiently high energy scale. Yukawa coupling matrix $\tilde{y}^{RL}_{2}$ that is introduced in Eq.~\eqref{eq:main_t_R_2} is predicted to be symmetric since it is proportional to the matrix $y$ up to a multiplicative constant. The model thus connects both proton decay physics and neutrino physics to the leptoquark physics that can be probed at the LHC. The phenomenology of this model that relies on higher-dimensional contributions to generate viable charged fermion masses has accordingly been extensively studied~\cite{Dorsner:2005fq,Dorsner:2005ii,Dorsner:2006hw}. (Available parameter space of the model is shown in Fig.~1 of Ref.~\cite{Dorsner:2006hw}.) The fact that the model relates neutrino masses to the LQ decay pattern has been exploited to show that LQ decay signatures depend on the possible neutrino mass hierarchies in a testable way~\cite{FileviezPerez:2008dw}.

Another possible class of $SU(5)$ models that might yield light scalar LQs is the one that resorts to $45$-dimensional scalar representation to address the observed difference between down-type quark masses and charged lepton masses in a manner initially proposed by Georgi and Jarlskog~\cite{Georgi:1979df}. Note that three out of six scalar LQs reside in $45$-dimensional representation as can be seen in Table~\ref{tab:LQs_GUT}. This representation has two renormalizable contractions with the SM fermions. These are $y'_{ij} \epsilon_{\alpha \beta \gamma \delta \eta} \mathbf{10}^{\alpha \beta}_i \mathbf{10}^{\gamma \delta}_j \mathbf{45}^{\eta \nu}_{\nu}$ and $y''_{ij} \mathbf{10}_i^{\alpha \beta} \overline{\mathbf{5}}_{j,\,\nu} \overline{\mathbf{45}}^{\nu}_{\alpha \beta}$, where $y'$ and $y''$ are Yukawa coupling matrices in flavor space. The former operator generates an antisymmetric contribution to the up-type quark masses~\cite{Dorsner:2006dj} whereas the latter one yields contributions to the down-type quark masses and charged lepton masses~\cite{Georgi:1979df}. Only recently it has been shown that two LQs --- $S_3$ and $\tilde{S}_1$ --- in $45$-dimensional representation might be light if one completely suppresses operator proportional to $y'$ or makes $y''$ symmetric in flavor space~\cite{Dorsner:2012nq}. (See Table~II in Ref.~\cite{Dorsner:2012nq} for explicit couplings of the LQs to fermions due to $SU(5)$ contractions.) In fact, it has been shown that phenomenologically viable scenario of $SU(5)$ unification with one $45$-dimensional scalar representation that was introduced in Ref.~\cite{Perez:2007rm} yields light $\tilde{S}_1$ with a mass around 1\,TeV~\cite{Dorsner:2009cu}. This regime when $\tilde{S}_1$ is accessible at the LHC was further investigated in Ref.~\cite{Dorsner:2010cu}.


The possibility that one could have a scalar LQ that is light within the context of $SO(10)$ unification has been pursued in Ref.~\cite{Senjanovic:1982ex}. It has been shown that a particular chain of symmetry breaking results in a collider accessible LQ. Again, successful unification predicts light leptoquark. Note that the $SO(10)$ framework involves two sets of vector LQs --- $V_2$ and $\tilde{V}_2$ --- that reside in $45$-dimensional representation of $SO(10)$. In view of our discussion on matter stability within the context of $SU(5)$ and flipped $SU(5)$ they both need to be superheavy in the most conservative scenario. The point is that if one wants to construct self-consistent model of unification with light vector LQs one needs to look beyond simple groups of $SU(5)$ or $SO(10)$. See, for example, Refs.~\cite{Frampton:1989fu,Frampton:1990hz,Frampton:1991ay} for explicit model realizations within an $SU(15)$ framework. 

$SO(10)$ gauge symmetry allows for multiple scalar LQs to be light if needed. For example, the $SU(5)$ operator $y''_{ij} \epsilon_{\alpha \beta \gamma \delta \eta} \mathbf{10}^{\alpha \beta}_i \mathbf{10}^{\gamma \delta}_j \mathbf{45}^{\eta \nu}_{\nu}$ that was discussed in the previous paragraph can be embedded into an $SO(10)$ operator $y_{ij} \mathbf{16}_i \mathbf{16}_j \mathbf{126}$, where three $16$-dimensional representations contain the SM fermions and three right-chiral neutrinos while $126$-dimensional representation is the relevant scalar representation containing LQs. The $SO(10)$ symmetry insures symmetricity of $y_{ij}$ in flavor space which might allow for $S_3$ and $\tilde{S}_1$ LQs that reside in $\mathbf{126}$ to be light. The absence of the ``diquark'' couplings for both $S_3$ and $\tilde{S}_1$ in $\mathbf{126}$ of $SO(10)$ is simply due to the underlying gauge symmetry of the model. 


The possibility to have light fields that are triplets of $SU(3)$ and singlets of $SU(2)$ of vector-like nature that could decay through leptoquark-type or ``diquark''-type couplings has been investigated within a supersymmetric context in Ref.~\cite{Kang:2007ib}. The authors use the language of $E_6$ to provide motivation for the origin of such fields and discuss at length potential signals of these states at the LHC.

Another $E_6$ inspired model --- the so-called Alternative Left--Right Symmetric Model --- has been studied in order to provide an explanation of the observed enhancement of the $\bar{B} \to D^{(*)} \tau \bar{\nu}$ rates with respect to the SM predictions via the scalar leptoquark exchange~\cite{Hati:2015awg}. The authors show compatibility of couplings that are required to explain experimental data on $\bar{B} \to D \tau \bar{\nu}$ and $\bar{B} \to D^{(*)} \tau \bar{\nu}$ with constraints originating from leptonic decays $D_s^+ \to \tau^+ \bar{\nu}$, $B^+ \to \tau^+ \bar{\nu}$, $D^+ \to \tau^+ \bar{\nu}$, and $D^0$--$\bar{D}^0$ mixing.


One can get light vector LQs in scenarios that do not exclusively rely on the use of simple groups. A prime example of potentially light vector LQ in the literature is $(\mathbf{3},\mathbf{1},5/3) \equiv U_1$. It is a vector LQ that is not dangerous for proton decay. Also, it is a part of the gauge sector of the Pati--Salam model~\cite{Pati:1974yy}. The mass of $U_1$ is thus associated with the scale of the Pati--Salam $SU(4)$ color gauge group breaking into the color group of the SM. The coupling strength of the Pati--Salam LQ to matter is consequently determined by the strength of the QCD coupling at the Pati--Salam symmetry breaking scale. See Eq.~\eqref{eq:main_U_1} for relevant couplings of $U_1$ to matter. The lower limit on the mass of this leptoquark has been discussed in Refs.~\cite{Deshpande:1982bf,Shanker:1982nd,Kuznetsov:1994tt}. 
An explicit model with light $U_1$ that is based on the SM gauge group extended with an $SU(4)$ gauge symmetry has been presented in Ref.~\cite{Blumhofer:1997vb}.

Two low scale quark--lepton unification scenarios that predict existence of both vector and scalar LQs have been presented in Refs.~\cite{Smirnov:1995jq,Perez:2013osa}. The unification scenarios are based on the $SU(4)_C \times SU(2)_L \times U(1)_R$ gauge group product. The individual analyses of these scenarios presented in Refs.~\cite{Popov:2005wz,Perez:2013osa} stipulate that scalar LQs could be accessible at the LHC and in the IceCube experiment~\cite{Ahrens:2002dv}. The scalar LQs in question are $SU(2)$ doublets $R_2$ and $\tilde{R}_2$. The model presented in Ref.~\cite{Perez:2013osa} also addresses the nature of neutrino mass that is of Majorana type through an inverse seesaw mechanism.


\subsection{Low-energy scenarios of neutrino mass }

LQs have been used to address the question of neutrino masses and associated mixing parameters within extensions of the SM that do not necessarily aim or lead to unification~\cite{Chua:1999si,Mahanta:1999xd,AristizabalSierra:2007nf,Dey:2008ht,Babu:2010vp,Kohda:2012sr,Angel:2013hla,Cai:2014kra,Sierra:2014rxa,Helo:2015fba,Pas:2015hca,Hagedorn:2016dze,Deppisch:2016qqd}. Namely, it has been stipulated that the neutrino masses are of the one-loop or two-loop origin, where some of the fields that participate in the loops are scalar LQs. This approach is used to naturally explain or justify observed smallness of neutrino masses. 

\begin{figure}[tbp]
\centering
\includegraphics[scale=0.9]{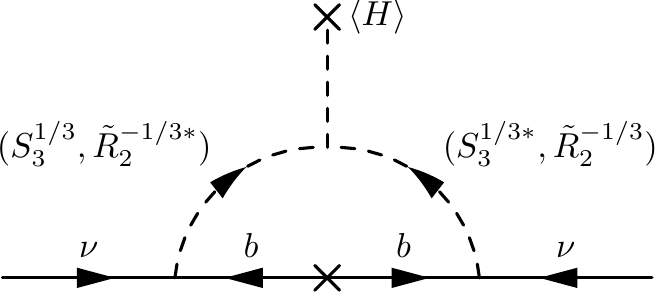}
\caption{\label{fig:neutrino_mass} One-loop diagram of neutrino mass generation. Scalar LQ fields that mix due to the couplings with the SM Higgs boson $H$ are bracketed. Flavor indices are omitted.}
\end{figure}

Let us outline the main features of this approach. We require the presence of $\tilde{R}_2$. More specifically, we need a term proportional to $\tilde{y}_2^{RL} \bar{d}_{R}\nu_{L}\tilde{R}_{2}^{-1/3}$, where we omit flavor indices. Note that $\tilde{R}_{2}^{-1/3}$ couples neutrino to the right-chiral down-type quark. The other necessary ingredient is the presence of either $S_1$ or $S_3$. Both of these LQs couple the leptonic doublet with the quark doublet representation. They thus couple neutrino to the left-chiral down-type quark. The required couplings are either $y_1^{LL} \bar{d}_{L}^{C} \nu_{L} S_{1}$ or $y_3^{LL} \bar{d}_{L}^{C} \nu_{L} S^{1/3}_{3}$. At this point it is sufficient to introduce mixing between $\tilde{R}_2$ and either $S_{1}$ or $S_{3}$ through the Higgs boson $H$ of the SM to generate neutrino mass(es) at the loop level. The particles in the loop are leptoquarks and the down-type quarks. The schematic depiction of this approach can be written as follows:
\begin{equation}
\tilde{y}_2^{RL} \bar{d}_{R}\nu_{L}\tilde{R}_{2}^{-1/3} \begin{array}{cccc}
\nearrow & \tilde{R}_{2} H S_1 & \longleftrightarrow & y_1^{LL} \bar{d}_{L}^{C} \nu_{L} S_{1}\\
\\
\searrow & \tilde{R}_{2} H S_3 & \longleftrightarrow & y_3^{LL} \bar{d}_{L}^{C} \nu_{L} S^{1/3}_{3} \\
\end{array}
\end{equation}
The one-loop diagram of neutrino mass generation that corresponds to the mixing between $\tilde{R}_2$ and $S_{3}$ is given in Fig.~\ref{fig:neutrino_mass}. The loop is closed through a mass insertion for the down-type quarks. To have a loop with the up-type quarks one would need to start with $R^{2/3}_2$ leptoquark. It is also possible to generate neutrino masses through the loops that involve leptoquarks of vector nature. This has been done in Ref.~\cite{Deppisch:2016qqd}. We discuss several explicit realizations of this approach in what follows. 

The one-loop scenario for neutrino masses that involves the $\tilde{R}_2$--$S_1$ mixing has been discussed in Ref.~\cite{Chua:1999si}. The proposed model is based on the supersymmetric SM with an explicit R-parity conservation. The non-supersymmetric neutrino mass model presented in Ref.~\cite{Mahanta:1999xd} studies potential contributions of three scalar LQs --- $S_3$, $S_1$, and $\tilde{R}_2$ --- towards neutrino masses. Again, the idea is to couple these LQs to the quark--lepton pairs and to introduce mixing between different LQs that is due to the coupling of $S_3$--$\tilde{R}_2$ and $S_1$--$\tilde{R}_2$ multiplet pairs to the Higgs boson of the SM. It is this mixing that generates small neutrino masses of the Majorana type through loop effects. In both cases the fermions in the loop are the down-type quarks. Note that the realization of proposed scenarios requires one to forbid ``diquark'' couplings to prevent rapid proton decay. The proposed scenario of Ref.~\cite{Mahanta:1999xd} has furthermore used experimental results on neutrino oscillations to constrain specific combinations of LQ couplings to quark--lepton pairs and to the Higgs boson as a function of LQ masses. 

Neutrino mass generation at the one-loop level by LQ exchange has also been studied in Ref.~\cite{AristizabalSierra:2007nf}. The authors consider all possible interactions of LQs with the Higgs boson to generalize the neutrino mass generation analysis. This approach also allows the up-type quarks in the loop, where one requires a presence of $R_2$ leptoquark to get the coupling $y^{RL}_2 \bar{u}^i_R \nu^j_L R_2^{2/3}$. Note that $R_2^{2/3}$ cannot mix directly with the Higgs and $S^{-2/3}_{3}$ component of $S_3$ but that can be overcome with the help of $\tilde{R}_2$. The proposed scenarios establish a link between the LQ decay pattern and Yukawa couplings that are responsible for the neutrino mass matrix. Consequentially, measured neutrino data are used to predict several decay branching ratios of LQ states that can be tested at the LHC. The authors of Ref.~\cite{AristizabalSierra:2007nf} single out large lepton flavor violating rates in muon and tau final states as the most interesting signatures of the proposed scenario. The study presents neutrino mass formulas containing all possible LQs in the loops taking into account both the down-type and up-type quark contributions.

One-loop model of Majorana neutrino mass that employs $S_3$ and $\tilde{R}_2$ LQs is presented in Ref.~\cite{Pas:2015hca}. The corresponding one-loop diagram of neutrino mass generation is given in Fig.~\ref{fig:neutrino_mass}. The model of Ref.~\cite{Pas:2015hca} uses the same LQ multiplets that help generate neutrino mass to address the observed anomaly in $R_K$~\cite{Hiller:2003js} by the LHCb Collaboration~\cite{Aaij:2014ora}.

\begin{figure}[tbp]
\centering
\includegraphics[scale=1.0]{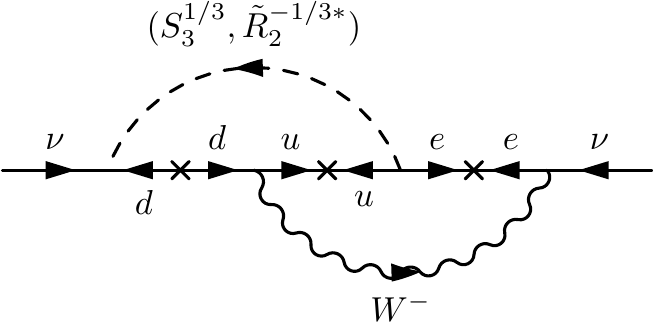}
\caption{\label{fig:neutrino_mass_b} Two-loop diagram of neutrino mass generation due to the LQ and $W$ boson exchange. Scalar LQ fields that mix due to the couplings with the SM Higgs boson are bracketed. Flavor indices are omitted.}
\end{figure}

Neutrino mass generation that relies on LQs can also be at the two-loop level~\cite{Dey:2008ht}. This approach can justify even better the observed smallness of the relevant mass parameters than the one-loop level approach. This particular possibility has been pursued in the supersymmetric version of the SM with the use of specific R-parity violating couplings~\cite{Dey:2008ht}. Another model of radiatively induced neutrino mass at the two-loop level that uses scalar LQs in the TeV range is given in Ref.~\cite{Babu:2010vp}. See Fig.~\ref{fig:neutrino_mass_b} for a representative two-loop diagram with the gauge boson in the loop. (There is one more contribution to the neutrino masses that involves the Higgs scalar in the loop. See Fig.~1 of Ref.~\cite{Babu:2010vp}.) The proposed LQs of the model --- $S_1$ and $\tilde{R}_2$ --- mix through the SM Higgs boson. They are accessible at the LHC and their decay branching ratios can probe neutrino oscillation parameters~\cite{Babu:2010vp}. The presence of these LQs can also substantially affect both the single and the pair Higgs productions as shown in Ref.~\cite{Enkhbat:2013oba}. The two-loop models of neutrino masses that use LQs that could be accessible at the LHC have also been studied in Refs.~\cite{Kohda:2012sr,Angel:2013hla}.

Recently, four systematic studies of radiatively induced neutrino mass models that also include the LQ loop scenarios have been conducted in Refs.~\cite{Cai:2014kra,Sierra:2014rxa,Helo:2015fba,Hagedorn:2016dze}. The first one places an emphasis on the predictions related to the LHC physics~\cite{Cai:2014kra}. The second one has studied in detail the two-loop realizations of the Majorana neutrino mass~\cite{Sierra:2014rxa}. The third work emphasizes the connection between double beta decay and neutrino mass models~\cite{Helo:2015fba} whereas the fourth one addresses the ability of the one- and two-loop neutrino mass models to be compatible with the gauge coupling unification. (For a mechanism that leads to the neutrinoless double beta decay through the exchange of either scalar or vector LQs see Ref.~\cite{Hirsch:1996ye}.)  


\subsection{Flavor models}

If the LQs do not couple to diquarks there is no eminent danger of proton decay. Also, the flavor constraints usually allow for light LQs if these couple diagonally and in chiral manner to the fermions in the mass eigenstate basis. The model that uses horizontal symmetry, i.e., symmetry that distinguishes fermion families, to justify particular set of couplings of a pair of the first generation LQs of type $S_1$ has been proposed in Ref.~\cite{Baver:1994hh}. It is a supersymmetric scenario that is based on a horizontal symmetry that consists of $Z_8 \times Z_7 \subset U(1)_{H_1} \times U(1)_{H_2}$ and three leptonic quantum numbers, where $U(1)_{H_1}$ and  $U(1)_{H_2}$ are QCD anomaly free groups. The same horizontal symmetry is used to address the hierarchy of quark mass matrices.

A possibility to discriminate between different models of flavor that reproduce the observed mixing parameters in the lepton sector through data on rare decays has been investigated in Ref.~\cite{Varzielas:2015iva}. The authors adapted frameworks with non-Abelian flavor symmetries that simultaneously yield experimentally viable leptonic mixing matrix and explain lepton-nonuniversality as observed through the  $R_K$ anomaly~\cite{Aaij:2014ora}. This is a bottom-up approach, where the pattern of the LQ couplings to matter manifests itself through the observable effects in lepton flavor violation and lepton-nonuniversality processes.

Another possible approach to construct the model of flavor for the leptoquarks is to apply the hypothesis of the Minimal Flavor Violation (MFV)~\cite{Chivukula:1987py,Buras:2000dm,D'Ambrosio:2002ex}. The idea is to build Yukawa couplings of leptoquarks out of the SM Yukawa coupling matrices. This is what was done in Ref.~\cite{Davidson:2010uu}, where the leptoquark Yukawa couplings were constructed on the premise of the MFV in order to be consistent with the bounds from rare physics processes. Once the LQ Yukawa couplings were constructed, i.e., fixed, and the mass of the LQ set to be just above the direct experimental search limits the authors of Ref.~\cite{Davidson:2010uu} investigated the most promising LQ search channels. Since the construction is not unique the authors considered several options and tabulated inferred limits for the given Yukawa coupling {\it ansatz}. 

There are three notable works on the application of the MFV hypothesis within the context of supersymmetric SM with R-parity violation~\cite{Nikolidakis:2007fc,Csaki:2011ge,Arcadi:2011ug}. Recall, there are three R-parity violating operators in the superpotential that can be written down if one wants to accommodate the minimal fermionic field content of the SM. These operators, written in terms of superfields, are
\begin{equation}
\label{eq:RPV_superpotential}
\lambda_{ijk} L_i L_j \bar{E}_k+\lambda'_{ijk} L_i Q_j \bar{D}_k+\lambda''_{ijk} \bar{U}_i \bar{D}_j \bar{D}_k,
\end{equation} 
where $i,j,k=1,2,3$ are flavor indices and $\lambda_{ijk}$, $\lambda'_{ijk}$, and $\lambda''_{ijk}$ are dimensionless couplings. Clearly, the second operator would yield the couplings of squarks with the quark--lepton pair once one writes down the interactions in terms of ordinary fields. More precisely, one gets $\lambda'_{ijk} L_i Q_j \bar{D}_k \supset \lambda'_{ijk} \bar{d}_{R}^{k}e_{L}^{i}\tilde{u}_L^{j}+\lambda'_{ijk} \bar{e}_{L}^{C\,i} u_{L}^{j} \tilde{d}_R^{k\,*}$, where $\tilde{u}_L^{j}$ and $\tilde{d}_R^{k}$ are squarks. These couplings can be directly mapped onto the leptoquark couplings. (See the first term in Eq.~\eqref{eq:main_t_R_2_a} and the second term in Eq.~\eqref{eq:main_S_1_a}.) Note also that the first two terms in Eq.~\eqref{eq:RPV_superpotential} violate $L$ whereas the third term violates $B$. If simultaneously allowed, these three terms would lead to rapid proton and bound neutron decays. The easiest way to prevent this from happening is to forbid all of these terms using R-parity. It has also been demonstrated that it is possible to suppress $B$ and $L$ violation to the extent that there is no problem with proton decay even if one does not demand strict R-parity conservation but instead resorts to the MFV hypothesis \cite{Nikolidakis:2007fc,Csaki:2011ge,Arcadi:2011ug}. The authors of Refs.~\cite{Nikolidakis:2007fc,Csaki:2011ge,Arcadi:2011ug} also addressed the issue of neutrino masses within this framework. 

There has also been an attempt to construct bottom-up scenario involving very light leptoquarks that is not affected by the constraints that originate from the flavor changing neutral currents. The idea is to make the neutral current interactions to be diagonal in a similar fashion as it is done with the Glashow-Iliopoulos-Maiani (GIM) mechanism. This approach calls for $n \times n$ scalar leptoquarks that couple universally to $n$ generations of the SM fermions in the weak basis, where the leptoquarks need to be degenerate in mass~\cite{Suzuki:1997ma}.




\section{Leptoquark effects on low-energy physics of flavor and CP violation}
\label{sec:FLAVOR}

LQs have renormalizable complex couplings to two SM fermions
and can thus be probed in low-energy experiments measuring rare flavor
changing and CP violating processes. The most stringent limits are
expected from observables to which the LQs contribute at tree-level
and involve both quarks and leptons: leptonic and semi-leptonic meson
decays, semi-leptonic $\tau$ decays, and $\mu$--$e$ conversion in
nuclei. Numerous constraints on such two-quark two-lepton effective
interactions have been analyzed in Ref.~\cite{Carpentier:2010ue}.
However, LQs can also contribute in loops, generating contributions to
electric and magnetic dipole moments of quarks and leptons, rare
radiative decays of mesons and leptons as well as neutral meson
anti-meson oscillations.

In principle more than one LQ state could exist at low energies. Low-energy observables can consequently receive simultaneous and
potentially interfering contributions from several LQs. Since the
resulting phenomenology becomes very model dependent, in the following
we focus on low-energy effects of single LQ exchange and derive limits
on the corresponding LQ couplings assuming they represent the only
significant contribution to the relevant observables beyond the SM. We
do not assume any flavor structure or alignment of the
LQ couplings with the SM Yukawas. The Minimal Flavor
Violation hypothesis has been applied to LQ scenarios in, for example, Refs.~\cite{Nikolidakis:2007fc,Davidson:2010uu,Csaki:2011ge,Arcadi:2011ug}.

Finally, LQs furnishing non-trivial $SU(2)$ representations can exhibit mass splittings among their weak multiplet mass eigenstates. Thus in principle their contributions to low-energy observables could be suppressed by varying mass scales. In practice however, sizable inter-multiplet mass splittings are severely constrained by electroweak precision observables (cf. Section~\ref{sec:Oblique}). In the remainder of this section we therefore neglect such effects and present all low-energy constraints as functions of a single LQ mass. Note however that most individual observables receive contributions from only one of the LQ multiplet components. Constraints derived from those observables can thus easily be rescaled, taking into account possible inter-multiplet LQ mass hierarchies.

In this section we will present in each of the following subsections basic
phenomenological tools that are needed to confront LQ predictions
and experimental data, where we will also comment on the current
experimental figures. First we will
consider LQ contributions in charged current processes and discuss
phenomenology of leptonic and semileptonic decays.
Next, we repeat similar analysis for neutral current semileptonic
processes, as well as for neutral meson mixing. In the last part we focus on charged lepton observables, such
as leptonic and semileptonic decays, lepton flavor violating (LFV) decays, $\mu$--$e$ conversion
on nuclei, lepton magnetic and
electric dipole moments.

\subsection{Semileptonic charged current processes}
The direct tree-level probes of the LQ effects
are the semileptonic charged current processes, whereas purely hadronic or leptonic
charged processes must proceed via higher perturbative order which
results in reduced sensitivity to LQ couplings.

Charged current processes can be induced by baryon number conserving
($F=0$) LQ states $R_2$, $\tilde R_2$, $U_3$, and $U_1$, whose common
feature is that they all contain a $2/3$ charged state. On the other
hand, the baryon number violating states $S_3$, $S_1$, $V_2$,
$\tilde V_2$ entail charged currents via exchange of $Q=1/3$
components~\cite{Dobrescu:2008er}. In the presence of one of the
abovelisted LQs the suitable generalization of the effective
weak Lagrangian is the following:
\begin{equation}
  \label{eq:ccLag}
  \begin{split}
  \mc{L}_{\bar u^i d^j \bar \ell \nu_k} = -\frac{4G_F}{\sqrt{2}}
  \Bigg[ &(V_{ij}  U_{\ell k}+g^{L}_{ij;\ell k})(\bar u_L^i \gamma^\mu  d_L^j) (\bar \ell_L \gamma_\mu
  \nu_L^k) \\
&+ g^{R}_{ij;\ell k}(\bar u_R^i \gamma^\mu  d_R^j) (\bar \ell_R \gamma_\mu
  \nu_R^k)\\
&+ g^{RR}_{ij;\ell k} (\bar u_R^i  d_L^j) (\bar \ell_R
  \nu_L^k) + h^{RR}_{ij;\ell k} (\bar u_R^i \sigma^{\mu\nu} d_L^j) (\bar \ell_R
  \sigma_{\mu\nu} \nu_L^k) \\
&+ g^{LL}_{ij;\ell k} (\bar u_L^i  d_R^j) (\bar \ell_L
  \nu_R^k) +  h^{LL}_{ij;\ell k}  (\bar u_L^i \sigma^{\mu\nu} d_R^j) (\bar \ell_L
  \sigma_{\mu\nu} \nu_R^k) \\
&+ g^{LR}_{ij;\ell k} (\bar u_L^i  d_R^j) (\bar \ell_R
  \nu_L^k) \\
&+ g^{RL}_{ij;\ell k} (\bar u_R^i  d_L^j) (\bar \ell_L
  \nu_R^k) \Bigg] + \mathrm{h.c.}.
  \end{split}
\end{equation}
In Eq.~\eqref{eq:ccLag} the
indices $i,j,\ell,k$ refer to the fermion mass eigenstates and $G_F$ is the Fermi coupling constant. In the SM
limit only the term proportional to $V_{ij} U_{\ell k}$
survives and results in lepton flavor universal process rates, when
these are summed over undetected neutrino flavor $k$. Modifications of
the left-handed currents, parameterized by $g^{L}$, are expected in the presence
of LQs that transform either as singlets or triplets under
$SU(2)$. On the other hand, the right-handed currents proportional
to $g^{R}$ are distinct signature of weak singlet LQ states. Further
operators involving chirality flipping currents are possible in the
presence of LQ states and are parameterized by couplings
$g^{XY}_{ij;\ell k}$, where $X$ and $Y$ refer to chiralities of the
up-type quark and charged lepton, respectively. (The chiralities of
the down-type quark and the neutrino are then determined as being
opposite to chiralities of the up-type quark and charged lepton,
respectively. E.g., $g^{RL}_{ij;\ell k}$ multiplies an operator that
is composed of right-handed up-type quark and left-handed charged
lepton.)  Pair of scalar and tensor couplings $(g^{RR}, h^{RR})$ has a common origin in a
single operator in the Fierzed basis that is characteristic of
non-chiral LQ scenarios. The same holds true for the pair ($g^{LL}, h^{LL}$). At
the matching scale the two couplings are related as
$g^{RR(LL)} = \pm 4 h^{RR(LL)}$, where minus sign applies for LQ
states with $|F|=2$. Finally, effective couplings $g^{LR}$
and $g^{RL}$ can be non-zero only when a vector LQ state is integrated
out.

The tree-level matching procedure is performed for each LQ at the
matching scale, which is here taken to be the mass $\mlq$ of the LQ state
being integrated out, and results in a subset of effective couplings
which are laid out in Table~\ref{tab:CCs}. The quoted expressions are
in accordance with the notation established in
Section~\ref{sec:INTRODUCTION}. Note that we use $v = 246\e{GeV}$ to
denote the vacuum expectation of the SM Higgs field.
The derivation of the expressions contained
in Table~\ref{tab:CCs} is straightforward: first one writes down the
tree-level LQ exchange amplitude in the leading order in
$q^2/\mlq^2$ expansion, where $q^\mu$ is the momentum flow through
the LQ propagator and $\mlq$ is the LQ mass. LQ exchanges lead to effective four-fermion
operators in the operator basis that is Fierzed with respect to
\eqref{eq:ccLag} and one has to be careful in the matching procedure
to take into account an extra minus sign that comes from different
ordering of fermionic operators acting on asymptotic states in the
LQ model and in the effective theory~\eqref{eq:ccLag} calculation. The most transparent procedure
that takes care of the sign due to fermionic permutations is to
Fierz-transform the fermionic operators on the level of effective
operators \eqref{eq:ccLag} and then perform the matching. Furthermore,
the amplitude for the LQ states with $|F|=2$ will contain currents with
charge-conjugated spinors. In such cases one first expresses the
relevant effective operator in terms of charge-conjugated fields and
only then applies the Fierz transformation.  To achieve that one invokes
a transpose relation of field bilinears:
\begin{equation}
\label{eq:TransposeRelation}
  \bar\psi_1\,(\gamma^\alpha \gamma^\beta \cdots
  \gamma^\omega)\,\psi^C_2 = \pm \overline{\psi_2}\, (\gamma^\omega \cdots
  \gamma^\beta \gamma^\alpha)\, \psi^C_1 .
\end{equation}
The plus (minus) sign in Eq.~\eqref{eq:TransposeRelation} applies for
even (odd) number of $\gamma$ matrices between the two fields.

The values of Wilson coefficients in Table~\ref{tab:CCs} are valid at the
scale $\mlq$. In order to re-sum potential large logarithmic effects due to the running of the QCD gauge coupling and to make predictions for experimentally measured quantities, they have to be run to the characteristic scales of the relevant hadronic
processes. The renormalization group effects of QCD multiplicatively
renormalize the scalar and tensor coefficients of the Lagrangian
\eqref{eq:ccLag}, while the vector currents are not renormalized. At
the leading logarithmic order the set of scalar couplings runs with
the anomalous dimension of the mass term:
\begin{equation}
\label{eq:RG-ST}
  \begin{split}
  g^{XY}_{ij;\ell k} (\mu) =&
  \left[\frac{\alpha_3(\mu)}{\alpha_3(m_{q^{f+1}})}\right]^{-\frac{\gamma_S}{2\beta_0^{(f)}}}
\cdots
\left[\frac{\alpha_3(m_b)}{\alpha_3(m_t)}\right]^{-\frac{\gamma_S}{2\beta_0^{(5)}}}
\left[\frac{\alpha_3(m_t)}{\alpha_3(\mlq)}\right]^{-\frac{\gamma_S}{2\beta_0^{(6)}}}
g^{XY}_{ij;\ell k} (\mlq).
  \end{split}
\end{equation}
Here $XY=LL,RR,LR,RL$ and $f$ is the number of quarks lighter than
scale $\mu$. The scalar anomalous dimension is $\gamma_S = -8$ and the
beta function coefficient reads $\beta_0^{(n)} = 11-2n/3$~\cite{Chetyrkin:1997dh}, where $n$ is the
number of active quark flavors. Analogous relation holds for the
tensor couplings $h^{RR,LL}_{ij;\ell k}$ with the appropriate change
$\gamma_S \to \gamma_T = 8/3$~\cite{Gracey:2000am}.
\begin{table}[tbp]
\centering
{\scriptsize
\begin{tabular}{|P{.75cm}||P{2.55cm}|P{1.35cm}|P{2.6cm}|P{2.6cm}|}
\hline
$S=0$ & $\frac{\mlq^2}{v^2}g^{L}_{ij;\ell k}$ & 
$\frac{\mlq^2}{v^2} g^{R}_{ij;\ell k}$ &
$\frac{\mlq^2}{v^2} \begin{pmatrix}
                                                           g^{RR}_{ij;\ell
                                                            k} \\ h^{RR}_{ij;\ell
                                                         k}\end{pmatrix}$
  &   $\frac{\mlq^2}{v^2} \begin{pmatrix} g^{LL}_{ij;\ell k} \\
    h^{LL}_{ij;\ell  k} \end{pmatrix}$
  \\
\hline
 $S_3$ & $-\frac{(y_3^{LL\dagger}
                           V^*)_{\ell i} (y_3^{LL}
                           U)_{jk}}{4}$
                     &  & & \\
$R_2$ & & & 
$\frac{(y_2^{RL} U)_{ik} y^{LR}_{2\,\ell j}}{16} \begin{pmatrix} 4 \\ 1 \end{pmatrix}$
& \\
$\tilde{R}_2$ & && &$\frac{(V \tilde
                     y_2^{\overline{LR}})_{ik} \tilde{y}_{2\,
                     j\ell}^{RL*}}{16} \begin{pmatrix} 4 \\ 1 \end{pmatrix}$
\\
$S_1$ 
& 
$\frac{(y_1^{LL} U)_{jk} (V^T y_1^{LL})_{i\ell }^*}{4}$
& 
$- \frac{y^{\overline{RR}}_{1\,jk} y^{RR\ast}_{1\, i\ell }}{4}$
&
$ \frac{(y_1^{LL}
        U)_{jk} y_{1\,i\ell }^{RR*}}{16} \begin{pmatrix} 4 \\ -1\end{pmatrix}$ 
&
$\frac{y_{1\,jk}^{\overline{RR}} (V^T y_1^{LL})_{i\ell }^*}{16}  \begin{pmatrix} 4 \\ -1\end{pmatrix}$ 
\\
\hline
\multicolumn{5}{c}{}\\
\hline
$S=1$& $\frac{\mlq^2}{v^2} g^{L}_{ij;\ell k} $ & $ \frac{\mlq^2}{v^2}g^{R}_{ij;\ell k} $ &  $ \frac{\mlq^2}{v^2}g^{LR}_{ij;\ell k}$ &  $\frac{\mlq^2}{v^2}g^{RL}_{ij;\ell k}$ \\
\hline 
 $U_3$ & $-\frac{(V x_3^{LL} 
                           U)_{ik} x_{3\,j\ell }^{LL*}}{2}$
                     &  & &\\
 $V_2$ & &  & $(x_2^{RL} U)_{jk}  
                           (V^T x_2^{LR})_{i\ell }$
                                                                                                                       &\\
 $\tilde V_2$ & &&& $-\tilde{x}^{RL*}_{2\,i\ell}
                    \tilde{x}_{2\,jk}^{\overline{LR}}$\\
 $U_1$ & 
$\frac{(V x_1^{LL}
         U)_{ik} x_{j\ell }^*}{2}$ &
$\frac{x_{1\,ik}^{\overline{RR}} x_{1\,j\ell }^{RR*}}{2}$   &
$-(V x_1^{LL} U)_{ik} x_{1\,j\ell }^{RR*}$ &
$-x_{1\,ik}^{\overline{RR}} x_{1\,j\ell}^{LL*}$ 
\\
\hline
\end{tabular}
}
\caption{\label{tab:CCs} Effective LQ charged-current
  couplings --- $g(\mlq)$ and $h(\mlq)$ --- defined by
  Eq.~\eqref{eq:ccLag} at the matching scale $\mlq$. $V$ and $U$ are
  CKM and PMNS matrices, respectively. The electroweak vacuum expectation value
  is $v=246\e{GeV}$.}
\end{table}

A comment related to the unobservability of PMNS matrix effects is in
order here. In Table~\ref{tab:CCs} we list the effective couplings for
neutrino mass eigenstates whereas experimentally one cannot differentiate
between neutrino species. We have consistently imposed a convention
where the couplings to the left-handed lepton doublets are defined in
the mass eigenstate basis of charged leptons. Left-handed neutrino in the final
eigenstate $k$ is thus always accompanied by $U_{\ell k}$, where $\ell$
refers to a charged lepton. The rates for $k=1,2,3$ are then summed
employing the unitarity of the PMNS matrix and the final rate is equal
to the prediction of a theory with massless neutrinos and no mixing,
i.e., $U_{ij}=\delta_{ij}$~\cite{Dorsner:2009cu}.

\subsubsection{$P \to \ell^- \bar \nu$}


The leptonic decays $P \to \ell^- \bar \nu$ of charged pseudoscalar mesons are among the most
sensitive probes of the presence of light LQs. The distinct helicity
suppression of the left-handed current operators in the SM and their
lepton universality make these decays very sensitive to
LQs. Experimentally these decays have been well measured for charged
pseudoscalars $P = \pi,
K, D, D_s$, and $B$ with $\ell = e,\mu$ and possibly also $\tau$
 in the case of decaying $D_s$ or $B$ meson. Measured values of
 the branching fractions follow the helicity suppression as predicted in
 the SM where partial branching fractions to individual lepton flavors are
 distributed proportionally to $m_\ell^2$~\cite{Agashe:2014kda}. This makes the
 decays of $B$ and $D$ mesons to light leptons very suppressed and
 currently only bounded from above, and these same bounds are
 restricting the parameter space of the LQ models that break the
 helicity suppression. These rare charged-current decays will be
 further probed in the forthcoming  Belle 2 experiment~\cite{Aushev:2010bq}.

On the theory side, the only hadronic input relevant to leptonic decays is a decay
constant, defined via
\begin{equation}
  \label{eq:DecayConst}
  \Braket{0 | \bar u_i \gamma^\mu \gamma_5 d_j | P(p)} =  i f_P
  p^\mu,
\end{equation}
where the flavor of the pseudoscalar is $P^-(\bar u^i d^j)$.
The matrix element of pseudoscalar density can be expressed, via
the equation of motion $i \slashed{D} \psi = m\psi$, by the derivative
of the axial current, $\bar u \gamma^5 d = -i\partial_\mu (\bar u \gamma^\mu \gamma_5
  d)/(m_u+m_d)$ and one can furthermore express
$\Braket{0 | \bar u_i \gamma_5 d_j | P(p)} =  -\frac{i f_P m_P^2}{m_{u^i}+m_{d^j}}$.
The decay amplitude is then
\begin{equation}
  \label{eq:CCLeptonicAmp}
\begin{split}
  \mc{A}_{P \to \ell \bar \nu^k} = \sqrt{2} G_F
  f_P m_P 
\Bigg\{ &\left[-\hat m_{\ell}(V_{ij} U_{\ell k}+g^L_{ij;\ell k}) +
  \frac{g^{RR}_{ij;\ell k}-g^{LR}_{ij;\ell k}}{\hat m_{u^i}+\hat
    m_{d^j}} \right] (\bar u_\ell P_L
    v_{\bar\nu})\\
 +&\left[\hat m_{\ell} g^R_{ij;\ell k} -
  \frac{g^{LL}_{ij;\ell k}-g^{RL}_{ij;\ell k}}{\hat m_{u^i}+\hat m_{d^j}} \right] (\bar u_\ell P_R
    v_{\bar\nu})  \Bigg\},
\end{split}
\end{equation}
where $\hat m_\ell = m_\ell/m_P$. The decay width reads
\begin{equation}
  \label{eq:CCLeptonicWidth}
  \begin{split}
\Gamma_{P \to \ell \bar \nu} = &\frac{G_F^2}{8\pi}f_P^2 m_P^3
  (1-\hat m_{\ell}^2)^2 \\
\times&\Bigg\{
 |V_{ij}|^2 \hat m_{\ell}^2 +  2 \hat m_{\ell} \re
\sum_k \left[V_{ij}^* U_{\ell k}^* \left(\hat m_{\ell} g^L_{ij;\ell k} -
  \frac{g^{RR}_{ij;\ell k}-g^{LR}_{ij;\ell k}}{\hat m_{u^i}+\hat m_{d^j}}\right)
\right]\\
&+\sum_k \left| \hat m_{\ell} g^L_{ij;\ell k} -
  \frac{g^{RR}_{ij;\ell k}-g^{LR}_{ij;\ell k}}{\hat m_{u^i}+\hat
    m_{d^j}}\right|^2 
+\sum_k \left| \hat m_{\ell} g^R_{ij;\ell k} -
  \frac{g^{LL}_{ij;\ell k}-g^{RL}_{ij;\ell k}}{\hat m_{u^i}+\hat m_{d^j}}\right|^2
\Bigg\}.
  \end{split}
\end{equation}
The summation over neutrino species $k$ eliminates all dependence on
the PMNS matrix $U$ when explicit expressions for effective couplings
are inserted into~\eqref{eq:CCLeptonicWidth}. In the decay width
expression~\eqref{eq:CCLeptonicWidth} it is apparent that LQ states
which introduce couplings other than $g^L$ or $g^R$ do not entail the
$m_\ell^2$ scaling of the width, which makes their parameter space
especially sensitive to the ratio of charged pion decays to different
charged leptons, i.e.,
$\pi^- \to e^- \bar \nu, \mu^- \bar\nu$.  The ratio
$R^\pi_{e/\mu} = \frac{\Gamma(\pi^- \to e^- \bar \nu)}{\Gamma(\pi^-
  \to \mu^- \bar \nu)}$
is testing the SM pattern of lepton flavor universality in the first
two generations of leptons, and its prediction is very precise due to
cancellation of hadronic and CKM element
uncertainties~\cite{Davidson:1993qk,Leurer:1993ap,Cirigliano:2007xi}:
\begin{equation}
\label{eq:Rpi_emu}
  R^{\pi(\mrm{SM})}_{e/\mu} =
  \frac{m_e^2(m_\pi^2-m_e^2)^2}{m_\mu^2(m_\pi^2-m_\mu^2)^2}
  (1+\Delta) = (1.2352 \pm 0.0001)\E{-4}.
\end{equation}
Here $\Delta$ parametrizes radiative corrections and the SM
prediction is in good agreement with the measured value
$R_\mrm{e/\mu}^{\pi(\mrm{exp})} = (1.2327\pm 0.0023)\E{-4}$~\cite{Agashe:2014kda}.  Scalar operators
that arise in the presence of non-chiral LQ states lift the helicity
suppression with the charged lepton mass, present in vector
operators. Early studies showed that non-chiral LQs can easily violate
the bound from leptonic decays of pions, $K$, and $D$
mesons~\cite{Davidson:1993qk,Leurer:1993ap,Leurer:1993em,Leurer:1993qx,Hewett:1997ce}.
In the kaon sector one defines a similar ratio $R_\mrm{e/\mu}^{K}$ that also turns out to be very
constraining~\cite{Fajfer:2015ycq}, where $R_\mrm{e/\mu}^{K(\mrm{exp})} = (2.488\pm
  0.010)\E{-5}$~\cite{Lazzeroni:2012cx} and $R_\mrm{e/\mu}^{K(\mrm{SM})} =(2.477\pm 0.001)\E{-5}$~\cite{Cirigliano:2007xi}. Leptonic decays of
$D_s \to \ell \bar \nu$ in presence of light scalar LQs have been studied
in~\cite{Dobrescu:2008er,Kronfeld:2008gu,Benbrik:2008ik,Dorsner:2009cu}.
Pion leptonic decays $\pi \to e \bar \nu, \mu \bar \nu$ have been
studied as subsidiary constraints in the context of LQ models
in Ref.~\cite{Benbrik:2010cf}. Ref.~\cite{Dorsner:2011ai} sets the various
flavor constraints on the Yukawa couplings of $\tilde S_1$, resulting in a
prediction of $B \to \tau \bar \nu$.  The authors of
Ref.~\cite{Dorsner:2013tla} made predictions for the
$B_c \to \tau \bar\nu$ branching fraction in relation to
$B \to D^{(*)} \tau \bar \nu$ in the presence of scalar $R_2$.

\subsubsection{$\tau \to \nu P$}
The amplitude for the $\tau \to \nu P$  decay is a crossed version of the one of $P \to \ell
\nu$ ($\ell = e, \mu$) with appropriate lepton flavor rearrangement. Considering
$\tau \to \nu P$ allows one to extend the lepton flavor universality
ratio with light pseudoscalar $P$ to the third lepton generation
\begin{equation}
R_{\tau/\mu}^P  = \frac{\Gamma(\tau^-\to P^- \nu)}{\Gamma(P^- \to \mu^- \bar \nu)},
\end{equation}
where $P=K,\pi$ and the amplitude for the $\tau$ semileptonic decay is just the crossed
amplitude for the $P \to \tau \bar \nu$ decay, given in Eq.~\eqref{eq:CCLeptonicAmp}:
\begin{equation}
  \label{eq:tau-Pnu}
\begin{split}
  \mc{A}_{\tau^+ \to P  \bar\nu^k} = \sqrt{2} G_F
  f_P m_P 
\Bigg\{ &\left[-\hat m_{\tau}(V_{ij} U_{\ell k}+g^{L}_{ij;\tau k}) +
  \frac{g^{RR}_{ij;\tau k}+g^{LR}_{ij;\tau k}}{\hat m_{u^i}+\hat
    m_{d^j}} \right] (\bar v_\tau P_L
    v_{\nu})\\
 +&\left[\hat m_{\tau} g^{R}_{ij;\tau k} -
  \frac{g^{LL}_{ij;\tau k}+g^{RL}_{ij;\tau k}}{\hat m_{u^i}+\hat m_{d^j}} \right] (\bar v_\tau P_L
    v_{\nu})  \Bigg\}.
\end{split}
\end{equation}
The $R_{\tau/\mu}^K$ has been combined with $R_{\mu/e}^K$ to derive
stringent constraints on the muon couplings of vector LQ
$U_3$~\cite{Fajfer:2015ycq}. In Ref.~\cite{Wise:2014oea} a connection
between neutrino interactions with light quarks and the ratio
$R_{\tau/\mu}^\pi$ has been established. Bounds on the LQ couplings of the same order of
magnitude are expected from $\tau \to \nu V$, where $V$ is a vector
meson.

\subsubsection{$P \to P' \ell^- \bar \nu$}


The effective Lagrangian~\eqref{eq:ccLag} in general modifies the SM
predictions for semileptonic decays
$P(p) \to P' (p') \ell^-(p_\ell) \bar \nu^k(k)$ or
$P(p) \to V (p',\epsilon) \ell^-(p_\ell) \bar \nu^k(k)$. For the
underlying quark flavor transition $d_j \to u_i$ the corresponding amplitude for the
pseudoscalar $P$ decay to another pseudoscalar $P'$ reads~\cite{Bobeth:2007dw}
\begin{equation}
  \begin{split}
\mc{A}_{P \to P' \ell^- \bar \nu_k} = -i\sqrt{2} G_F
&\Big[
\mc{A}_{ij;\ell k}^V \bar u_\ell \slashed{p} v_\nu + \mc{A}_{ij;\ell k}^A \bar u_\ell \slashed{p} \gamma_5 v_\nu \\
&+ (\mc{A}_{ij;\ell k}^{S}  +\mc{A}_{ij;\ell k}^{T}  \cos
\theta) \bar u_\ell v_\nu \\
&+ (\mc{A}_{ij;\ell k}^{P}+\mc{A}_{ij;\ell k}^{T5} \cos\theta) \bar u_\ell \gamma_5 v_\nu
\Big].
  \end{split}
\end{equation}
The Dirac spinors correspond to the charged lepton, $u_\ell = u_\ell(p_\ell)$, and
antineutrino, $v_\nu = v_\nu(k)$. Angle $\theta$ is between $P$ and
$\ell^-$ in the rest frame of $\ell \bar \nu^k$ system.
The $q^2$ dependent functions, where $q = p_\ell + k$, are expressed via hadronic form
factors and the effective couplings
\begin{equation}
\label{eq:Amps-SL}
  \begin{split}
    \mc{A}_{ij;\ell k}^V &=(V_{ij} U_{\ell k} + g^L+g^R)f_+(q^2) + 4(h^{RR} + h^{LL}) \frac{m_\ell}{m+m'} f_T(q^2),\\
    \mc{A}_{ij;\ell k}^A &=(-V_{ij} U_{\ell k} - g^L+g^R)f_+(q^2) + 4(-h^{RR} + h^{LL}) \frac{m_\ell}{m+m'} f_T(q^2),\\
    \mc{A}_{ij;\ell k}^S &=(g^{LR} + g^{RR}+g^{RL}+g^{LL})\frac{m^2-m'^2}{2(m_d -
      m_u)} f_0 (q^2) \\
    &+ \frac{m_\ell}{2}(V_{ij} U_{\ell
      k}+g^L+g^R)\left[\frac{m^2-m'^2}{q^2}(f_0(q^2)-f_+(q^2))-f_+(q^2)\right]\\
&- 2(h^{RR}+h^{LL}) m_\ell^2 \frac{f_T(q^2)}{m+m'} \left(1+\frac{m^2-m'^2}{q^2} \right),\\
    \mc{A}_{ij;\ell k}^P &=(-g^{LR} - g^{RR}+g^{RL}+g^{LL})\frac{m^2-m'^2}{2(m_d -
      m_u)} f_0 (q^2) \\
    &+ \frac{m_\ell}{2}(-V_{ij} U_{\ell
      k}-g^L+g^R)\left[\frac{m^2-m'^2}{q^2}(f_0(q^2)-f_+(q^2))-f_+(q^2)\right]\\
&- 2(-h^{RR}+h^{LL}) m_\ell^2 \frac{f_T(q^2)}{m+m'} \left(1+\frac{m^2-m'^2}{q^2} \right),\\
    \mc{A}_{ij;\ell k}^T &= 2(h^{RR}+h^{LL}) \frac{f_T(q^2)}{m+m'} \lambda^{1/2}
    \beta_\ell(q^2)^2,\\
    \mc{A}_{ij;\ell k}^{T5} &= 2(-h^{RR}+h^{LL}) \frac{f_T(q^2)}{m+m'} \lambda^{1/2}
    \beta_\ell(q^2)^2.
  \end{split}
\end{equation}
The pseudoscalar masses have been denoted by $m$ and $m'$, while
$m_{u,d}$ are the masses of valence quarks taking part in interaction. In order to
ease readability we have suppressed subscripted flavor indices
$ij;\ell k$ on the couplings $g$ and $h$. We have also employed
standard notation
\begin{equation}
  \label{eq:lambda}
  \begin{split}
\lambda(x,y,z) = (x+y+z)^2-4(xy+yz+zx)
  \end{split}  
\end{equation}
and $\beta_\ell(q^2) = \sqrt{1-m_\ell^2/q^2}$. The hadronic form
factors that enter in amplitudes~\eqref{eq:Amps-SL} are defined as:
{\footnotesize
\begin{equation}
\label{eq:ffs}
  \begin{split}
\Braket{P' (p')| \bar u \gamma^\mu  d| P(p)} &=
f_+(q^2)\left[(p+p')^\mu - \frac{m^2 - m'^2}{q^2} q^\mu \right]
+ f_0 (q^2) \frac{m^2 - m'^2}{q^2}  q^\mu ,\\
\Braket{P (p')| \bar u \sigma^{\mu  \nu} (1 \pm \gamma_5) d| P(p)} &= i \frac{f_T(q^2)}{m +m'} \left[ (p+p')^\mu  q^\nu - (p+p')^\nu  q^\mu \pm i \epsilon^{\mu \nu \alpha \beta} (p+p')_\alpha   q_\beta \right].
  \end{split}
\end{equation}
}
The form factors can be either extracted experimentally in a related
process that is presumably free of LQ contributions
or calculated using a nonperturbative QCD method,
e.g., lattice QCD. For recent lattice QCD results for $D$ or $B$ form factors c.f.~\cite{Aoki:2013ldr}.

For the scalar operators we have used the identity 
\begin{equation}
\Braket{P'(p') | \bar u d | P(p)} = \frac{q_\mu}{m_d - m_u}
\Braket{P'(p') | \bar u \gamma^\mu d | P(p)}  
\end{equation}
that follows from the equation of motion. The decay spectrum
distribution is given by the amplitudes
$\mc{A}_{ij;\ell k}^{V,A,S,P,T,T5}$ and can be found in
Refs.~\cite{Bobeth:2007dw, Fajfer:2015mia}.  Branching fractions
themselves provide a good handle only on the LQ Yukawas since they are
competing with large tree-level SM contributions that suffer from
uncertainties in the hadronic matrix elements and from experimental
errors. Again, the lepton universality ratios
$\Gamma(P \to P' \ell_1 \nu)/\Gamma(P \to P' \ell_2 \nu)$ are superior
in precision over individual branching fractions and spectra. A
notable example is the semileptonic lepton flavor universality~(LFU)
ratio
$R_{D} \equiv \Gamma(\bar B \to D \tau \bar\nu)/\Gamma(\bar B \to D
\ell \bar\nu)$
comparing the $\tau$ and light lepton couplings with large precision.
It has been shown that scalar~\cite{Tanaka:2012nw,Dorsner:2013tla,
  Sakaki:2013bfa, Sakaki:2014sea, Alonso:2015sja,
  Freytsis:2015qca,Bauer:2015knc,Li:2016vvp} as well as vector
LQs~\cite{Freytsis:2015qca,Alonso:2015sja,Fajfer:2015ycq,Barbieri:2015yvd}
may significantly alter this LFU ratio. The world averages of $R_D$ and
the analogous observable defined for the $B \to D^* \ell \bar\nu$
transition $R_{D^*}$ calculated in the winter 2016 HFAG
edition~\cite{Amhis:2014hma} agree with the SM at the $4\,\sigma $
level, as shown in Eqs.~\eqref{eq:RDexp} and \eqref{eq:RDtheory}.

It has been suggested in
Ref.~\cite{Bauer:2015knc} that a minimal extension of the SM with
$S_1$ scalar LQ of mass of the order of $1$\,TeV can provide an
explanation of $R_{D^{(*)}}$ in addition to observables $R_K$ and
muon $(g-2)$, to be discussed in the following sections. Scalar LQ
$S_1$ couples with order one generation-diagonal couplings to quarks
and leptons that are doublets of $SU(2)$ of the SM. The presence of
$S_1$ allows the authors of Ref.~\cite{Bauer:2015knc} to address the
$R_{D^{(*)}}$ anomalies at the tree-level, while the $R_K$
contributions are generated at loop level. Note that $S_1$ could have
``diquark'' couplings and accordingly mediate proton decay.
The flavor structure of this proposal in view of current LHC constraints from the
$8$\,TeV and $13$\,TeV data sets has been investigated in
Ref.~\cite{Dumont:2016xpj}. The authors of Ref.~\cite{Dumont:2016xpj}
have, furthermore, analyzed in great detail prospects to test this
particular $S_1$ scenario at $14$\,TeV run at the LHC. 
A possible resolution of $R_K$ and $R_{D^{(*)}}$ puzzles through the
use of vector LQ $U_3$ has been proposed in
Ref.~\cite{Fajfer:2015ycq}. One particularly simple phenomenologically viable leptoquark model
that features dark matter coannihilation through the leptoquark
mediation has been presented in Ref.~\cite{Baker:2015qna}. The authors
establish connection between the visible and invisible sectors in the
form of $R_2$ leptoquark with very specific Yukawa couplings (and mass) that satisfy constraints of both direct and indirect detection origin.
Finally, flavor models that can provide an explanation of the observed enhancement of the $\bar{B} \to D^{(*)} \tau \bar{\nu}$ rates with respect to the SM predictions through scalar and/or vector LQ(s) have been pursued in Ref.~\cite{Freytsis:2015qca}.


\subsubsection{$P \to V \ell^- \bar \nu$}


As an example of semileptonic decays with vector mesons in the final state 
we shall use the decays $\bar B \to D^{*} \ell \bar \nu$. The most
interesting experimental observable is the analogue of $R_D$,
but defined for vector meson in the final state:
\begin{equation}
R_{D^*} = \frac{{\cal  B} (\bar B \to D^* \tau^- \bar \nu) }{{\cal  B} (\bar B \to D^* \ell^- \bar \nu) }
\label{RX}
\end{equation}
with $\ell= e,\mu$. The HFAG averages over LHCb and $B$ factories'
results and finds~\cite{Amhis:2014hma}
\begin{equation}
  \label{eq:RDexp}
  \begin{split}
    R_{D} &= 0.397 \pm 0.068,\\
    R_{D^*} &= 0.316 \pm 0.026,
  \end{split}
\end{equation}
where we have summed the statistical and systematical errors. Taking
into account the correlation between the two ratios $-0.26$ the
agreement with the SM is at the $4\,\sigma$ level~\cite{Na:2015kha,Fajfer:2012vx}:
\begin{equation}
  \label{eq:RDtheory}
  \begin{split}
    R_{D}^\mrm{SM} &= 0.300 \pm 0.008,\\
    R_{D^*}^\mrm{SM} &= 0.252 \pm 0.003.
  \end{split}
\end{equation}

In the case of $\bar B\to D^* \tau \bar\nu$ decays the helicity amplitudes
formalism is a convenient description that is commonly
used~\cite{Fajfer:2012vx,1602.03030,1602.07671,Sakaki:2013bfa,1302.1042}. First one needs to parametrize matrix elements of the
weak currents for the $B \to D^*$ transition.  Standard
parametrization of vector and axial hadronic matrix elements for the
$\bar B \to D^*$ transition is given by~\cite{Isgur:1990kf}
{\small
\begin{align}
\label{one}
\Braket{D^*(k,\epsilon)| \bar{c}\gamma_\mu b|\bar B(p) } &=\frac{2 i V(q^2)}{m_B+m_{D^*}}\;\epsilon_{\mu\nu\alpha\beta}
\epsilon^{*\nu}p^\alpha k^\beta ,
\\
\label{two}
\Braket{ D^*(k,\epsilon)|\bar{c}\gamma_\mu\gamma_5 b|\bar B(p) }
&=2m_{D^*}\,A_0(q^2)\frac{\epsilon^*\cdot q}{q^2}q_\mu
\\
&\quad + (m_B+m_{D^*})\,A_1(q^2)\,\left(\epsilon^*_\mu - \frac{\epsilon^*\cdot q}{q^2}q_\mu
\right) \nonumber\\
&\quad - A_2(q^2)\,\frac{\epsilon^*\cdot q}{m_B+m_{D^*}}\left( (p+k)_\mu 
-\frac{m_B^2-m_{D^*}^2}{q^2}q_\mu\right), \nonumber
\end{align}}
where $q = p - k$. The matrix element of the tensor operator
is parameterized as~\cite{Isgur:1990kf}:
\begin{equation}
  \label{eq:ffNPDsTens}
\begin{split}
\Braket{ D^{*}\left(k,\epsilon\right)|\bar c\sigma_{\mu\nu}b|\bar
  B\left(p\right) }& = \frac{\epsilon^*\cdot
                           q}{(m_B+m_{D^*})^2}T_0(q^2)\varepsilon_{\mu\nu\alpha\beta}p^{\alpha}k^\beta \\
&\quad+T_1(q^2)\varepsilon_{\mu\nu\alpha\beta}p^\alpha\epsilon^{*\beta}+T_2(q^2)\varepsilon_{\mu\nu\alpha\beta}k^\alpha\epsilon^{*\beta},
\end{split}
\end{equation}
and can be related to
  \begin{equation}
\label{eq:ffNPDsTens5}
\begin{split}
\Braket{ D^{*}\left(k,\epsilon\right)|\bar
  c\sigma_{\mu\nu}\gamma_5\,b|\bar
  B(p) }&=\frac{i\,\epsilon^*\cdot
                            q}{(m_B+m_{D^*})^2}T_0(q^2) \left(p_\mu
                            k_\nu-k_\mu p_\nu\right)\\
&\quad+i\,T_1(q^2)\left(p_\mu\epsilon^*_\nu-\epsilon^*_\mu p_\nu\right)\\
&\quad+i\,T_2(q^2)\left(k_\mu\epsilon^*_\nu-\epsilon^*_\mu k_\nu\right).
\end{split}
\end{equation}
The $\bar B\to D^{(*)}\tau\bar\nu$ amplitude can be written as:
{\normalsize
  \begin{equation}
\label{eq:ampBDtaunu}
\begin{split}
 \mathcal{A}_{\bar B\to D^{(*)}\tau\bar\nu}=-\sqrt{2}G_F \Big\{&H_{V}^\mu \bar u \gamma_\mu (1-\gamma_5)v+H_{S}\bar u  (1-\gamma_5) v
 \\
&+H_T^{\mu\nu}\bar u  \sigma_{\mu\nu} (1-\gamma_5)v\Big\},
\end{split}
  \end{equation}}
with $H_V^\mu$, $H_S$ and $H_T^{\mu\nu}$: 
\begin{equation}
\begin{split}
&H_V^\mu=\left(V_{cb}+g^L \right) \langle  D^{*}|\bar{c}\gamma^\mu b |B\rangle
- \left(V_{cb} + g^L\right)\langle  D^{*}|\bar{c}\gamma^\mu\gamma_5 b |B\rangle,\\
&H_S=(g^{RR}+g^{LR} ) \langle D^{*}|\bar c b |B \rangle-(g^{RR}- g^{LR} )\langle D^{*}|\bar c\gamma_5  b |B\rangle,\\
&H_T^{\mu\nu}=h^{RR} \langle D^{*}| \bar c\sigma^{\mu\nu}(1-\gamma_5)b
  |B\rangle .\label{eq:Hampsdef}
\end{split}
\end{equation}
The functions $H_{V,S,T}$  can be projected onto different angular-momentum states of the lepton and antineutrino specified by its polarization vectors $\epsilon^\mu(\lambda)$ and using the completeness relation 
$g_{\mu\nu}=\sum g_{mn}\epsilon_\mu(m)\epsilon_\nu^*(n)$. The projections of $H_V^\mu$ and $H_T^{\mu\nu}$
define the helicity amplitudes,
\begin{equation}
H_\lambda=H_V^{\mu}\epsilon^*_\mu(\lambda),\hspace{0.5cm}H_{\lambda\lambda^\prime}=H_T^{\mu\nu}\epsilon^{*}_\mu(\lambda)\epsilon^{*}_\nu(\lambda^\prime),  \label{eq:helicityDef}
\end{equation}
while $H_S$ contributes only to the $\lambda=t$ component. The tensor helicity amplitudes are antisymmetric with respect to the
exchange of indices, $H_{\lambda^\prime\lambda}=-H_{\lambda\lambda^\prime}$.
For the lepton--antineutrino pair, polarization vectors are 
$\epsilon^\mu(\pm)=(0,\pm1,-i,0)/\sqrt{2}$, $\epsilon^\mu(0)=(|\vec k|,0,0,-q^0)/\sqrt{q^2}$ and 
$\epsilon ^\mu(t)=(q^0,0,0,-|\vec k|)/\sqrt{q^2}$. The $D^*$
polarizations are given as
$\epsilon_{D^*}^\mu(\pm)=(0,\mp 1,-i,0)/\sqrt{2}$ and
$\epsilon_{D^*}^\mu(0)=(|\vec{k}|,0,0,E_{D^*})/m_{D^*}$.
The helicity amplitudes calculated 
 in the $B$-rest frame are:
{\small
\begin{align}
&H_\pm=-(V_{cb}+g^L  )(m_B+m_{D^*})A_1(q^2)\pm(V_{cb}+ g^L -g^R)\frac{2m_B|\vec k|}{m_B+m_{D^*}}V(q^2),\nonumber\\
&H_0=-\frac{(V_{cb}+g^L )}{2 m_{D^*}\sqrt{q^2}}\left[\left(m_B+m_{D^*}\right)\left(m_B^2-m_{D^*}^2-q^2\right)A_1(q^2)
-\frac{4m_B^2|\vec k|^2}{m_B+m_{D^*}}A_2(q^2)\right],\nonumber\\
&H_t=-(V_{cb}+g^L )\frac{2m_B|\vec k|}{\sqrt{q^2}}A_0(q^2),\hspace{0.5cm}H_S=(g^{LR} - g^{RR} )\frac{2m_B|\vec k|}{m_b+m_c}A_0(q^2),\nonumber\\
&H_{\pm0}=\pm\frac{ih^{RR} }{2\sqrt{q^2}}\left[(m_B^2-m_{D^*}^2\pm2m_B|\vec k|)(T_1(q^2)+T_2(q^2))+q^2(T_1(q^2)-T_2(q^2))\right],\nonumber\\
&H_{\pm t}=\frac{ih^{RR}}{2\sqrt{q^2}}\left[(m_B^2-m_{D^*}^2\pm2m_B|\vec k|)(T_1(q^2)+T_2(q^2))+q^2(T_1(q^2)-T_2(q^2))\right],\nonumber\\
&H_{+-}=-H_{t0}=i h^{RR}\left[\frac{m_BE_{D^*}}{m_{D^*}}T_1(q^2)+m_{D^*}T_2(q^2)+\frac{m_B^2|\vec k|^2}{m_{D^*}(m_{D^*}+m_B)^2}T_0(q^2)\right].\label{eq:helicityBDV}
\end{align}}

By separating the contributions to the decay rate due to different
$D^{(*)}$ and $\tau$ helicity states, the $\bar B\to D^{(*)}\tau\bar\nu$
decay rates $\Gamma_{\tilde \lambda,\lambda_\tau}$ take the following form:
\begin{equation}
\label{eq:BDtaunurate}
\begin{split}
\frac{d^2\Gamma_{B,+}}{dq^2d(\cos\theta_\tau)}&=\frac{G_F^2}{256\pi^3}\frac{|\vec{k}|}{m_B^2}\left(1-\frac{m_\tau^2}{q^2}\right)^2\\
&\times
\left\{2\cos\theta^2_\tau\,\Gamma_{0+}^0+2\Gamma_{0+}^t+2\cos\theta_\tau\,\Gamma_{0+}^I+
\sin\theta^2_\tau(\Gamma_{++}+\Gamma_{-+})\right\}, \\
\frac{d^2\Gamma_{B,-}}{dq^2d(\cos\theta_\tau)}&=\frac{G_F^2}{256\pi^3}\frac{|\vec{k}|}{m_B^2}\left(1-\frac{m_\tau^2}{q^2}\right)^2\\
&\times
\left\{2\sin\theta^2_\tau\Gamma_{0-}+(1-\cos\theta_\tau)^2\Gamma_{+-}+(1+\cos\theta_\tau)^2\Gamma_{--}\right\}.
\end{split}
\end{equation}
For the decay into a $\tau(1/2)$ and a longitudinal $D^{(*)}$ there are three
contributions stemming from the longitudinal ($\Gamma_{0+}^0$) and time-like ($\Gamma_{0+}^t$) components of the lepton--antineutrino state and from
their interference ($\Gamma_{0+}^I$). Finally, the $\Gamma^{(X)}_{\tilde \lambda,\lambda_\tau}$ are functions of the helicity amplitudes:
\begin{equation}
\label{eq:Gammashel}
\begin{split}
\Gamma_{0+}^0&=\left|2i\sqrt{q^2}\left(H_{+-}+H_{0t}\right)-m_\tau
               H_0\right|^2, \\
\Gamma_{0+}^t&=\left|m_\tau H_t+\sqrt{q^2}H_S\right|^2,\\
\Gamma_{0+}^I &=2{\rm Re}\left[(2i\sqrt{q^2}\left(H_{+-}+H_{0t}\right)-m_\tau H_0)(m_\tau H_t+\sqrt{q^2}H_S)^*\right],\\
\Gamma_{++}&=\left|m_\tau
             H_+-2i\sqrt{q^2}\left(H_{+t}+H_{+0}\right)\right|^2,\\
\Gamma_{-+}&=\left|m_\tau H_--2i\sqrt{q^2}\left(H_{-t}-H_{-0}\right)\right|^2,\\
\Gamma_{0-}&=\left|\sqrt{q^2}H_0-2im_\tau\left(H_{+-}+H_{0t}\right)\right|^2,\\
\Gamma_{+-}&=\left|\sqrt{q^2} H_+-2im_\tau\left(H_{+t}+H_{+0}\right)\right|^2,\\
\Gamma_{--}&=\left|\sqrt{q^2} H_--2im_\tau\left(H_{-t}-H_{-0}\right)\right|^2.
\end{split}
\end{equation}
Evaluation of the above observables requires the knowledge of form
factors which can be related to a single function in the
heavy quark limit as described in~\cite{Isgur:1990kf,Fajfer:2012vx,1602.03030,1602.07671, Sakaki:2013bfa,1302.1042}.


The number of observables in the case $\bar B \to D^* \tau \bar\nu$ is much
larger than in the case of $\bar B \to D \tau \bar \nu$ as discussed
already by many authors \cite{Fajfer:2012vx,1602.03030,1602.07671,Sakaki:2013bfa,Freytsis:2015qca,1302.1042,Alonso:2015sja}. The
opening angle asymmetry between $D^*$ and $\tau$ probes right-handed currents~\cite{Fajfer:2012vx,Sakaki:2013bfa}
while $\tau$ helicity-dependent decay rates have also been shown to
deviate from the SM prediction~\cite{Tanaka:2010se,Fajfer:2012vx}.
In particular, the $q^2$ distributions have been shown to be able to
distinguish between different LQ scenarios~\cite{Sakaki:2013bfa}. The
Belle analysis of decay kinematics~\cite{Abdesselam:2016cgx} managed to exclude a portion
of parameter space  $R_2$ scalar LQ scenario favored by $R_{D^{(*)}}$
deviation. Based
on current results on $R_{D}$ and $R_{D^*}$, the authors of
Ref.~\cite{Freytsis:2015qca} have found best fit point for the
coupling $g^L = 0.18\pm 0.04$.
The invariant dilepton mass distribution can
be obtained after performing integration over the angle $\theta_\tau$
in \eqref{eq:BDtaunurate}, as discussed in
Refs.~\cite{Fajfer:2012vx,1602.07671,Freytsis:2015qca,1302.1042,Sakaki:2014sea}.

\subsection{Rare meson decays}
\begin{figure}[!htbp]
  \centering
  \begin{tabular}{lcr}
  \includegraphics[width=0.3\textwidth]{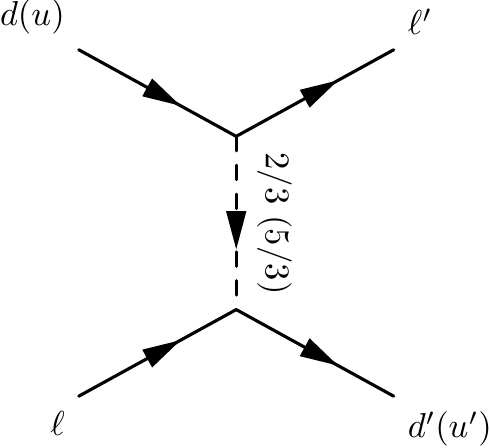} && \includegraphics[width=0.3\textwidth]{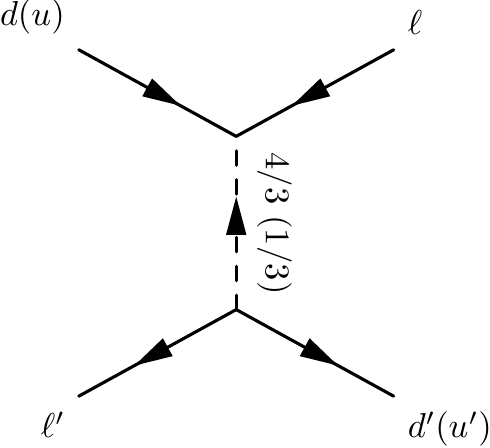}
    \end{tabular}
    \caption{Diagrams of rare meson decays of flavor
      $(\bar d' d)(\bar \ell' \ell)$ and
      $(\bar u' u)(\bar \ell' \ell)$ induced by LQs. The $F=0$
      LQs contribute as shown in the left-hand
side diagram, $|F|=2$ as shown in the right-hand diagram. Propagators
are labeled by charge of the relevant components of the LQ.}
\label{fig:LQRareDiag}
\end{figure}

The mechanisms of LQ contributions to rare meson decays differ
qualitatively from the mechanism of rare SM decays. By rare decays we
mean here decays with neutral quark and lepton currents
$(\bar q' q) (\bar \ell' \ell)$ or $(\bar q' q) (\bar \nu' \nu)$ that are induced exclusively via loop
diagrams and are suppressed by the unitarity of the CKM matrix (GIM
mechanism) or are even absent (lepton flavor violation) in the SM. The competing LQ amplitudes can be
induced at the tree-level and present a very serious constraint on the flavor changing
LQ couplings.
 Tree-level LQ amplitudes are exemplified in Fig.~\ref{fig:LQRareDiag}
for decays that involve charged leptons.

The fact that tree-level LQ diagrams, as shown in Fig.~\ref{fig:LQRareDiag}, contribute to dimension-6
operators with chiral fermions motivates the following convention for 
$\bar q^i q^j \bar\ell \ell^{\prime}$ effective Lagrangian:
\begin{equation}
  \label{eq:RareLeff}
  \begin{split}
  \mc{L}_{\bar q^i q^j \bar \ell \ell^\prime} = -\frac{4G_F}{\sqrt{2}}
  &\Bigg[ c^{LL}_{ij;\ell \ell^\prime}(\bar q_L^i \gamma^\mu  q_L^j) (\bar \ell_L \gamma_\mu
  \ell_L^{\prime}) 
+ c^{RR}_{ij;\ell \ell^\prime}(\bar q_R^i \gamma^\mu  q_R^j) (\bar \ell_R \gamma_\mu
  \ell^\prime_R)\\
&+c^{LR}_{ij;\ell \ell^\prime}(\bar q_L^i \gamma^\mu  q_L^j) (\bar \ell_R \gamma_\mu
  \ell_R^{\prime}) 
+ c^{RL}_{ij;\ell \ell^\prime}(\bar q_R^i \gamma^\mu  q_R^j) (\bar \ell_L \gamma_\mu
  \ell^\prime_L)\\
&+ g^{RR}_{ij;\ell \ell^\prime} (\bar q_R^i  q_L^j) (\bar \ell_R
  \ell_L^\prime) + h^{RR}_{ij;\ell \ell^\prime} (\bar q_R^i \sigma^{\mu\nu} q_L^j) (\bar \ell_R
  \sigma_{\mu\nu} \ell^\prime_L) \\
&+ g^{LL}_{ij;\ell \ell^\prime} (\bar q_L^i  q_R^j) (\bar \ell_L
  \ell_R^\prime) +  h^{LL}_{ij;\ell \ell^\prime}  (\bar q_L^i \sigma^{\mu\nu} q_R^j) (\bar \ell_L
  \sigma_{\mu\nu} \ell_R^\prime) 
\\
 &+ g^{LR}_{ij;\ell \ell^\prime} (\bar q_L^i  q_R^j) (\bar \ell_R
   \ell_L^\prime) + g^{RL}_{ij;\ell \ell^\prime} (\bar q_R^i  q_L^j) (\bar \ell_L
   \ell_R^\prime)
  \Bigg] + \mathrm{h.c.}.
  \end{split}
\end{equation}
The above operator basis is analogous to the charged-current
semileptonic basis~\eqref{eq:ccLag} with an exception of coefficients
$c^{LR}$, $c^{RL}$ whose counterparts are not present in the
charged-current effective Lagrangian. In order to connect to large
body of phenomenological literature on rare leptonic and semileptonic
meson decays, we provide below the mapping from the above introduced ``chiral basis'' of
operators~\eqref{eq:RareLeff} to the basis commonly used
in rare $B$ meson decays~\cite{Buchalla:1995vs}:
\begin{equation}
\label{eq:RareLeffStandard}
  \begin{split}   
  \mc{L}_{\bar q^i q^j \bar\ell\ell'} = -\frac{4G_F}{\sqrt{2}} \lambda_q
  \Bigg[&C_7^{ij} \mc{O}^{ij}_7 + C_{7^\prime}^{ij}    \mc{O}^{ij}_{7^\prime}
  +\sum_{X=9,10,S,P}
  \left(C_X^{ij;\ell\ell'} \mc{O}_X^{ij;\ell\ell'} + C_{X^\prime}^{ij;\ell\ell'}
    \mc{O}_{X^\prime}^{ij;\ell\ell'}\right) \\
   &+ C^{ij;\ell\ell'}_T
  \mc{O}^{ij;\ell\ell'}_T + C_{T5}^{ij;\ell\ell'}
  \mc{O}^{ij;\ell\ell'}_{T5}\Bigg] + \mrm{h.c.}.
  \end{split}
\end{equation}
The following operators are all susceptible to LQ effects
\begin{equation}
  \label{eq:RareOps}
    \begin{split}
 { \cal O}^{ij}_7 &= \frac{e m_{q^j}}{(4\pi)^2}\, (\bar q^i \sigma_{\mu\nu} P_R q^j)
  \,F^{\mu\nu}\,, \qquad \quad \  {\cal O}_S^{\ell\ell'} = \frac{e^2}{(4\pi)^2}\, (\bar q^i P_R q^j)
  (\bar\ell  \ell'),\\
 {\cal O}_9^{ij;\ell\ell'} &= \frac{e^2}{(4\pi)^2}\, (\bar q^i \gamma^\mu P_L q^j)
  (\bar\ell \gamma_\mu \ell') \,, \qquad \quad {\cal O}_P^{ij;\ell\ell'} = \frac{e^2}{(4\pi)^2}\, (\bar q^i P_R q^j)
  (\bar\ell \gamma_5 \ell') ,\\
  {\cal O}_{10}^{ij;\ell\ell'} &= \frac{e^2}{(4\pi)^2}\, (\bar q^i \gamma^\mu P_L q^j)
  (\bar\ell \gamma_\mu \gamma_5 \ell'),\qquad
  {\cal O}^{ij;\ell\ell'}_T = \frac{e^2}{(4\pi)^2}\, (\bar q^i \sigma_{\mu\nu} q^j)
  (\bar\ell \sigma^{\mu\nu}\ell'),\\
 {\cal O}^{ij;\ell\ell'}_{T5} &= \frac{e^2}{(4\pi)^2}\, (\bar q^i \sigma_{\mu\nu} q^j)
  (\bar\ell \sigma^{\mu\nu} \gamma_5\ell').
\end{split}
\end{equation}
The set of operators with ``primes'', $X'=7',9',10',S',P'$, are
related to the ``unprimed'' set by switching the roles of $P_L$ and
$P_R$. The standard operator basis is suited to the SM loop
contributions and contains an overall CKM factor $\lambda_q$, which
reads $\lambda_q = V_{qj} V_{qi}^*$ in the case of
$d_j \to d_i \ell^- \ell^{\prime+}$ processes while for
$u_j \to u_i \ell^- \ell^{\prime+}$ one has
$\lambda_q = V_{jq}^* V_{iq}$. The conventions follow the structure of
effective Hamiltonian in rare $B$ decays~\cite{Buchalla:1995vs}, where
the dominant SM contribution is from the top quark and one puts
$q = t$. For the short-distance contributions to rare charm decays one
usually sets $q=b$. The basis commonly used in rare kaon decays
differs from the above
convention~\cite{Pich:2014zta,Mescia:2006jd}. The definition of
$\mc{O}_7^{ij}$ includes the external quark mass which is taken to be
the mass of the decaying quark ($m_{q^j}$ in the case of $q^j \to q^i$
transition). The prescription for mapping from the chiral to the
standard basis of Wilson coefficients is
\begin{equation}
  \label{eq:12}
  \begin{split}
    C_{9,10} &= \frac{2\pi}{\aem \lambda_q} \left(c^{LR} \pm
      c^{LL}\right),\qquad  C_{9',10'} = \frac{2\pi}{\aem \lambda_q} \left(c^{RR} \pm
      c^{RL}\right),\\
    C_{S,P} &= \frac{2\pi}{\aem \lambda_q} \left(g^{LL} \pm
      g^{LR}\right),\qquad  C_{S',P'} = \frac{2\pi}{\aem \lambda_q} \left(g^{RL} \pm
      g^{RR}\right),\\
    C_{T,T5} &= \frac{2\pi}{\aem \lambda_q} \left(h^{LL} \pm h^{RR}\right),
  \end{split}
\end{equation}
with flavor indices $ij;\ell \ell'$ implied.

The result of tree-level matching of a single LQ scenario to the
effective Lagrangian at the scale $\mlq$ is given in the form of Wilson
coefficients collected in Table~\ref{tab:RareLep}. The running of
the Wilson coefficients from high scale $\mlq$ down to the typical energy
scale of a physical process is governed by Eqs.~\eqref{eq:RG-ST} for
coefficients $g^{LL,RR,LR,RL}$ and $h^{LL,RR}$. At one-loop level the
contributions to the electromagnetic penguin operators
$\mc{O}^{ij}_7,\mc{O}_{7^\prime}^{ij}$ may appear as well, but these
have negligible effects compared to the Wilson coefficients that
appear at the tree-level (cf.~\cite{Becirevic:2015asa}).

\begin{table}[!htbp]
{\normalsize
  \centering
 \begin{tabular}{||p{0.4cm}|p{4.3cm}|p{6.0cm}||}
\hline\hline
LQ              & $d_j \to d_i \ell^- \ell^{\prime+}$ decays & $u_j \to u_i \ell^- \ell^{\prime+}$ decays\\\hline
$S_3$ & $c^{LL} = -\tfrac{v^2}{2\mlq^2}  y_{3\,j\ell'}^{LL}
        y^{LL*}_{3\,i\ell}$ & $c^{LL} = -\tfrac{v^2}{2\mlq^2}   (V^T
                              y_3^{LL})_{j\ell'} (V^T y_3^{LL})^*_{i\ell}$
\\\hline
    $R_2$ & 
$c^{LR} = \tfrac{v^2}{4\mlq^2} y_{2\,\ell j}^{LR} y^{LR*}_{2\,\ell' i}$& 
$c^{LR} = \tfrac{v^2}{4\mlq^2} (y_2^{LR} V^\dagger)_{\ell  j}(y_2^{LR}V^\dagger)_{\ell' i}^*$\newline
$c^{RL} = \tfrac{v^2}{4\mlq^2} y_{2\,i\ell'}^{RL} y_{2\,j\ell}^{RL*}$\newline
$g^{LL} = 4 h^{LL} = -\tfrac{v^2}{4\mlq^2} y^{RL*}_{2\,j\ell} (y_2^{LR} V^\dagger)^*_{\ell' i}$\newline
$g^{RR} = 4 h^{RR} = -\tfrac{v^2}{4\mlq^2} y_{2\,i\ell'}^{RL} (y_2^{LR} V^\dagger)_{\ell  j}$
\\\hline
    $\tilde R_2$ & $c^{RL} = \tfrac{v^2}{4\mlq^2} \tilde
                   y_{2\,i\ell'}^{RL}  \tilde y_{2\,j\ell}^{RL*}$&
 \\\hline
    $\tilde S_1$ & 
$c^{RR} = -\tfrac{v^2}{4\mlq^2} \tilde y_{1\,j\ell'}^{RR} \tilde y_{1\,i\ell}^{RR*}$& \\\hline
    $S_1$ && 
  $c^{LL} =-\tfrac{v^2}{4\mlq^2} (V^T y_1^{LL})_{j\ell'} (V^T y_1^{LL})_{i \ell}^*$\newline
  $c^{RR} = -\tfrac{v^2}{4\mlq^2} y_{1\,j\ell'}^{RR} y_{1\,i\ell}^{RR*}$\newline
  $g^{LL} = -4 h^{LL} = \tfrac{v^2}{4\mlq^2} y_{1\,j\ell'}^{RR}  (V^T y_1^{LL})_{i\ell}^*$\newline
  $g^{RR} = -4 h^{RR} = \tfrac{v^2}{4\mlq^2} (V^T y_1^{LL})_{j\ell'} y^{RR*}_{1\,i\ell} $
\\\hline \hline
    $U_3$ &
    $c^{LL} =  \tfrac{v^2}{2\mlq^2} x^{LL}_{3\,i\ell'} x_{3\,j\ell}^{LL*}$& 
    $c^{LL} =  \tfrac{v^2}{\mlq^2} (V x_3^{LL})_{i\ell'} (V x_3^{LL})_{j\ell}^*$                 
\\\hline
    $V_2$ & 
     $c^{LR} =-\tfrac{v^2}{2\mlq^2} x_{2\,j\ell'}^{LR} x_{2\,i\ell}^{LR*}$\newline 
     $c^{RL} =  -\tfrac{v^2}{2\mlq^2} x_{2\,j\ell'}^{RL}  x_{2\,i\ell}^{RL*}$ \newline            
     $g^{LR} = \tfrac{v^2}{2\mlq^2} x_{2\,j\ell'}^{RL} x_{2\,i\ell}^{LR*}$ \newline 
     $g^{RL} = \tfrac{v^2}{2\mlq^2} x_{2\,j\ell'}^{LR} x_{2\,i\ell}^{RL*}$&
$c^{LR} = -\tfrac{v^2}{2\mlq^2} (V^T x_2^{LR})_{j\ell'} (V^T x_2^{LR})_{i\ell}^*$
 \\\hline
    $\tilde V_2$ && 
      $c^{RL} = -\tfrac{v^2}{2\mlq^2} \tilde{x}_{2\,j\ell'}^{RL} \tilde{x}_{2\,i\ell}^{RL*}$
 \\\hline
    $\tilde U_1$ && 
      $c^{RR} = \tfrac{v^2}{2\mlq^2} \tilde{x}_{1\,i\ell'}^{RR}  \tilde{x}_{1\,j\ell}^{RR*}$
\\\hline
    $U_1$ &  
    $c^{LL} = \tfrac{v^2}{2\mlq^2} x_{1\,i\ell'}^{LL}  x_{1\,j\ell}^{LL*}$\newline 
            $c^{RR} = \tfrac{v^2}{2\mlq^2}  x_{1\,i\ell'}^{RR}  x_{1\,j\ell}^{RR*}$ \newline 
            $g^{LR} =-\tfrac{v^2}{\mlq^2} x_{1\,i\ell'}^{LL}    x_{1\,j\ell}^{RR*}$ \newline 
            $g^{RL} = -\tfrac{v^2}{\mlq^2}  x_{i\ell'}^{RR} x^{LL*}_{1\,j\ell}$&
    \\
\hline  \hline
  \end{tabular}
}
  \caption{Tree-level Wilson coefficients of LQ models in rare
    semileptonic decays. Values quoted are valid at the matching
    scale taken to be the LQ mass $\mlq$. We have not explicitly written
    the lepton and quark flavor indices on the Wilson coefficients, as introduced in
    the operator basis. For example, a table entry for $c^{LL}$ stands for
    $c^{LL}_{ij;\ell \ell'}$. $V$ and $U$ are
  CKM and PMNS matrices, respectively. The electroweak vacuum expectation value
  is $v=246\e{GeV}$.}
  \label{tab:RareLep}
\end{table}

\subsubsection{$P \to \ell^- \ell^{\prime+}$}
The leptonic decay width of a neutral meson to two charged leptons of
same flavor depends on the Wilson coefficients $C^{ij;\ell \ell}_{10^{(\prime)}}$,
$C^{ij;\ell \ell}_{S^{(\prime)}}$, $C^{ij;\ell \ell}_{P^{(\prime)}}$:
\begin{equation}
  \begin{split}   
\Gamma_{P \to \ell^+ \ell^-} =
f_{P}^2 m_{P}^3& \frac{G_F^2 \aem^2}{64 \pi^3}  |\lambda_q|^2   \beta_\ell(m_{P}^2)  \left[ \
\frac{m_{P}^2}{(m_{q^i} + m_{q^j})^2}\Big| C^{ij;\ell \ell}_S -C^{ij;\ell \ell}_{S^\prime} \Big|^2  \beta_\ell(m_P^2)^2 \right.  \\ 
 & \left. +  \ \Big| \frac{m_{P}}{m_{q^i} + m_{q^j}} \left( C^{ij;\ell \ell}_P -C^{ij;\ell \ell}_{P^\prime}
   \right) + 2 \frac{m_\ell}{m_{P}} \left( C^{ij;\ell \ell}_{10} - C^{ij;\ell \ell}_{10^\prime} \right) \Big|^2 \right],
  \end{split}
\end{equation}
where $\beta_\ell(q^2) = \sqrt{1-4m_\ell^2/q^2}$ and $m_{q^j}$,
$m_{q^j}$ are the masses of the valence quarks in the pseudoscalar
meson $P$.  The SM contributes only to the lepton flavor conserving
coefficient $C_{10}^{ij;\ell\ell}$
that enters with a helicity suppression factor $m_\ell^2$
in decay width, which is not present for scalar or pseudoscalar
coefficients. Experimental limits from $P \to \ell^+ \ell^-$ are thus
mostly sensitive to LQ scenarios that generate
$C^{ij;\ell \ell}_{S^{(\prime)}}$,
$C^{ij;\ell \ell}_{P^{(\prime)}}$~\cite{Becirevic:2012fy}. The most
notable constraining decay modes are $\mrm{Br}(K_L \to \mu^+\mu^-)$,
$\mrm{Br}(D^0 \to e^+ e^-, \mu^+ \mu^-)$, whose experimental results or
upper bounds are difficult to interpret since the SM long-distance
contributions are not well known (see e.g.~\cite{Burdman:2001tf} for
charm and \cite{D'Ambrosio:1997jp} for $K_L$ decays). Thus, with lowering
upper experimental bounds in the near future one should improve the SM
theoretical predictions for these modes in order to gain more
information on the short distance physics.
On the other hand, $B$ meson decays, $B_{(s)}^0 \to e^+e^-, \mu^+ \mu^-, \tau^+ \tau^-$, can be much more reliably
predicted in the SM with the only hadronic uncertainty coming from
the rather well known decay constants $f_{B_{(s)}}$. Presently, only the
$B_s \to \mu^+ \mu^-$ has been observed to agree with the SM with
branching fraction $(2.8^{+0.7}_{-0.6})\E{-6}$ reported by LHCb, CMS~\cite{CMS:2014xfa},
and reconfirmed by ATLAS~\cite{Aaboud:2016ire}, whereas decay modes
with $e^+e^-$ pair are helicity
suppressed, $B^0_d$ decays are Cabibbo suppressed, and modes with $\tau^+ \tau^-$
are more difficult on the experimental side.

While the lepton conserving decay modes $P \to \ell^+ \ell^-$ do not
depend on $C_9^{ij;\ell\ell}$ since the leptonic vector current
$\bar \ell \gamma^\mu \ell$ is conserved\footnote{Effectively, the
  interaction term is proportional to $f_P (\bar \ell(x) \gamma^\mu
  \ell(x)) (\partial_\mu P(x))$ that is proportional to lepton current
divergence after integrating by parts.}, the LFV decay
modes $P \to \ell \ell'$ depend further on
$C^{\ell \ell^\prime}_9$~\cite{Becirevic:2016zri} due to non-conserved
vector current and are furthermore theoretically cleaner than the corresponding
lepton flavor conserving modes. Leptonic decays have been studied
in~\cite{Bobeth:2011st} as well as in Ref.~\cite{Dighe:2012df}, where
decay width of $B_s \to \tau \tau$ was related to the absorptive
mixing amplitude in $B_s$-$\bar B_s$ system. Bounds on LFV decays are
important in situations where an LQ state is interacting with different
generations of leptons,
cf. Refs.~\cite{Davidson:1993qk,Smirnov:2007hv,Raidal:2008jk,Carpentier:2010ue,Dorsner:2011ai}.


\subsubsection{$P \to P'(V) \ell^- \ell^{\prime+}$}
\label{sec:P-Pll}
Decay width expressions for the three body decay $P \to P' \ell^+ \ell^-$ can be
found for the case of $B$ meson decays in, e.g.,~\cite{Bobeth:2007dw,
  Becirevic:2012fy}:
\begin{equation}
  \label{eq:Pllspectrum}
   \frac{ d^2 \Gamma_{P \to P' \ell^+ \ell^-} (q^2, \cos \theta) }{dq^2 d \cos \theta } =  a_\ell \qq + b_\ell \qq \cos \theta + c_\ell \qq \cos^2 \theta,
\end{equation}
where $q^2$ is the lepton pair invariant mass and $\theta$ the angle
between $\ell^-$ and decaying mother particle $P$ momenta in the rest frame
of leptons. The $q^2$ dependent angular coefficients are expressed as
\begin{equation}
\label{coeff_BKell}
\begin{split}
a_\ell \qq &= \mc{C} (q^2)\Big[  q^2 \left( \beta_{\ell}^2(q^2) \lvert F_S(q^2)\rvert^2 + \lvert F_P(q^2) \rvert^2 \right)  +
 \frac{\lambda \qq }{4}  \left( | F_A (q^2)\rvert^2 +  \lvert F_V(q^2) |^2 \right)    \\ 
 & \quad + 4 m_{\ell}^2 m_P^2 | F_A(q^2) |^2+2m_{\ell}  \left( m_P^2 -m_{P'}^2 +q^2\right) \re\left( F_P(q^2) F_A^{\ast}(q^2) \right)\Big], \\ 
b_\ell \qq &= 2\  {\cal C}(q^2)\Big\{q^2 \left[  \beta_{\ell}^2(q^2)  \re  \left( F_S(q^2) F_T^{\ast}(q^2) \right) + \re \left(F_P(q^2) F_{T5}^{\ast}(q^2)  \right)\right] \\
 & \qquad\qquad\quad +    m_{\ell}  \Big[\sqrt{\lambda \qq } \beta_{\ell}(q^2) \re
 \left( F_S(q^2) F_V^{\ast}(q^2)\right)\\
& \qquad \qquad \qquad\qquad+\left(m_P^2 - m_{P'}^2+q^2 \right) \re \left(F_{T5}(q^2) F_A^{\ast}(q^2) \right) \Big] \Big\} , \\
c_\ell \qq &= {\cal C}(q^2)\Big[ q^2 \left( \beta_{\ell}^2(q^2) \lvert
  F_T(q^2) \rvert^2 + \lvert F_{T5}(q^2) \rvert^2 \right) \\
&\qquad\qquad - \frac{\lambda \qq}{4} \beta_{\ell}^2(q^2) \left( \lvert F_A(q^2) \rvert^2 +  \lvert F_V(q^2) \rvert^2 \right)  \\ 
 & \qquad\qquad+ 2 m_{\ell} \sqrt{\lambda \qq} \beta_{\ell}(q^2) \re \left( F_T(q^2) F_V^{\ast}(q^2) \right) \Big],
\end{split}
\end{equation}
where $\mc{C}(q^2)= (G_F^2 \aem^2 |\lambda_q |^2)/(512 \pi^5 m_P^3)
\beta_{\ell}(q^2)   \sqrt{\lambda(q^2)}$ and $\lambda(q^2) \equiv
\lambda(q^2,m_P^2,m_{P'}^2)$, as defined in
Eq.~\eqref{eq:lambda}. The Wilson coefficients and hadronic form
factors enter the decay rate in following combinations:
\begin{equation}
  \begin{split}  
F_V \qq  &=   \left( C_9^{ij;\ell\ell} + C_{9'}^{ij;\ell\ell} \right)
f_+ \qq + \frac{2 m_{q^j}}{m_P +m_{P'}} \left(C_7^{ij}  + C_{7^\prime}^{ij}  + \frac{4 m_{\ell}}{m_{q^j}}  C_T^{ij;\ell\ell} \right)  f_T \qq ,  \\ 
F_A \qq    &=  \left( C_{10}^{ij;\ell\ell} +C_{10^\prime}^{ij;\ell\ell} \right)  f_+ \qq ,   \\
 F_S \qq &=     \frac{m_P^2 -m_{P'}^2}{2m_{q^j}}  \left( C_S^{ij;\ell\ell} + C_{S^\prime}^{ij;\ell\ell} \right) f_0 \qq,   \\
F_P \qq  &=  \frac{m_P^2 -m_{P'}^2}{2m_{q^j}} \left( C_P^{ij;\ell\ell} + C_{P^\prime}^{ij;\ell\ell} \right)  f_0 \qq   \\
&- m_{\ell} \left( C_{10}^{ij;\ell\ell}+C_{10^\prime}^{ij;\ell\ell} \right) \left[  f_+ \qq - \frac{m_P^2-m_{P'}^2}{q^2} \left(f_0  \qq - f_+ \qq \right) \right]  ,\\
F_T \qq &=  \frac{2 \sqrt{\lambda \qq} \beta_{\ell}(q^2)}{m_P + m_{P'}}  C_T^{ij;\ell\ell} f_T\qq,\\
 F_{T5} \qq  &=  \frac{2 \sqrt{\lambda \qq} \beta_{\ell}(q^2)}{m_P +
   m_{P'}}  C_{T5}^{ij;\ell\ell} f_T\qq . 
  \end{split}
\end{equation}
The hadronic form factors are defined as in Eqs.~\eqref{eq:ffs}.
Interplay of
leptonic and semileptonic $P \to P' \ell^+ \ell^-$ decays together
constrain the couplings of LQ in the $b \bar s$ systems very
effectively~\cite{Saha:2010vw,Dorsner:2011ai,Kosnik:2012dj,Mohanta:2013lsa,Sahoo:2015fla,Sahoo:2015wya}. Whereas
the leptonic $P \to \ell^+ \ell^-$ are sensitive to differences 
between a particular Wilson coefficient and its chirally flipped
counterpart, e.g., $C_{10}^{ij;\ell\ell}-C_{10'}^{ij;\ell\ell}$, the $P \to P' \ell^+ \ell^-$ are sensitive to sums of the
two coefficients. In the three-body decay hadronic uncertainties
due to local (form factor) and non-local hadronic contributions are hindering
the access to the LQ couplings and thus decay rates
do not present the most stringent and reliable constraints. These obstacles can be 
avoided by resorting to LFU ratios, as in e.g. $R_K = \Gamma(B
\to K \mu^+ \mu^-)/ \Gamma(B \to K e^+ e^-)$, $q^2 \in [1,6]\,\mrm{GeV}^2$, where experimental and
theoretical uncertainties are significantly reduced. In the SM the LFU
violation induced by lepton masses is practically negligible, $R_K^\mrm{SM}
= 1.0003(1)$~\cite{Hiller:2003js} and is $2.6\,\sigma$ lower than the
LHCb measured value
\begin{equation}
  \label{eq:RKexp}
  R_K^\mrm{LHCb} = 0.745^{+0.090}_{-0.074} \pm 0.036.
\end{equation}
The statistical error of $R_K$ will be reduced in the ongoing LHC run
2 at LHCb and it will be further probed by the Belle 2
experiment. Generalization of $R_K$ to other channels ($B \to K^*
\ell^+ \ell^-$, $B_s \to \phi \ell^+ \ell^-$) and/or kinematical regions of
$q^2$ would allow to confirm the excess and to learn more about the
type of New Physics (NP) responsible for LFU violation. The (partial) decay spectra
have also been well measured for $B \to K$ decays at low- and
high-$q^2$ kinematical regions and their branching ratios are of the
order $10^{-7}$. The bounds on the LFV decay modes of $B \to P \ell
\ell'$ are bounded to be below $\sim 10^{-8}$ when only light leptons are
involved and $\sim 10^{-5}$ for modes involving $\tau$~\cite{Agashe:2014kda}.

For recent works addressing the $R_K$ observable
cf.~\cite{Hiller:2014ula,Hiller:2014yaa,Varzielas:2015iva,Becirevic:2015asa,Barbieri:2015yvd,Alonso:2015sja,Pas:2015hca,Gripaios:2014tna,Bauer:2015knc}.
Two simple extensions of the SM that aim at addressing the $R_K$
anomaly through physics of LQs have been presented in
Ref.~\cite{Hiller:2014yaa}. First extension calls for the SM with one
scalar LQ that transforms as $\tilde{R}_2$, where the LQ in question
couples to electrons. Second possibility is to introduce $S_3$ scalar LQ that
couples to muons. Note that the authors of Ref.~\cite{Hiller:2014yaa}
consider very specific sets of couplings of LQs to matter that prevent
one from identifying them to be of the first, second or third
generation type. Scalar LQ of $S_3$ type that arises as a
pseudo-Goldstone boson of the strong dynamics has been used to address
the $R_K$ anomaly in Ref.~\cite{Gripaios:2014tna}. New physics
contributions due to exchange of scalar LQs $R_2$ and $\tilde{R}_2$
towards rare decays of $B$ mesons have been studied in
Ref.~\cite{Sahoo:2015wya}.  The $R_K$ anomaly has also been addressed
through physics of scalar LQ $\tilde R_2$ in
Ref.~\cite{Becirevic:2015asa}.

Finally, decays with a vector meson $P \to V \ell^+ \ell^-$ in the final
state are effectively 4-body decays that offer many independent
angular observables~\cite{Altmannshofer:2008dz} and these observables
can be used to cross-check scenarios that fit well with the $R_K$
measurements. Result of the global fit of the effective
coefficients~\eqref{eq:RareLeffStandard} to the $b\to s \ell^+ \ell^-$
data~\cite{Descotes-Genon:2015uva} also strongly disfavors the SM
hypothesis and is in line with the deviation observed in $R_K$. 
The fit unambiguously prefers that $C_9$ should deviate from its SM value.
The global deviations observed in $b \to s \ell^+ \ell^-$ have recently been interpreted in terms of a
vector LQ model~\cite{Fajfer:2015ycq,Barbieri:2015yvd}. One of the
recent attempts to resolve the problems of the SM with lepton flavor
universality ratios $R_K$ and $R_D^{(*)}$ is by introducing the LQ
state $S_1$. The LQ state $S_1$ modifies lepton universality ratio
$R_D^{(*)}$ at tree-level and the rare decays' ratio $R_K$ only at
loop level, thus mimicking the SM loop suppression pattern of the two
observables~\cite{Bauer:2015knc}.

Rare charm decays offer less stringent constraints due to large
hadronic uncertainties of the relevant SM contributions.  They have
been studied in
Refs.~\cite{Burdman:2001tf,Dorsner:2009cu,deBoer:2015boa,Fajfer:2015mia}.

\subsubsection{$P \to V \gamma$ and $P \to X \gamma$}
Rare radiative decays, e.g., $B \to X_s \gamma$, correspond to penguin
diagrams with a virtual LQ and a lepton and are suppressed by the
electromagnetic coupling and a loop factor
$\aem/(4\pi) \sim 10^{-3}$ at the amplitude level.  Most notable
decays are $B \to K^* \gamma$, $B \to X_s \gamma$, and $D \to \rho \gamma$
with well measured branching fractions, but loop suppressed
sensitivity to the LQ couplings. Generally, same couplings are much
better constrained through tree-level contributions of processes with
$\gamma$ replaced by $\ell^+ \ell^-$ pair. Thus, for the 
LQ contributions to rare processes with flavor structure $\bar q^i q^j \bar\ell
\ell^\prime$, the electromagnetic penguin insertions $\bar q^i q^j \gamma$ are
negligible except in cases when tree-level contributions to $\bar
q^i q^j \bar \ell \ell^\prime$ vanish. In the case of $S_1$ scalar LQ
the $\bar s b \ell^+ \ell^-$ processes are mediated by box
diagrams while semileptonic decays $b \to c \ell \bar \nu$
occur at tree-level. Such mimicry of the SM pattern of the relative
suppression of the two types of processes enables one to address the $R_K$
and $R_{D^{(*)}}$ puzzles simultaneously~\cite{Bauer:2015knc}.

\subsubsection{$P \to P' \nu\bar \nu$, $P \to V \nu\bar\nu$} 
\begin{figure}[!htbp]
  \centering
  \begin{tabular}{lcr}
  \includegraphics[width=0.3\textwidth]{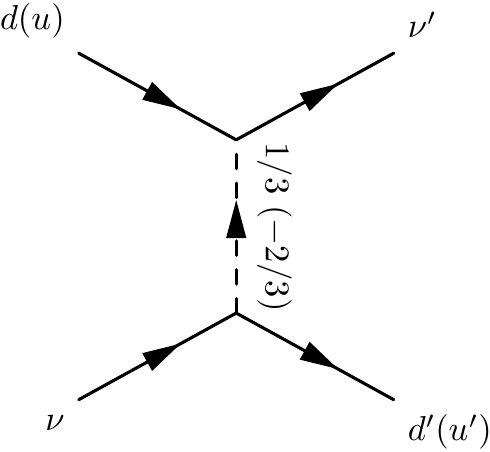} && \includegraphics[width=0.3\textwidth]{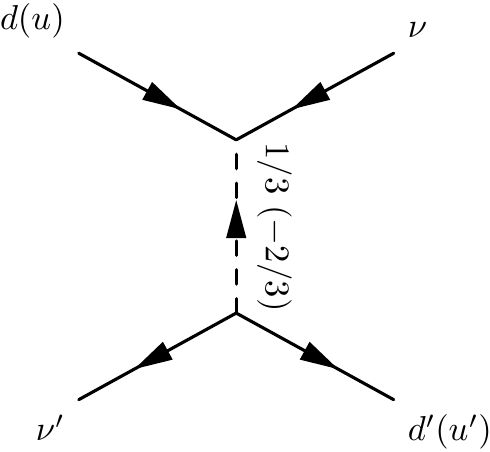}
    \end{tabular}
    \caption{ Diagrams of rare meson decays of flavor
      $(\bar d' d)(\bar \nu \bar\nu)$ and
      $(\bar u' u)(\bar \nu \bar\nu)$ induced by LQs. The $F=0$
      $(|F|=2)$ LQs contribute as shown in the left-hand
      (right-hand) side diagram. Propagators are labeled by charge of
      the relevant components of the LQ.  }
\label{fig:LQRareDiagNu}
\end{figure}

Rare decays with neutral quark current and a neutral $\bar\nu \nu$
pair in the final state are in close correspondence to decays with the same
quark flavors and $\bar\ell^{\prime +} \ell^-$ in the final
state. Namely, in all LQ models where lepton doublets are involved
both of these processes are relevant and their couplings differ only
because of the PMNS mixing. The relevant effective
Lagrangian of LQ contributions to $P \to P'(V) \nu\bar \nu$ introduced
in~\cite{Buras:2014fpa} and extended to account for a possibility of
right-handed neutrino involved in LQ interactions reads:
\begin{equation}
  \label{eq:P-Pnunu}
  \begin{split}   
\mc{L}_\mrm{eff}^{\bar q^i q^j \bar \nu \nu' } = \sqrt{2}G_F
&\Big[
c^{LL}_{ij;\nu \nu^\prime} (\bar q^i_L \gamma_\mu  q_L^j)(\bar \nu_L
\gamma^\mu \nu^\prime_L)  
+c^{RR}_{ij;\nu \nu^\prime} (\bar q^i_R \gamma_\mu  q_R^j)(\bar \nu_R
\gamma^\mu \nu^\prime_R)  \\
&+c^{LR}_{ij;\nu \nu^\prime} (\bar q^i_L \gamma_\mu  q_L^j)(\bar \nu_R
\gamma^\mu \nu^\prime_R)  
+c^{RL}_{ij;\nu \nu^\prime} (\bar q^i_R \gamma_\mu  q_R^j)(\bar \nu_L
\gamma^\mu \nu^\prime_L)  \\
&+g^{LL}_{ij;\nu \nu^\prime} (\bar q^i_L q_R^j)(\bar \nu_L
\nu^\prime_R) +  h^{LL}_{ij;\nu \nu^\prime} (\bar q^i_L
\sigma^{\mu\nu}q_R^j)(\bar \nu_L \sigma_{\mu\nu}
\nu^\prime_R) \\
&+g^{RR}_{ij;\nu \nu^\prime} (\bar q^i_R q_L^j)(\bar \nu_R
\nu^\prime_L) +  h^{RR}_{ij;\nu \nu^\prime} (\bar q^i_R
\sigma^{\mu\nu}q_L^j)(\bar \nu_R \sigma_{\mu\nu}
\nu^\prime_L) \\
&+g^{LR}_{ij;\nu \nu^\prime} (\bar q^i_L q_R^j)(\bar \nu_R
\nu^\prime_L)+g^{RL}_{ij;\nu \nu^\prime} (\bar q^i_R q_L^j)(\bar \nu_L
\nu^\prime_R)
\Big].
  \end{split}
\end{equation}
The Lagrangian \eqref{eq:P-Pnunu} is written in the fermion mass
eigenstate basis.  The experiments do not detect neutrinos directly
and can measure the inclusive width summed over all neutrino species:
\begin{equation}
\Gamma(P \to P' \nu \bar \nu) \equiv \sum_{\nu,\nu^\prime} \Gamma(P
\to P'  \nu \bar \nu^\prime).
\end{equation}
Thus the above observable receives contributions from lepton flavor
conserving as well as LFV amplitudes~\cite{Buras:2014fpa,Fajfer:2015ycq}.
The effective coefficients of \eqref{eq:P-Pnunu} at scale $\mlq$ are
summarized
in Table~\ref{tab:RareNu}. (See Fig.~\ref{fig:LQRareDiagNu}.)

For the $B \to K \bar \nu \nu$ decays 
the main contribution is due to the top quark dominance in the SM
and the effective contribution to the ``left'' Wilson coefficient is
$c^{LL,\mrm{SM}}_{sb;\nu \nu^\prime} = \delta_{\nu
  \nu^\prime}\tfrac{\aem \lambda_t}{\pi} C^\mrm{SM}_L$,
where $C^\mrm{SM}_L = -6.35$~\cite{Buras:2014fpa}. The corresponding
SM prediction is
$\mrm{Br}(B^+ \to K^+ \bar \nu \nu) = (4.0\pm 0.5)\E{-6}$, while the
current experimental upper bound on this decay is
$1.6\E{-5}$~\cite{Lees:2013kla}.  
Belle 2 experiment plans to probe the SM prediction of $B\to K^* \nu \bar\nu$ with 30\%
error~\cite{Lutz:2013ftz}.

The effective Lagrangian of rare
decays $K \to \pi \bar \nu \nu$, on the other hand, gets SM
contributions of box diagrams with charm/top quark and charged
leptons. The charm part of the effective coefficient is sensitive to
the mass of charged lepton and contributes about $30\%$ of the total
branching ratio in $K^+ \to \pi^+ \bar\nu \nu$ but has no effect in
$K_L \to \pi^0 \bar \nu \nu$~\cite{Buchalla:1998ba,Buras:2006gb,Brod:2010hi}. The dominant lepton flavor universal part of the
Wilson coefficient, stemming from the top quark penguins and boxes is
equal as in the case of $B \to K \bar \nu \nu$, namely
$C^\mrm{SM}_L = -6.35$~\cite{Brod:2010hi}, while the predicted
branching fractions in the SM are
$\mrm{Br}(K^+ \to \pi^+ \bar \nu \nu)^\mrm{SM} = (7.8 \pm 0.8)\E{-11}$
and
$\mrm{Br}(K_L \to \pi^0 \bar \nu \nu)^\mrm{SM} = (2.4 \pm
0.4)\E{-11}$. The charged mode has been observed by the E949 experiment,
$\mrm{Br}(K^+ \to \pi^+ \bar \nu \nu) = (1.7\pm
1.1)\E{-10}$~\cite{Artamonov:2008qb},
whereas the neutral mode has been bounded from above,
$\mrm{Br}(K_L \to \pi^0 \bar \nu \nu) < 2.6\E{-8}$ at $90\%$C.L., by the E391a
experiment~\cite{Ahn:2009gb}. The NA62 plans to measure $K 
\to \pi \nu \bar \nu$ decays with $10\%$ error~\cite{Palladino:2016pdq}
while $K_L \to \pi^0 \nu \bar \nu$ will be sought after in KOTO experiment~\cite{Shiomi:2014sfa}.


\begin{table}[!htbp]
{\scriptsize
  \centering
\begin{tabular}{||P{0.3cm}|p{5.2cm}|p{5.1cm}||}
\hline\hline
LQ              & $d_j \to d_i \nu \bar\nu^\prime$ decays & $u_j \to u_i \nu \bar\nu^\prime$ decays\\\hline
 $S_3$ & $c^{LL} = \tfrac{v^2}{2\mlq^2} (y_3^{LL} U)_{j\nu'}  (y_3^{LL} U)^*_{i\nu}$& 
 $c^{LL} = \tfrac{v^2}{\mlq^2} (V^T y_3^{LL} U)_{j\nu'} (V^T y_3^{LL} U)^*_{i\nu}$
\\\hline
 $R_2$ & &$c^{RL} = -\tfrac{v^2}{2\mlq^2}  (y_2^{RL} U)_{i\nu'}  (y_2^{RL} U)^*_{j\nu}$
\\\hline
    $\tilde R_2$ &
 $c^{RL} = -\tfrac{v^2}{2\mlq^2}  (\tilde y_2^{RL} U)_{i\nu'} (\tilde y_2^{RL} U)^*_{j\nu}$\newline
$c^{LR} = -\tfrac{v^2}{2\mlq^2}    \tilde y_{2\,i\nu'}^{\overline{LR}}
                   \tilde y^{\overline{LR}*}_{2\,j\nu}$\newline
$g^{RR} = 4 h^{RR} = -\tfrac{v^2}{2\mlq^2}
          (\tilde y_{2}^{RL} U)_{i\nu'}  \tilde y^{\overline{LR}*}_{2\,j\nu}$\newline
$g^{LL} = 4 h^{LL} = -\tfrac{v^2}{2\mlq^2} \tilde
                   y_{2\,i\nu'}^{\overline{LR}} (\tilde y_2^{RL}U)_{j\nu}^*  $ &
$c^{LR} = -\tfrac{v^2}{2\mlq^2} (V \tilde y_2^{\overline{LR}})_{i\nu'} (V y_2^{\overline{LR}})^*_{j\nu}$
\\\hline
    $S_1$ & 
$c^{LL} = \tfrac{v^2}{2\mlq^2} (y_1^{LL} U)_{j\nu'} (y_1^{LL} U)^*_{i\nu}$\newline
$c^{RR} = \tfrac{v^2}{2\mlq^2}    y_{1\,j\nu'}^{\overline{RR}} y^{\overline{RR}*}_{1\,i\nu}$\newline
$g^{RR} = -4 h^{RR} = \tfrac{v^2}{2\mlq^2}   (y_1^{LL} U)_{j\nu'} y^{\overline{RR}*}_{1\,i\nu}$\newline
$g^{LL} = -4 h^{LL} = \tfrac{v^2}{2\mlq^2} y_{1\,j\nu'}^{\overline{RR}} (y_1^{LL} U)^*_{i\nu}$
&
\\\hline
 $\bar S_1$ & & $c^{RR} = \tfrac{v^2}{2\mlq^2} \bar
                y_{1\,j\nu'}^{\overline{RR}} \bar y_{1\,i\nu}^{\overline{RR}*}$
\\\hline\hline
    $U_3$ & $c^{LL} = -\tfrac{2v^2}{\mlq^2}  (x_3^{LL} U)_{i\nu'}   (x_3^{LL} U)^*_{j\nu}$ &
$c^{LL} = -\tfrac{v^2}{\mlq^2} (Vx_3^{LL} U)_{i\nu'} (V x_3^{LL} U)^*_{j\nu}$
\\\hline
    $V_2$ & $c^{RL} = \tfrac{v^2}{\mlq^2}(x_2^{RL} U)_{j\nu'} (x_2^{RL} U)^*_{i\nu}$
                                   &
\\\hline
    $\tilde V_2$ & &
      $c^{RL} = \tfrac{v^2}{\mlq^2} (\tilde x_2^{RL} U)_{j\nu'}(\tilde x_2^{RL}U)^*_{i\nu}$\newline
       $c^{LR} = \tfrac{v^2}{\mlq^2} (V^T \tilde x_2^{\overline{LR}})_{j\nu'}  (V^T \tilde x_2^{\overline{LR}})^*_{i\nu}$\newline
     $g^{RL} = \tfrac{2v^2}{\mlq^2} (V^T \tilde x_2^{\overline{LR}})_{j\nu'}    (\tilde x_2^{RL}U)^*_{i\nu}$\newline
   $g^{LR} = \tfrac{2v^2}{\mlq^2} (\tilde x_2^{RL} U)_{j\nu'} (V^T \tilde x_2^{\overline{LR}})^*_{i\nu}$
\\\hline
    $U_1$ & &
                     $c^{LL} = -\tfrac{v^2}{\mlq^2}  (Vx_1^{LL} U)_{i\nu'} (Vx_1^{LL}U)^*_{j\nu}$\newline
                     $c^{RR} = -\tfrac{v^2}{\mlq^2}  x_{1\,i\nu'}^{\overline{RR}} x^{\overline{RR}*}_{1\,j\nu}$\newline
                     $c^{LR} = \tfrac{2v^2}{\mlq^2}  (Vx_1^{LL}U)_{i\nu'} x^{\overline{RR}*}_{1\,j\nu}$\newline
                     $c^{RL} = \tfrac{2v^2}{\mlq^2}  x_{1\,i\nu'}^{\overline{RR}} (Vx_1^{LL}U)^*_{j\nu}$
 \\\hline  
    $\bar U_1$ & 
                     $c^{RR} = -\tfrac{v^2}{\mlq^2}  x^{RR}_{1\,i\nu'}
            x^{RR*}_{1\,j\nu}$ &
 \\\hline    \hline
  \end{tabular}
}
  \caption{Tree-level Wilson coefficients of LQ models in rare
    decays $q^j \to q^i \nu \bar \nu'$. Values quoted are valid at the matching
    scale taken to be the LQ mass $\mlq$. The flavor indices have been
    suppressed. For example, entry $c^{LL}$ stands for
    $c^{LL}_{ij;\nu \nu^\prime}$. $V$ and $U$ are
  CKM and PMNS matrices, respectively. The electroweak vacuum expectation value
  is $v=246\e{GeV}$.}
  \label{tab:RareNu}
\end{table}
Constraints from $B \to X_s \nu \bar\nu$ have been first derived
in~\cite{Grossman:1995gt} while recent detailed studies of LQ contributions to
$B \to K^{(*)} \nu \bar\nu$ can be found in
Refs.~\cite{Buras:2014fpa,Sahoo:2015fla,Becirevic:2015asa}. The scalar LQ state $S_3$ as
an explanation of the $R_K$ ratio~\eqref{eq:RKexp} has been constrained by $B
\to K^{(*)} \nu\bar\nu$ and $K^+\to \pi^+ \nu\bar\nu$ decay widths in Ref.~\cite{Gripaios:2014tna}.


\subsection{Neutral meson anti-meson oscillations}


The LQ induced $M \overline M$ mixing amplitudes originate from diagrams
with LQ state with $F=0$ or $|F|=2$, accompanied by leptons in the
box. For the $|F|=2$ states which always couple to two quarks, additional box
diagrams with virtual quarks are possible. The $|F|=2$ states na\"ively
seem to mediate meson mixing at tree-level, however 
the diquark couplings ($z_{ij}$) are always antisymmetric in flavor space, as shown in
Section~\ref{sec:INTRODUCTION} (also cf.~\cite{Davies:1990sc}). One can
easily see that mixing amplitudes vanish for antisymmetric diquark Yukawas.

With LQ couplings to lepton and quark fields as the main focus of this
review, proton decay constraints force the diquark couplings of any 
$|F|=2$ LQ state to be negligible. For an alternative approach where
the diquark aspect of $|F|=2$ LQ state has been considered
addressing neutral meson mixing constraints see~\cite{Dorsner:2010cu}.

\begin{figure}[!htbp]
  \centering
  \includegraphics[scale=0.35]{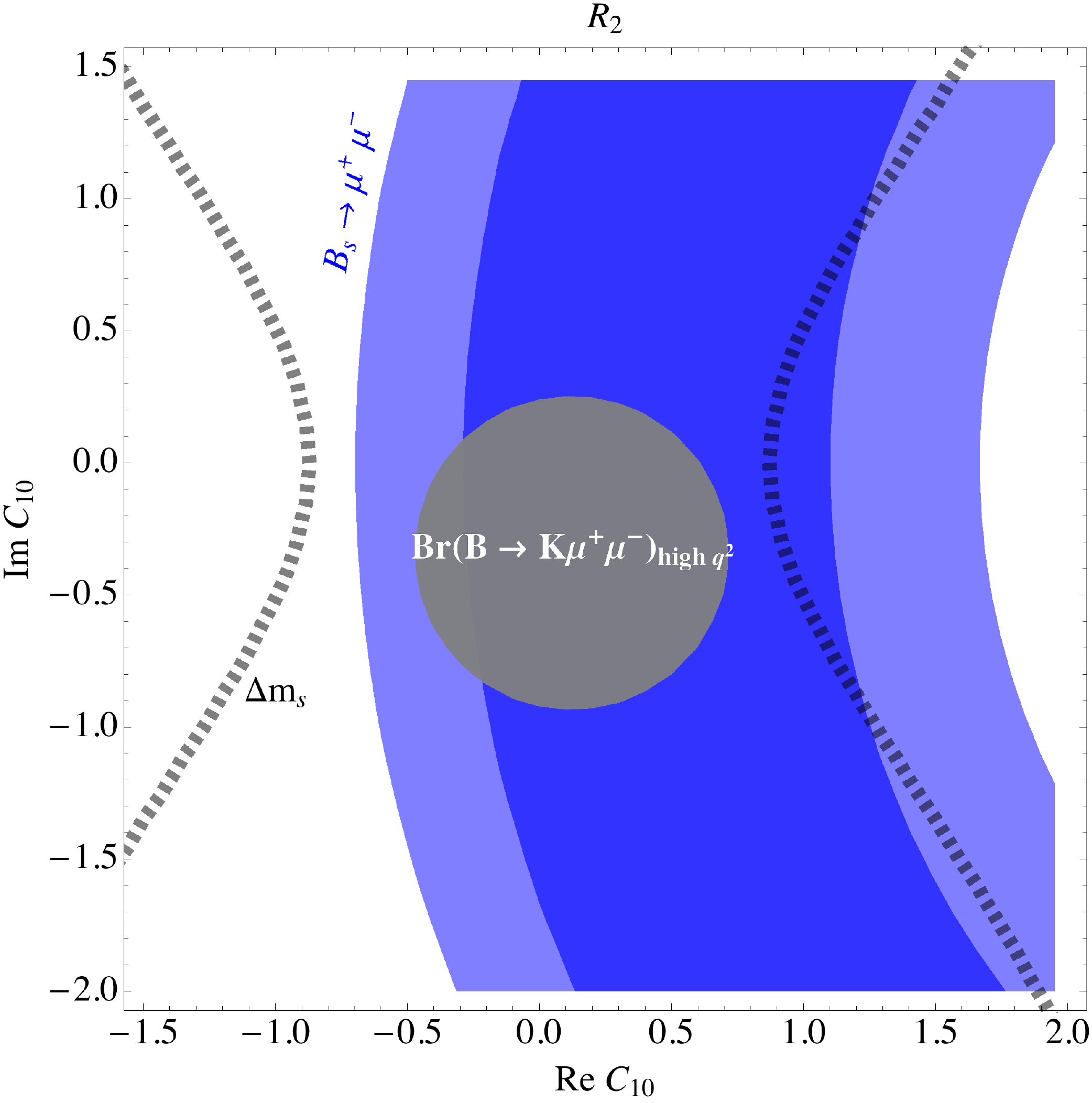}
  \caption{Constraints in the complex plane of Wilson coefficient
    $C_{10}^{sb;\mu\mu}$, for a $R_2$ scalar LQ scenario where
    $C_{10} = C_{9}$. Gray circle corresponds to $2\,\sigma$ agreement
    with $B\to K \mu^+ \mu^-$ decay rate in region
    $q^2 \in [15,22]\e{GeV}^2$, dark and light blue regions are $1$
    and $2\,\sigma$ regions due to $B_s \to \mu^+ \mu^-$ branching
    ratio. Taken at mass $\mlq = 20\e{TeV}$, the agreement with mass
    splitting $\Delta m_s$ at $1\,\sigma$ prefers the parameter space
    between the dashed curves.}
  \label{fig:R2}
\end{figure}

The interaction Lagrangian of any scalar LQ can be split down to
couplings to left- and right-handed quarks of definite charge:
\begin{equation}
  \label{eq:MMbarLagS}
  \mc{L} = \bar q^i \left[l_{ij} P_R + r_{ij} P_L \right] \ell^j \,S + \mrm{h.c.},
\end{equation}
where $q^i$ denotes quarks of definite charge ($2/3$ or $-1/3$), $\ell^j$ stands for
charged leptons or neutrinos or their charge conjugate (charges $-1,0,1$). Couplings $l$
and $r$ can be straightforwardly obtained from defining Lagrangians
of each scalar LQ presented in Section~\ref{sec:INTRODUCTION}.
The
effective Lagrangian in the approximation where we only keep the LQ
mass finite reads
\begin{equation}
  \label{eq:MMbarLeffS}
  \begin{split}   
    \mc{L}_\mrm{eff} = \frac{-1}{128 \pi^2 \mlq^2} \big[&(l
    l^\dagger)_{ji}^2 (\bar q^j \gamma^\mu P_L q^i) (\bar q^j
    \gamma_\mu P_L q^i) +(r r^\dagger)_{ji}^2 (\bar q^j \gamma^\mu P_R
    q^i) (\bar q^j
    \gamma_\mu P_R q^i)\\
    & - 4 (l l^\dagger)_{ji} (r r^\dagger)_{ji} (\bar q^j 
    P_L q^i) (\bar q^j P_R q^i) \big].
  \end{split}
\end{equation}
As an example we demonstrate complementarity of constraints coming from rare decays and mass
splitting in meson mixing constraints for the case of $R_2$ scalar LQ
state. For simplicity we consider only two non-zero Yukawa couplings,
$y^{LR}_{2\,\mu s}$ and $y^{LR}_{2\,\mu b}$, which drive the effects in
$B_s \to \mu^+ \mu^-$, $B \to K^{(*)} \mu^+ \mu^-$, and also
$B_s$--$\bar B_s$ mixing. Looking into Tab.~\ref{tab:RareLep} we find
that the effective coefficient $c_{LR}$ (equivalent to $C_9 = C_{10}$
Wilson coefficients in the standard basis)
scales as $y^{LR*}_{2\,\mu s}
y^{LR}_{2\,\mu b}/\mlq^2$, whereas in Eq.~\eqref{eq:MMbarLeffS} the
operator  $(\bar s \gamma^\mu P_L b)^2$ is multiplied by  $(y^{LR*}_{2\,\mu s}
y^{LR}_{2\,\mu b})^2/\mlq^2$. Thus the two constraints depend on the
same Yukawas but scale differently with mass. The effective coupling in the mixing
scales as $\mc{A}_\mrm{mix} \sim \mc{A}_\mrm{decay} \mlq^2$, thus
making the constraint on $\Delta m_s$, the mass splitting in the $B_s$
system, more constraining for heavy $R_2$. In Fig.~\ref{fig:R2} we
show an interplay of constraints coming from the measured branching fraction of
$B_s \to \mu^+ \mu^-$ and the measured branching fraction of
$B^+ \to K^+\mu^+ \mu^-$ in the region $q^2 \in [15,22]\e{GeV}^2$~\cite{Aaij:2014pli}, as
has been done in Ref.~\cite{Becirevic:2015asa} for the $\tilde R_2$ state. The
two constraints are invariant with respect to the mass of
leptoquark. However, for high masses, here taken to be $\mlq= 20\e{TeV}$,
the constraint from $\Delta m_s$ is becoming relevant and disallows
large real parts of the Wilson coefficient $C_{10}$, as shown by the
dashed lines in Fig.~\ref{fig:R2}. For the SM prediction, we employed
the value provided in the lattice study~\cite{Bazavov:2016nty},
$\Delta m_s= (19.6\pm 1.6)\e{ps}^{-1}$, while the world experimental
average was taken to be $\Delta m_s = (17.76 \pm 0.02)\e{ps}^{-1}$~\cite{Amhis:2014hma}.

Specification of the vector LQ interactions suited for the study of
neutral meson mixing is completely analogous to Eq.~\eqref{eq:MMbarLagS}:
\begin{equation}
  \label{eq:MMbarLagV}
  \mc{L} = \bar q^i \gamma^\mu \left[l_{ij} P_L + r_{ij} P_R \right] \ell^j \,V_\mu + \mrm{h.c.},
\end{equation}
with coupling matrices which can be straightforwardly related to
LQ couplings specified in Section~\ref{sec:INTRODUCTION}.
Here $V_\mu$ is an eigencharge component of a vector LQ
multiplet. However, the effect of vector LQs in loop processes is
model dependent to a large extent. Simple extensions of the SM with a
massive vector LQ with explicit mass term are not renormalizable and are
thus dependent on the details of the ultraviolet (UV) completion~\cite{Leurer:1993qx}. In particular, the
longitudinal component of the virtual massive vector is not damped
sufficiently in the UV region. In order to render the setup renormalizable
one should formulate a theory with an extended gauge symmetry where the vector LQ is
a gauge boson associated with part of this symmetry. The gauge symmetry is
then broken via the Higgs mechanism which provides the mass for the vector LQ. The
theory in the broken phase would then be renormalizable. Another
consequence in such a setting is that coupling matrices $l_{ij}$ and
$r_{ij}$ of Eq.~\eqref{eq:MMbarLagV} must be proportional to some
unitary matrices, if we assume that the gauge symmetry is blind to
the generation indices of fermions. The matrices $l_{ij}$ and
$r_{ij}$ are proportional to $\delta_{ij}$ only in the flavor basis whereas in the mass
basis the identity matrix is transformed to a product of unitary rotations connecting
flavor and mass eigenstates.

Some aspects of the spontaneously broken extended gauge symmetry are independent of the
UV details. Indeed, in the unitary gauge the relevant interaction
terms reduce to \eqref{eq:MMbarLagV} and
the vector propagator carries three physical polarizations,
  $-i  (g^{\mu\nu} - k^\mu
      k^\nu/\mlq^2)/(k^2 -\mlq^2)$,
while all other degrees of freedom that depend on the UV completion decouple.
In the case of chiral vector LQs the unitarity of the coupling
matrix $l_{ij}$ (or $r_{ij}$)  is
sufficient to eliminate all UV divergences in neutral meson mixing
amplitudes. For the vector LQ with couplings $l_{ij}$ the box
amplitudes obtained in the unitary gauge match onto the effective
Lagrangian:
\begin{equation}
  \label{eq:MMbarLeffV}
  \begin{split}
        \mc{L}_\mrm{eff} &= \frac{1}{32 \pi^2 m_V^2}
        \sum_{\ell,\ell'} (l_{i\ell}
    l^*_{j\ell}) (l_{i\ell^\prime}
    l^*_{j\ell^\prime})  S(x_\ell, x_{\ell'})(\bar q^j \gamma^\mu P_L q^i) (\bar q^j
    \gamma_\mu P_L q^i), \\
    x_\ell &= m_\ell^2/m_V^2,
\end{split}
\end{equation}
where $S_0$ is the Inami--Lim function~\cite{Inami:1980fz}:
\begin{equation}
       S_0(x,y) = xy\left[\frac{(x^2-8x+4)\log x}{4(x-1)^2 (x-y)} +  (x \leftrightarrow
         y)\right] - \frac{3 xy}{4(x-1)(y-1)}.
\end{equation}
Here the double unitarity sum over leptons $\ell$, $\ell^\prime$ has been invoked to cancel
all the divergent terms and to impose boundary conditions $S_0(0,y) = S_0(x,0) = 0$. The part of the amplitude that survives must
thus have non-zero and non-degenerate lepton masses. This GIM-like
suppression is completely analogous to the GIM mechanism in the SM but
contrary to the situation in the SM where large top quark mass typically
enhances the amplitudes, here the final amplitudes are small due to
tiny ratio $m_\ell^2/m_V^2$. In the case of non-chiral vector LQs, lepton helicity
flips are allowed in which case the corresponding divergent parts each
comes with a lepton mass in front, invalidating the GIM cancellation of
divergences.  

The cleanest observable offered by the $K \bar K$ system is the
indirect CP violating parameter
$\epsilon = (2.228\pm 0.011)\E{-3}$~\cite{Agashe:2014kda}. Mass
splitting $\Delta m_K$ and $\epsilon'/\epsilon$ are more prone to long
distance QCD effects~\cite{Agashe:2014kda}. The properties of the
$D \bar D$ system have been precisely measured and
the fitted value of mass and width differences, both hard to
predict precisely, are in the ballpark of the SM values. The indirect CP
violation parameters $|p/q|$ and
$\sin \phi$~\cite{Gedalia:2009kh,Dorsner:2010cu} are still consistent
with null value~\cite{Amhis:2014hma}. In the $B^0_{(s)}$ sector the indirect
CP violation as well as meson mass splittings are well
measured~\cite{Agashe:2014kda} and dominated by the short distance
contributions. Mass splittings in this case can be also precisely predicted thanks
to the advances in the lattice QCD predictions of relevant bag
parameters and meson decay constants~\cite{Aoki:2013ldr}.

Approximate bounds on vector LQs from neutral meson mixing
have been studied in Ref.~\cite{Davidson:1993qk}. The lifetime
difference in the $B_s$--$\bar B_s$ system has been discussed
in Refs.~\cite{Dighe:2007gt, Dighe:2010nj}. Constraints stemming from the $D^0$--$\bar D^0$
system properties have been derived in Ref.~\cite{Golowich:2007ka}. An approach to
mixing and other loop mediated amplitudes without the assumption of
unitarity of the vector LQ couplings has been pursued in
Ref.~\cite{Barbieri:2015yvd}.

\subsection{Rare processes of charged leptons}
LQ states contribute to rare charged lepton processes by forming a
loop with quarks and LQs. For the scalar LQs we employ
the Lagrangian already introduced in the subsection on neutral meson
mixing, Eq.~\eqref{eq:MMbarLagS}:
\begin{equation}
  \label{eq:RareLeptonLag}
  \mc{L} = \bar q^i \left[l_{ij} P_R + r_{ij} P_L \right] \tilde \ell^j \,S + \mrm{h.c.}.
\end{equation}
Recall that $q^i$ is a quark field whereas $\tilde\ell^j$ stands for
a charged lepton $\tilde \ell^j = \ell^j$ (charge-conjugated charged
lepton $\tilde \ell^j = (\ell^{j})^C$) in the case of scalar $S$
belonging to an $F=0$ ($|F|=2$) LQ multiplet. For the processes with
charged leptons it is convenient to separate the $F=0$ and $|F|=2$ cases
and rewrite the Lagrangian \eqref{eq:RareLeptonLag} for both cases in 
terms of definite lepton  charge:
\begin{equation}
\label{eq:RareLeptonLag2}
  \begin{split}  
    \mc{L}^{F=0} &= \bar q^i \left[l_{ij} P_R + r_{ij} P_L \right] \ell^j \,S + \mrm{h.c.},\\
    \mc{L}^{|F|=2} &= \overline {q^{i,C}} \left[l^*_{ij} P_L + r^*_{ij} P_R \right] \ell^j \,S^* + \mrm{h.c.}.
  \end{split}
\end{equation}
Note that the couplings $l_{ij}$ and $r_{ij}$ are defined as couplings
to left- and right-handed quarks, respectively, irrespective of the value of $F$.

\subsubsection{$\ell \to \ell' \gamma$}
The experimental searches for $\ell \to \ell' \gamma$ have resulted in
upper bounds on these decays' branching ratios that severely constrain
LFV parameters of NP models. Among the strictest upper bounds ever
experimentally obtained is the bound on $\mrm{Br}(\mu \to e \gamma) <
4.2\E{-13}\,@\,90\%\mrm{CL}$
that has been obtained at the MEG experiment~\cite{TheMEG:2016wtm}.
The MEGII experiment plans to further lower this bound by another
order of magnitude~\cite{Baldini:2013ke}. LFV decays of the $\tau$
lepton are considerably weaker --- $\mrm{Br}(\tau \to \mu \gamma) <
4.4\E{-8}\,@\,90\%\mrm{CL}$ and $\mrm{Br}(\tau \to e \gamma) <
3.3\E{-8}\,@\,90\%\mrm{CL}$ --- as determined by the BaBar Collaboration~\cite{Aubert:2009ag}.


\begin{figure}[!htbp]
  \centering
  \begin{tabular}{lcr}
    \includegraphics[width=.33\textwidth]{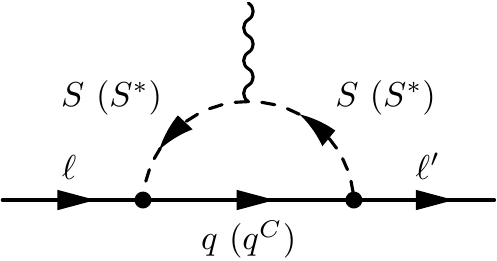}&&
     \includegraphics[width=.33\textwidth]{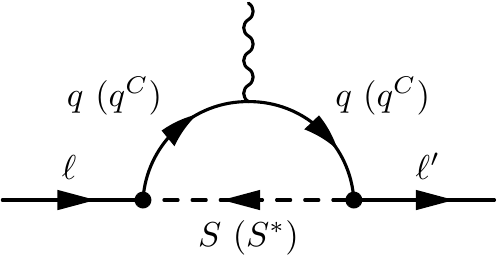}\\\\
    \includegraphics[width=.33\textwidth]{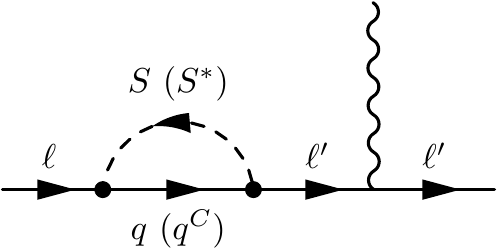}&&
 \includegraphics[width=.33\textwidth]{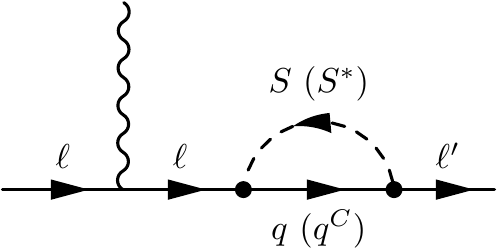}
  \end{tabular}
  \caption{One loop scalar LQ contributions to $\ell \to \ell' \gamma$ and to
    dipole moments of leptons.}
  \label{fig:l-lgamma}
\end{figure}
\label{sec:tamuga}
The effective Lagrangian for $\ell \to \ell' \gamma$ respecting gauge invariance of quantum electrodynamics
can be written as:
\begin{equation}
  \label{eq:l-lprimegamma}
 \mc{L}_\mrm{eff}^{\ell \to \ell^\prime \gamma} = \frac{e}{2}\
 \bar\ell' i\sigma^{\mu\nu} F_{\mu\nu}  \left(\sigma^{\ell\ell'}_L P_L + \sigma^{\ell\ell'}_R P_R \right) \ell.
\end{equation}
For definiteness let us mention that the charge sign convention used
throughout this section for the electromagnetic (EM) vertex is such that the tree-level
Lagrangian for lepton--photon interaction is
$e \bar \ell A_\mu \gamma^\mu \ell$
(cf. Appendix~\ref{sec:APPENDIX_C}). The radiative LFV decay width
following from Eq.~\eqref{eq:l-lprimegamma} is
\begin{equation}
  \Gamma(\ell \to \ell' \gamma) = \frac{\aem m_\ell^3 \left(1-
    m_{\ell'}^2/m_\ell^2\right)^3}{4} \left(|\sigma^{\ell\ell'}_L|^2 + |\sigma^{\ell\ell'}_R|^2 \right).
\end{equation}
Taking the Lagrangian for $F=0$ LQ states in~\eqref{eq:RareLeptonLag2} as a starting point
one finds the contribution of the diagrams in Fig.~\ref{fig:l-lgamma}
amounts to:
\begin{equation}
  \label{eq:llgamma-loops}
  \begin{split}   
  \sigma^{\ell\ell'}_L = \frac{iN_c}{16\pi^2 \mlq^2} \sum_q \Bigg\{&\left(l_{q\ell'}^*
    l_{q\ell} m_\ell + r_{q\ell'}^*
    r_{q\ell} m_{\ell'} \right) \left[Q_S f_S(x_q) - f_F(x_q)\right]
  \\
  &+ l_{q\ell'}^*r_{q\ell} m_q \left[Q_S g_S(x_q)-g_F(x_q)\right]\Bigg\},\\
  \sigma^{\ell\ell'}_R = \frac{iN_c}{16\pi^2 \mlq^2} \sum_q \Bigg\{&\left(r_{q\ell'}^*
    r_{q\ell} m_\ell + l_{q\ell'}^*
    l_{q\ell} m_{\ell'}\right) \left[Q_S f_S(x_q) - f_F(x_q)\right] \\
&+ r_{q\ell'}^*
    l_{q\ell} m_q \left[Q_S g_S(x_q)-g_F(x_q)\right]\Bigg\}.
  \end{split}
\end{equation}
In the above expressions $\sigma^{\ell\ell'}_{L,R}$ are proportional to the number of
colors, $N_c = 3$, and are summed over all virtual quark flavors
$q$. Charge $Q_S$ is defined as the charge of the scalar LQ field $S$ in
the Lagrangians~\eqref{eq:RareLeptonLag}, \eqref{eq:RareLeptonLag2}. The
loop functions of the argument $x_q = m_q^2/\mlq^2$ are:
\begin{equation}
\label{eq:llgamma-func}
  \begin{split}
  f_S(x) &= \frac{x+1}{4(1-x)^2} + \frac{x \log x}{2(1-x)^3} \sim
  \frac{1}{4},\\
  f_F(x) &= \frac{x^2-5x-2}{12(x-1)^3} + \frac{x \log x}{2(x-1)^4} \sim
  \frac{1}{6},\\
 g_S(x) &= \frac{1}{x-1}  -\frac{\log x}{(x-1)^2} \sim -\log x,\\
g_F(x) &= \frac{x-3}{2(x-1)^2} + \frac{\log x}{(x-1)^3} \sim -\log x.
  \end{split}  
\end{equation}
The above expressions agree with the formulas presented in Ref.~\cite{Lavoura:2003xp}.
In Eqs.~\eqref{eq:llgamma-func} the limiting behavior of the
functions is indicated when $x$ becomes small. Note that in such a limit
the contribution of a chiral LQ with charge $Q_S=2/3$ becomes
negligible due to cancellation between the terms with $f_S$ and $f_F$.
Eqs.~\eqref{eq:llgamma-func} have been derived for the $F=0$ case and
are easily adapted to the $|F|=2$ case by flipping the scalar charge,
$Q_S \to -Q_S$ (but with $Q_S$ still defined as charge of the field $S$), and applying $l_{q\ell} \to r_{q\ell}^*$, $r_{q\ell} \to l_{q\ell}^*$.

Rare radiative processes with LFV have been studied for vector and
scalar LQs with unitary coupling matrices in
Ref.~\cite{Gabrielli:2000te}. Constraints on the scalar leptoquarks
from LFV radiative decays were also tackled in the literature in Refs.~\cite{Benbrik:2010cf,Dorsner:2015mja}.

\subsubsection{Anomalous magnetic moments}
Virtual corrections due to LQ states can modify the tree-level electromagnetic interactions of charged leptons
$\ell$. At the level 
of the $\ell(p) \to \ell(p') \gamma^*(q,\epsilon)$  amplitude one
has~\cite{Jegerlehner:2009ry}:
\begin{equation}
  ie  \bar u_\ell(p')\left[ \gamma^\mu - \frac{a_\ell}{2m_\ell}
    i\sigma^{\mu\nu}q_\nu \right]  u_\ell(p) \epsilon_\mu^*,\qquad q_\mu = (p-p')_\mu.
\end{equation} 
The gyromagnetic ratio $g_\ell$ is then obtained from the relation
$a_\ell(q^2 \to 0) = (g_\ell-2)/2$. At the effective Lagrangian level
$a_\ell$ corresponds to the following interacting Lagrangian:
\begin{equation}
\label{eq:magmom}
\mc{L}_{a_\ell} = e \bar \ell \left(\gamma_\mu A^\mu + \frac{a_\ell}{4m_\ell} \sigma_{\mu\nu}F^{\mu\nu}\right) \ell,
\end{equation}
where $F_{\mu\nu} = \partial_\mu A_\nu - \partial_\nu A_\mu$. The
terms in the brackets are independent of electric charge convention,
while the overall sign reflects the choice of the covariant derivative
for lepton: $D_\mu = \partial_\mu - i e A_\mu$. 
From the amplitude~\eqref{eq:l-lprimegamma} adapted to the
$\ell = \ell'$ case one can extract
$a_\ell = i m_\ell (\sigma_L + \sigma_R)$.  Scalar LQ contributions to
lepton anomalous magnetic moments have been known for some
time~\cite{Djouadi:1989md}. All $F=0$ scalar LQs contribute as
\begin{equation}
\label{eq:g-2}
  \begin{split}
  a_\ell = -\frac{N_c m_\ell}{8\pi^2 \mlq^2} \sum_q \Bigg[&m_\ell
    \left(|l_{q\ell}|^2 + |r_{q\ell}|^2 \right) \left(Q_S f_S(x_q) -
      f_F(x_q)\right) \\
 &+m_q \re (r_{q\ell}^*    l_{q\ell})  \left(Q_S g_S(x_q)-g_F(x_q)\right)  \Bigg].
  \end{split}
\end{equation}
The above expression is valid for LQ with $F=0$ and can be recast to the
$|F|=2$ case by changing the sign before $Q_S$. Eq.~\eqref{eq:g-2}
reveals that a chiral LQ charge-eigenstate contributes to $a_\ell$
with a definite sign depending only on $Q_S$. However, complete LQ
multiplets could have contributions from different $Q_S$ and in such
cases the sign may depend on the interplay between Yukawa couplings of
differently charged states. We do not consider contributions of vector
LQ states towards $a_\ell$ since their contributions largely depend on
the UV details of the theory.

Currently the most precise experimental result on the magnetic moment of
the muon is due to the E821 experiment performed at
BNL~\cite{Bennett:2004pv,Bennett:2006fi} and amounts to
$a_\mu^\mrm{exp} = 1.16592080(63)\E{-3}$. Comparing it to the
contemporary state-of-the-art SM predictions,
$a_\mu^\mrm{SM} = 1.16591803(70)\E{-3}$~\cite{Agashe:2014kda} (see
also \cite{Davier:2010nc}) the difference between the two numbers, $\Delta a_\mu \equiv
a_\mu^\mrm{exp} - a_\mu^\mrm{SM} = (2.8\pm 0.9)\E{-9}$, entails a
strong constraint on the couplings of LQ states to muon. Next
generation experiments will reduce the error in $a_\mu^\mrm{exp}$ by a
factor of 4~\cite{Hertzog:2015jru}.

The contributions of LQs to $a_\ell$ have been presented in
Refs.~\cite{Couture:1995he,Cheung:2001ip, Chakraverty:2001yg, Mahanta:2001yc,
  Queiroz:2014zfa}. The connection between $a_\mu$ and the electric dipole of
the muon, in light of the BNL measurement, has been
advocated in~\cite{Chakraverty:2001yg}. The interplay between LFV lepton
decays have been studied in Refs.~\cite{arXiv:0807.4240,Benbrik:2010cf}.
 Bounds on the LQ couplings,
albeit weak, have been also derived from the measurements of neutrino
magnetic moments in Ref.~\cite{Chua:1998yk}.

\subsubsection{$\ell \to \ell' \ell' \ell''$}
The four-lepton lepton flavor violating amplitude is mediated by box
diagrams involving leptoquarks and quarks corresponding to diagrams of
neutral meson mixing via LQ turned inside out.  The amplitudes of
scalar LQ mediation scale
as $\sim y^4/\mlq^2$, where $y$ is the relevant LQ coupling, whereas
for the vector LQ the contribution could be suppressed by a GIM-like
mechanism~\cite{Davidson:1993qk}. Formulas
for the neutral meson mixing amplitudes, i.e.,~\eqref{eq:MMbarLeffS}
and \eqref{eq:MMbarLeffV}, can be straightforwardly adapted to cover
the $\ell \to \ell' \ell' \ell''$ processes. The upper bounds stemming
from $\mu \to 3e$ decay on the relevant sets of couplings
characteristic of box diagrams have been derived
in~\cite{Davidson:1993qk}, with $Z$ and $\gamma$ penguin
contributions neglected. The box diagram of $\mu \to 3 e$ is a product
of LQ couplings present in $\mu \to e \gamma$ and the LQ coupling to electrons
and quarks. On the other hand, the photon penguin diagram
contribution to $\ell \to \ell'\ell' \ell''$ have similar structure as
$\ell \to \ell' \gamma$ amplitudes and have been shown to be important
in the case of unitary coupling matrices in
Ref.~\cite{Gabrielli:2000te}.

The muon decay $\mu \to 3 e$ is very strongly bounded 
with branching ratio below $1.0\E{-12}$ at 90\% C.L.~\cite{Bellgardt:1987du}.
Bounds on many different decay modes of the $\tau$ lepton are considerably weaker
with branching fractions below $\sim 10^{-8}$~\cite{Agashe:2014kda}.

\subsubsection{$\mu$--$e$ conversion in nuclei}
Competitive lepton flavor violation constraints on LQ interactions can be obtained from searches for
$\mu$--$e$ conversion in nuclei. Currently the strongest bounds on
BR$_{\mu e} \equiv \Gamma_{\text{conversion}}/
\Gamma_{\text{capture}}$ were set by the SINDRUM Collaboration from
experiments on titanium with BR$^{(\text{Ti})}_{\mu e} < 4.3 \times
10^{-12}$~\cite{Dohmen:1993mp} and gold target setting BR$_{\mu
  e}^{(\text{Au})} < 7 \times 10^{-13}$~\cite{Bertl:2006up}, both at
90\%C.L.  The $\mu$--$e$ conversion bounds are expected to be further
improved in the future by several orders of magnitude. The COMET~\cite{Krikler:2015msn} and DeeMe~\cite{Nguyen:2015vkk} experiments currently under construction are expected to improve the current SINDRUM bounds by an order of magnitude. Furthermore, according to
proposals~\cite{Ankenbrandt:2006zu,Knoepfel:2013ouy,Bartoszek:2014mya} and~\cite{P21Jparc, P20Jparc}, one
can expect a sensitivity of few times $10^{-17}$ at Mu2e or even $10^{-18}$ by the
PRISM/PRIME experiment.

To get the constraint in the $\mu$--$e$ channel from these experiments,
one needs to compute the relevant transition matrix elements in different nuclei. A
detailed numerical calculation has been carried out
in Ref.~\cite{Kitano:2002mt} and we use their Eq.~(14) to
calculate the desired conversion rates. (For more recent evaluation of
uncertainties due to nuclear matrix element see Ref.~\cite{Cirigliano:2009bz}.) They receive both tree-level and 
one-loop
LQ contributions.  Using the notation of~\cite{Kitano:2002mt} we thus
have
\begin{align}
  \label{eq:conversion-rate}
  \Gamma_{\text{conversion}} &= 2 G_F^2 
  \left| A_R^* D +  (2g_{LV}^{(u)} + g_{LV}^{(d)}) V^{(p)} + (g_{LV}^{(u)} + 2 g_{LV}^{(d)}) V^{(n)}  \right. \nonumber\\
  &\qquad\qquad\left.+  (G_S^{(u,p)} g_{LS}^{(u)} + G_S^{(d,p)} g_{LS}^{(d)}+ G_S^{(s,p)} g_{LS}^{(s)}) S^{(p)}  \right. \nonumber\\
  &\qquad\qquad\left. + (G_S^{(u,n)} g_{LS}^{(u)} + G_S^{(d,n)} g_{LS}^{(d)} + G_S^{(s,n)} g_{LS}^{(s)}) S^{(n)} \right|^2 \nonumber\\
  &\quad+ (L\leftrightarrow R) \,,
\end{align}
where $G_S^{(u,p)} = G_S^{(d,n)} = 5.1$, $G_S^{(d,p)} = G_S^{(u,n)} = 4.3$, and $G_S^{(s,p)} = G_S^{(s,n)} = 2.5$~\cite{Kosmas:2001mv}, and  one can identify $g_{RV,LV}^{(q)} = 4(c_{qq;e\mu}^{RR,LR}+c_{qq;e\mu}^{RL,LL})$,  $g_{RS,LS}^{(q)} = 4(g_{qq;e\mu}^{RR,LR}+g_{qq;e\mu}^{RL,LL})$ and $8 G_F m_\mu A_{R,L} / \sqrt{2} = e \sigma^{e\mu}_{R,L}$. Note that quark tensor ($h^{RR,LL}_{qq;e\mu}$), pseudoscalar density ($g_{qq;e\mu}^{RR,LR}-g_{qq;e\mu}^{RL,LL}$) and axial current ($c_{qq;e\mu}^{RR,LR}-c_{qq;e\mu}^{RL,LL}$) operators do not contribute to the coherent conversion processes and their effects are usually neglected altogether. 

\begin{table}
{\small
  \begin{center}
  \begin{tabular}{lccc} \hline
    Nucleus                        &$D [m_\mu^{5/2}]$  & $V^{(p)} [m_\mu^{5/2}]$ & $V^{(n)} [m_\mu^{5/2}] $\\ \hline
    $\text{\text{Ti}}^{48}_{22}$    & 0.0864 &              0.0396    &             0.0468  \\
    $\text{\text{Au}}^{197}_{79}$ &  0.189 &              0.0974
                                                                                 &             0.146    \\ 
&&&\\
  & $S^{(p)} [m_\mu^{5/2}]$ & $S^{(n)} [m_\mu^{5/2}] $ &
                                                         $\Gamma_{\text{capture}}[10^6\e{s}^{-1}]$
    \\ \hline
 $\text{\text{Ti}}^{48}_{22}$ &   0.0368 & 0.0435            &  2.59\\
 $\text{\text{Au}}^{197}_{79}$& 0.0614 &  0.0918         &  13.07\\
\hline
\end{tabular}
  \end{center}
}
  \caption{Data taken from Tables I and VIII of~\cite{Kitano:2002mt}.}
  \label{TabTiAuData}
\end{table}
Since both Ti and Au processes feature the same flavor structure, the differences between
them are entirely determined by the numerical factors quoted in
Table~\ref{TabTiAuData}.  The Ti parameters entering in the conversion rate are on average a
factor $\sim 2.5$ smaller than the ones for Au, whereas the capture rates differ by a factor $\sim 5$. Accordingly the difference between these branching ratios is a factor of $\sim 2$. Since the current experimental Au bound is a factor of six stronger, the most stringent constraints can be derived from this target nucleus, in particular in absence of accidental cancellations one obtains $\sigma^{e\mu}_{L,R} < 1/(2\times 10^8\,\rm TeV)$, $c^{XY}_{uu;e\mu} < 2 \times 10^{-8}$, $c^{XY}_{dd;e\mu} < 3 \times 10^{-8}$, $g^{XY}_{uu;e\mu} < 2 \times 10^{-8}$, $g^{XY}_{dd;e\mu} < 2 \times 10^{-8}$ and $g^{XY}_{ss;e\mu} < 3 \times 10^{-8}$ resulting in very stringent constraints on products of LQ interactions coupling light quarks to electrons and muons.

\subsubsection{$\tau \to \ell' P$, $\tau \to \ell' V$, $\tau \to \ell'
  X$}
These decays are uniquely sensitive to Yukawas with light quarks
and the tau lepton. Limits on their branching fractions have been set by the B factories
and are expected to be further extended in the near future by the upcoming Belle-II experiment. As an
example, consider the width of the pionic channel in the model with
$\tilde S_1$, studied in~\cite{Dorsner:2011ai}:
\begin{equation}
  \label{eq:tauwidth}
  \Gamma_{\tau \to \ell \pi^0} = \frac{\left|x_{d \ell} x_{d \tau}\right|^2}{2048\pi} \frac{f_\pi^2 m_\tau^3}{\mlq^4}
  \left[1-3 m_\ell^2/m_\tau^2 - 2 m_\pi^2/m_\tau^2 \right].
\end{equation}
The experimental upper bounds for these decays were obtained at the B factories and
are currently of the order $10^{-7}$ constraining the product of
couplings at $|x_{d \ell} x_{d \tau}| \lesssim 0.1$ at $\mlq = 1$\,TeV.

\subsection{Electric dipole moments and leptoquark contributions}
Many New Physics models can contribute to electric dipole moments
(EDMs) of quarks and leptons. (For an excellent review
see Ref.~\cite{Pospelov:2005pr}.) Only upper bounds of the electric dipole
moments are established experimentally. The strongest constraint has
been obtained for the EDM of neutron~\cite{Baker:2006ts}, the
$^{199}$Hg atom~\cite{Griffith:2009zz} and the ThO
molecule~\cite{Baron:2013eja}. Among leptons the most robust limit is for the electron EDM. It reads $|d_e|< 10.5 \times 10^{-28}$\,$e$\,cm~\cite{Hudson:2011zz,Jung:2013mg}. This bound then gives very strong constraints on the imaginary part of New Physics
couplings. The upper bounds of EDM for
$\mu$ and $\tau$ and $s$, $c$, $b$, and $t$ quarks are orders of
magnitude smaller than the abovementioned limits. LQs might contribute to the electric dipole moments of
quarks and leptons even at the one-loop level as pointed out
in~\cite{Arnold:2013cva}.  The scenario is possible if there are two
different couplings of an LQ to a quark and a lepton. Namely, an LQ
coupling to both left- and right-handed quarks can produce an EDM of a
lepton at the one-loop level due to a chirality flip on the internal
quark line. Of course, such a scenario can only be efficient for heavy
quarks (in particular the top) in the loop. For the same reason one
loop LQ contributions to quark EDMs, with leptons running in the loop
are severely suppressed and generically ignored. Leptoquarks might
contribute to the electric dipole moments of quarks and leptons even
on the one-loop level as pointed out in Ref.~\cite{Arnold:2013cva}. The scenario
is possible if there are two different couplings of the leptoquark to
quark and lepton. Namely, for the lepton dipole moment if leptoquark
couples to both left- and right-handed quarks, at one-loop to the
chirality flip, one can produce electric dipole moment at one
loop. with the quark and lepton with two different couplings.
Of course, such scenario can be efficient if the top quark is in the loop. 
Although such scenario is applicable to all leptonic generations, it makes sense to determine electron electric dipole moment, due to rather week bounds on muon and tau EDM. 

As noticed by the authors of Ref.~\cite{Arnold:2013cva} EDM of the
electron is found to be in the case of $R_2$ scalar leptoquark 
\begin{equation}
\label{EDM-e1}
d_e = \frac{3 m_t}{16 \pi^2 \mlq^2} \im [(y_{2}^{LR}
V^\dagger)_{13} y_{2\,31}^{RL} ]   f(m_t^2/\mlq^2 )
\end{equation}
with the function $f$ defined as
$f(x)=(9-53 x +60 x^2 -40 x^3 +2 ( 2+5 x) \log x )/(x-1)^3$.  The
upper bound on $|d_e|< 2.0 \times 10^{-28}$ $e\e{cm}$ gives constraints on
the imaginary part of the relevant couplings
\begin{equation}
\im [(y_{2}^{LR} V^\dagger)_{13} y_{2\,31}^{RL} ] < 2.2 \times 10^{-11}.
\label{edm-7/6}
\end{equation} 
Among all other leptoquarks we find that only $S_1$ scalar LQ state
has two different couplings leading to the electric dipole moment
\begin{equation}
d_e = \frac{ m_t}{64\pi^2 \mlq^2} \im [y_{1\,31}^{RR} (V^T y_1^{LL}
)^*_{31} ]   g(m_t^2/\mlq^2 )
\label{EDM-e2}
\end{equation}
with $g(x)=(-5+8x - 3x^2 +2(x-2) \log x ) /(x-1)^3$. The ensuing bound is
\begin{equation}
\im [y_{1\,31}^{RR} (V^T y_1^{LL}
)^*_{31} ]  < 9.2  \times 10^{-12}.
\label{edm-1/3}
\end{equation} 
The LQs might contribute to EDMs at two-loop level through
contributions reviewed in~\cite{Pospelov:2005pr} and recently
in~\cite{Cirigliano:2016njn}.  Such extensive analyses of LQs
contributions at two-loop levels were not performed yet.


\section{Top quark, electroweak precision and Higgs physics constraints}
\label{sec:EWK&HIGGS}

Measurements involving top quarks, electroweak gauge bosons and the Higgs boson typically proceed at much larger energy scales than precision flavor and CP violation probes. Nonetheless, existing direct constraints on LQ masses that are discussed in Section~\ref{sec:COLLIDERS} typically justify the use of effective field theory methods in this regime, i.e., the relevant expansion $(v/m_\mathrm{LQ})^n$ is expected to be convergent and well approximated by the first non-vanishing terms.

A special role is played by the electroweak precision observables which are uniquely sensitive to possible mass splittings within $SU(2)$ LQ multiplets. In particular, the stringently constrained oblique corrections are discussed in Section~\ref{sec:Oblique}.  

Finally, special attention needs to be paid to the role of scalar LQs in Higgs physics. In particular, these states can generically couple to the Higgs boson through the scalar potential which contains gauge invariant marginal operator of the form $\Phi^\dagger \Phi H^\dagger H$, where $\Phi$ is the LQ field and $H$ is the SM Higgs doublet. The so-called ``Higgs portal'' operator, after electroweak symmetry breaking, induces corrections to the masses of weak $\Phi$ components ($\Phi_i$) as well as the couplings to the physical Higgs boson ($\Phi_i^\dagger \Phi_i h$), which in turn can affect loop induced Higgs production and decay processes. Such effects can be probed with the  Higgs signal strength measurements at hadron and lepton colliders. In Section~\ref{sec:lqhiggs} we review the impact of  scalar LQs on Higgs boson phenomenology. We show how the currently available LHC Higgs data constrain the relevant parameter space and discuss possible improvements with future-projected data. 



\subsection{Leptoquarks in processes of top quarks}

LQs coupling to the third generation can leave observable imprints in processes involving top quarks. Most of the general considerations presented in the opening paragraphs of Section~\ref{sec:FLAVOR} apply here as well. At the tree level, LQs (assumed heavier than the top quark) can contribute to effective four-fermion interactions in Eq.~\eqref{eq:RareLeff}
where the relevant LQ contributions  can be read from Table~\ref{tab:RareLep}. 
Their implications for top quark phenomenology have been studied recently in Ref.~\cite{Davidson:2015zza}. As reviewed in Section~\ref{sec:FLAVOR} most stringent existing constraints on such LQ couplings can actually be derived from low-energy phenomenology. In particular LFV rare meson and charged lepton decays receive relevant contributions from LQ interactions with top quark at the one-loop or even two-loop level.

In  top quark phenomenology, indirect LQ effects can be probed using both top quark decays and production. In particular, tree-level LQ contributions can induce LFV three-body decays of top quarks, which can be searched for using the large top quark data sets at high energy hadron (or future lepton) colliders.  Alternatively, indirect effects of LQ's coupling simultaneously to top quarks, electrons and light quarks can be searched for in $e p$ colliders (see also Section~\ref{sec:ep-colliders}). We review both directions in the following subsections. 
On the other hand, LQ contributions to radiative top quark decays and single top quark production at hadron colliders are necessarily loop suppressed and thus offer limited sensitivity to the relevant LQ couplings. Especially since they have to compete against sizable CKM allowed SM contributions to the two-body dominated top quark decay width and hadronic single top production, respectively. 

\subsubsection{Lepton flavor violating top quark decays}

The large top quark event samples produced at high energy hadron colliders in principle allow to probe very rare top quark decay modes. The dominant tree-level effects of LQ's to the rare LFV top decays $t \to \ell_i^+ \ell_j^- u_k$ can be summarized as
\begin{align}
\Gamma(t \to \ell_i^+ \ell_j^- u_k)  &= \frac{G_F^2 m_t^5}{192 \pi^3} \sum_{X,Y=L,R}  \biggr[  |c_{kt;ij}^{XY}|^2 + \nonumber \\
& + \frac{1}{4} \left( | g_{kt;ij}^{X\neq Y}|^2 +  |\hat g_{kt;ij}^{X}|^2 +  |\hat h_{kt;ij}^{X}|^2 - 2 {\rm Re} \{\hat g_{kt;ij}^{X} \hat h_{kt;ij}^{X*} \} \right) \biggr]\,,
\end{align} 
where we have neglected all fermionic masses except of the top quark and introduced $\hat h_{kl;ij}^X \equiv  8  h_{kl;ij}^{XX}$ and $\hat g_{kl;ij}^{X} \equiv  g_{kl;ij}^{XX} + 4 h_{kl;ij}^{XX}$. Currently no dedicated searches exist for such signatures, but a recent sensitivity study~\cite{Davidson:2015zza} drawing comparisons with a search for related rare top quark decays $t \to q Z$~\cite{Chatrchyan:2013nwa} estimates the sensitivity of 
\begin{equation}
\mathcal B(t\to q \mu^\pm e^{\mp}) < 6.3 (1.2) \times 10^{-5} ~ {\rm @ \, 95\% \, C.L.}\,,
\end{equation}
with 20\,fb${}^{-1}$ of 8\,TeV (100\,fb${}^{-1}$ of 13\,TeV) LHC data, corresponding to a final sensitivity of 13\,TeV LHC run to $|c^{XY}_{qt;e\mu}|, |c^{XY}_{qt;\mu e}| <0.06$ and $|\hat h^{X}_{qt;e\mu}|$, $|\hat h^{X}_{qt;\mu e}|$, $|\hat g^X_{qt;e\mu}|$, $|\hat g^X_{qt;\mu e}| < 0.1$. 

\subsubsection{Single top production at $e\,p$ colliders}

The HERA electron--proton collider at DESY had performed searches for single top quarks in the final state. It collided protons with positrons (or electrons) at a center of mass energy of $319$\,GeV. The H1 Collaboration at HERA actually observed a few events with energetic isolated leptons and missing energy, consistent with $e^\pm p \to t e^\pm + X$ followed by a leptonic decay of the top quark~\cite{Aktas:2003yd,Aaron:2009vv}. However, this signal was not confirmed by the ZEUS experiment~\cite{Chekanov:2003yt}, and neither Collaboration had a signal consistent with hadronic top decays. Both Collaborations thus set bounds on $\sigma(e^\pm p \to e^\pm t X)$. In particular H1 obtained a somewhat weaker bound due to their observed excess
\begin{equation}
\sigma(e^\pm p \to e^\pm t + X) < 0.3 \, {\rm pb \, @ 95\% C.L.}\,. 
\end{equation}
LQ interactions with the flavor structure$(\bar e \ell)(\bar u t)$ and $(\bar \ell e)(\bar u t)$ could respectively mediate $e^-p \to \ell^- t X$ and $e^+p \to \ell^+ t X$, where $\ell = e, \mu, \tau$. In particular, the relevant $e^\pm u \to \ell^\pm t$ partonic production cross-section can be written as
\begin{align}
\hat \sigma(e^\pm u \to \ell^\pm t) &=  \frac{(1-r_{t/\hat s})^2}{192 \pi \hat s r^2_{t/\hat s}} \sum_{X,Y=R,L}\left[ 12 |c_{u t; e\ell }^{XX}|^2  + {\rm Re}(\hat h_{ u t; e\ell}^{XX}\hat g_{ u t; e\ell}^{X})  \left(1 - {r_{t/\hat s}}{}\right) +\right.\nonumber\\
&\left. + \left (4 |c_{u t; e\ell }^{X\neq Y}|^2 + |\hat h_{ u t; e\ell}^{X\neq Y}|^2 + |\hat h_{ u t; e\ell}^{XX}+\hat g_{u t; e\ell }^{X}|^2\right )  \left(1 + \frac{r_{t/\hat s}}{2}\right)   \right]\,,
\end{align}
where $r_{t/\hat s} \equiv m_t^2/\hat s$ and $\hat s$ is the partonic invariant mass squared.
After integrating over the relevant parton luminosities and accounting for experimental efficiency and acceptance effects of the H1 analysis~\cite{Aktas:2003yd,Aaron:2009vv}, some approximate estimates of the resulting bounds on the effective LFV four-fermion interactions involving the top quark have been derived in~\cite{Davidson:2015zza}, in particular $|c_{ ut; e\mu}^{XX}| \lesssim 0.22$, $|c_{ut; e\mu }^{X\neq Y}| \lesssim 0.33$, $|\hat h_{ ut; e\mu}^{XY}|, |\hat g_{ut; e\mu }^{X}| \lesssim 0.66$.



\subsection{Leptoquarks and precision electroweak measurements}

\subsubsection{Atomic parity violation}

Stringent constraints on LQ couplings to the first generation fermions can be derived from atomic parity violation (APV) measurements in ${}^{133}{\rm Cs}$ (see for example~\cite{Langacker:1990jf}).
The relevant LQ contribution occurs at the tree level and can be parameterized as
\begin{equation}
\mathcal{L}_{PV} = \frac{G_F}{\sqrt{2}} \bar e \gamma^\mu \gamma^5 e ~(\delta{C}_{1u} \bar u \gamma_\mu u + \delta{C}_{1d} \bar d \gamma_\mu d)~,
\label{eq:apv-lag}
\end{equation}
where the effective $\delta{C}_{1u}$ and $\delta{C}_{1d}$ coefficients receive the matching contribution from a LQ propagation.  A generic LQ contribution can easily be obtained by comparing the effective Lagrangian in Eq.~\eqref{eq:apv-lag} to the one in Eq.~\eqref{eq:RareLeff},
\begin{equation}
\delta{C}_{1q} = c^{LL}_{qq;ee}-c^{LR}_{qq;ee}+ c^{RL}_{qq;ee}-c^{RR}_{qq;ee} ~,
\label{eq:apv-matching}
\end{equation}
where $q=u,d$, and using the results of the matching calculations reported in Table~\ref{tab:RareLep}.
The effective $\delta{C}_{1u}$ and $\delta{C}_{1d}$ coefficients modify the nuclear weak charge as~\cite{Gresham:2012wc} 
\begin{equation}
\delta Q_{W}(Z,N)=-2(2Z+N)\delta{C}_{1u}-2(Z+2N)\delta{C}_{1d}\,,
\end{equation}
where the measured value deviates from the SM expectation by $1.5\,\sigma$~\cite{Dzuba:2012kx}
\begin{equation}
Q_{W}-Q_{W}^{SM}\equiv\delta Q_{W}=0.65(43),
\end{equation}
which implies
\begin{equation}
| \delta{C}_{1u(d)} | \lesssim 10^{-3}~,
 \label{e3PV}
\end{equation}
assuming only one of the two contributions is present at a given moment.
These bounds can be easily translated for the specific LQ by using Eq.~\eqref{eq:apv-matching} and Table~\ref{tab:RareLep}. Similar bounds have been derived in Refs.~\cite{Baker:2015qna,Dorsner:2014axa}.

Before we conclude this section it is worth to point out that it is not possible to have a flat direction in $\delta{C}_{1u}-\delta{C}_{1d}$ that is induced by a single LQ representation except for $R_2$ and $V_2$. (See Table~\ref{tab:RareLep} for more details.)




\subsubsection{Oblique corrections}
\label{sec:Oblique}

The effects of new heavy particles, which contribute to $W$ and $Z$ boson self-energies at one-loop level, can be constrained from oblique electroweak parameters $S$ and $T$~\cite{Peskin:1991sw} (or equivalently~\cite{Altarelli:1990zd}). A recent global fit of the electroweak precision data~\cite{Baak:2014ora} reports
\begin{equation}
S=0.05\pm0.11~, \;\; T= 0.09 \pm 0.13~,
\end{equation} 
with the correlation coefficient of $+0.90$. 
Scalar LQ representation, with a mass splitting in the multiplet, can induce large corrections to these parameters. For example, the contribution of a scalar LQ doublet representation to oblique parameters is~\cite{Keith:1997fv,Froggatt:1991qw}
\begin{equation}
\begin{split}
\Delta S &= \frac{-Y}{2\pi} \log \frac{m_1^2}{m_2^2}~,\\
\Delta T &= \frac{3}{16 \pi m_Z^2 s_W^2 c_W^2} \left( m_1^2+m_2^2-\frac{2 m_1^2 m_2^2}{m_1^2-m_2^2}  \log \frac{m_1^2}{m_2^2}\right)~,
\end{split}
\end{equation}
where $Y$ is the weak hyper-charge, $m_1$ and $m_2$ are the masses of the two isospin components, $m_Z$ is the $Z$ boson mass and $s_W=\sqrt{1-c_W^2} \equiv \sin \theta_W$ is the sine of the weak mixing angle. Defining $\Delta m = m_2 - m_1$ and assuming $\Delta m \ll m_1,m_2$, we can expand the above relations
\begin{equation}
\begin{split}
\Delta S &= \frac{Y}{\pi} \frac{\Delta m}{m_1}+\mathcal{O}\left(\frac{\Delta m^2}{m_1^2}\right)~,\\
\Delta T &= \frac{\Delta m^2}{m_Z^2 \pi \sin^2 (2\theta_W) } \left(1+ \mathcal{O}\left(\frac{\Delta m^2}{m_1^2}\right)\right)~.
\end{split}
\end{equation}
In other words, the dominant constraint on the heavy LQ doublet is due to the $T$ parameter which depends on the mass splitting ($\Delta m$) and does not depend on the LQ mass ($m_1$).
Using the experimental data reported above, we find a robust limit on the mass splitting
\begin{equation}
|\Delta m | < 53\,\textrm{GeV} \; \; \; \textrm{at} ~ 95\% \textrm{C.L.}~.
\label{eq:ewpoLIM}
\end{equation}
Consider an operator of the form, $\mathcal{L} \supset -\lambda_S (S^\dagger H) (H^\dagger S)$, where $S$ is the LQ doublet representation and $SU(2)$ contraction singlets are separated with brackets. After electroweak symmetry breaking, the operator induces mass splitting in the doublet, in particular, $\Delta m = - \lambda_S v^2 / (4 m_1)$ where $v$ is the electroweak condensate. If no other source of mass splitting is present, the bound from Eq.~\eqref{eq:ewpoLIM} translates into a bound on the above ``Higgs portal'' operator ($\lambda_S$).


\subsubsection{Non-oblique corrections: $Z\to b \bar b$ }

LQ induced effects in precision electroweak measurements can also
show up as vertex corrections to $V \bar f f$ interactions, where
$V$ is an electroweak gauge boson while $f$ is a SM fermion. Along these lines, authors in Ref.~\cite{Mizukoshi:1994zy} perform a global fit to LEP data including oblique and vertex corrections induced by LQs. The constraints on LQs from $Z$ decays have also been considered in Ref.~\cite{Carpentier:2010ue}.
More general study of the impact of heavy vectors and scalars on the electroweak data, including the cases of scalar and vector LQs, has been performed in Refs.~\cite{delAguila:2010mx,deBlas:2014mba}.
 Also, Refs.~\cite{Alcaraz:2006mx, Schael:2013ita} report bounds on LQ interactions from LEP-II $e^+ e^- \to f \bar f$ cross section and distribution measurements.

To illustrate the size of the effect, we consider the impact of the scalar LQ representation $R_2 
\equiv (\mathbf{3},\mathbf{2},7/6)$ on $Z \bar b b$ effective coupling discussed in Ref.~\cite{Dorsner:2013tla}, updating the numerical analysis. We adopt the following parametrization 
of the effective interaction
\begin{equation}
\label{eq:Zbbdef}
\mc{L}_{Zb\bar b} = \frac{g_2}{c_W} Z^\mu 
    \bar b \gamma_\mu \left[(g_L^b + \delta g_L^b) P_L +  (g_R^b +
      \delta g_R^b) P_R \right] b\,~,
\end{equation}
where $SU(2)$ coupling is denoted by $g_2$, $c_W$ is the cosine of the
weak mixing angle and $P_{L,R} = (1 \pm \gamma_5)/2$ are the chiral
projectors. 
In the SM at tree-level, the couplings are $g_L^{b0} = -1/2+s_W^2/3$ and $g_R^{b0} = s_W^2/3$. 
Going beyond  tree-level, the higher-order electroweak corrections get the largest contribution from top quark in the loop.

As far as $R_2$ representation is concerned, the $R_2^{2/3}$ component couples to $b$-$\tau$ pair proportionally to $y_{2\,33}^{LR}$, see  Eq.~\eqref{eq:main_R_2_a}. This contributes to the $Z \to b \bar b$ amplitude at the one-loop level (order $|y_{2\,33}^{LR}|^2$) and thus allows one to constrain it directly. The LQ correction to the left-handed coupling is
\begin{align}
  \label{eq:dgLb}
  \delta g_L^b(y_{2\,33}^{LR}) = \frac{|y_{2\,33}^{LR}|^2}{384 \pi^2} \left[ g_0(x) + s_W^2 g_2(x) \right]\,,
\end{align}
where $x = m_{R_2}^2/m_Z^2$, and we take $m_Z = 91.2$\,GeV and $s_W^2
= 0.231$. The asymptotic expansion of functions $g_0(x)$ and $g_2(x)$ at large $x$ is
\begin{align}
  g_0(x) &\simeq -\frac{2}{3x}\,,\nonumber\\
   g_2(x) &\simeq \frac{8}{x} (\log x+2/9+i \pi )\,,\nonumber
\end{align}
while their full expressions can be found in Ref.~\cite{Dorsner:2013tla}.

On the other hand, the LEP experiment measured precisely the decay modes of $Z$ bosons to $f \bar f$ pairs. 
A recent global fit to electroweak data~\cite{Ciuchini:2013pca}, finds the coupling shifts to be
\begin{align}
  \label{eq:dgb}
  \delta g_L^b = (0.0016 \pm 0.0015) \,,\qquad \delta g_R^b = (0.019 \pm
  0.007)~,
\end{align}
with the correlation coefficient $\rho=0.82$~\cite{Ciuchini:2013pca}. A slight tension  with respect to the SM prediction (above $2\,\sigma$ significance) is observed. Although $\delta g_L^b$ develops an
imaginary part due to on-shell $\tau$ leptons, the constraint from the
electroweak fit in Eq.~\eqref{eq:dgb} is sensitive to the interference term
between the approximately real $g_L^b$ and complex $\delta
g_L^b(y_{2\,33}^{LR})$, and therefore only the real part of $\delta
g_L^b(y_{2\,33}^{LR})$ enters the prediction. This implies a very weak constraint
on the LQ coupling, for example, $m_{R_2}=500\,$GeV gives $|y_{2\,33}^{LR}| \lesssim 10$.



\subsection{Leptoquarks and Higgs physics}
\label{sec:lqhiggs}

Scalar LQs with masses around electroweak scale and  tree level couplings to the Higgs boson ($S_i^\dagger S_i h$), can significantly modify the Higgs boson production and decay processes that are induced at one-loop level in the SM.  Examples of such are the dominant single and double Higgs boson production mechanisms, $gg\to h$ and $gg\to hh$, as well as experimentally interesting Higgs decays, $h \to \gamma \gamma$ and $h\to Z\gamma$. In these processes,  loop diagrams with sufficiently light LQs can easily compete with the SM contributions and lead to $\mathcal{O}(1)$ effect in physical observables such as the total cross sections and partial decay widths.

Assuming the different weak components ($S_i$) to be degenerate, as expected from the electroweak precision constraints, the contribution of a single colored scalar representation ($S$) to loop induced processes can be parameterized in terms of the relevant coupling and mass. The Higgs portal interaction is
\begin{equation}
\mathcal L_{} \ni -\lambda_S (S_{i a}^{\dagger}S_{i a}) (H_j^{\dagger}H_j) =  -\lambda_{S} v ~ S_{i a}^\dagger S_{i a} h + \ldots\,,
\label{eq:2Phi}
\end{equation}
where the sum over weak ($i,j$) and color ($a$) indices is shown explicitly. In other words, the effects of such interactions are described in terms of two parameters only. These are the coupling constant $\lambda_S$ and the common mass of the weak components $\mlq$. The electroweak condensate is $v=246.22$\,GeV.

\subsubsection{Loop-induced Higgs processes}

Single LQ representation $S$ interacting with the Higgs through Eq.~\eqref{eq:2Phi}, modifies the partial decay width for $h\to\gamma\gamma$ as~\cite{Djouadi:2005gj,Carena:2012xa,Chang:2012ta,Dorsner:2012pp,Agrawal:1999bk,Enkhbat:2013oba,Gori:2013mia}
\begin{equation}
\label{eq:1higgsLQ}
\Gamma_{h\to\gamma\gamma}=\frac{G_F \aem^{2}m_{h}^{3}}{128\sqrt{2}\pi^{3}}\left|\mathcal A_{1}({x}_{W})+
\frac{4}{3}\mathcal A_{1/2}({x}_{t})+\underset{i}{\sum}\frac{\lambda_{S}}{2}\frac{v^{2}}{m_{\rm{LQ}}^{2}}
d(r_{S})Q_{S_i}^{2}\mathcal A_{0}({x}_{s})\right|^{2},
\end{equation}
where the sum is taken over all weak components of the $SU(2)$ multiplet and $d(r_{S})$ and  $Q_{S_i}$ 
are the dimension of the color representation of $S$, and the electric charges of the weak $S_i$ components, 
respectively. Also, $G_F$ is the Fermi coupling constant, $m_h$ is the mass of the Higgs boson, $\aem=e^2/(4\pi)$ is the fine structure constant and
${x}_{i}=m_{h}^{2}/(4m_{i}^{2})$ for $i=W,t,s$. The relevant one-loop functions are given by
\begin{eqnarray}
\mathcal A_{1}({x})&=&-\left(2{x}^{2}+3{x}+3(2{x}-1)f({x})\right){x}^{-2},\\
\mathcal A_{1/2}({x})&=&2\left({x}+({x}-1)f({x})\right){x}^{-2},\\
\mathcal A_{0}({x})&=&-\left({x}-f({x})\right){x}^{-2},\\
f({x})&=&\left\{ \begin{array}{cc}
\arcsin^{2}\sqrt{{x}} & {x}\leq1\\
-\frac{1}{4}\left(\log\frac{1+\sqrt{1-{x}^{-1}}}{1-\sqrt{1-{x}^{-1}}}-i\pi\right)^{2} & {x}>1
\end{array}\right..
\end{eqnarray}
The first two terms in Eq.~\eqref{eq:1higgsLQ} are the SM contribution from the $W$ boson loop interfering destructively with the subdominant top quark loop. The numerical values of the relevant SM loop functions evaluated for  $m_h=125$\,GeV, are $\mathcal A_{1}({x}_{W})=-8.3$, $\mathcal A_{1/2}({x}_{t})=1.4$. On the other hand, the new physics loop function in the limit of heavy LQ is $\mathcal A_{0}({x}_{s}\to 0) \to 1/3$. This limit is a good approximation for the interesting LQ mass range. For instance, $m_{\rm{LQ}}=200$\,GeV implies $\mathcal A_{0}({x}_{s}) = 0.35$. To sum up, the modified $h\to \gamma \gamma$ partial decay width normalized to the SM prediction is 
given by
\begin{equation}
\frac{\Gamma_{h\to\gamma\gamma}}{\Gamma_{h\to\gamma\gamma}^{\textrm{SM}}}=|\kappa_\gamma|^2,\;\; \textrm{where}\;\;
\kappa_\gamma=1-0.026~ \lambda_{S}\frac{v^{2}}{m_{\rm{LQ}}^{2}} ~ d(r_{S}) \underset{i}{\sum} Q_{S_i}^{2}~,
\label{eq:AAmodif}
\end{equation}
where the $\kappa_\gamma$ parameter can be extracted from fits to LHC Higgs data.

Similarly, the gluon-gluon fusion production of the Higgs boson is affected by the presence of LQ representation $S$ with  interactions given by Eq.~\eqref{eq:2Phi}. The parton level leading order $gg\to h$ cross section at partonic center of mass energy ($\sqrt{\hat s}$) is~\cite{Djouadi:2005gj,Chang:2012ta}
\begin{equation}
\hat \sigma_{gg\to h}=\sigma_{0}m_{h}^{2}\delta(\hat{s}-m_{h}^{2}),
\end{equation}
where
\begin{equation}
\sigma_{0}=\frac{G_{\mu}\alpha_{3}^{2}}{128\sqrt{2}\pi}\left|\frac{1}{2}\mathcal A_{1/2}({x}_{t})+\underset{i}{\sum}
\frac{\lambda_{S}}{2}\frac{v^{2}}{m_{\rm{LQ}}^{2}}C(r_{S})\mathcal A_{0}({x}_{s})\right|^{2}.
\end{equation}
Here, $\alpha_3$ is the strong coupling constant and $C(r_{S})$ is the index of color representation of $S$ which for the color triplet reads $C(\mathbf 3)=1/2$. The first term represents the dominant SM contribution from the top quark loop. Total hadronic cross section ($\sigma_{ggF}$) is obtained from the partonic one ($\hat \sigma_{gg\to h}$) after appropriate convolution with the partonic distributions functions.
That is, the total modified gluon-gluon fusion production cross section normalized to the SM prediction is
\begin{equation}
\frac{\sigma_{ggF}}{\sigma_{ggF}^{\textrm{SM}}}=|\kappa_g|^2,\;\; \textrm{where}\;\;
\kappa_g=1+0.24 ~ \lambda_{S}\frac{v^{2}}{m_{\rm{LQ}}^{2}} ~ N_{S_i} C(r_{S})~,
\label{eq:ggmodif}
\end{equation}
and $N_{S_i}$ is the number of $S_i$ components in the weak multiplet $S$.
As pointed out before, we consider heavy enough LQ such that the loop function $\mathcal A_{0}$ is in the 
decoupling limit, $\mathcal A_{0}({x}_{s}\to 0) \to 1/3$. In general, the gluon-gluon fusion production receives large corrections from higher order QCD effects~\cite{Heinemeyer:2013tqa}, however, the ratio in Eq.~\eqref{eq:ggmodif} is found to be less sensitive~\cite{Gori:2013mia}.

As a final remark, combining Eq.~\eqref{eq:AAmodif} with Eq.~\eqref{eq:ggmodif} we find
\begin{equation}
\kappa_g=1-9.4 (\kappa_\gamma -1) \frac{N_{S_i} C(r_{S})}{d(r_{S}) \underset{i}{\sum} Q_{S_i}^{2}}~,
\end{equation}
showing explicitly the correlation in the two observables induced by a single LQ representation. Interestingly, the correlation depends on the quantum numbers of the representation only. The correlated effects are expected to take place in other relevant loop induced Higgs processes, such as $h \to Z\gamma$ and $g g \to h h$. In principle, the global approach to Higgs measurements should allow for disentangling among different representation hypotheses.

\begin{table}
\begin{centering}
\begin{tabular}{ccc}
\hline
$SU(3) \times SU(2) \times U(1)$ &  SYMBOL  & $\lambda_S (v/m_{\rm{LQ}})^2$  \tabularnewline
\hline
$(\overline{\mathbf{3}},\mathbf{3},1/3)$ & $S_3$ & $0.02 \pm 0.25$ \tabularnewline 
$(\mathbf{3},\mathbf{2},7/6)$ & $R_2$ &  $-0.06 \pm 0.31$ \tabularnewline
$(\mathbf{3},\mathbf{2},1/6)$ & $\tilde{R}_2$  &  $0.11 \pm 0.38$  \tabularnewline
$(\overline{\mathbf{3}},\mathbf{1},4/3)$ & $\tilde{S}_1$ &   $-0.14 \pm 0.58$ \tabularnewline
$(\overline{\mathbf{3}},\mathbf{1},1/3)$ & $S_1$ &   $0.25 \pm 0.74$ \tabularnewline
$(\overline{\mathbf{3}},\mathbf{1},-2/3)$ &  $\bar{S}_1$ &   $0.16 \pm 0.78$ \tabularnewline
\hline
\end{tabular}
\par\end{centering}
\caption{The limits on the ``Higgs portal'' interactions ($\lambda_S v^2 /m_{\rm{LQ}}^2$ combination) from the latest available LHC Higgs data for several LQ representations.}
\label{tab:csresults}
\end{table}

\subsubsection{The Higgs fit}

The combination of Higgs coupling measurements performed by ATLAS and CMS experiments is reported in Refs.~\cite{ATLAS-CONF-2015-044, CMS:2015kwa}. Among other analyses, the Collaborations perform a fit assuming that there is no large modification of the tree level Higgs couplings and that the new physics gives a non-standard contribution to the loop-induced $h g g$ and $h \gamma \gamma$ couplings. The Collaborations report the combined best fit, $68\%$, and $95\%$ C.L.\ intervals in $(\kappa_{\gamma}\equiv \sqrt{\Gamma_{h\to\gamma\gamma}/\Gamma^{SM}_{h\to\gamma\gamma}},\kappa_{g}\equiv \sqrt{\Gamma_{h\to g g}/\Gamma^{SM}_{h\to g g}} )$ plane. Digitizing the plot and parametrizing the likelihood with the Gaussian form we find
\begin{equation}
\kappa_\gamma=1.045\pm0.090~, \;\; \kappa_g = 1.017\pm0.094~,
\label{eq:LHChigg}
\end{equation}
with the correlation coefficient $\rho=-0.35$.

Referring to the discussion from the previous section, the dominant impact of LQs on Higgs data is expected to be the modification of the loop-induced Higgs couplings. Thus, the results of the fit reported above can be used to constrain the size of LQ interactions to the Higgs boson via the Higgs portal operator. In Table~\ref{tab:csresults} we show the preferred range for $\lambda_S (v/m_{\rm{LQ}})^2$ for all scalar LQ representations.

\begin{figure}
\centering
\includegraphics[scale=0.7]{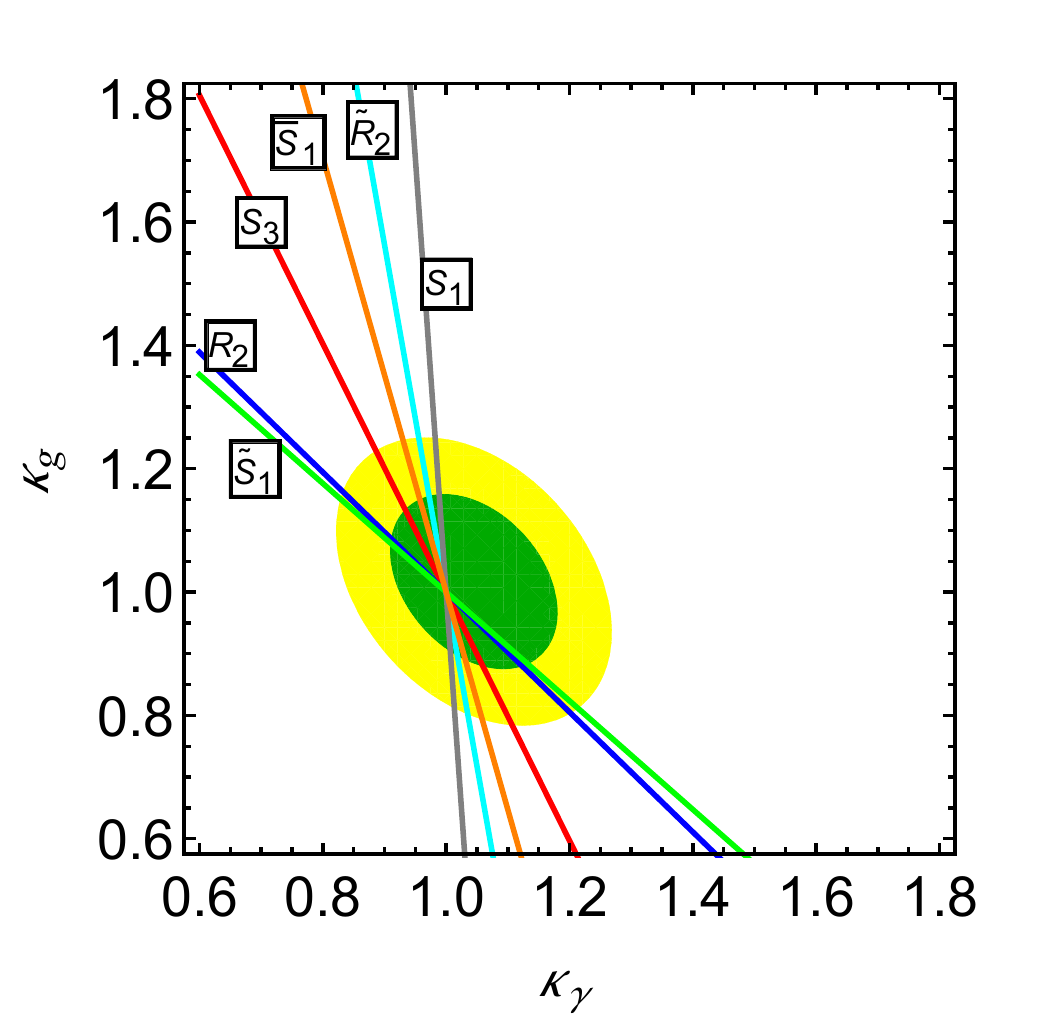} \;\;
\caption{The preferred region from the fit to the latest LHC Higgs signal strength measurements at $68\%$ and $95\%$ C.L.\ in $(\kappa_{\gamma}\equiv \sqrt{\Gamma_{h\to\gamma\gamma}/\Gamma^{SM}_{h\to\gamma\gamma}},~\kappa_{g}\equiv \sqrt{\Gamma_{h\to g g}/\Gamma^{SM}_{h\to g g}} )$ plane is shown in green and yellow, respectively. The correlations induced by single LQ representations are shown in various curves.}
\label{fig:coloredplot}
\end{figure}

Furthermore, a single LQ representation induces a correlation among $\kappa_\gamma$ and $\kappa_g$ parameters. The correlation curves in $(\kappa_\gamma,\kappa_g)$ plane are shown in Fig.~\ref{fig:coloredplot} for several LQ representations. The constraints from the LHC reported in Eq.~\eqref{eq:LHChigg}, are shown in green and yellow for $68\%$ and $95\%$ C.L.s regions.

Let us summarize the results of the fit. In particular, the decoupling limit to the SM ($\lambda_S \to 0$ and (or)  $m_{\rm{LQ}} \to \infty$) is compatible with the current Higgs data. Assuming $\lambda_S \sim 1$, masses around electroweak condensate scale are being probed. The constraints are mainly driven by the measurements of the gluon-gluon fusion production. The approximate relation derived from Eq.~\eqref{eq:ggmodif},
\begin{equation}
\delta \kappa_g = 0.24 N_{S_i} C(r_S) ~ \delta\left(\lambda_S \frac{v^2}{m_{\rm{LQ}}^2} \right)~,
\end{equation}
allows us to estimate the expected precision achievable in future. In particular, the projections for Higgs couplings determinations at HL-LHC with full statistics of $3000$\,fb$^{-1}$ of data are computed in~\cite{Dawson:2013bba,CMS:2013xfa}. The precision on $\kappa_g$ and $\kappa_\gamma$ is expected to be $3$--$5\%$ and $2$--$5\%$ per-experiment, respectively. The two numbers in the estimates are given for two different assumptions on the theoretical uncertainty, namely, if it stays the same or is reduced by half. To conclude, compared to the present precision on $\lambda_S \frac{v^2}{m_{\rm{LQ}}^2}$, the situation is expected to improve by a factor of $4.4$--$2.7$ at the HL-LHC.

\subsubsection{Predictions for $h\to Z\gamma$}

Higgs decay to $Z$ and $\gamma$ is  generated at one-loop level in the SM. Being extremely suppressed, this decay has not yet been observed by the experimental collaborations. The present limits from ATLAS~\cite{Aad:2014fia} and CMS~\cite{Chatrchyan:2013vaa} experiments are $|\kappa_{Z\gamma}|<3.3$ and $|\kappa_{Z\gamma}|<3.1$ at $95\%$~C.L., respectively. However, once it is measured, it will constitute a complementary probe of new physics compared to $h\to \gamma \gamma$.
The partial decay width for $h\to Z\gamma$ in the presence of a scalar LQ representation $S$ is given by~\cite{Carena:2012xa}
\begin{equation}
\begin{split}
&\Gamma_{h\to Z\gamma} = \frac{G_F^2 m_W^2 \aem}{64\pi^4} m_h^3\left( 1-\frac{m_Z^2}{m_h^2} \right)^3 \times \\
&|{\cos\theta_{W}\mathcal C_{1}({x}^{-1}_{W},y_{W})+\frac{2(1-(8/3)\sin^{2}\theta_{W})}{\cos\theta_{W}}\mathcal C_{1/2}({x}^{-1}_{t},y_{t})} \\
&+{\frac{\sin\theta_{W}}{2} \sum_i{\frac{\lambda_{S}v^2}{m_{\rm{LQ}}^{2}}} g_{Z S_i S}d(r_{S})Q_{S_i} \mathcal C_{0}({x}^{-1}_{s},y_{s})} |^{2},
\end{split}
\label{hzgamma}
\end{equation}
where $y_{i}\equiv{4m_{i}^{2}}/{m_{Z}^{2}}$ and the coupling of
$S$ to $Z$ boson is given in units of $e>0$, that is $g_{Z sS_iS_i}=2({I_{S_i}^{3}-Q_{S_i}\sin^{2}\theta_{W}})/{\sin2\theta_{W}}$ where $\theta_W$ is the weak mixing angle. Also,  $I_{S_i}^{3}$ represents the value of the weak isospin of $S_i$ and in Eq.~\eqref{hzgamma} we sum over all $i$ within the given weak multiplet $S$.  The relevant one-loop functions are given by:
\begin{align*}
\mathcal C_{0}(x,y) =& I_{1}(x,y), \\
\mathcal C_{1}(x,y) =& 4(3-\tan^{2}\theta_{W})I_{2}(x,y)\\&+\left((1+2x^{-1})\tan^{2}\theta_{W}-(5+2x^{-1})\right)I_{1}(x,y),  \\
\mathcal C_{1/2}(x,y) =& I_{1}(x,y)-I_{2}(x,y), 
\end{align*}
where
\begin{align*}
I_{1}(x,y) =& \frac{x y}{2(x-y)}+\frac{x^{2} y^{2}}{2(x-y)^{2}}\left(f(x^{-1})-f(y^{-1})\right)\\
&+\frac{x^{2}y}{(x-y)^{2}}\left(g(x^{-1})-g(y^{-1})\right),\\
I_{2}(x,y) =& -\frac{x y}{2(x-y)}\left(f(x^{-1})-f(y^{-1})\right),\\
g(x) =& \sqrt{x^{-1}-1}\arcsin\sqrt{x},\;\;\;\;x\geq1.
\end{align*}
The SM loop functions induced by the $W$ boson and top quark are $5.8$ and  $-0.3$, respectively.
Using the above relations, we find the deviations in $h\to Z\gamma$ partial decay width, normalized to the SM prediction, to be
\begin{equation}
\frac{\Gamma_{h\to Z \gamma}}{\Gamma_{h\to Z\gamma}^{SM}}=|\kappa_{Z \gamma}|^2,\;\; \textrm{where}\;\;
\kappa_{Z \gamma}=1+0.018  \lambda_{S} \frac{v^{2}}{m_{\rm{LQ}}^{2}}  d(r_{s}) \underset{i}{\sum} Q_{S_i}(I^3_{S_i}-0.23Q_{S_i}) .
\label{eq:gagamodif}
\end{equation}
The correlation in $h\to Z \gamma$ and $h\to \gamma \gamma$ due to single LQ representation is given by
\begin{equation}
\kappa_{Z \gamma}=1-0.69 (\kappa_{\gamma} -1) \frac{\underset{i}{\sum} Q_{S_i}(I^3_{S_i}-0.23Q_{S_i})}{ \underset{i}{\sum} Q_{S_i}^{2}}.
\end{equation}
This relation can, in principle, be used to discriminate among different representation hypotheses.

\subsubsection{Flavor violating Higgs boson decays}

\label{sec:LQ}
\begin{figure}[t]
\centering
\centering\begin{tabular}{ccc}
\includegraphics[scale=.8]{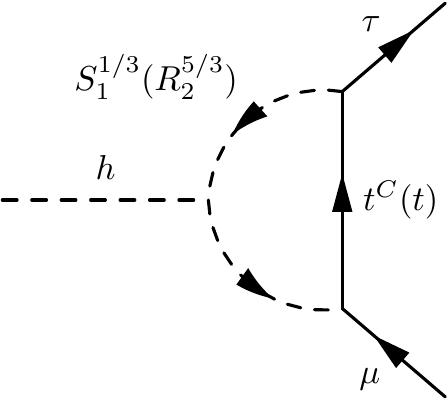}  && \includegraphics[scale=.8]{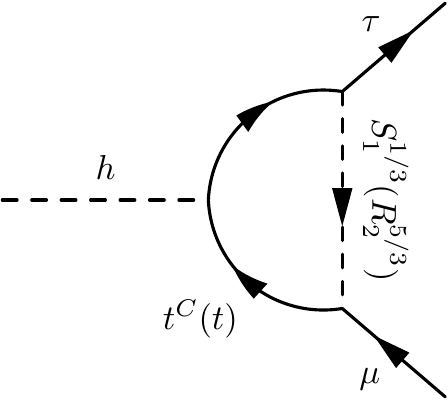}\\\\
\includegraphics[scale=.9]{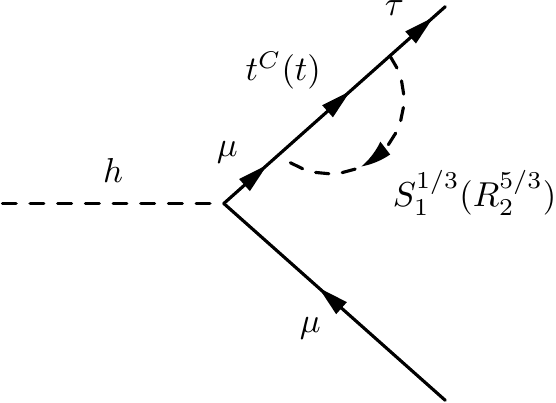}  && \includegraphics[scale=.9]{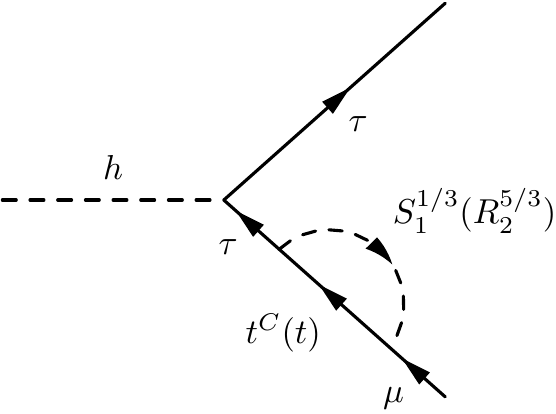}\\
\end{tabular}
\caption{Feynman diagrams for LQ contribution to $h\to \tau \mu$~\cite{Dorsner:2015mja}.}
\label{fig:LQdiag}
\end{figure}
Flavor violating (FV) Higgs interactions are negligible in the SM. Finding these at the LHC would be an indisputable evidence of the physics beyond the SM. Particular attention is given to $t\to h q$, with $q=u,c$ and $h \to \tau \ell$, with $\ell = e, \mu$. The reason is that the signal rates expected from new physics can easily be within the reach of the LHC and yet in agreement with the low-energy data. See, for example, Refs.~\cite{Harnik:2012pb,Blankenburg:2012ex,Greljo:2014dka,Goudelis:2011un,McKeen:2012av,Arhrib:2012mg,Kopp:2016rzs,Buschmann:2016uzg}.

Interestingly enough, the CMS Collaboration has recently reported $2.4\,\sigma$ excess in the search for $h\to \tau \mu$~\cite{Khachatryan:2015kon}. The best fit for the branching ratio of $h\to\tau \mu $ (more precisely,  $h \to \tau^- \mu^+ + \tau^+ \mu^-$), assuming SM Higgs production, is found to be 
\begin{equation}
\mathcal{B}(h\to \tau \mu)=\left(0.84^{+0.39}_{-0.37}\right) \% \,.
\label{eq:Br}
\end{equation}
A scalar LQ can, in principle, induce $h \to \tau \mu$ decay via quark--LQ penguin diagrams~\cite{Dorsner:2015mja,Baek:2015mea}. The helicity flip is required on one of the fermion lines in these diagrams, thus, only LQs that couple to top quark and charged leptons are suitable candidates to accommodate  the excess. Furthermore, both left and right chiralities of leptons and top quark are required to couple to the LQ. The relevant diagrams are shown in Fig.~\ref{fig:LQdiag}. These involve the contribution from the top-LQ coupling, as well as, the ``Higgs portal'' operator (see Eq.~\eqref{eq:2Phi}).

\begin{figure}
\begin{centering}
\includegraphics[scale=0.43]{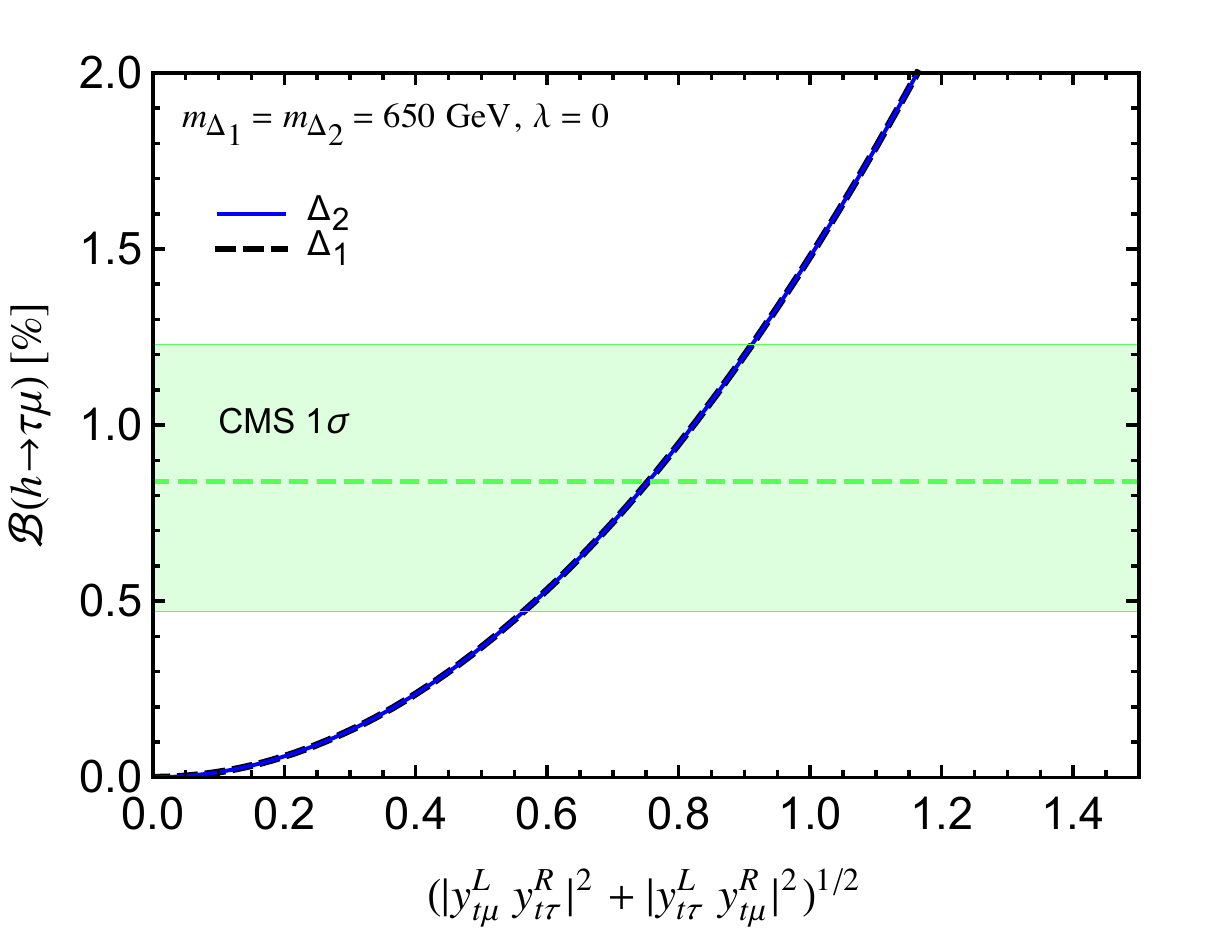} \;
\includegraphics[scale=0.45]{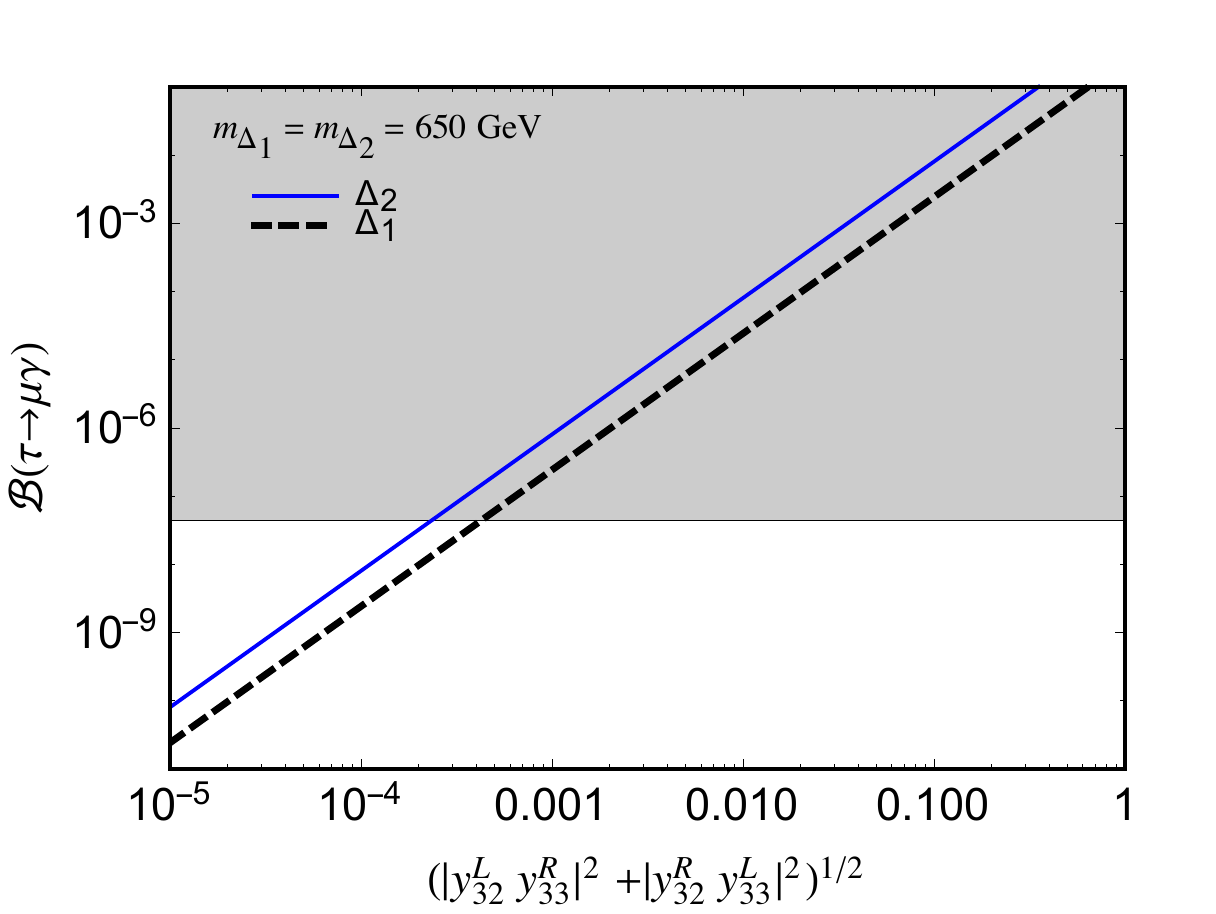}
\par\end{centering}
\caption{Higgs branching ratio to $\tau\mu$ final state in the presence of the scalar LQs (left panel). The corresponding $\tau\to\mu\gamma$ constraints (right panel). See Ref.~\cite{Dorsner:2015mja} for details.}
\label{fig:-plot-2}
\end{figure}

Let us discuss the impact of the particular LQ representation $S_1=(\overline{\mathbf{3}},\mathbf{1},1/3)$.
The interactions of $S_1$ with the SM fermions are given in Eq.~\eqref{eq:main_S_1}. The relevant interactions for this purpose, after expanding the $SU(2)$ indices, are
\begin{equation}
\label{eq:main_1}
\mathcal{L}_{S_1} \supset y^L_{ij}\bar{u}_{L}^{i} e_{L}^{C\,j} S_1^* + y^R _{ij}\bar{u}_{R}^{i} e_{R}^{C\,j} S_1^* +\textrm{h.c.}~,
\end{equation}
where $y^{L*} \equiv V^T y_1^{LL}$ and $y^{R*} \equiv  y_1^{RR}$ from Eq.~\eqref{eq:main_S_1}.
All fields in the above Lagrangian (up quarks, charged leptons and LQ) are specified in the mass eigenstate basis. Finally, the effective $h$--$\tau$--$\mu$ couplings at one-loop are obtained by an explicit computation of the diagrams shown in Fig.~\ref{fig:LQdiag} (see Ref.~\cite{Dorsner:2015mja})
\begin{align}
y_{\tau\mu \,(\mu\tau)}  &= -\frac{N_c}{16\pi^2} \frac{m_t}{v} g_1(\lambda, m_{S_1}) \,y^{R}_{t \mu} y^{L*}_{t \tau} \,\,\left (y^{R}_{t \tau} y^{L*}_{t \mu}\right)\,,
\end{align}
where $N_c =3$ is the number of colors and $\lambda$ is the Higgs portal coupling. The relevant loop function is
\begin{equation}
\begin{split}
g_1(\lambda,m_{S_1}) &=(m_{S_1}^{2}+m_{t}^{2})C_{0}(0,0,m_{h}^{2},m_{t}^{2},m_{S_1}^{2},m_{t}^{2})+B_{0}(m_{h}^{2},m_{t}^{2},m_{t}^{2})\\
&\phantom{=}-B_{0}(0,m_{t}^{2},m_{S_1}^{2})+\lambda v^{2}C_{0}(0,0,m_{h}^{2},m_{S_1}^{2},m_{t}^{2},m_{S_1}^{2})\,,
\end{split}
\end{equation}
where $B_0$ and $C_0$ are the well-known Passarino--Veltman functions. Finally, the  $h \to \tau \mu$ partial decay width is given by
\begin{equation}
\label{eq:LQ1-BR}
\Gamma(h\to\tau\mu)=\frac{9 m_{h}m_{t}^{2}}{2^{13}\pi^{5}v^{2}}\left(\left|y_{t\mu}^{L}y_{t\tau}^{R}\right|^{2}+\left|y_{t\tau}^{L}y_{t\mu}^{R}\right|^{2}\right)\left|g_1(\lambda,m_{S_1})\right|^{2}\,.
\end{equation}
Shown in Fig.~\ref{fig:-plot-2} (left panel) is the expected Higgs branching ratio to $\tau\mu$ final state, $\mathcal{B}(h\to\tau\mu$), as a function of the LQ--fermion couplings assuming the total Higgs decay width to be SM-like. The mass of $S_1$ LQ is set to $650$\,GeV, consistent with current direct search limits at the LHC~\cite{ATLAS:2013oea,CMS:2014gha} and the Higgs portal coupling is turned off. The CMS measurement is shown in the same plot.

Conclusively, to fit the CMS excess, the relevant LQ couplings are required to be $\mathcal{O}(1)$. On the other hand, the same state also contributes to the $\tau \to \mu \gamma$ with the same coupling combinations present in $h \to \tau \mu$. Referring to Section~\ref{sec:tamuga} for an extended discussion of the effect, we find
\begin{equation}
 \mathcal{B}(\tau \to \mu \gamma) = \frac{\aem m_\tau^3}{2^{12} \pi^4 \Gamma_\tau}\,\frac{m_t^2}{m_{S_1}^4} \, h_1(x_t)^2 \, \left(\left|y_{t\mu}^{L}y_{t\tau}^{R}\right|^{2}+\left|y_{t\tau}^{L}y_{t\mu}^{R}\right|^{2}\right)\,.
\end{equation}
where
\begin{equation}
h_1(x) = \frac{7-8x +x^2+(4+2x)\ln x}{(1-x)^3}\,,
\label{eq:48}
\end{equation}
with $x_t = m_t^2/m_{S_1}^2$.

Shown in Fig.~\ref{fig:-plot-2} (right) is $\mathcal{B}(\tau\to \mu \gamma)$ as a function of LQ--fermion couplings again for $m_{S_1} = 650\,\mathrm{GeV}$. The parameter space colored in gray is excluded by experiment. To sum up, a single LQ representation cannot simultaneously explain the CMS $h\to \tau \mu$ excess while still remaining in agreement with low-energy data from LFV $\tau$ decays. (For a possibility to address the $h \to \tau^\mp \mu^\pm$ excess with two LQs --- $S_1$ and $R_2$ --- while avoiding the $\tau \to \mu \gamma$ constraint through cancellation of relevant contributions see Ref.~\cite{Cheung:2015yga}.)

It is illustrative to study the impact of a non-zero Higgs portal coupling in the model with $S_1$ LQ representation. Again we take $m_{S_1}=650$\,GeV for the numerical benchmark. The loop function dependence on the portal coupling in the numerical form is given by
\begin{equation}
g_1(\lambda, m_{S_1} = 650\,\mathrm{GeV})=-(0.26+0.12 \lambda)\,.
\end{equation}
In principle, a positive large $\lambda$ could reduce the required size of the LQ-fermion couplings and yield sizable $h\to \tau \mu$ rates while keeping $\tau \to \mu \gamma$ under control. This, however, turns out to be disfavored by the Higgs data. 
As discussed before, the Higgs portal coupling is constrained from the global fit to Higgs data mainly due to its contributions to $h\to\gamma\gamma$ and to gluon-gluon fusion Higgs production. We can directly use the limit derived in Table~\ref{tab:csresults} which translates into $g_1(\lambda, m_{S_1})= 0.6\pm 0.9$. This, in turn, is not enough to make $\left|y_{t\mu}^{L}y_{t\tau}^{R}\right|^{2}+\left|y_{t\tau}^{L}y_{t\mu}^{R}\right|^{2}$ below $\tau \to \mu \gamma$ constraints and yet keep $h\to\tau\mu$ at the observed level.

Let us conclude this section with  an explicit example of a phenomenological viable model in which $\tau\to \mu \gamma$ is fine-tuned to zero, while accommodating the $h\to\tau\mu$ excess. Following Ref.~\cite{Dorsner:2015mja}, extra scalar LQ $S_1=(\overline{\mathbf{3}},\mathbf{1},1/3)$ and a vector-like top quark partner $T'_L, T'_R=(\bf{3},\bf{1},2/3)$ are added to the SM. The top partner mixes with the SM top quark.  After rotating to the mass basis, $\tau\to\mu\gamma$ decay amplitude will receive two competing contributions from both, $t$ and $T$, at one loop level. This scenario requires large amount of fine-tuning in $\tau\to \mu \gamma$, while keeping $h\to\tau\mu$ at the observed level.


\subsection{Diphoton decays of heavy scalars in presence of leptoquarks}

After analyzing the first data sets from $13$\,TeV proton--proton collisions at the LHC, both ATLAS and CMS Collaborations have reported a tantalizing excess in the diphoton invariant mass spectrum around $m_S \sim 750$\,GeV~\cite{ATLAS-CONF-2015-081, CMS:2015dxe, ATLAS:Moriond, CMS:Moriond, CMS:2016owr}. Interpreting the excess in terms of a promptly produced s-channel resonance ($S$), the significance of the observed excess and the corresponding signal cross-section depend somewhat on the assumed width ($\Gamma_S$) and spin of the resonance. Even more importantly, the absence of significant deviations in this invariant mass region in 8\,TeV di-photon analyses singles out gluon fusion and/or heavy quark annihilation as the dominant production mechanisms, since these are expected to be sufficiently enhanced at higher collision energies~\cite{Franceschini:2015kwy}.  


For a narrow scalar (or pseudoscalar) resonance, produced dominantly via gluon-gluon fusion, the combined fit to 8\,TeV and 13\,TeV results yields a signal strength at 13\,TeV of~\cite{Kamenik:2016tuv}
\begin{equation}
\mu_{\rm{13\,TeV}} = (3.2\pm 0.7)\,\rm{fb}~,
\end{equation}
and roughly twice this value in the case of a wide resonance with $\Gamma_S/m_S \sim 0.06$ as preferred by the ATLAS results~\cite{ATLAS-CONF-2015-081}. 

The situation prompts a brief discussion on possible LQ effects in radiative decays of heavy color and EM neutral scalar resonances. Focusing for concreteness on  the observed excess, such a  resonance should couple to photon and gluon field strengths, $F_{\mu\nu}$ and $G^{A}_{\mu\nu}$, $A=1,\ldots,8$, via effective interactions~\cite{Buttazzo:2015txu,Franceschini:2015kwy},
\begin{equation}
\mathcal{L}^{\rm eff} \supset c_{G} \frac{\alpha_3}{12 \pi m_S} S  \, G_{\mu \nu}^A G^{A,\mu\nu} + c_{\gamma \gamma} \frac{\aem}{4 \pi m_S} S F_{\mu \nu} F^{\mu \nu}~.
	\label{eq:effLagrMass}
\end{equation}
Following analogous discussions for the Higgs boson, the relevant effective interactions can be generated at one-loop level by new heavy colored and charged particles with the renormalizable couplings to the resonance. Vector LQs that appear in the solutions of $B$ physics anomalies~\cite{Murphy:2015kag} or diboson excess~\cite{deBlas:2015hlv} have been advocated to provide enough contribution to fit the signal. Scalar LQs, on the other hand, have been suggested for the similar purpose in Refs.~\cite{Bauer:2015boy,Chao:2015nac,Hati:2016thk,Dorsner:2016ypw}. The self-consistency analysis of two particular scalar LQ scenarios has been subsequently presented in Ref.~\cite{Staub:2016dxq}. The scenarios under scrutiny comprise a model with a single LQ~\cite{Bauer:2015boy} and a model with two LQs~\cite{Chao:2015nac}. Diphoton excess and $B$ physics anomalies as unavoidable manifestations of the new strongly interacting sector with vector-like confinement, have been discussed in Ref.~\cite{Buttazzo:2016kid}. The main contribution is due to the lowest-lying resonances spectrum, pseudo Nambu-Goldstone bosons and vectors, which necessarily contain scalar and vector LQs, respectively.


\section{Collider Searches}
\label{sec:COLLIDERS}
The most direct way to probe for LQs is to produce them at colliders and to investigate their subsequent decay patterns, i.e., decay channels and associated branching fractions. LQs have accordingly been searched for and studied in the context of $e^+ e^-$, $e p$, $p \bar{p}$ and $p p$ colliders. We opt to present these studies in this particular order concentrating especially on the LHC case. The $e \gamma$ and $\gamma p$ collider signatures will also be discussed for completeness. However, before we turn our attention towards issues that depend on intricacies of particular production mechanisms we address the question of LQ decay that is common for all colliders. 


\subsection{Leptoquark decays}
Once produced LQs should decay. The most general expression for the width $\Gamma_S$ of the two-body decay of a scalar leptoquark $S$ of mass $m_{\mathrm{LQ}}$ into a quark $q$ of mass $m_q$ and a lepton $l$ of mass $m_l$ is
\begin{equation}
\Gamma_S=|y|^2 \frac{\sqrt{m^4_{\mathrm{LQ}}+m^4_l+m^4_q-2 m^2_{\mathrm{LQ}}m^2_l-2 m^2_{\mathrm{LQ}}m^2_q-2 m^2_q m^2_l}}{16 \pi m_{\mathrm{LQ}}^3} (m^2_{\mathrm{LQ}}-m^2_l-m^2_q), 
\end{equation}
where $y$ denotes particular Yukawa coupling of the scalar LQ to a given quark--lepton pair in the mass-eigenstate basis in accordance with our notation in Section~\ref{sec:INTRODUCTION}. (Recall, our normalization is such that $\mathcal{L} \supset y  \, \bar q \, \ell \, S$.) The two-body decay width $\Gamma_V$ of a vector leptoquark $V$ is $\Gamma_V=(2/3) \Gamma_S$.

If we neglect lepton and quark masses altogether the partial decay width for scalar (vector) LQ simplifies to
\begin{equation}
\label{eq:LQ_gamma}
\Gamma_S=\frac{|y|^2}{16 \pi}m_{\mathrm{LQ}} \qquad \qquad \left(\Gamma_V=\frac{|x|^2}{24 \pi}m_{\mathrm{LQ}}\right), 
\end{equation}
where $x$ denotes particular coupling strength of the vector LQ interaction with a given quark--lepton pair.
 
Total decay width is simply the sum of all partial widths. The decay length is thus inversely proportional to both the sum of squares of relevant couplings and the LQ mass. This usually results in the following implicit assumption when it comes to LQ decays at colliders. Namely, the coupling(s) $y$ ($x$) are taken to be sufficiently large to insure prompt decay of a given scalar (vector) LQ in the detector. One can estimate that this puts a lower bound of about $|y| \sim |x| \geq 10^{-7}$ for $m_{\mathrm{LQ}}=500$\,GeV. This, again, is an underlying assumption of all LQ searches to date.

One also expects the strength of the couplings of leptoquarks to the SM fermions to be bounded from above purely on theoretical grounds. Namely, a requirement that a self-consistent theory be perturbative places an upper bound on Yukawa (gauge) couplings that is of the order of $4 \pi$ ($\sqrt{4 \pi}$). The perturbativity bound is clearly not very stringent, especially when compared with existing bounds from flavor physics, and we do not refer to it in the rest of the discussion.

Let us now mention one model building issue that is related to LQ decay. This concerns all those LQs that couple to a quark multiplet and/or a lepton multiplet that transforms non-trivially under $SU(2)$ factor of the SM gauge group. Let us assume, for example, that LQ $R_2$ has only one Yukawa coupling different from zero. (Recall, $R_2$ is an $SU(2)$ doublet. It thus needs to couple to another doublet to yield a gauge singlet. See, for example, Eqs.~\eqref{eq:main_R_2} and \eqref{eq:main_R_2_a}.) Let that coupling be $y^{RL}_{2\,11}=y$, i.e., $y^{RL}_{2\,ij}=\delta_{1i}\delta_{1j}y$ and $y^{LR}_{2\,ij}=0$ for $i,j=1,2,3$, in order to have $R_2^{5/3}$ to be purely of the first generation kind. The decay width of $R_2^{5/3}$ is then $\Gamma(R_2^{5/3})=(|y|^2m_{\mathrm{LQ}})/(16 \pi)$, whereas the other charged component of $R_2$, i.e., $R_2^{2/3}$, has $\Gamma(R_2^{2/3})= \left(|y|^2m_{\mathrm{LQ}} \sum_i |U|^2_{1i}\right)/(16 \pi)=\Gamma(R_2^{5/3})$. We see that this simple ansatz predicts, through unitarity of PMNS matrix, equal decay widths for both $R_2^{5/3} \rightarrow  e u$ and $R_2^{2/3} \rightarrow  \nu u$ channels, where $u$ is an up quark.

It is often assumed that LQ under investigation has two distinct types of decay channels. One possible signature corresponds to the LQ decay into a charged lepton of particular flavor ($l$) and a jet ($j$) and the other signature is due to the LQ decay into a neutrino ($\nu$) and a jet. It is furthermore customary to denote corresponding branching ratios as $\beta$ and $(1-\beta)$, respectively. This assumption needs to be handled with care. For example, $S^{1/3}_3$ can simultaneously couple to $lj$ and $\nu j$ pairs. (See Eq.~\eqref{eq:main_S_3_a}.) However, these two sets of couplings are related via CKM and PMNS mixing matrices. In other words, $\beta$ is fixed for a given Yukawa coupling ansatz. The same is true for $S_1$, $U^{2/3}_3$, and $U_1$ LQs. Experimental observation of the branching fractions of the LQ decays into these final states might thus allow for determination of flavor mixing parameters at low energies in a similar manner to what was initially proposed for $B$ and $L$ violating LQ decays in proton decay experiments~\cite{DeRujula:1980qc}. 

LQs that can have both $lj$ and $\nu j$ decay signatures but with couplings that are not correlated through the known mixing parameters are $\tilde{R}_{2}^{2/3}$, $\tilde{R}_{2}^{-1/3}$, $R_{2}^{2/3}$, $V_{2}^{1/3}$, and $V_{2}^{-2/3}$, where we consider only those cases when $\nu$ corresponds to neutrinos of left-chirality.

Note that it is entirely possible to have LQs produced in ultrahigh-energy neutrino interactions. These LQs could then be accessible to neutrino telescopes. This idea has been investigated in several works~\cite{Berezinsky:1985yw,Robinett:1987ym,Doncheski:1997it,Carena:1998gd,Anchordoqui:2006wc}. For example, an analysis of the possibility to probe parameter space of second and third generation LQs at the IceCube experiment~\cite{Ahrens:2002dv} has been put forth in Ref.~\cite{Anchordoqui:2006wc}. We will not comment further on this possibility but turn to discuss topics that are related to the collider specific physics instead.   


\subsection{$e \gamma$ collider signatures}

Production of LQs has been discussed within the context of $e \gamma$ machines~\cite{Hewett:1989cs,Rizzo:1990hx,CiezaMontalvo:1992bs,Nadeau:1993zv,Eboli:1993qx,Belanger:1993zi,Doncheski:1993ni,Doncheski:1994cr,Cuypers:1995ax,Aliev:1996qj,Doncheski:1996dp,Doncheski:1996un}. For example, the possibility to have single production of vector LQs through the process $\gamma e^- \rightarrow \mathrm{LQ} q$ was investigated in Ref.~\cite{CiezaMontalvo:1992bs} for the case of LQ of electric charge $|Q|=2/3$. This LQ needs to couple to an electron in order to be produced and accordingly couples to the down-type quarks $d$, $s$, and $b$. 

\begin{figure}[tbp]
\centering
\includegraphics[scale=0.8]{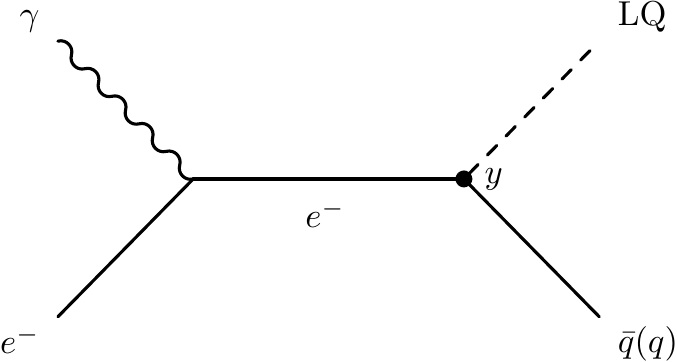}
\includegraphics[scale=0.8]{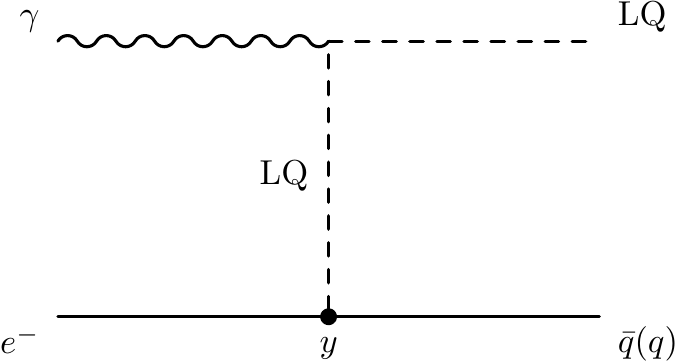}\newline
\newline
\includegraphics[scale=0.8]{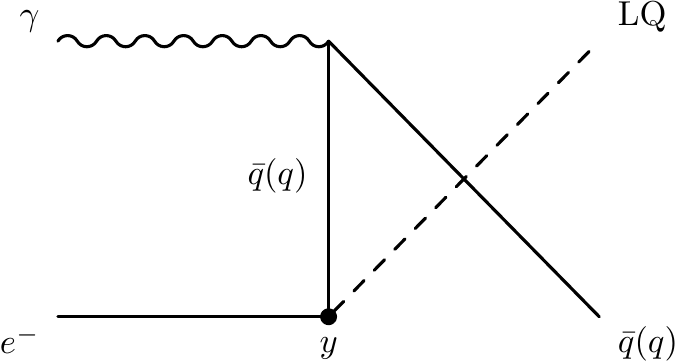}
\caption{\label{fig:e_gamma} Diagrams for a single LQ production at $e \gamma$ collider. $y$ represents relevant Yukawa coupling. }
\end{figure}

The possibility for the detection of scalar LQs at
$e \gamma$ colliders through the single production mechanism has been presented in Ref.~\cite{Nadeau:1993zv}. The authors of Ref.~\cite{Nadeau:1993zv} have shown that the detection of real LQs would be possible even for the case when the LQ masses are as high as the center-of-mass energy of the collider and Yukawa couplings are as small as $10^{-5}$. This has been done considering two particular values --- $\sqrt{s}=500$\,GeV and $\sqrt{s}=1$\,TeV --- for the center-of-mass energy. For relevant differential cross-section for the real scalar LQ production as a function of the LQ electric charge see Eq.~(2) of Ref.~\cite{Nadeau:1993zv}. We reproduce in Fig.~\ref{fig:e_gamma} diagrams for single LQ production at $e \gamma$ machine considered in Ref.~\cite{Nadeau:1993zv}, where the final state is $q$ ($\bar{q}$) for the LQ with $F=0$ ($F=-2$). 

There has been one experimental search at LEP $e^+ e^-$ collider that was performed with the OPAL detector for first generation scalar and vector LQs that relied on the process $eq \rightarrow \mathrm{LQ} \rightarrow e q (\nu q)$, where the quark in the initial state originated from a photon that in turn originated from the beam electron. The search was performed at the center-of-mass energy $\sqrt{s_{ee}}$ of $189$\,GeV with the $160$\,pb$^{-1}$ of data collected~\cite{Abbiendi:2001aj}. Under the assumption that the coupling of the LQ is larger than $\sqrt{4 \pi \aem}$ the OPAL Collaboration established lower limits on the LQ mass that ranged from $121$\,GeV to $175$\,GeV ($149$\,GeV to $188$\,GeV) as a function of the branching ratio $\beta$ of the scalar (vector) LQ state. 

\subsection{$\gamma p$ collider signatures}
The leptoquark production has been studied within the context of $\gamma p$ colliders~\cite{Atag:1994hk,Atag:1994np,Aliev:1995xm,Aydin:2000yn}. For example, pair (single) production of scalar leptoquarks in the collisions of photons and proton beams has been investigated in Ref.~\cite{Atag:1994hk} (Ref.~\cite{Atag:1994np}). The production of both scalar and vector LQs in the same environment has been pursued in Ref.~\cite{Aliev:1995xm}. Also, the possibility to test R-parity violation at TeV scale $\gamma p$ colliders has been investigated in Ref.~\cite{Aydin:2000yn}.

\subsection{$e^+ e^-$ collider signatures}

\begin{figure}[tbp]
\centering
\includegraphics[scale=0.8]{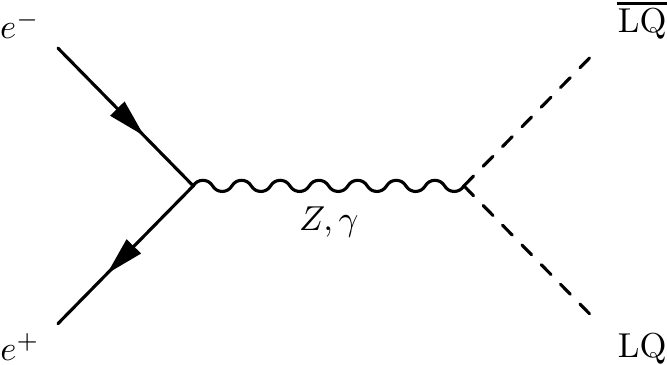} \hspace{1cm}
\includegraphics[scale=0.8]{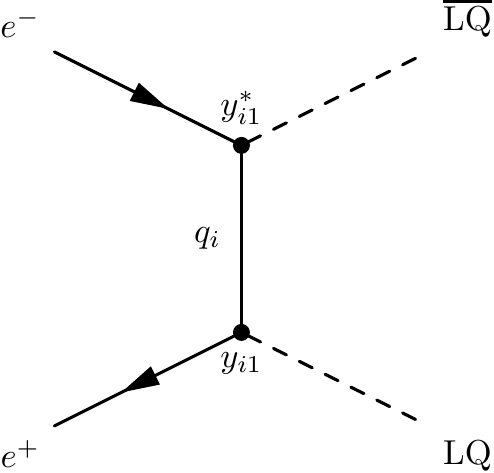}\newline
\newline
\includegraphics[scale=0.8]{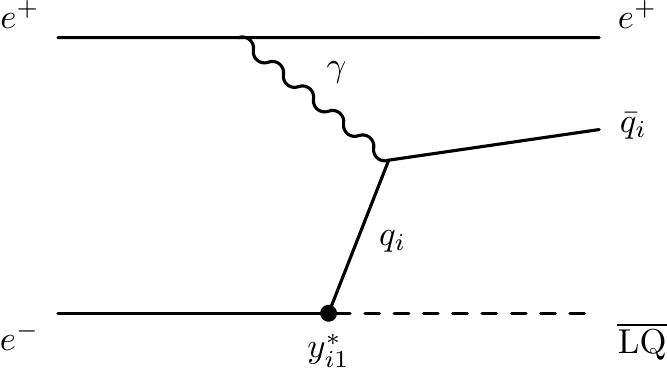}
\caption{\label{fig:LQ_e_p} Feynman diagrams relevant for a pair production (top row) and a single production (bottom row) of scalar LQs through $e^+ e^-$ annihilation. Here, $y_{i1}$, $i=1,2,3$, represents appropriate Yukawa coupling of a quark ($q_i$) and an electron with an LQ.}
\end{figure}

Production of scalar and vector LQs through $e^+ e^-$ annihilation has a long history~\cite{Mohapatra:1984aq,Hewett:1987bh,Tracas:1988pm,Hewett:1989cs,Schaile:1987va,Djouadi:1991us,Blumlein:1991gw,CiezaMontalvo:1992bs,Blumlein:1992ej,Choudhury:1994he,Blumlein:1994tu,Aliev:1996qj,Blumlein:1996gb}. Relevant diagrams for a pair production of scalar LQs are shown in the top row of Fig.~\ref{fig:LQ_e_p}. The pair production is driven by the LQ couplings to electroweak gauge bosons ($\gamma$ and $Z$) as well as Yukawa couplings of LQ to matter. The underlying assumption that a leptoquark couples to an electron is a necessity for the latter production scenario. Note that the term ``electron'' is used generically to refer to both electrons and positrons, if not otherwise stated.

To obtain Feynman diagrams relevant for a pair production of vector LQs through  $e^+ e^-$ annihilation one needs to replace scalar LQs with vector LQs and Yukawa couplings $y$ with couplings $x$ in Fig.~\ref{fig:LQ_e_p}. 

The most general expressions for the pair production cross sections and differential distributions for both scalar and vector LQs can be found in Ref.~\cite{Blumlein:1992ej} under the assumption that the LQ couplings are family diagonal. (See also Refs.~\cite{Schaile:1987va,Djouadi:1991us}.) The contribution that corresponds to photon exchange was derived in Ref.~\cite{Cabibbo:1961sz}. One possible effect of the LQ exchange in $e^+ e^-$ annihilation is on the production of quark-antiquark pairs. This effect has been studied in Ref.~\cite{Kalinowski:1997fk}. LQs can also be produced singly at $e^+ e^-$ colliders. For the studies that investigate this particular possibility see, for example, Refs.~\cite{Belanger:1993zi,Papadopoulos:1997mi}. 

The possibility to test the LQ Yukawa couplings to fermion fields in the context of an $e^- e^-$ mode of a linear collider through  production of non-conjugate LQ pairs has been investigated in Ref.~\cite{Cuypers:1996xg}. For typical Feynman diagrams for production of the non-conjugate leptoquark pairs in $e^- e^-$ collisions see, for example, Fig.~1 of Ref.~\cite{Cuypers:1996xg}.

Leptoquarks can contribute towards the process $e^+ e^- \rightarrow q \bar{q}$ through a $t$-channel exchange. See, for example, Fig.~1 of Ref.~\cite{Berger:1996gm} for the relevant Feynman diagram. It is thus possible to use this process, i.e., the associated overall rate and the angular distribution, to look for the LQ presence indirectly. This possibility has been studied within a context of the $e^+ e^-$ and $\mu^+ \mu^-$ machines in Ref.~\cite{Berger:1996gm}. The authors of Refs.~\cite{Berger:1996gm,Rindani:2004ue} also investigated a possibility to use polarized beams to probe the chiral nature of the LQ couplings.


\subsubsection{Searches at $e^+ e^-$ colliders}

First experimental results on the LQ production at $e^+ e^-$ colliders have been generated at the storage ring PETRA (Positron-Elektron-Tandem-Ring-Anlage) with CELLO~\cite{Behrend:1986jz} and JADE~\cite{Bartel:1987de} detectors. For example, the CELLO Collaboration performed a search for light LQs decaying into either a muon or a neutrino and a quark. The lack of evidence for pair production of LQs allowed the CELLO Collaboration to exclude the LQ masses between $7$ and $20.5$\,GeV~\cite{Behrend:1986jz}. These results have been followed with the searches for LQs in $e^+ e^-$ annihilations at TRISTAN~\cite{Kim:1989qz} (The Transposable Ring Intersecting Storage Accelerator in Nippon) using the AMY detector. The center-of-mass energies that were used to collect data at TRISTAN varied from $50$\,GeV to $60.8$\,GeV. These should be compared with the center-of-mass energies at PETRA that were around $35$\,GeV.

LQs have been subsequently searched for at the LEP $e^+ e^-$ collider by the ALEPH~\cite{Decamp:1991uy}, DELPHI~\cite{Abreu:1991df,Abreu:1993bc,Abreu:1998fw}, L3~\cite{Adeva:1991dv,Adriani:1993gk}, and OPAL~\cite{Alexander:1991sv,Abbiendi:1999wk,Abbiendi:2001sw} Collaborations. 

The LEP limits on the LQ parameter space were primarily generated by considered LQ pair production.
For example, a search performed by OPAL detector for pair-produced LQs at center-of-mass energies between 189 and 209\,GeV and with the total integrated luminosity of 596\,pb$^{-1}$, yielded robust limits on scalar and vector LQs with $m_{\mathrm{LQ}} \lesssim 100$\,GeV, depending on the specific LQ representation. (For details see Table~12 of Ref.~\cite{Abbiendi:2003iv}.) DELPHI, however, did considered single LQ production. The representative diagram for this process is shown in the bottom row of Fig.~\ref{fig:LQ_e_p}. For example, the DELPHI detector was used to look for events with one jet and at most one isolated lepton using the LEP-2 data corresponding to an integrated luminosity of $47.7$\,pb$^{-1}$. The data were collected at a center-of-mass energy of $183$\,GeV. The analysis produced the set of upper limits on Yukawa couplings of scalar and vector LQs as a function of $m_{\mathrm{LQ}}$. These were presented in Fig.~4 of Ref.~\cite{Abreu:1998fw}. Also, the lower limits were placed on the mass of a first generation LQs for a coupling parameter $y=\sqrt{4 \pi \aem}$ assuming branching ratios to charged leptons and quarks to be 1 and 0.5. These limits are presented in Table 2 of Ref.~\cite{Abreu:1998fw}. The most stringent limit on $m_{\mathrm{LQ}}$ of scalar (vector) LQ at 95\% confidence level was $161$\,GeV ($171$\,GeV) for $|Q|=1/3,5/3$ and $\beta=1$, where $Q$ is electric charge and $\beta$ branching ratio.

                      
\subsection{$e p$ collider signatures}
\label{sec:ep-colliders}

\begin{figure}[tbp]
\centering
\includegraphics[scale=0.7]{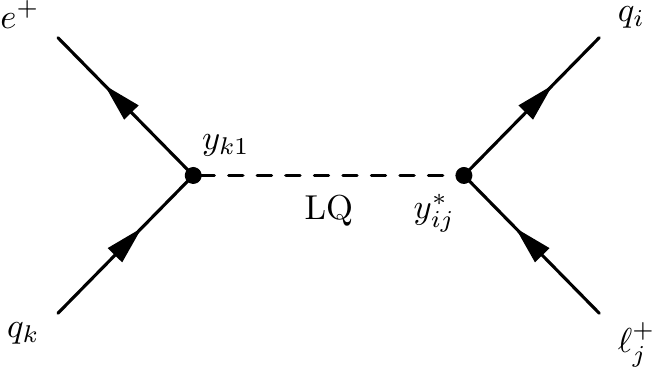} \hspace{1cm}
\includegraphics[scale=0.7]{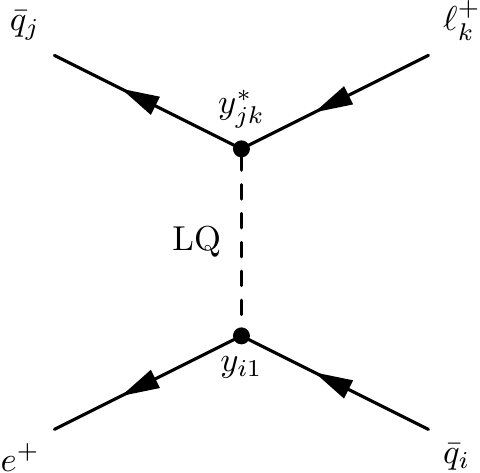}
\caption{\label{fig:ep_PRODUCTION} Feynman diagrams relevant for the $(3B+L)=0$ scalar LQ searches at $e p$ colliders. An $s$-channel resonant production diagram is shown in the left panel and a $u$-channel LQ exchange diagram is presented in the right panel. Here, $y_{ij}$, $i,j=1,2,3$, represents appropriate quark--lepton--LQ Yukawa coupling.}
\end{figure}

LQs in lepton--quark collisions have been studied extensively in the past. See, for example, Refs.~\cite{Wudka:1985ef,Buchmuller:1986zs,Dobado:1987pj,Gunion:1987ge,Grifols:1987ah,Blumlein:1994qd,Ilyin:1995jv,Kunszt:1997at,Plehn:1997az,Friberg:1997nn}. Feynman diagrams that are relevant for the search for scalar LQs at $e p$ colliders at the leading order are shown in Fig.~\ref{fig:ep_PRODUCTION}. Scalar LQ can be produced through an $s$-channel process that is shown in the left panel of Fig.~\ref{fig:ep_PRODUCTION}. It can also be exchanged in a $u$-channel process that is presented in the right panel of Fig.~\ref{fig:ep_PRODUCTION}. Note that the diagrams in Fig.~\ref{fig:ep_PRODUCTION} correspond to the LQ models involving either $\tilde{R}_2$ or $R_2$. To switch to the diagrams relevant for the $(3B+L)=-2$ LQs it is sufficient to make the following exchange: $q_i \leftrightarrow \bar{q}_i$.

Contact interactions of the form $eeqq$ within the context of $e p$ collider that are due to the LQ exchange have been studied in Refs.~\cite{Barger:1997nf,DiBartolomeo:1997mb,Cheung:1998gg,Zeppenfeld:1998un,Zarnecki:1999je,Zarnecki:2000qc,Cheung:2001wx}. Some of these studies have also included the LEP and Tevatron data as well as other relevant low-energy measurements to place a constraint on the strength of these interactions. The possibility to use the effect of the virtual leptoquark exchange on charged and neutral current processes in the context of the $e p$ collider to search for leptoquarks has been studied in Ref.~\cite{Doncheski:1992ds}.


\subsubsection{Searches at $e p$ collider}

There are numerous searches for the LQs at the HERA $e p$ collider that were conducted by the H1~\cite{Abt:1993nr,Ahmed:1994td,Aid:1995kf,Aid:1995wd,Adloff:1999tp,Adloff:2000dp,Adloff:2001cp,Aktas:2005pr,Aaron:2011zz,Collaboration:2011qaa} and ZEUS~\cite{Derrick:1993by,Chekanov:2003af,Abramowicz:2012tg} Collaborations. Clearly, an LQ needs to couple to an electron to be produced in the first place. However, the LQ decay and/or the LQ exchange can result, in principle, in leptons of all flavors as indicated in Fig.~\ref{fig:ep_PRODUCTION}. Here we summarize existing searches that are based on $e^+ p$ and $e^- p$ scattering data at HERA. 

The limits on the parameter space of first generation scalar and vector LQs were presented in Ref.~\cite{Aktas:2005pr}. The search was performed using the $e^+ p$ and $e^- p$ data corresponding to a total integrated luminosity of $117$\,pb$^{-1}$. The data were collected by the H1 experiment at HERA between 1994 and 2000. The study looked for a signal of direct or indirect LQ production in data samples with a large transverse momentum final state electron or with large missing transverse momentum. Due to the lack of signal LQs with masses of up to $275$--$325$\,GeV were ruled out assuming LQ couplings to matter of electromagnetic strength, i.e., $y=0.3 (\approx \sqrt{4 \pi \aem})$. 

The H1 experiment did not only look for first generation LQs but also for events with a high transverse momentum $\mu$ or $\tau$ to cover the mixed fermion generation scenario~\cite{Adloff:1999tp}. The analysis in question used data collected from 1994 to 1997 that amounted to an integrated luminosity of $37$\,pb$^{-1}$. Scalar LQs were excluded for masses up to $275$\,GeV for a coupling of electromagnetic strength. The study also excluded any LQ type up to $400$\,GeV for couplings of order one. Constraints were also set on the product of Yukawa coupling involving $\mu$ or $\tau$ with Yukawa coupling involving the electron as a function of the LQ mass~\cite{Adloff:1999tp}. Final results were presented in the form of tabulated constraints --- upper limits --- on $(y_{i1} y_{jk})/m_{\mathrm{LQ}}^2$, where $k=2,3$ and $i,j=1,2$ for real couplings. (See Fig.~\ref{fig:ep_PRODUCTION} for notation.) 

Another analysis of production of second and third generation LQs was also performed by the H1 Collaboration~\cite{Aaron:2011zz}. This study  excluded existence of LQs produced in $e^+ p$ and $e^- p$ collisions with a coupling strength of $y=0.3$ to an electron--quark pair and decaying with the same coupling strength to a muon--quark or a tau--quark pair at 95\% confidence level up to LQ masses of $712$ and $479$\,GeV, respectively. See Tables 1, 2, 3, and 4 of Ref.~\cite{Aaron:2011zz} for comprehensive list of all results. For example, the ratio $(y_{11} y_{22})/m_{\mathrm{LQ}}^2$ was constrained to be below $0.7$ in units of TeV$^{-2}$ at 95\%\,C.L.\ for LQ $S^L_{1/2}$. All other bounds on $(y_{i1} y_{jk})/m_{\mathrm{LQ}}^2$ were of the same order of magnitude. Here, $k=2,3$ and $i,j=1,2$.

A search for resonances that could decay into $e j$ and $\nu j$ was performed with the ZEUS detector at HERA operating at center-of-mass energies of $300$\,GeV and $318$\,GeV~\cite{Chekanov:2003af}. The analysis that was based on an integrated $e^+ p$ ($e^- p$) luminosity of $114.8$\,pb$^{-1}$ ($16.7$\,pb$^{-1}$) found no evidence for such a resonance. For example, the study excluded LQs with masses of up to $290$\,GeV for Yukawa coupling of $0.1$~\cite{Chekanov:2003af}. (For the search for LQs with the ZEUS detector that was based on $26$\,nb$^{-1}$ of integrated luminosity see analysis presented in Ref.~\cite{Derrick:1993by}.)

The most stringent constraints on parameter space of first generation LQs in $e p$ collisions at HERA came from the study of full H1 data sample~\cite{Collaboration:2011qaa}. An integrated luminosity of $446$\,pb$^{-1}$ was used to exclude first generation LQs with masses up to $800$\,GeV at 95\% confidence level for LQ couplings of electromagnetic strength. The ZEUS detector, on the other hand, performed a search for first generation LQs using a total of $498$\,pb$^{-1}$ of data to set limits on the LQ Yukawa coupling as a function of its mass. The data in question originated from the 1994--2000 and 2003--2007 collection periods. LQs with a coupling of electromagnetic strength were excluded for masses up to $699$\,GeV~\cite{Abramowicz:2012tg}.

Before we conclude discussion of experimental searches at $e p$ colliders it is worth mentioning that both H1 Collaboration~\cite{Adloff:1997fg} and ZEUS~\cite{Breitweg:1997ff} reported in 1997 an anomalous measurement while studying $e p$ scattering at high values of squared four-momentum transfer denoted $Q^2$. For example, the ZEUS detector at HERA observed two events for $Q^2 > 35,000$\,GeV$^2$ while $0.145 \pm 0.013$ events were expected. The sample under consideration was based on the reaction $e^+\,p \rightarrow e^+\,X$ with a $20.1$\,pb$^{-1}$ of data collected during the years 1994 through 1996. For $Q^2$ below $15,000$\,GeV$^2$ the observed signal at ZEUS detector was in good agreement with the SM expectations. This anomaly has accordingly attracted a lot of attention with one of potential physical explanations being that of an $s$-channel LQ production. See, for example, Refs.~\cite{Choudhury:1997dt,Altarelli:1997ce,Dreiner:1997cd,Doncheski:1997fp,Blumlein:1997fm,Kalinowski:1997fk,Babu:1997jr,Hewett:1997ce,Leontaris:1997fq,Kunszt:1997at,Kon:1997qj,Barbieri:1997zn,Montvay:1997dk,King:1997wf,Belyaev:1997yw,Dutta:1997us,Altarelli:1997qu,Babu:1997tx,Ellis:1997tz,Caravaglios:1997tk,Giusti:1997tw,Kim:1997in,Joshipura:1997yb,Lola:1997dm,Cao:1997gi,Keith:1997fv,Deshpande:1997pg,Cheung:1997jt,Hewett:1997ba,Altarelli:1997eb,Frampton:1997in,Joshipura:1997vw} for studies/reviews directly related to or motivated by aforementioned anomaly.
  

\subsection{$p \bar{p}$ and $p p$ collider signatures}

LQs are colored fields. They could thus be copiously produced at hadron machines. Moreover, the fact that they decay into (charged) leptons makes the detection of their signatures feasible. We start by discussing production mechanisms at leading order.

\begin{figure}[tbp]
\centering
\includegraphics[scale=0.75]{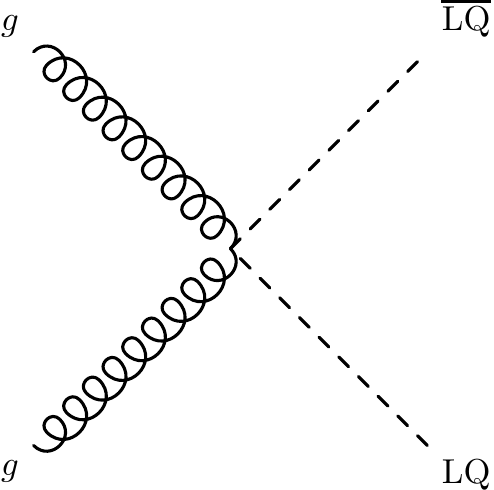} \hspace{1cm} \includegraphics[scale=0.75]{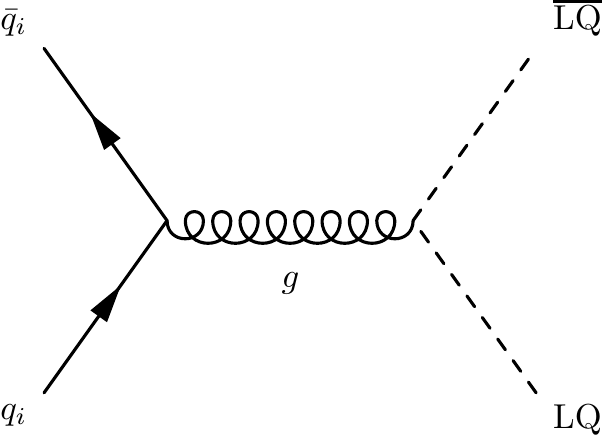}\newline
\newline
\includegraphics[scale=0.75]{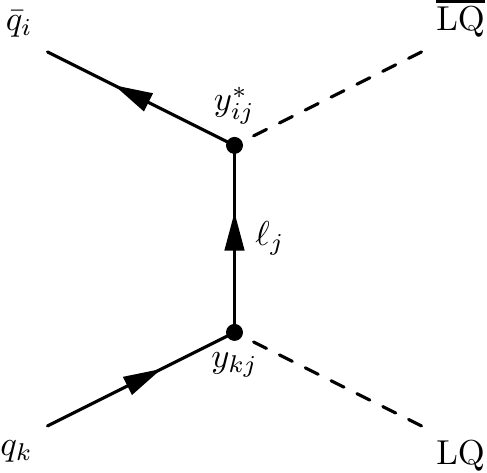}
\caption{\label{fig:PAIR_PRODUCTION} Feynman diagrams relevant for a pair production of scalar LQs at hadron colliders. Representative diagram for a gluon--gluon fusion (quark--antiquark annihilation) process is shown in the upper left (right) panel. The diagram in the lower panel represents a $t$-channel production mechanism. Here, $y_{ij}$, $i,j=1,2,3$, represents appropriate Yukawa coupling of a quark ($q_i$) and a lepton ($l_j$) with an LQ.}
\end{figure}

Feynman diagrams that depict the scalar LQ pair production at hadron colliders at the leading order are shown in Fig.~\ref{fig:PAIR_PRODUCTION}.
The QCD diagrams that contribute to an LQ pair production are numerous. We opt to show one representative diagram for gluon-gluon fusion and one diagram that corresponds to a quark-antiquark annihilation process in the upper left and right panels of Fig.~\ref{fig:PAIR_PRODUCTION}, respectively. There is, on the other hand, only one type of the Yukawa coupling contribution to the LQ pair production and it corresponds to a $t$-channel process shown in the lower panel of Fig.~\ref{fig:PAIR_PRODUCTION}. The amplitude that corresponds to a $t$-channel is proportional to a product of two relevant Yukawa couplings. This makes it especially relevant in the limit of large Yukawas. 

If we assume that the scalar LQ pair production is purely QCD driven and completely neglect $t$-channel contribution there are only two distinct partonic level processes that should be taken into consideration. The first corresponds to the gluon--gluon fusion ($g\,g \rightarrow \mathrm{LQ} \overline{\mathrm{LQ}}$) and the second process is due to the quark--antiquark annihilation ($q \bar{q} \rightarrow \mathrm{LQ} \overline{\mathrm{LQ}}$). Representative diagram for the former (latter) process is shown in the upper left (right) panel of Fig.~\ref{fig:PAIR_PRODUCTION}. The partonic level contributions towards total cross section~\cite{Grifols:1981aq,Antoniadis:1983hz,Eichten:1984eu,Altarelli:1984ve,Dawson:1983fw,Blumlein:1996qp} read
\begin{align}
\label{eq:pair_gg}
\hat{\sigma}[g g \rightarrow \mathrm{LQ} \overline{\mathrm{LQ}}]= &  \frac{\alpha_3^2 \pi}{96 \hat{s}} \left[\beta (41-31 \beta^2)+(18 \beta^2-\beta^4-17)\log\frac{1+\beta}{1-\beta}\right],\\
\label{eq:pair_qq}
\hat{\sigma}[q \bar{q} \rightarrow \mathrm{LQ} \overline{\mathrm{LQ}}]= &  \frac{2 \alpha_3^2 \pi}{27 \hat{s}} \beta^3,
\end{align}
where $\alpha_3 \equiv g_3^2/(4 \pi)$ is strong coupling, $\hat{s}$ stands for the square of invariant mass of the relevant process, $\beta=\sqrt{1-4 m_\mathrm{LQ}^2/\hat{s}}$, and $m_\mathrm{LQ}$ is the LQ mass. 

We show, in Table~\ref{tab:Tevatron}, total cross section for pair production of scalar LQs using Eqs.~\eqref{eq:pair_gg} and~\eqref{eq:pair_qq} at $\sqrt{s}=1.8$\,TeV proton--antiproton center-of-mass energy for different LQ masses. To obtain numerical results for the case of $p \bar{p}$ collider we resort to MSTW 2008 parton distribution functions~\cite{Martin:2009iq} at the leading order and set $\alpha_3(m_Z) = 0.13939$. These results are relevant for Tevatron and can be directly compared with results presented in Table~1 of Ref.~\cite{Kramer:1997hh}. For an early study of the experimental signal of scalar leptoquarks at the $p \bar{p}$ collider see Ref.~\cite{Mohapatra:1984aq}.
\begin{table}[tbp]
\centering
\begin{tabular}{|c|ccc|}
\hline
$m_{\mathrm{LQ}}$\,[GeV] & $\sigma_{gg}$\,[pb] & $\sigma_{q\overline{q}}$\,[pb] & $\sigma_{\mathrm{tot}}$\,[pb] \\
\hline 
$150$ & 0.255 & 0.733 & 0.988 \\
$175$ & 0.073 & 0.311 & 0.384 \\
$200$ & 0.023 & 0.137  & 0.160 \\
$250$ & 0.003 & 0.029 & 0.031 \\
$300$ & 0.000 & 0.006 & 0.006 \\
\hline 
\end{tabular}
\caption{\label{tab:Tevatron} Total cross section $\sigma_{\mathrm{tot}}$ at the leading order for the QCD driven LQ pair production with a breakdown of individual contributions for $\sqrt{s}=1.8$\,TeV center-of-mass energy at $p \bar{p}$ collider.}
\end{table} 
Clearly, at high enough LQ masses the QCD driven total cross section at $p \bar{p}$ colliders is dominated by the quark--antiquark annihilation processes.

The dominant contribution at $p p$ colliders, if we assume that Yukawa couplings are reasonably small, comes from the gluon--gluon fusion processes. To quantify that we present, in Table~\ref{tab:LHC}, QCD driven total cross section at the leading order for pair production of scalar LQs at $\sqrt{s}=14$\,TeV proton--proton center-of-mass energy for different LQ masses. Here we have in mind the Large Hadron Collider. (For the early studies of the signal and backgrounds that are relevant for the scalar leptoquark pair production at the LHC see Refs.~\cite{Dion:1997jw,Abdullin:1999im,Mitsou:2004hm}.) Values presented in the second and third column of Table~\ref{tab:LHC} are obtained integrating over partonic level contributions given in Eqs.~\eqref{eq:pair_gg} and~\eqref{eq:pair_qq}, respectively. To do that we again use MSTW 2008 parton distribution functions~\cite{Martin:2009iq} at the leading order. Note that the importance of quark--antiquark annihilation grows with the LQ mass at $p p$ collider. For example, it amounts to one third (sixth) of gluon--gluon fusion contribution at $m_\mathrm{LQ}=1.1$\,TeV ($m_\mathrm{LQ}=700$\,GeV).
\begin{table}[tbp]
\centering
\begin{tabular}{|c|ccc|}
\hline
$m_{\mathrm{LQ}}$\,[TeV] & $\sigma_{gg}$\,[pb] & $\sigma_{q\overline{q}}$\,[pb] & $\sigma_{\mathrm{tot}}$\,[pb] \\
\hline 
$0.4$   & 1.4073 & 0.1297 & 1.5370 \\
$0.5$   & 0.3904 & 0.0461 & 0.4365 \\
$0.6$   & 0.2586 & 0.0187 & 0.1480 \\
$0.7$   & 0.0484 & 0.0084 & 0.0568 \\
$0.8$   & 0.0198 & 0.0040 & 0.0238 \\
$0.9$   & 0.0087 & 0.0020 & 0.0107 \\
$1.0$   & 0.0040 & 0.0010 & 0.0051 \\
$1.1$   & 0.0020 & 0.0006 & 0.0025 \\
\hline 
\end{tabular}
\caption{\label{tab:LHC} Total cross section $\sigma_{\mathrm{tot}}$ at the leading order for the scalar LQ pair production with a breakdown of individual contributions for $\sqrt{s}=14$\,TeV center-of-mass energy at $p p$ collider.}
\end{table} 
These cross sections can be compared with the total cross sections presented in Table~1 of Ref.~\cite{Kramer:2004df}. (Note that units of [fb] should be replaced with [pb] in Table 1 of Ref.~\cite{Kramer:2004df}.)


The cross sections for a pair production of vector leptoquarks at hadron machines can be found in Refs.~\cite{Hewett:1993ks,CiezaMontalvo:1993hp,Djouadi:1995es,Blumlein:1996qp,Dion:1996rv}. Note that the calculation requires certain assumptions if the underlying theory for the origin of vector leptoquarks is not specified. For example, if these are elementary vector bosons associated with the gauge symmetry then the relevant couplings of the pair of vector LQs to a single gluon or a pair of gluons are completely fixed. If that is not the case one can introduce additional parameters to describe the strength of the relevant couplings~\cite{Hewett:1993ks,Djouadi:1995es,Blumlein:1996qp,Dion:1996rv,Rizzo:1996ry,Hewett:1997ce}. For example, it is common to introduce a single dimensionless parameter $\kappa$ that is defined through the following interaction term
\begin{align}
\mathcal{L} \supset & - i g_3 \kappa \Phi_\mu^\dagger G^{\mu \nu} \Phi_\nu,
\end{align}  
where $\Phi_\nu$ stands for all vector leptoquarks and $G^{\mu \nu}$ represents the gluon field strength tensor. The relevant production cross-section for a pair of vector leptoquarks at the LHC as a function of $\kappa$ is shown in Fig.~1 of Ref.~\cite{Rizzo:1996ry}. The same reference contains the summary of search reaches for vector leptoquarks at hadron colliders through pair production. A compilation of cross-sections for the pair production of both scalar and vector leptoquarks at Tevatron and the LHC can be found in Ref.~\cite{Blumlein:1998ym}. 

If we allow for non-negligible Yukawa couplings of the LQ to $u$, $d$ or $s$ quark the pair production is not driven solely by QCD anymore. In fact, the LQ pair production cross section exhibits three distinct regions as a function of the LQ Yukawa coupling(s). For small enough Yukawa couplings the production is purely QCD driven. For intermediate values of Yukawa couplings there exists a region where a negative interference between the QCD diagrams and the $t$-channel diagram shown in Fig.~\ref{fig:PAIR_PRODUCTION} decreases the total cross section by up to 15\% depending on the quark type ($u$, $d$, or $s$), mass of the LQ and strength of the coupling. Finally, there is a region of the large Yukawa coupling, where the $t$-channel contribution does not only dominate the QCD driven one but significantly enhances the total cross section. We will discuss these features in more detail later on when we address Yukawa dependence of the cross section for the single LQ production.

Single LQ production is very relevant at hadron colliders~\cite{Hewett:1987yg,Eboli:1987vb,DeMontigny:1989yd,Ohnemus:1994xf,Blumlein:1997fm,CiezaMontalvo:1998sk,Belyaev:1998ki,Eboli:1999ye,Belyaev:2005ew}. Moreover, the importance of a combined analysis of the LQ single and pair production has been demonstrated in Ref.~\cite{Eboli:1997fb} (Ref.~\cite{Eboli:1999ye}) for the LHC (Tevatron) case. These works have partially motivated subsequent study presented in Ref.~\cite{Belyaev:2005ew} that produced the combined analysis of the LQ single and pair production at the level of detector simulation for the LHC. 

\begin{figure}[tbp]
\centering
\includegraphics[scale=0.7]{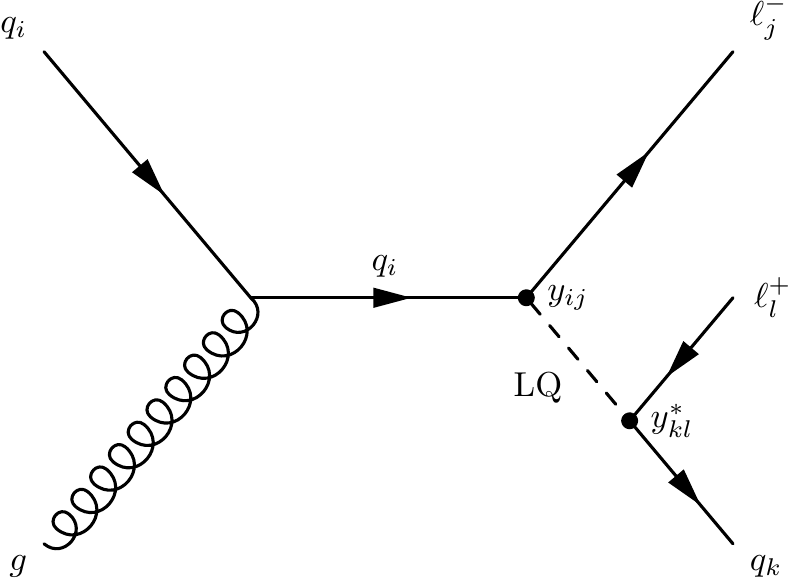}\newline
\newline
\includegraphics[scale=0.7]{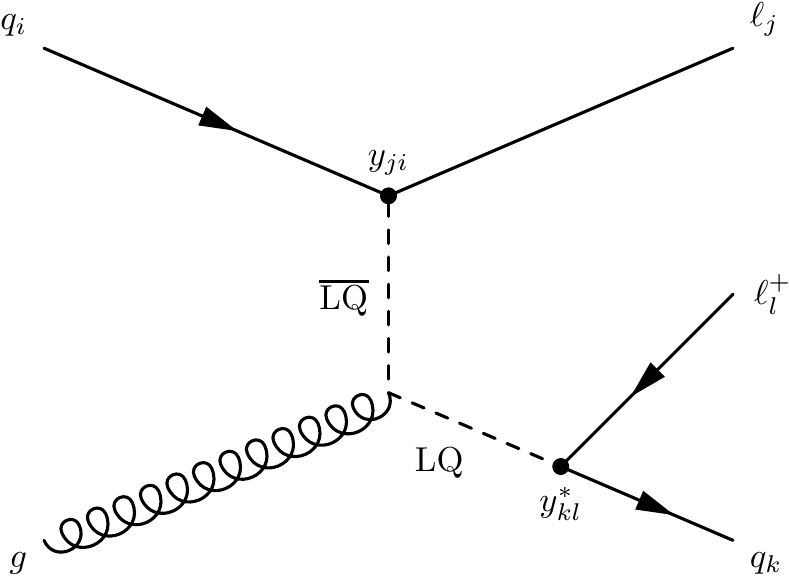}\hspace{1cm}
\includegraphics[scale=0.7]{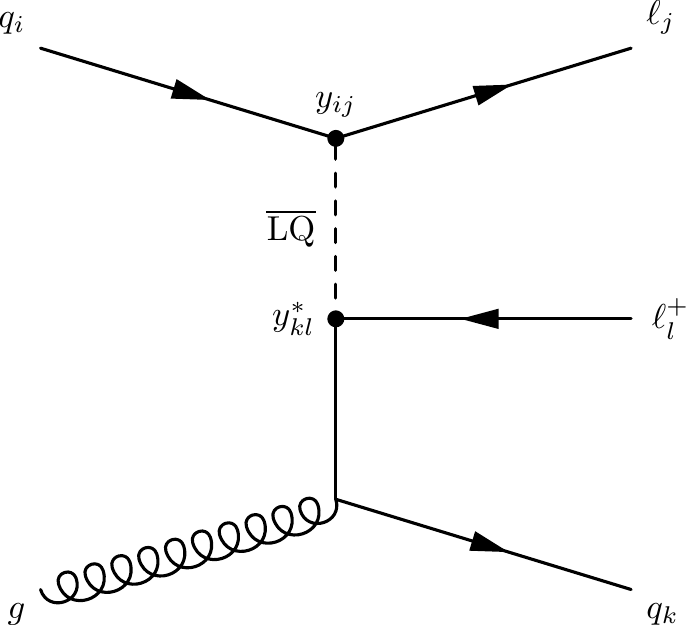}
\caption{\label{fig:SINGLE_PRODUCTION} Complete set of the leading order Feynman diagrams for a single LQ production at hadron colliders and subsequent decay with $l_j \overline{l}_l q_k$ final state. Here, $y_{ij}$, $i,j=1,2,3$, represents appropriate Yukawa coupling of a quark ($q_i$) and a lepton ($l_i$) with LQ.}
\end{figure}

We show complete set of the leading order Feynman diagrams that are relevant for a single LQ production/exchange and subsequent decay at hadron colliders that yields two leptons and one jet in the final state in Fig.~\ref{fig:SINGLE_PRODUCTION}. The diagrams shown in Fig.~\ref{fig:SINGLE_PRODUCTION} are an $s$-channel diagram (upper panel), a $t$-channel diagram (lower left panel), and a non-resonant diagram (lower right panel). We use generic symbols for all the fields including the LQs. 
If one is interested purely in single LQ production at hadron machines one should disregard a non-resonant diagram in Fig.~\ref{fig:SINGLE_PRODUCTION}. The parton level differential cross section for the scalar (vector) leptoquark production through $g q \rightarrow \mathrm{LQ} l$ can be found in Refs.~\cite{Hewett:1987yg,Eboli:1987vb,Dobado:1988wi,Barger:1988pm} (Ref.~\cite{Hewett:1997ce}). 

Single LQ production is particularly important in the regime of large LQ masses and/or large couplings. To demonstrate that we compare in what follows the scalar LQ pair production cross section with the single production cross section for the case of proton--proton collisions at $\sqrt{s}=8$\,TeV center-of-mass energy. To that end we, for simplicity, assume that scalar LQs couple to the SM fermions without cross-generational mixing. This ansatz would correspond to the case when $y_{ij}=y_i\delta_{ij}$, $i,j=1,2,3$, in Figs.~\ref{fig:PAIR_PRODUCTION} and~\ref{fig:SINGLE_PRODUCTION}. We parametrically write both pair and single production cross sections as   
\begin{align}
\label{eq:pair_X}
\sigma_{\textrm{pair}}(y_i,m_\mathrm{LQ})= &a_0(m_\mathrm{LQ})+a_2(m_\mathrm{LQ}) |y_i|^2+a_4 (m_\mathrm{LQ}) |y_i|^4,\\
\label{eq:single_X}
\sigma_{\textrm{single}}(y_i,m_\mathrm{LQ})= &a(m_\mathrm{LQ}) |y_i|^2,
\end{align}
where coefficients $a_0(m_\mathrm{LQ})$, $a_2(m_\mathrm{LQ})$, $a_4(m_\mathrm{LQ})$, and $a(m_\mathrm{LQ})$ depend solely on the LQ mass. (Clearly, Eqs.~\eqref{eq:pair_X} and~\eqref{eq:single_X} are useful only if $y_i$ represents Yukawa coupling of LQ to $u$, $d$ or $s$ quark due to proton composition at partonic level.) We can now calculate cross sections for 
$p p \rightarrow \overline{\mathrm{LQ}} l_i$, $p p \rightarrow \mathrm{LQ}  \overline{l_i}$, and $p p \rightarrow \mathrm{LQ}  \overline{\mathrm{LQ}}$ for several $m_\mathrm{LQ}$ choices while setting $y_i$ to one, i.e, $|y_i|=1$., to find the functional dependence of $a_0(m_\mathrm{LQ})$, $a_2(m_\mathrm{LQ})$, $a_4(m_\mathrm{LQ})$, and $a(m_\mathrm{LQ})$ by simple interpolation. To do that we implement two LQ models introduced in sections~\ref{R_2} and~\ref{R_2_tilda} in FeynRules (v1.6.16)~\cite{Christensen:2008py}. In particular, we look at the production of $R_2^{5/3}$ assuming that only $y^{RL}_{2\,11} \neq 0$ in Eq.~\eqref{eq:main_R_2_a} and the production of $\tilde{R}_2^{2/3}$ assuming that either $\tilde{y}^{RL}_{2\,11}$ or $\tilde{y}^{RL}_{2\,22}$ in Eq.~\eqref{eq:main_t_R_2_a} is different from zero. We thus take $R_2^{5/3}$ to couple only to an electron $e$ and an up quark $u$. $\tilde{R}_2^{2/3}$, on the other hand, is assumed to couple only to a down quark $d$ and an electron $e$ or a strange quark $s$ and a muon $\mu$. 

\begin{figure}[tbp]
\centering
\includegraphics[scale=0.69]{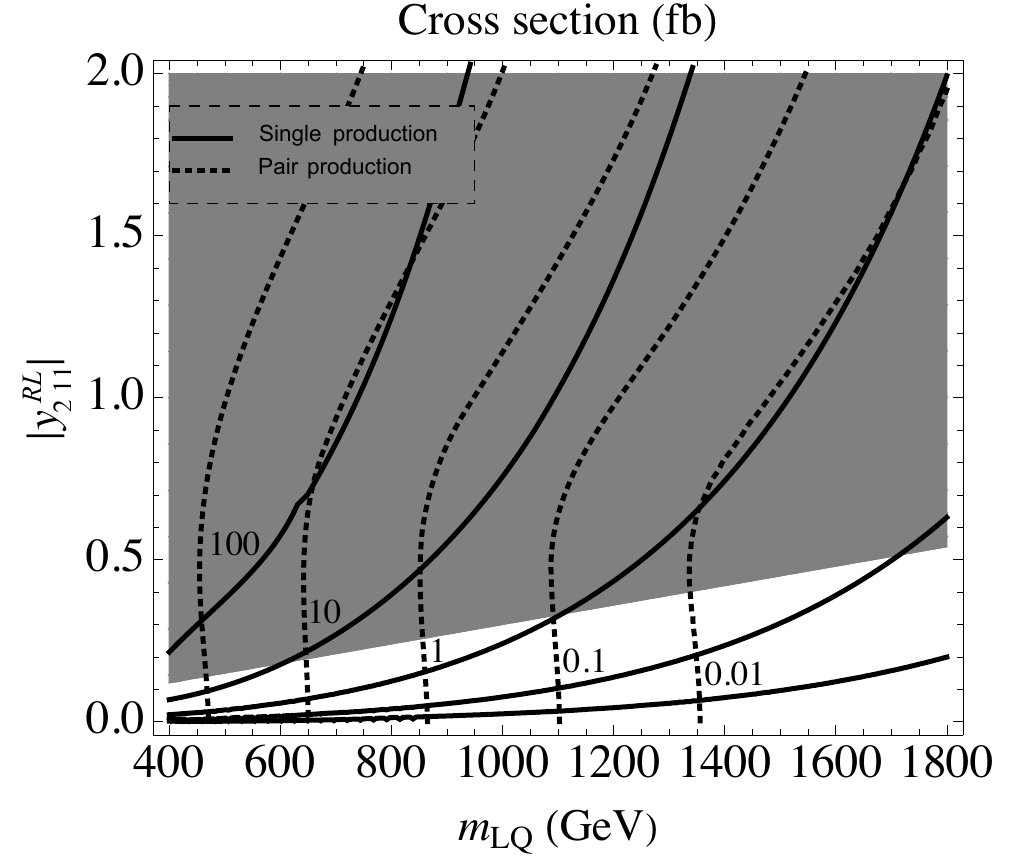}
\includegraphics[scale=0.69]{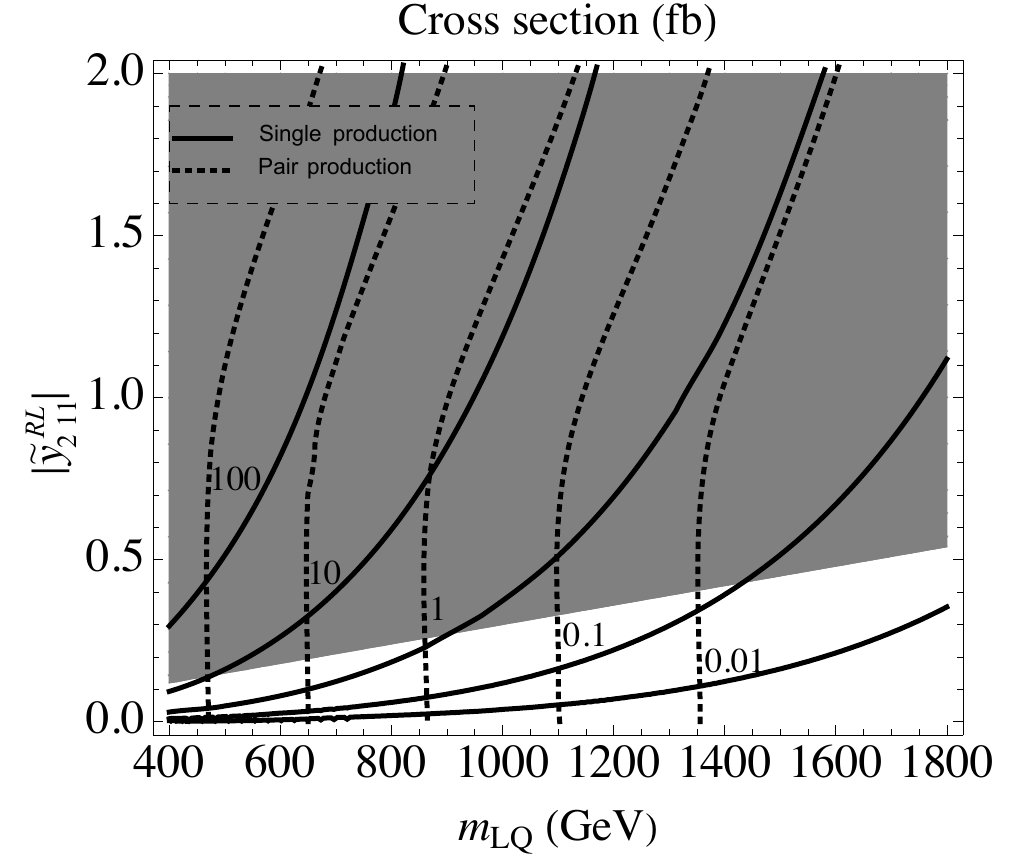}
\includegraphics[scale=0.69]{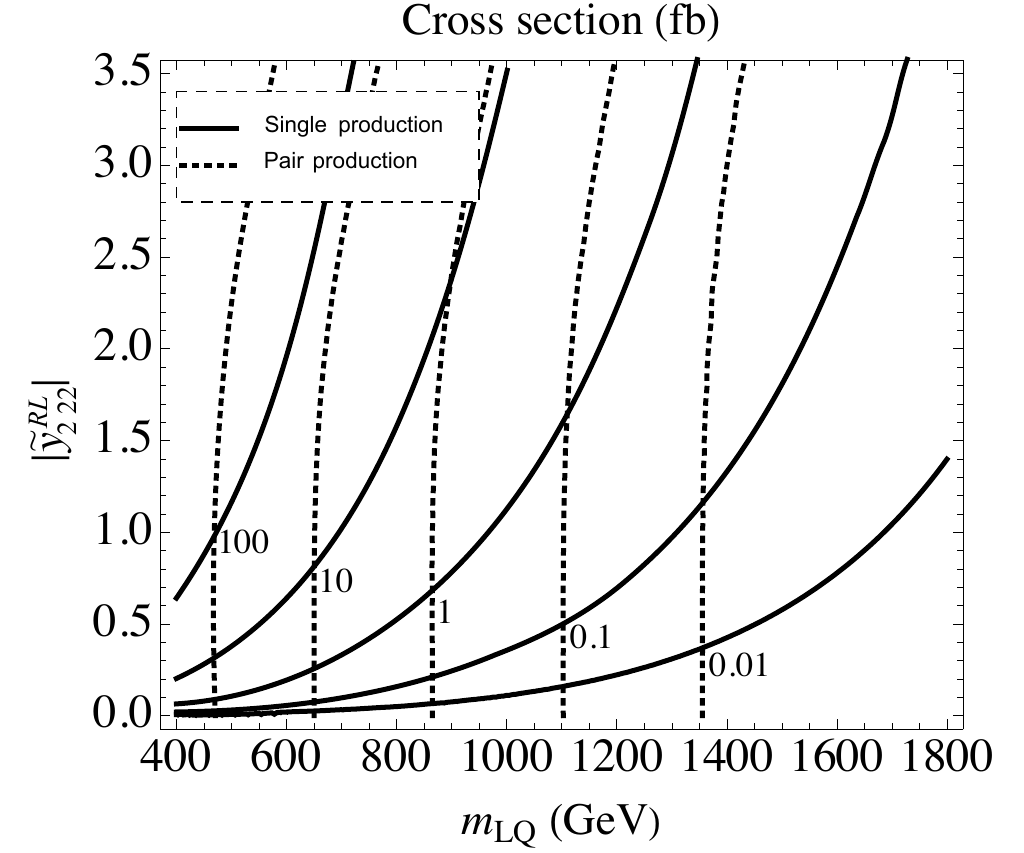}
\caption{\label{fig:plot-1} Contours of constant leading order cross sections for single LQ production (solid lines)
and the LQ pair production (dotted lines) at $\sqrt{s}=8$\,TeV center-of-mass
energy in proton--proton collisions in the $(m_\mathrm{LQ},|y_i|)$ planes, where $|y_i|=|y^{RL}_{2\,11}|, |\tilde{y}^{RL}_{2\,11}|, |\tilde{y}^{RL}_{2\,22}|$. The region shown in gray is excluded from the atomic parity violation constraints.}
\end{figure}

The leading order production cross sections are calculated using 
MadGraph~5 (v1.5.11)~\cite{Alwall:2011uj}. We take $\sqrt{s}=8$\,TeV for proton--proton center-of-mass energy and fix renormalization 
($\mu_R$) and factorization ($\mu_F$) scales to $\mu_R=\mu_F=m_\mathrm{LQ}/2$. The final results for the $y^{RL}_{2\,11}$, $\tilde{y}^{RL}_{2\,11}$, and $\tilde{y}^{RL}_{2\,22}$ scenarios are shown in Fig.~\ref{fig:plot-1} in terms of contours of constant cross 
section in the $(m_\mathrm{LQ},|y_i|)$ plane, where $|y_i|=|y^{RL}_{2\,11}|, |\tilde{y}^{RL}_{2\,11}|, |\tilde{y}^{RL}_{2\,22}|$. The contours of constant cross section for the single LQ production (LQ pair production) are shown in solid (dotted) lines in Fig.~\ref{fig:plot-1}.

The LQ pair production is QCD driven for small Yukawa couplings. This behavior corresponds to a region where dotted lines run vertically in Fig.~\ref{fig:plot-1}. However, in the large coupling regime, the total cross section can be significantly enhanced by the $t$-channel contribution. The LQ single production, on the other hand, vanishes in the zero coupling limit. However, due to the final state phase space, it drops less rapidly with larger LQ masses compared to the LQ pair production. In other words, the contributions from the single LQ production mechanism become increasingly important at larger LQ masses. The relative strengths of two cross sections depend on the parton distribution functions of the initial state partons. The largest (smallest) effect for the fixed value of appropriate Yukawa coupling is seen in the $y^{RL}_{2\,11}$ ($\tilde{y}^{RL}_{2\,22}$) panels in Fig.~\ref{fig:plot-1}. To sum all this up, the contributions from the single LQ production mechanism might have significant impact on the LQ phenomenology at the LHC.


There are two recasts of experimental searches by the CMS Collaboration for pair production of LQs at the LHC~\cite{Dorsner:2014axa,Mandal:2015vfa} that demonstrate the ability of the LHC to probe flavor structure of LQ models through inclusion of the single LQ production. (For the early study of the discovery potential for pair produced scalar LQs at the LHC of the CMS (ATLAS) detector see Ref.~\cite{Abdullin:1999im} (Ref.~\cite{Mitsou:2004hm}).) These recasts combine reported data on the pair production searches of LQs with expected signals for single LQ production in order to obtain collider based limits on relevant Yukawa couplings as a function of the LQ mass. The first recast~\cite{Dorsner:2014axa} concerns the CMS Collaboration search for the second generation scalar LQs based on $19.6$\,fb$^{-1}$ of data at $\sqrt{s}=8$\,TeV proton--proton center-of-mass energy~\cite{CMS:zva}. The second recast~\cite{Mandal:2015vfa} addresses the CMS search for the pair production of the first generation scalar LQs in $p p$ collisions at $\sqrt{s} = 8$\,TeV~\cite{CMS:2014qpa}. 


The recast presented in Ref.~\cite{Dorsner:2014axa} allows for second generation LQ $\tilde{R}^{2/3}_2$ to couple purely to a strange quark and a muon via coupling $\tilde{y}^{RL}_{2\,22}$ to stay within the single generation LQ paradigm but allows for arbitrary strength of the Yukawa coupling itself. The study disregards non-resonant diagram in Fig.~\ref{fig:SINGLE_PRODUCTION}. Also, the analysis is based purely on the leading order calculations.
The recast adopts the same preselection and final cuts as used in Ref.~\cite{CMS:zva}. The CMS analysis uses the following three variables to optimize the signal for the first (second) generation leptoquark search through the pair production: (i) the invariant mass of the dielectron (dimuon) pair, (ii) the scalar sum of the transverse momenta of 
the two leading $p_T$ electrons (muons) and the two leading $p_T$ jets, and (iii) the smallest of the two electron--jet (muon--jet) invariant masses that minimizes the LQ mass difference. The main results of the recast of the second generation LQ search is the exclusion region in parameter space spanned by $\tilde{y}^{RL}_{2\,22}$ and $m_\mathrm{LQ}$ that is shown in 
in Fig.~5 of Ref.~\cite{Dorsner:2014axa}. The limit on the LQ mass is more stringent for larger values of $\mu$--$s$--$\mathrm{LQ}$ coupling.


One expects that the single LQ production contribution will be even more pronounced for the first generation case due to particularities of proton composition whenever LQ couples to either $u$ or $d$ quarks. This has indeed been confirmed by the recast~\cite{Mandal:2015vfa} of the CMS search~\cite{CMS:2014qpa} for the pair production of the first generation scalar LQs in $p p$ collisions at $\sqrt{s} = 8$\,TeV. The authors of Ref.~\cite{Mandal:2015vfa} have produced exclusion limits on Yukawa coupling $y$ of the first generation LQ for two different LQ scenarios as a function of $m_{\mathrm{LQ}}$ considering possible coexistence of $eejj$ and $e \nu jj$ final states. This has been accomplished by overlying the signal of the single LQ production and subsequent decay over the pair production. See, for example, Fig.~3 of Ref.~\cite{Mandal:2015vfa} for quantitative results.    

\begin{figure}[tbp]
\centering
\includegraphics[scale=0.6]{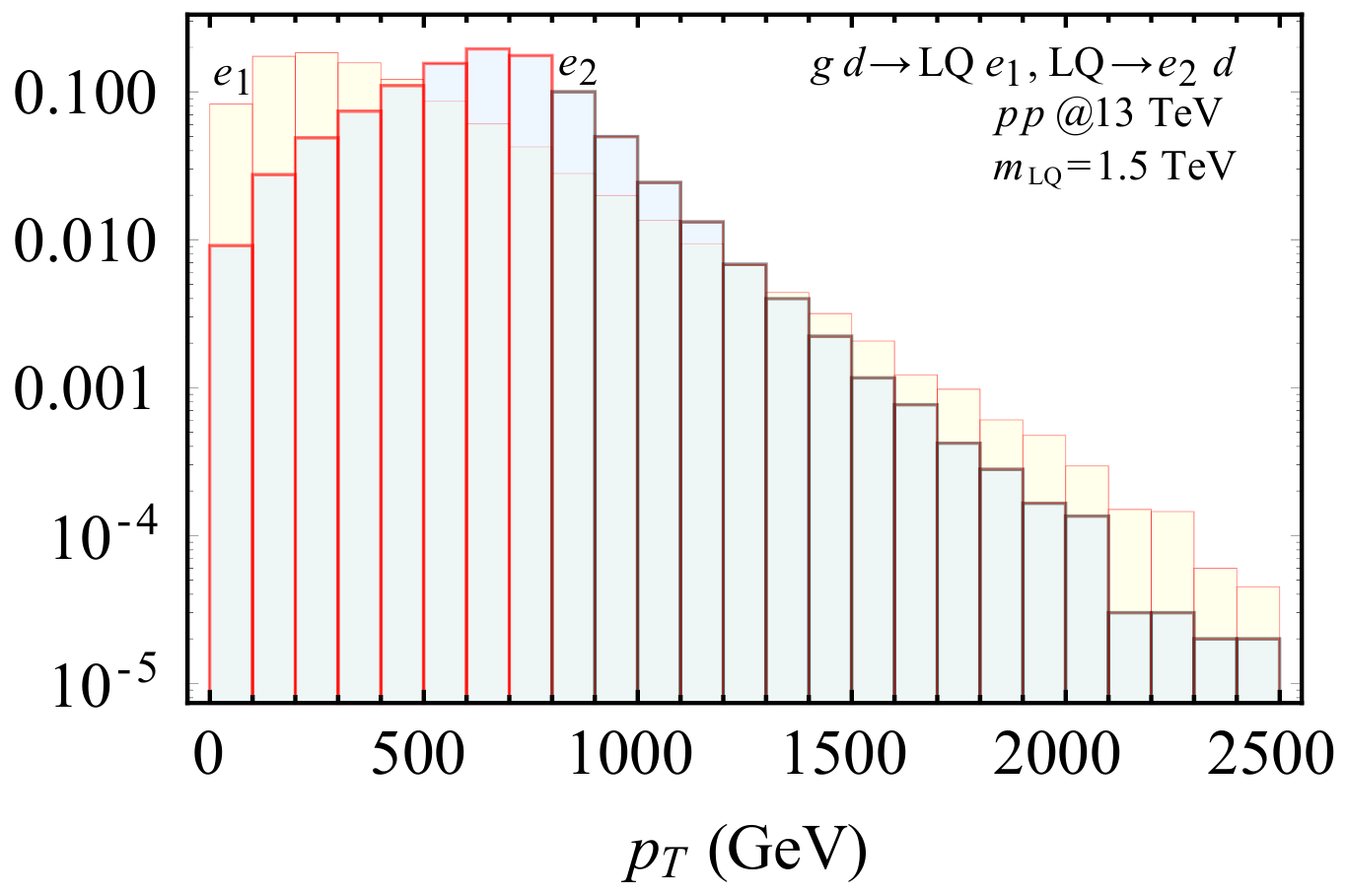}
\includegraphics[scale=0.6]{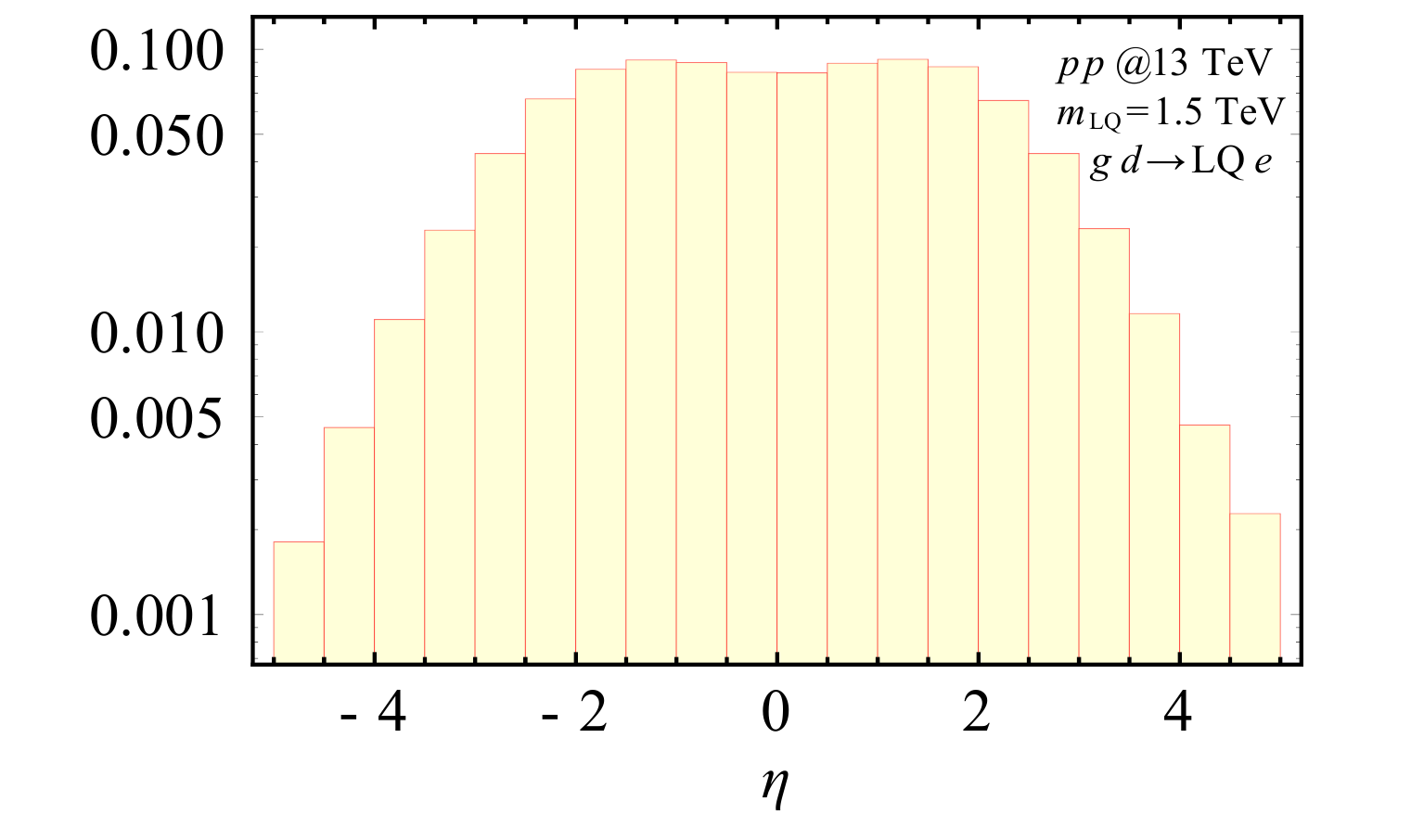}
\caption{\label{fig:plot-5} Upper panel: $p_T$ distributions of electrons in the final state of the single LQ production process at the LHC. $e_2$ represents electron originating from the LQ decay. The LQ mass is taken to be $m_\mathrm{LQ}=1.5$\,TeV. Lower panel: Pseudorapidity ($\eta$) distribution of the first generation scalar LQ in the single LQ production process at the LHC.}
\end{figure}

The non-resonant contribution towards the single LQ production is always subleading. If we accordingly neglect non-resonant contribution one of the two leptons in the final state of the single LQ production processes would always originate from LQ. It is expected that this lepton should be the leading one of the two in the $p_T$ parameter space, where $p_T$ is a transverse momentum. We accordingly present in Fig.~\ref{fig:plot-5} an example of $p_T$ distributions of two charged leptons in the final state of the single LQ production process at the LHC. Electron $e_1$ originates from the production of first generation scalar LQ whereas electron $e_2$ comes from the actual LQ decay. Our simulation is based on $m_\mathrm{LQ}=1.5$\,TeV for $p p$ collisions at $13$\,TeV. Clearly, $p_T$ distribution of $e_2$ that is rendered in thick lines in Fig.~\ref{fig:plot-5} peaks at much higher values of transverse momentum when compared to the $e_1$ distribution. We also show the pseudorapidity ($\eta$) distribution of the first generation scalar LQ in the single LQ production process at the LHC in Fig.~\ref{fig:plot-5} to show that LQ should occupy region of small values of $\eta$. Pseudorapidity $\eta$ is defined as $\eta=-\ln[\tan(\theta/2)]$, where $\theta$ is the angle between the particle momentum and the beam axis. Both panels in Fig.~\ref{fig:plot-5} are based on simulation of the signal for first generation LQ that couples exclusively to a $d$ quark and an electron $e$. 

The next-to-leading order QCD corrections to the LQ pair production are shown to substantially enhance 
the tree level cross section for pair production of LQs at the LHC~\cite{Kramer:2004df}. That work has built upon previous study that investigated pair production of scalar LQs at the Tevatron~\cite{Kramer:1997hh} at the next-to-leading order. The main result of the analysis is the total cross section including next-to-leading order QCD corrections and associated $K$-factors for $p p \rightarrow \mathrm{LQ}+\overline{\mathrm{LQ}}+X$ at the LHC energy of $\sqrt{s} = 14$\,TeV. These are summarized in Table~1 of Ref.~\cite{Kramer:2004df}. Here, $K$-factor is simply defined as $K=\sigma_\mathrm{NLO}/\sigma_\mathrm{LO}$, where $\sigma_\mathrm{LO}$ ($\sigma_\mathrm{NLO}$) represents leading (next-to-leading) order cross section. The $K$-factor was found to vary from $K=1.48$ for $m_\mathrm{LQ}=200$\,GeV to $K=2.07$ for $m_\mathrm{LQ}=2$\,TeV~\cite{Kramer:2004df}. 

Pair production of the first generation scalar LQs at the LHC to next-to-leading order accuracy using the MC@NLO formalism (with parton shower matching) has been investigated in Ref.~\cite{Mandal:2015lca}. The authors have evaluated pair production cross sections and associated $K$-factors at center-of-mass energy of $\sqrt{s} = 14$\,TeV. These can be found in Table 1 of Ref.~\cite{Mandal:2015lca} for two different sets of parton distribution functions. One can compare these results with results presented in Ref.~\cite{Kramer:2004df}. We show central values of one set of cross sections and $K$-factors from Ref.~\cite{Mandal:2015lca} in Table~\ref{tab:NLO}. These values are based on CTEQ6(M/L1) parton distribution functions~\cite{Pumplin:2002vw}.
\begin{table}[tbp]
\centering
\begin{tabular}{|c|cc|}
\hline
$m_{\mathrm{LQ}}$\,[TeV] &  $\sigma_\mathrm{NLO}$\,[fb] & $K$-factor \\
\hline 
$0.2$ & $7.434 \times 10^4$ & $1.483$ \\
$0.4$ & $2.223 \times 10^3$ & $1.586$ \\
$0.6$ & $2.220 \times 10^2$ & $1.641$ \\
$0.8$ & $3.810 \times 10^{1}$ & $1.724$ \\
$1.0$ & $8.322 \times 10^{0}$ & $1.748$ \\
$1.2$ & $2.202 \times 10^{0}$ & $1.807$ \\
$1.4$ & $6.505 \times 10^{-1}$ & $1.870$ \\
$1.6$ & $2.084 \times 10^{-1}$ & $1.923$ \\
$1.8$ & $7.089 \times 10^{-2}$ & $1.979$ \\
$2.0$ & $2.506 \times 10^{-2}$ & $2.047$ \\
\hline 
\end{tabular}
\caption{\label{tab:NLO} Total cross section $\sigma_{\mathrm{NLO}}$ and the associated $K$-factor at $\sqrt{s}=14$\,TeV center-of-mass energy at the LHC.}
\end{table} 

Single LQ productions and decays have also been a target of the next-to-leading order study~\cite{Hammett:2015sea}. (See, for example, Fig.~\ref{fig:SINGLE_PRODUCTION} for diagrams of relevant leading order contributions and Figs.~2, 3, and 4 of Ref.~\cite{Hammett:2015sea} for representative diagrams of next-to-leading order corrections to the LQ production and decay processes.) It has been conclusively demonstrated that the higher order corrections are significant for production of both scalar LQ ($R_2$) and vector LQ ($U_1$). These can amount to a 100\% correction with regard to the leading order result. The core process under investigation was $u+g \rightarrow e^-+e^+ + u$ ($d+g \rightarrow e^-+e^+ + d$) for scalar (vector) scenario. The authors of Ref.~\cite{Hammett:2015sea} have, in particular, investigated the effectiveness of the narrow-width approximation in the context of the next-to-leading order approach. This assumption has proved to be very reliable in the case of the scalar LQ but not so accurate in the case of the vector LQ. The authors attributed this to the fact that one cannot neglect the interference between the different helicity states of the intermediate vector LQs. Relevant results for the leading and next-to-leading order cross sections are summarized in Tables 1 through 7 of Ref.~\cite{Hammett:2015sea}. 

Another potentially relevant analysis that can in principle be carried out at the LHC is a search for the indirect LQ effects in the tails of the distributions, such as, dilepton invariant mass in $p p\to \ell \ell$~\cite{Davidson:2014lsa,Bessaa:2014jya}. For initial study on sensitivity of both $e^+ e^-$ colliders and hadron-hadron colliders to the indirect effects of virtual scalar LQs see Ref.~\cite{Dreiner:1987de}. Contact interactions of the form $llqq$ within the context of $p p$ and $p \bar{p}$ colliders that are due to the LQ exchange have been studied, for example, in Ref.~\cite{Bhattacharyya:1994yc}. The experimental constraints on contact interactions from the LHC have been used to place bounds on the strength of the leptoquark couplings to matter in Ref.~\cite{Wise:2014oea} for three concrete scenarios when only one leptoquark --- $S_1$, $R_2$, or $\tilde{R}_2$ --- is present at the given time.

For a particular study of the leptoquarks decaying to a top quark and a charged lepton of any flavor at hadron colliders see Ref.~\cite{Davidson:2011zn}. The authors investigated two distinct scenarios. First (second) scenario concerned the LQ decays with the LQ mass assumed to be below (above) the top quark mass at Tevatron (LHC). In the Tevatron case the authors produced a limit using existing data whereas in the LHC case the analysis resulted in a sensitivity study.


\subsubsection{Searches at $p \bar{p}$ colliders}


First results on the LQ parameter space that were inferred from the $p \bar{p}$ collider data came from the CERN proton--antiproton collider SPS. In the proton--antiproton regime the SPS collider was aptly referred to as Sp$\bar{\mathrm{p}}$S. For example, UA2 detector at the CERN Sp$\bar{\mathrm{p}}$S collider was investigating pair production of first generation LQs with their subsequent decays into a quark and either an electron or an electron neutrino. (A possibility to study single production of scalar LQs at Sp$\bar{\mathrm{p}}$S has been investigated in Ref.~\cite{Dobado:1988wi}.) The center-of-mass collision energy was $\sqrt{s} = 630$\,GeV and the sample was based on an integrated luminosity of $13$\,pb$^{-1}$. The lack of signal allowed the UA2 Collaboration to establish a lower limit on the mass of first generation LQs at $67$\,GeV ($76$\,GeV) at 95\% C.L.\ assuming that the branching ratio for a scalar LQ to decay into a quark and an electron was at 50\% (100\%)~\cite{Alitti:1991dn}. The LQ production and the LQ decay analyses at UA2 were based on the leading order contributions using results of Ref.~\cite{Hewett:1987yg}. (For the study of higher order QCD corrections that would increase the predicted production cross section at Sp$\bar{\mathrm{p}}$S see Ref.~\cite{deMontigny:1989kw}.)

LQs have been searched for at the Fermilab Tevatron, a $p \bar{p}$ collider, by the D\O\ \cite{Abachi:1993hr,Abachi:1995ik,Abbott:1997mq,Abbott:1997pg,Abbott:1998rb,Abbott:1999bc,Abbott:1999ka,Abazov:2001mx,Abazov:2001ic,Abazov:2004mk,Abazov:2006vc,Abazov:2006wp,Abazov:2006ej,Abazov:2007bsa,Abazov:2008jp,Abazov:2008at,Abazov:2008np,Abazov:2009ab,Abazov:2010wq,Abazov:2011qj} and CDF~\cite{Abe:1993ac,Abe:1995fj,Abe:1996dn,Abe:1997rc,Abe:1998it,Abe:1998bc,Abe:1998iq,Affolder:2000ny,Acosta:2004zb,Abulencia:2005ua,Acosta:2005ge,Aaltonen:2007rb} Collaborations. D\O\ has published results of a search for the pair production of first generation scalar LQs in the $eejj$ channel using a data set amounting to $123$\,pb$^{-1}$ of data that were collected during the $1992$--$1996$ time period~\cite{Abbott:1997mq}. A lower limit was consequently set on the mass of a first generation scalar LQ of $225$\,GeV. The underlying assumption was that the hypothetical LQ had a 100\% branching fraction to $eq$ pair. The CDF Collaboration has also presented the results of a search for first generation scalar LQs using $(110 \pm 7)$\,pb$^{-1}$ of data~\cite{Abe:1997rc}. It excluded scalar LQs with mass less than $213$\,GeV at 95\%\,C.L.\ for a branching ratio into $eq$ equal to 1. The results of the CDF and D\O\ Collaborations have then been combined to yield the most stringent Tevatron limit on the mass of first generation scalar LQ~\cite{GrossoPilcher:1998qt}. The combined lower limit on $m_\mathrm{LQ}$ from the two experiments is $242$\,GeV. D\O\ Collaboration has also searched for scalar LQ pair production in the $e u \nu_e d$ final state with $5.4$\,fb$^{-1}$ of data recorded at $\sqrt{s}=1.96$\,TeV. The lack of a signal led to a lower limit on $m_\mathrm{LQ}$ of $326$\,GeV for $\beta=0.5$ at the 95\%\,C.L.~\cite{Abazov:2011qj}.

D\O\ has also reported results of a search for pair production of second generation scalar LQs that was based on a data set  produced in $p \bar{p}$ collisions at the center-of-mass energy of $1.96$\,TeV amounting to $1.0$\,fb$^{-1}$ of integrated luminosity~\cite{Abazov:2008np}. It was assumed that the LQ in question could decay to both $\mu q$ and $\nu q$ pairs. Since D\O\ did not notice any excess over the SM prediction the Collaboration has set lower limits on the LQ mass of $316$\,GeV ($270$\,GeV) for $\beta = 1$ ($0.5$). 

Finally, D\O\ has also investigated pair production of third generation LQs that decay exclusively to a $\tau$ neutrino and a $b$ quark using the $5.2$\,fb$^{-1}$ data set~\cite{Abazov:2010wq}. Lack of any excess allowed the Collaboration to set a lower limit on the LQ mass of $247$\,GeV at 95\%\,C.L. 



\subsubsection{Searches at $p p$ collider}

We summarize here current status of experimental searches for LQs at the LHC that were conducted by the CMS and ATLAS Collaborations. We again note that the terms ``electron'', ``muon'', and ``tau'' are used generically to refer to corresponding particles and anti-particles, if not otherwise stated, in what follows.
 
The ATLAS experiment has performed a search for first generation scalar LQs in $p p$ collisions at $\sqrt{s}=7$\,TeV using $1.03$\,fb$^{-1}$ of data~\cite{Aad:2011ch}. The events under investigation had at least two jets and either two oppositely charged electrons or one electron and missing energy. The observed event yields were in agreement with the background expectations due to the SM physics. The ATLAS experiment has accordingly excluded first generation scalar LQs with masses below $660$\,GeV ($607$\,GeV) for $\beta=1$ ($\beta=0.5$) at 95\%\,C.L. (We always denote with $\beta$ the LQ decay rate into a charged lepton and a quark.) 

The ATLAS Collaboration has also conducted a search for the production of second generation scalar LQs with a total of $1.03$\,fb$^{-1}$ integrated luminosity of proton--proton collision data produced at $\sqrt{s} = 7$\,TeV at the LHC. This time the final states that were looked at consisted of either two oppositely charged muons and at least two jets or a muon plus missing transverse momentum and at least two jets. Since the observed event yield was in agreement with the expectations for the SM physics generated background the ATLAS Collaboration has excluded the production of second generation scalar LQs for $m_\mathrm{LQ} < 685$ (594) GeV for a branching ratio of $\beta=1.0$ ($\beta=0.5$) at 95\% confidence level~\cite{ATLAS:2012aq}.

The CMS Collaboration has also presented a search for the pair production of first and second generation scalar LQs at 
center-of-mass energy $\sqrt{s} = 7$\,TeV~\cite{Chatrchyan:2012vza}. (For the individual searches for the pair production of first and second generation scalar LQs at $\sqrt{s} = 7$\,TeV see Refs.~\cite{Khachatryan:2010mp,Chatrchyan:2011ar} and Ref.~\cite{Khachatryan:2010mq}, respectively.) The data sample that was used for analysis corresponded to an integrated luminosity of $5.0$\,fb$^{-1}$. The events under investigation involved at least two jets and either two charged leptons of the same flavor or a single charged lepton and missing transverse energy. We will not discuss initial selection criteria but will again mention variables that are used for event selection optimization. These are the invariant mass of the dilepton pair ($M_{l l}$), the scalar sum of the transverse momenta ($S^{ll}_T=p_T(l_1)+p_T(l_2)+p_T(j_1)+p_T(j_2)$) of the two leading $p_T$ leptons ($l_1$ and $l_2$) and the two leading $p_T$ jets ($j_1$ and $j_2$), and the smallest of the two lepton--jet invariant masses that minimizes the LQ mass difference ($M^\mathrm{min}_{lj}$). The outcome of the analysis was an exclusion of scalar LQ pair production at the 95\%\,C.L.\ for masses below $830$\,GeV ($840$\,GeV) for the first (second) generation LQs assuming $\beta = 1$. The exclusion changed to $m_{\mathrm{LQ}} < 640$\,GeV ($m_{\mathrm{LQ}} < 650$\,GeV)  for the first (second) generation LQs for $\beta = 0.5$~\cite{Chatrchyan:2012vza}.

The ATLAS Collaboration has performed a search for third generation scalar LQs through pair production using an integrated luminosity of $4.7$\,fb$^{-1}$ of data. These data were recorded at $\sqrt{s}= 7$\,TeV with the ATLAS detector~\cite{ATLAS:2013oea}. The underlying assumption was that LQs decay to a $\tau$ and a $b$ with a branching fraction equal to 100\%, i.e., $\beta=1$. The fact that no deviation was observed with respect to the SM expectation allowed the ATLAS Collaboration to establish a lower limit on the third generation scalar LQ masses of $534$\,GeV at 95\% confidence level~\cite{ATLAS:2013oea}. This limit was improved upon by the CMS Collaboration~\cite{Khachatryan:2014ura}. Namely, a search for pair production of third-generation scalar LQs that decay exclusively into $\tau$ leptons and $b$ quarks resulted in lower limits on the LQ masses of $740$\,GeV at 95\% confidence level. The CMS search was based on a data sample of proton--proton collisions at $\sqrt{s} = 8$\,TeV corresponding to an integrated luminosity of $19.7$\,fb$^{-1}$. 
The CMS Collaboration has also conducted a search for pair production of third generation scalar LQs of electric charge $-1/3$ that decay exclusively into $\tau$ leptons and $t$ quarks~\cite{Khachatryan:2015bsa} using the data set corresponding to integrated luminosity of $19.7$\,fb$^{-1}$ that was taken at a center-of-mass energy of $\sqrt{s} = 8$\,TeV. The events that were studied contained an electron or a muon, a hadronically decaying $\tau$, and two or more jets. The fact that the  observations were found to be consistent with the SM predictions allowed for exclusion of third generation LQs up to a mass of $685$\,GeV at 95\,\% C.L. 

Single production of scalar LQs in $p p$ collisions at $\sqrt{s} = 8$\,TeV was studied with the CMS detector~\cite{CMS:2015kda,Khachatryan:2015qda}. Namely, the CMS Collaboration has conducted a search for production of first (second) generation scalar LQs with a final state comprising two electrons (muons) and one jet. The data set that was used in analysis corresponded to an integrated luminosity of $19.6$\,fb$^{-1}$. Note that the event selection for the single production of scalar LQs utilizes the invariant mass of the dilepton pair ($M_{l l}$) and the scalar sum of the transverse momenta ($S_T=p_T(l_1)+p_T(l_2)+p_T(j_1)$) of the two leading $p_T$ leptons ($l_1$ and $l_2$) and the leading $p_T$ jet ($j_1$). The two leptons were required to be of opposite electric charges. Due to agreement between observed candidate events and expected background events the CMS Collaboration produced exclusion limits on $m_{\mathrm{LQ}}$ as a function of Yukawa coupling $y$. We reproduce in Table~\ref{tab:CMS_LIMITS} the content of Table~2 of Ref.~\cite{CMS:2015kda} that shows relevant exclusion limits on the LQ mass.
\begin{table}[tbp]
\centering
\begin{tabular}{|ccc|}
\hline
Yukawa coupling $y$ & LQ type & Excluded mass (GeV) \\
\hline 
$0.4$ & 1st gen. & $895$ \\
$0.6$ & 1st gen. & $1260$ \\
$0.8$ & 1st gen. & $1380$ \\
$1.0$ & 1st gen. & $1730$ \\
$1.0$ & 2nd gen. & $530$ \\
\hline 
\end{tabular}
\caption{\label{tab:CMS_LIMITS} Excluded masses for single LQ production search as a function of Yukawa coupling and LQ type at 95\% confidence limit. Results are taken from Ref.~\cite{CMS:2015kda}.}
\end{table} 

\begin{figure}[tbp]
\centering
\includegraphics[scale=0.6]{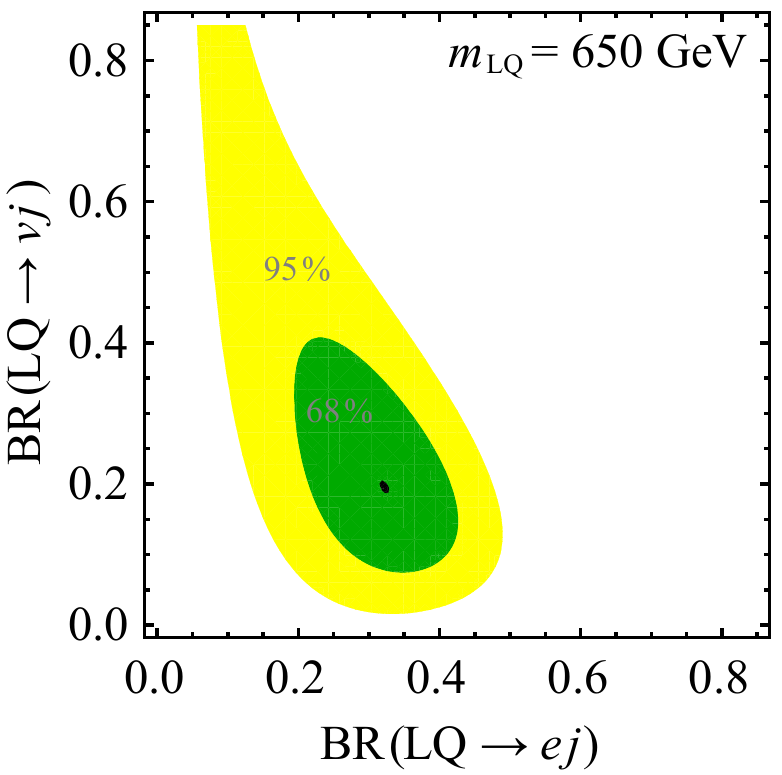}
\caption{\label{fig:plot-6} The best fit of the CMS observed anomaly for $m_{\mathrm{LQ}}=650$\,GeV.}
\end{figure}

There were also two searches for pair-production of first generation~\cite{CMS:2014qpa} and second generation scalar LQs~\cite{CMS:zva} that were performed by the CMS Collaboration using proton--proton collision data at $\sqrt{s} = 8$\,TeV. The final states under investigation were required to contain two electrons and at least two jets or an electron, a neutrino, and at least two jets in case of the first generation scalar LQ search~\cite{CMS:2014qpa}. The second generation search required charged leptons in the final state to be muons. The data in both instances corresponded to an integrated luminosity of $19.6$\,fb$^{-1}$ and were collected by the CMS detector at the LHC. Although the CMS Collaboration has excluded first generation scalar LQs with masses less than $1005$\,GeV ($845$\,GeV) for $\beta= 1$ ($\beta= 0.5$)~\cite{CMS:2014qpa} it has observed, in both channels, a significant excess in the final selection optimized for LQs with a mass of $650$\,GeV. We show in Fig.~\ref{fig:plot-6} best fit of the observed anomaly assuming that the branching fractions of LQ to go into $ej$ or $\nu j$ are not related to each other. We also fix mass of LQ to be $650$\,GeV. Obviously, if the anomaly is real the LQ should decay into something else rather than the first generation charged leptons. The most obvious solution is to have an LQ that can also couple to muons (and/or taus) but the constraints coming from the search for the second generation LQs do not allow for this possibility. Namely, the CMS Collaboration has excluded second generation scalar LQs with masses less than $1070$\,GeV ($785$\,GeV) for $\beta = 1$ ($\beta = 0.5$)~\cite{CMS:zva}.  

This anomaly has accordingly attracted attention of the high energy physics community and there were several proposals put forth to explain the observed discrepancy between the experimental signal and the SM expectations~\cite{Bai:2014xba,Chun:2014jha,Queiroz:2014pra,Dobrescu:2014esa,Allanach:2014nna,Dhuria:2015hta,Dutta:2015dka,Evans:2015ita,Berger:2015qra}. For example, the authors of Ref.~\cite{Queiroz:2014pra} have constructed a scenario that was able to account for the mild excess in the $e e j j$ and $e \nu j j$ channels that was observed by the CMS search for first generation LQs~\cite{CMS:2014qpa} through very specific  combination of branchings and kinematics. (The same scenario was used to accommodate the CMS observed $\ell^+\ell^- jj E\!\!\!\!/_{\rm T}$ excess~\cite{CMS:2014jfa} in Ref.~\cite{Allanach:2015ria}. However, the explanation of the $\ell^+\ell^- jj E\!\!\!\!/_{\rm T}$ excess has required use of different part of parameter space with regard to the one used to address excesses in the $e e j j$ and $e \nu j j$ channels.) The study of Ref.~\cite{Dutta:2015dka}, on the other hand, has investigated the correlation between the $e e j j$ and $e \nu j j$ CMS excess and the resonant enhancement in the very high energy shower event rates at the IceCube experiment within a context of a simple LQ scenario. (The production and subsequent decay of scalar LQs --- $R_2$ and $\tilde{R}_2$ --- within a framework of unification model based on the $SU(4)_C \times SU(2)_L \times U(1)_R$ gauge group product at IceCube experiment has been discussed in Ref.~\cite{Dey:2015eaa}.)

The CMS Collaboration has provided an update on the search for pair-production of first generation and second generation scalar LQs that was performed by using proton--proton collision data at $\sqrt{s} = 8$\,TeV~\cite{Khachatryan:2015vaa}. This time an integrated luminosity of $19.7$\,fb$^{-1}$ was used to exclude first generation scalar LQs with masses less than $1010$\,GeV ($850$\,GeV) for $\beta = 1.0$ ($\beta =0.5$), where $\beta$ represents the branching fraction of an LQ decaying to a charged lepton and a quark. The second generation scalar LQ limit, on the other hand, is $m_{\mathrm{LQ}}>1080$\,GeV ($m_{\mathrm{LQ}}>760$\,GeV) for $\beta = 1.0$ ($\beta =0.5$). The final states under consideration were required to contain either two charged leptons and two jets, or one charged lepton, one neutrino and two jets.

\begin{figure}[tbp]
\centering
\includegraphics[scale=0.26]{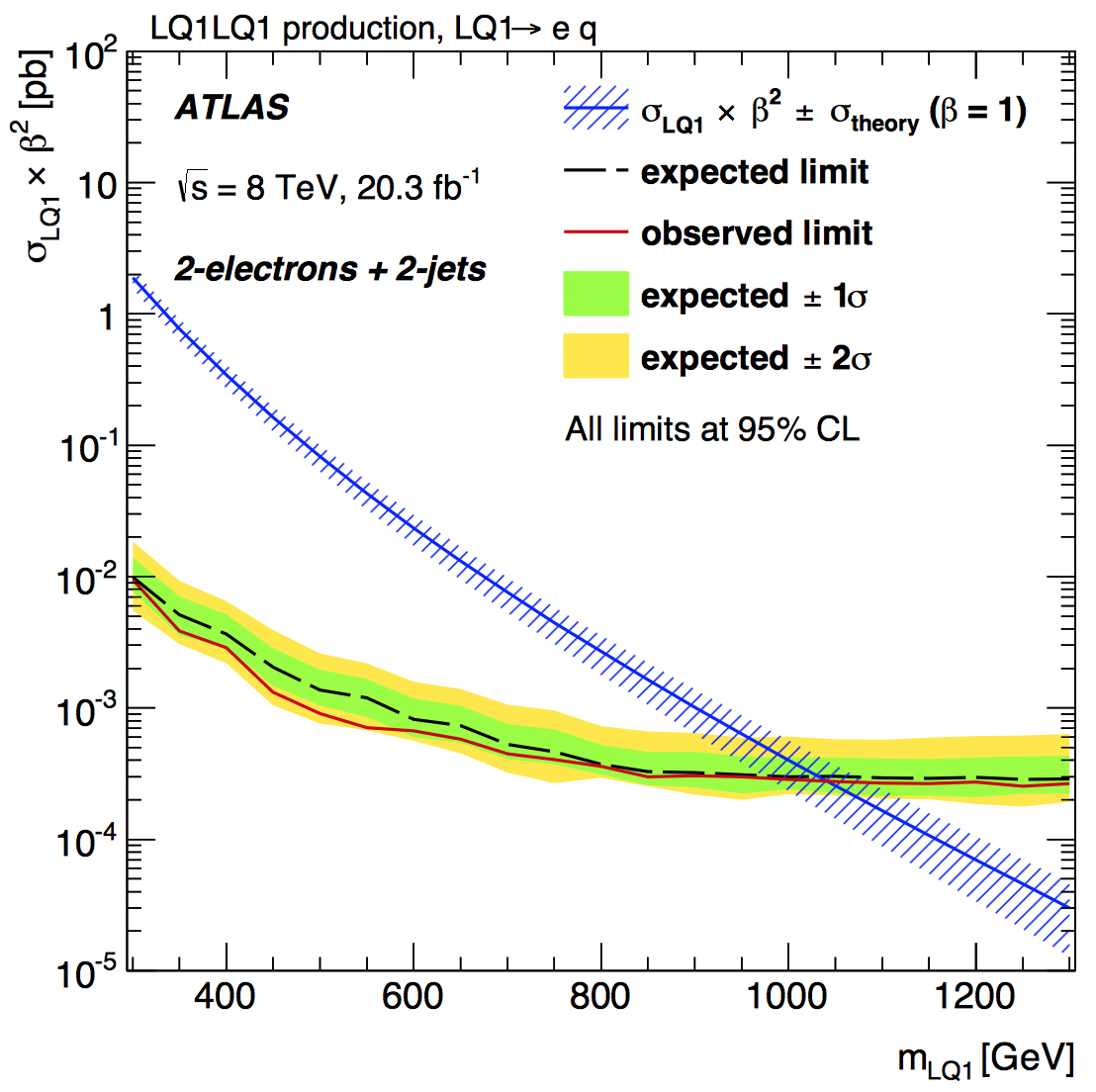}
\includegraphics[scale=0.3]{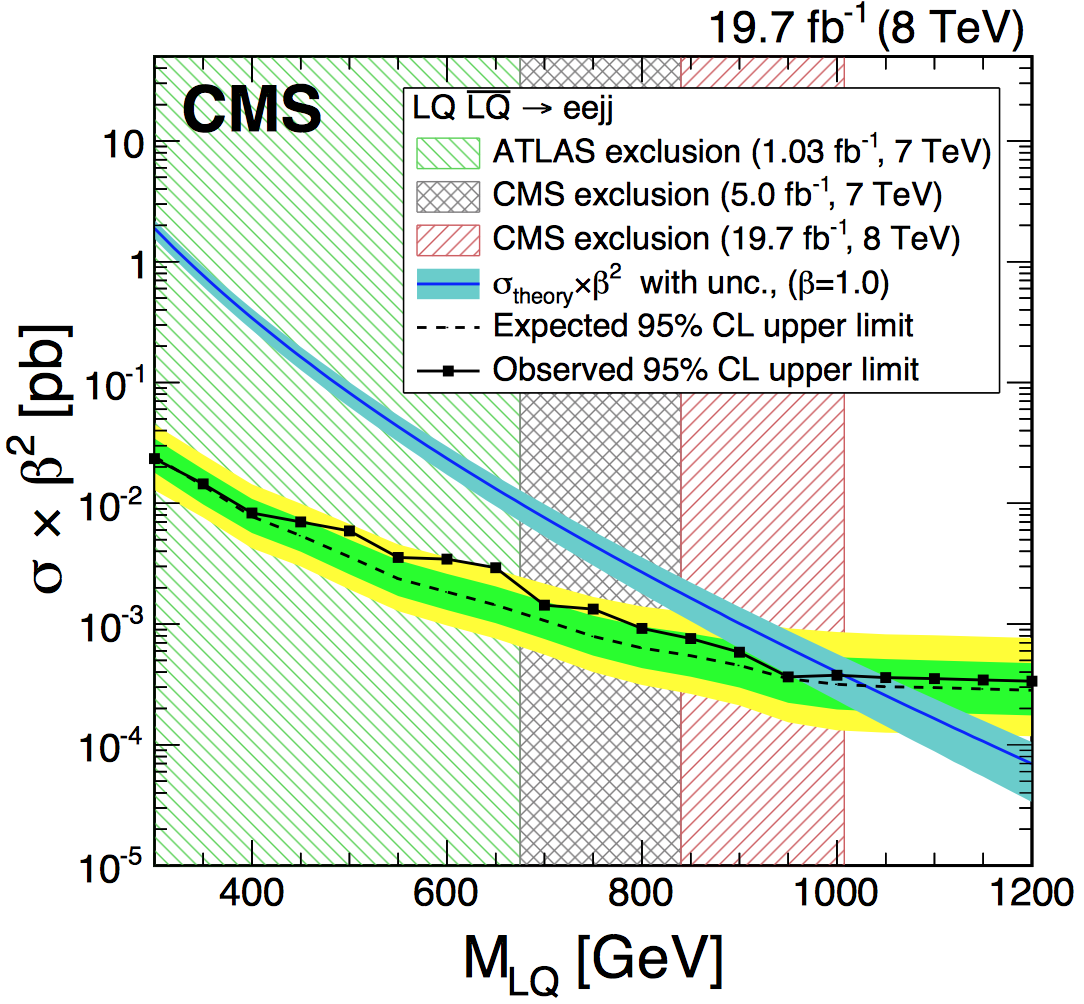}
\caption{\label{fig:plot-exclusions-atlas-cms} Exclusion limits on the LQ pair production cross section times the LQ branching ratio to $e$--jet final state from ATLAS (left plot~\cite{Aad:2015caa}) and CMS (right plot~\cite{Khachatryan:2015vaa}).}
\end{figure}

Finally, the ATLAS Collaboration has recently reported a search for pair produced scalar LQs in $8$\,TeV $p p$ collisions with full $20$\,fb$^{-1}$ of luminosity~\cite{Aad:2015caa}. Stringent exclusion limits at $95\,\%$\,C.L.\ are set on the first ($m_{\mathrm{LQ}}>1.05$\,TeV) and second ($m_{\mathrm{LQ}}>1.00$\,TeV) generation LQs. Shown in Fig.~\ref{fig:plot-exclusions-atlas-cms} (left plot) are the expected and observed limits on $\sigma \times \beta^2$ for LQ pair production with subsequent LQ decay to $e$-jet final state. Blue line, on the other hand, represents the predicted cross section (for $\beta=1$) in QCD. It is interesting to notice that the recent ATLAS analysis rules out the slight excess seen in the CMS search (Fig.~\ref{fig:plot-exclusions-atlas-cms} (right plot)) for $m_{\mathrm{LQ}}$ around $650$\,GeV.


\section{Conclusions}
\label{sec:CONCLUSIONS}

Physics of LQs is a very rich and mature subject. At the same time, it is still a rapidly evolving field on both experimental and theoretical fronts. The experimental constraints on these hypothetical particles that can turn quarks into leptons and vice versa are set to be improved in current, upcoming and future high energy and precision experiments. And, our ability to provide theoretical predictions of their presence goes hand in hand with the experimental quest.

The main focus of this review was on the phenomenology of light LQs that are relevant for precision experiments and particle colliders. We accordingly presented tools that can enable one to extract constraints on the LQ interactions with matter from precision low energy observables such as electric and magnetic dipole moments of SM fermions, atomic parity violation, neutral meson mixing, $\pi$, $K$, $B$, and $D$ meson decays to name but a few examples. We also addressed complementary constraints that originate from electroweak precision measurements, top, and Higgs physics. We finally discussed and summarized the current status of direct LQ searches at colliders.

Direct searches for LQs have pushed model independent lower limits on their masses up to a TeV scale. For example, the CMS (ATLAS) Collaboration has excluded first generation scalar LQs with masses less than $1010$\,GeV ($1050$\,GeV) with the $8$\,TeV $p p$ collision data. And, these limits are expected to become even more stringent in the near future with the LHC run II and HL-LHC data. Indirect searches are also reaching unprecedented levels of accuracy and precision, cornering the available parameter space of LQ couplings to matter. 

The hope is that LQs will eventually be experimentally observed. A leptoquark discovery would certainly point towards unification of quarks and leptons. It might happen that the LQ detection  could be further correlated with other observed phenomena that are not well accommodated within the SM. Here we primarily refer to neutrino masses, anomalies in flavor physics, dark matter, and potential anomalies in collider physics. This, in our view, would nicely dovetail with the observed unification of weak and electromagnetic interactions and potentially usher in a new era in particle physics.

\appendix

\renewcommand*{\thesection}{\Alph{section}}

\section{The leptoquark--Higgs couplings}
\label{sec:APPENDIX_A}
We summarize below the couplings of leptoquarks with the SM Higgs boson $H \equiv ({\bf 1},{\bf 2},-1/2)$. We assume that only one leptoquark is present at a given time. (See Table~\ref{tab:LQs} for the LQ multiplet assignments under the SM gauge group of $SU(3) \times SU(2) \times U(1)$. See Section~\ref{subsec:NOMENCLATURE} for relevant nomenclature.) The following two operators can be written in a compact form~\cite{Hirsch:1996qy}
\begin{align}
\mathcal{L} \supset &-(\eta_\Phi m_\Phi^2+\lambda_\Phi H^\dagger H) (\Phi^\dagger \Phi),
\end{align}
where $\Phi$ stands for all leptoquarks and $\eta_\Phi=1$ ($\eta_\Phi=-1$) for scalar (vector) leptoquarks. It is understood that the $(\Phi^\dagger \Phi)$ contraction is in the $SU(3)$, $SU(2)$, and Lorentz space when appropriate. There is an additional operator for the $SU(2)$ doublet leptoquarks $\tilde{R}_2$, $R_2$, $\tilde{V}_2$, and $V_2$ that reads
\begin{align}
\mathcal{L} \supset &-\lambda'_\Phi H^{\dagger \,a} \Phi^a \Phi^{\dagger\,b} H^b,
\end{align}
where we only specify contraction in the $SU(2)$ space.
The operator relevant for the $SU(2)$ triplet leptoquarks $S_3$ and $U_3$ is
\begin{align}
\mathcal{L} \supset &-\lambda''_\Phi H^{\dagger \,a} (\tau^i \Phi^i)^{ab} (\tau^j \Phi^j)^{\dagger\,bc} H^c,
\end{align}
where $\tau^i$ and $\tau^j$ are Pauli matrices with $i,j=1,2,3$.
Finally, there is one more operator for $\tilde{R}_2$ that reads~\cite{Arnold:2012sd}
\begin{align}
\mathcal{L} \supset &+h_\Phi \epsilon_{\alpha \beta \gamma} \epsilon^{ab} \epsilon^{cd}  H^{a} \tilde{R}_{2\,\alpha}^{b} \tilde{R}_{2\,\beta}^{c} \tilde{R}_{2\,\gamma}^{d},
\end{align}
where $\alpha, \beta, \gamma=1,2,3$ ($a, b, c, d=1,2$) are $SU(3)$ ($SU(2)$) indices.

\section{The leptoquark gauge couplings}
\label{sec:APPENDIX_C}
The purpose of this appendix is to state definitely the gauge couplings of leptoquarks. First, we choose the gauge covariant derivative acting on the LQ fields to be
\begin{equation}
  \label{eq:D}
D_\mu = \partial_\mu + i g_1 Y B_\mu + i g_2 I^k W^k_\mu + i g_3
\frac{\lambda^A}{2} G^A_\mu  ,
\end{equation}
where $Y$ is the LQ hyper-charge, $I^k$ are the $SU(2)$
representation matrices, depending on the $SU(2)$ assignment of the LQ state. For $SU(2)$ singlets $I^k = 0$, $k=1,2,3$, while for doublets $I^k =
\tau^k/2$, where $\tau^k$ are the standard Pauli matrices. For the weak
triplets $(I^k)_{ij} = -i \epsilon_{kij}$, $i,j,k=1,2,3$. Finally, all LQ states
transform as fundamental representation under $SU(3)$ thus $\lambda^A$, $A=1,\ldots,8$, in Eq.~\eqref{eq:D} are the Gell-Mann matrices. After the electroweak
symmetry breaking we have $e = g_2 \sin\theta_W >0$ and $g_1/g_2 = \tan \theta_W$.

For a scalar leptoquark, labeled here as $\phi$ all the gauge interactions are
contained within the gauge invariant kinetic term:
\begin{equation}
\mc{L} \supset  (D_\mu \phi)^\dagger (D^\mu \phi).
\end{equation}

For a vector leptoquark, $V^\mu$, one has to replace derivatives with
covariant derivative inside the canonic kinetic
term $-\tfrac{1}{2} V^{\mu \nu\,\dagger} V_{\mu\nu}$, where $V_{\mu\nu}
= \partial_\mu V_\nu-\partial_\nu V_\mu$. We define $D_{\mu\nu} \equiv
D_\mu V_\nu - D_\nu V_\mu$ and write
\begin{equation}
\mc{L} \supset -\frac{1}{2}   D^{\mu \nu\,\dagger} D_{\mu\nu} +
\sum_{i=1,2,3} \kappa_i V^{\mu\,\dagger} F^{(i)}_{\mu\nu} V^{\nu}.
\end{equation}
The sum in the last term that runs over all factors --- $U(1)$, $SU(2)$, and $SU(3)$ for
$i=1,2,3$, respectively --- in the direct
product of the SM gauge group
introduces three new arbitrary dimensionless parameters
$\kappa_i$. Here $F^{(1)}_{\mu\nu}$ corresponds to the $U(1)$ gauge
boson field strength $B_{\mu\nu}$, $F^{(2)}_{\mu\nu}$ to $W_{\mu\nu}$,
and $F^{(3)}_{\mu\nu}$ to $G_{\mu\nu}$~(cf.~\cite{Gabrielli:2015hua} and references therein).

\section{The leptoquark--quark--lepton couplings}
\label{sec:APPENDIX_B}
We summarize below the couplings of scalar leptoquarks $S_3$, $R_2$, $\tilde{R}_2$, $\tilde{S}_1$, $S_1$, and $\bar{S}_1$ to the quark--lepton pairs. (See Table~\ref{tab:LQs} for the LQ multiplet assignments under the SM gauge group of $SU(3) \times SU(2) \times U(1)$.)
\begin{align}
\nonumber
\mathcal{L} \supset &+y^{LL}_{3\,ij}\bar{Q}_{L}^{C\,i,a} \epsilon^{ab} (\tau^k S^k_{3})^{bc} L_{L}^{j,c}-\\
\nonumber
&-y^{RL}_{2\,ij}\bar{u}_{R}^{i} R_{2}^{a}\epsilon^{ab}L_{L}^{j,b}+y^{LR}_{2\,ij}\bar{e}_{R}^{i} R_{2}^{a\,*}Q_{L}^{j,a}-\\
\nonumber
&-\tilde{y}^{RL}_{2\,ij}\bar{d}_{R}^{i}\tilde{R}_{2}^{a}\epsilon^{ab}L_{L}^{j,b}+\tilde{y}^{\overline{LR}}_{2\,ij}\bar{Q}_{L}^{i,a} \tilde{R}_{2}^{a}\nu_{R}^{j}\\
\nonumber
&+\tilde{y}^{RR}_{1\,ij}\bar{d}_{R}^{C\,i} \tilde{S}_{1} e_{R}^{j}+\\
\nonumber
&+y^{LL}_{1\,ij}\bar{Q}_{L}^{C\,i,a} S_{1} \epsilon^{ab}L_{L}^{j,b}+y^{RR}_{1\,ij}\bar{u}_{R}^{C\,i} S_{1} e_{R}^{j}+y^{\overline{RR}}_{1\,ij}\bar{d}_{R}^{C\,i} S_{1} \nu_{R}^{j}+\\
\label{eq:main_scalars}
&+\bar{y}^{\overline{RR}}_{1\,ij}\bar{u}_{R}^{C\,i} \bar{S}_1 \nu_{R}^{j}+\textrm{h.c.}
\end{align}

We summarize below the couplings of vector leptoquarks $U_3$, $V_2$, $\tilde{V}_2$, $\tilde{U}_1$, $U_1$, and $\bar{U}_1$ to the quark--lepton pairs.
\begin{align}
\nonumber
\mathcal{L} \supset&+x^{LL}_{3\,ij}\bar{Q}_{L}^{i,a} \gamma^\mu (\tau^k U^k_{3,\mu})^{ab}  L_{L}^{j,b}+\\
\nonumber
&+x^{RL}_{2\,ij}\bar{d}_{R}^{C\,i} \gamma^\mu V^{a}_{2,\mu} \epsilon^{ab}L_{L}^{j,b} + x^{LR}_{2\,ij}\bar{Q}_{L}^{C\,i,a} \gamma^\mu \epsilon^{ab} V^b_{2,\mu} e_{R}^{j}+\\
\nonumber
&+\tilde{x}^{RL}_{2\,ij}\bar{u}_{R}^{C\,i} \gamma^\mu \tilde{V}^{b}_{2,\mu} \epsilon^{ab}L_{L}^{j,a} +\tilde{x}^{\overline{LR}}_{2\,ij}\bar{Q}_{L}^{C\,i,a} \gamma^\mu \epsilon^{ab} \tilde{V}^{b}_{2,\mu}  \nu_{R}^{j}+\\
\nonumber
&+\tilde{x}^{RR}_{1\,ij}\bar{u}_{R}^{i} \gamma^\mu \tilde{U}_{1,\mu} e_{R}^{j}+\\
\nonumber
&+x^{LL}_{1\,ij}\bar{Q}_{L}^{i,a} \gamma^\mu U_{1,\mu} L_{L}^{j,a} + x^{RR}_{1\,ij}\bar{d}_{R}^{i} \gamma^\mu U_{1,\mu} e_{R}^{j}+ x^{\overline{RR}}_{1\,ij}\bar{u}_{R}^{i} \gamma^\mu U_{1,\mu} \nu_{R}^{j}+\\
\label{eq:main_vectors}
&+x^{\overline{RR}}_{1\,ij}\bar{d}_{R}^{i} \gamma^\mu \bar{U}_{1,\mu} \nu_{R}^{j} +\textrm{h.c.}
\end{align}


\section*{Acknowledgements}
N.K.\ would like to thank Olcyr Sumensari for insightful discussions
regarding flavor aspects of this work. J.F.K.\ would like to thank the CERN TH Department for hospitality while this work was being completed. This work has been supported in part by
Croatian Science Foundation under the project 7118, the Slovenian
Research Agency (research core funding No.\ P1-0035), and the Swiss
National Science Foundation (SNF) under contract 200021-159720.

~

~

~

~

\bibliography{references}

\begin{thebibliography}{100}
\bibitem{CMS:2015kwa}
C.~Collaboration, {Measurements of the Higgs boson production and decay rates
  and constraints on its couplings from a combined ATLAS and CMS analysis of
  the LHC $p p$ collision data at $\sqrt{s} = 7$ and 8\,TeV
  (CMS-PAS-HIG-15-002)}.
\end{thebibliography}

\end{document}